\documentclass[a4paper, 11pt, titlepage, twoside]{report}
\pdfoutput=1

\usepackage[latin1]{inputenc}    
\usepackage{amsmath,amssymb}
\usepackage{graphicx}   
\usepackage{here}
\usepackage{mathrsfs} 
\usepackage{array}

\usepackage[a4paper,twoside, hmarginratio={3:2}]{geometry}

\usepackage[Lenny]{fncychap}

\usepackage{fancyhdr} 
\renewcommand{\chaptermark}[1]{\markboth{\textbf{Chapter \thechapter}:\ \textbf{#1}}{}}

\fancypagestyle{plain}{%
\fancyhf{} 
\fancyfoot[RO]{\textbf{\thepage}}
\fancyfoot[LE]{\textbf{\thepage}}
 
}
\usepackage[nottoc]{tocbibind}

\usepackage{psfrag}   

\usepackage[small,bf]{caption}

\usepackage{palatino}
\usepackage{pxfonts}

\usepackage[backref,
            linktocpage,
            colorlinks=true,
            pdftitle={Global versus Local Aspects of Critical Collapse},
            pdfauthor={Michael Puerrer},
            pdfsubject={Spherically Symmetric Gravitational Collapse of a Massless Scalar
            Field in Compactified Bondi Coordinates and Horizon-penetrating Double-null
            Coordinates},
            pdfkeywords={General Relativity, Numerical Relativity, 
                         Critical Collapse, Quasinormal Modes},
            plainpages=false
            ]{hyperref}

\newcommand{\forceemptypage}{
  \newpage
  \thispagestyle{empty}
  \mbox{}
  \newpage
}

\newcommand{\real}{\mathrm{Re}\ }
\newcommand{\imag}{\mathrm{Im}\ }

\newcommand{\bigO}{\mathcal O}
\newcommand{\pd}[2]{\frac{\partial #1}{\partial #2}}
\newcommand{\npd}[2]{(\partial #1/\partial #2)}
\newcommand{\Vr}{\frac{V}{r}}
\def\NP{c_{\scriptscriptstyle\text{NP}}}
\newcommand{\Scri}{{\mathscr I}}
\newcommand{\Scrh}{{\mathscr H}}
\newcommand{\mEXT}{m_{\scriptscriptstyle\text{EXT}}}
\newcommand{\mSSH}{m_{\scriptscriptstyle\text{SSH}}}
\newcommand{\SSH}{{\scriptscriptstyle\text{SSH}}}
\newcommand{\mB}{m_{\scriptscriptstyle\text{B}}}
\newcommand{\mBH}{m_{\scriptscriptstyle\text{BH}}}

\newcommand{\uB}{u_{\scriptscriptstyle\text{B}}}
\newcommand{\tauB}{\tau_{\scriptscriptstyle\text{B}}}
\newcommand{\pdrt}[1]{\frac{\partial #1}{\partial r}\bigg\vert_t}
\newcommand{\pdrts}[1]{\frac{\partial #1}{\partial r}\vert_t}
\newcommand{\dbar}[1]{\bar {\bar #1}}

\DeclareFontFamily{OT1}{rsfs}{}
\DeclareFontShape{OT1}{rsfs}{m}{n}{ <-7> rsfs5 <7-10> rsfs7 <10->
rsfs10}{} \DeclareMathAlphabet{\mycal}{OT1}{rsfs}{m}{n}

\def\cf{cf.\hbox{}}
\def\ie{i.e.\hbox{}}

\def\defn#1{``#1''}
\def\BH{{\scriptscriptstyle\text{BH}}}
\def\CS{{\scriptscriptstyle\text{C}}}
\def\BS{{\scriptscriptstyle\text{B}}}

\def\init{{\text{init}}}
\def\MS{{\scriptscriptstyle\text{MS}}}
\def\total{{\text{total}}}
\def\clap#1{\hbox to 0em{\hss{#1}\hss}}

\def\alignflush#1{%
\rlap{$\displaystyle #1$}\phantom{\displaystyle G_{\theta\theta}}}
\newcommand{\two}{{\scriptscriptstyle(2)}}
\newcommand{\quabtwo}{\square_{{}^\two\! g}}
\newcommand{\quabtwoT}{\tilde{\square}_{{}^\two\! g}}

\newcommand{\sdetgtwo}{\sqrt{-{}^\two\!g}}

\newcommand{\ebeta}{e^{2\beta}}

\newcommand{\half}{\frac{1}{2}}

\numberwithin{equation}{section}    

\begin{document}


\pagenumbering{alph}

\begin{titlepage}
\begin{center}
\vspace*{1cm}
\noindent
\parbox{\textwidth}{\huge\bf \center{Global versus Local Aspects\\ of Critical Collapse}}

\vspace{2cm}
Dissertation\\
zur Erlangung des Akademischen Grades\\
``Doktor der Naturwissenschaften''\\
an der\\
Fakult\"at f\"ur Physik\\
der Universit\"at Wien\\
\vspace{2,5cm}
eingereicht von\\
\vspace{1cm}
Michael P\"urrer\\
\vspace{1cm}
unter der Betreuung von\\
Univ.-Prof. Dr. Peter C. Aichelburg\\
\vspace{2cm}
an der\\
Fakult\"at f\"ur Physik\\

\vspace{1cm}
Wien, im Sommersemester 2007
\end{center}
\end{titlepage}


\forceemptypage

\pagenumbering{roman}

\chapter*{Acknowledgements}
\addcontentsline{toc}{chapter}{Acknowledgements}

I would like to express my sincere thanks to my advisor, Prof. Peter C. Aichelburg, 
for making this thesis possible financially, 
the questions and all the discussions we had, and his careful reading of the manuscript.
Special thanks to Sascha Husa for first arousing my interest in critical collapse and
for the rewarding teamwork on our paper, the invitations to Golm and his 
hospitality and support.
Part of this thesis relies on an extension of the DICE code which was originally 
developed for $\sigma$-model collapse by a team comprising Sascha Husa, Christiane Lechner, 
Jonathan Thornburg, Prof. Peter C. Aichelburg and myself.
I am very grateful to Prof. Piotr Bizon for always sharing interesting results and
for his helpful suggestions 

It is a pleasure for me to thank my colleges and friends Roland, Michael and Patrick for 
sharing an office with me and all the good times we had together:
Roland for his enthusiasm for all things mathematical and for sticking to his
upper Austrian dialect even in the heat of discussions.
His jokes and stories are one of a kind and a boon to every party and his guitar-playing
is not be missed, either ;-)
Michael, for his good spirits and advice, the Heimfestln, the movie sessions and
the memorable trip to the W\"orthersee.
Patrick, for the badminton battles, his power-adventures as a DM and for just being
a cool guy.

My friend Mark, who never shared an office with me, for his stunning knowledge of physics 
and science, in general, the Badminton sessions and his hospitality in Potsdam.
My friend Gerald for the great music we played together and the lovely trips to Perugia and
to the Aosta valley.

I would like to express my gratitude to my girlfriend Sabine for her encouragement and
understanding and simply for her wonderful presence in my life.
Last but not least, I would like to thank my parents, Karin and Ernst, and my sister 
Brigitte, for their support and all the good times we had together.

This work was supported by the Austrian Fonds zur F\"orderung der wissenschaftlichen
Forschung (FWF) (projects P15738 and P19126).


\forceemptypage

\chapter*{Abstract}
\markboth{\textbf{Abstract}}{}
\addcontentsline{toc}{chapter}{Abstract}

In this thesis I study the dynamics of a collapsing scalar 
field coupled to Einstein's equations. In this model, evolution of initial
data leads to one of two possible endstates: formation of a black hole or  
dispersion to flat space.
At the threshold between black hole formation and dispersion the behavior of the
system is characterized by so-called \emph{critical phenomena}: scaling, self-similarity 
and universality. These features of \emph{critical collapse} are numerically investigated 
from both local and global points of view.

On the one hand, only a small region of spacetime close to 
the origin is relevant for the dynamics of critical collapse.
On the other hand, it is also possible for a distant observer to analyze the radiation 
signal emitted by the collapsing matter field. 
In the framework of \emph{characteristic evolution}, such observers can be modelled
by employing radial compactification on outgoing null cones, so that the 
numerical grid extends to future null infinity. One may then extract global 
properties such as the Bondi mass and the news function.

We study the threshold behavior by numerical evolution of one parameter families
of initial data. The parameter is fine-tuned to the threshold via bisection.
In the evolution of such \emph{near-critical} data, the solution approaches the self-similar 
\emph{critical solution} for some time. We find that the \emph{critical exponent} that 
characterizes the discretely self-similar solution can be extracted both locally, in 
the self-similar region and globally, e.g. from the news function. In this sense, 
self-similarity is observable from future null infinity.

For late times in near-critical evolutions, we see a residual mass concentrated 
outside of the self-similar region and conjecture that it originates from 
backscattering of outgoing radiation during the collapse. 
The fate of this mass is unclear. If, in supercritical evolutions,
this mass were to fall into the black hole, the black hole mass would be finite, 
no matter how fine-tuned the initial data.

For subcritical evolutions we have numerically analyzed the exponents of \emph{power law 
tails} and have found agreement with analytical calculations for radiation along null infinity 
and along timelike lines.
We argue that for astrophysical observers the relevant falloff rate is that of
future null infinity.

We have also investigated the behavior of \emph{quasinormal modes} (QNM) in near-critical 
evolutions. Although the perturbation theory underlying QNMs requires a fixed black
hole background, we have found a surprising correlation between the radiation signal 
with the period of the first QNM determined by the time-dependent Bondi mass.
In this context, we have also been able to verify a stationarity criterion based on
QNMs for the Vaidya metric, which models the time-dependent Schwarzschild background.


\tableofcontents

\forceemptypage

\listoffigures

\listoftables

\forceemptypage

\pagenumbering{arabic}


\chapter{Introduction}
\label{ch:intro}

\markright{\textbf{Introduction}}

In the context of general relativity, consider the collapse of a spherical shell of 
matter under its own weight. The dynamics of this process, as modelled by the coupled
Einstein and matter field equations, can be understood intuitively in terms of the 
competition between gravitational attraction and repulsive internal forces 
(due, for instance, to kinetic energy or pressure).
Typically, such an isolated system ends up in one of three distinct states.
If the initial configuration is dilute, then the repulsive forces will dominate and
the collapsing matter will implode through the center and disperse, leaving flat
space behind.
If, on the other hand, the density of the initial configuration is sufficiently large,
some fraction of the initial mass will form a black hole. 
In some matter models it is also possible to form stable stars, but,
for the sake of simplicity, we will disregard this possibility in the following.
Critical gravitational collapse occurs when the attracting and repulsive forces governing
the dynamics of the collapse process are almost in balance, or, in other words, the
initial configuration is near the threshold of black hole formation.

Critical phenomena in gravitational collapse have been originally discovered in the
seminal numerical investigations of scalar field collapse by Choptuik \cite{Choptuik92,Choptuik93}.
Using sophisticated numerical techniques, Choptuik investigated the threshold of black 
hole formation for the self-gravitating massless scalar field in spherical symmetry.
He evolved one-parameter families of initial data that interpolate between black
hole formation and dispersion and fine-tuned the initial data parameter, $p$, through a
bisection search to its critical value, $p^*$, where a black hole is just formed.
Choptuik was able to give convincing evidence that black holes of arbitrarily
small mass can be created.
Moreover, he discovered the following surprising phenomena:
The black hole mass depends on the initial data parameter via a simple power law
\begin{equation}
  m_\text{BH} \propto \left( p - p^* \right)^\gamma,
\end{equation}
for $p \gtrsim p^*$.
All near-critical solutions approach a discretely self-similar solution at intermediate times.
This so-called ``critical solution'' or ``Choptuon'' is characterized by a constant $\Delta$.
These phenomena and the ``critical exponents'' $\gamma$ and $\Delta$ are independent of
the family of initial data. Therefore, critical phenomena are universal within a given model.
In the dynamical systems picture, the phase space (or space of initial data) of 
this system is divided into basins of attraction, with black holes and Minkowski space
as attractors.

Hamad\'e and Stewart \cite{Stewart96} have found numerical evidence, that the
critical solution contains a naked singularity which can be seen at future
null infinity. 
The problem has also been studied extensively from an analytic point of view by
Christodoulou \cite{Christodoulou86,Christodoulou87,Christodoulou91,Christodoulou94}.
In particular, he was able to prove that the space of regular initial 
data that lead to naked singularities has measure zero \cite{Christodoulou99},
so that the appearance of the naked singularity is non-generic.
Similar critical solutions -- exhibiting (continuous or discrete) self-similarity --
have also been found for several other types of matter fields, and have been
constructed directly in several cases \cite{Gundlach95,Gundlach97f,Gundlach96a,Lechner2001,LechnerPhD}.

In this thesis we present further numerical studies of spherically symmetric 
scalar field critical collapse.
We extend previous investigations by focussing on global aspects
of this problem, and, for the most part, use a compactified evolution scheme 
which includes null infinity on our numerical grid.
The motivation is twofold: The main goal of our investigation was to gain 
an understanding of local versus global issues in critical collapse.
In particular, we try to address questions like:
What is the role of asymptotic flatness for critical collapse 
(e.g. the critical solution, the ``Choptuon'' is self-similar, 
and thus not asymptotically flat)?
How would hypothetical detectors of radiation observe the dynamics 
close to criticality?
How can we understand the way null infinity approximates observers at large 
distances in this simple but nontrivial setup?
The second motivation is to test numerical algorithms which are based on
compactification methods in a situation that is very demanding on accuracy.
We will argue that at least in the model considered here, global methods
do not cause a significant penalty in accuracy, but simplify the
interpretation of certain results.

In the current work, we refer to critical collapse phenomena as
``critical collapse at the threshold of apparent horizon formation'' to avoid
possible misunderstandings, since critical collapse is essentially a quasilocal
phenomenon and the standard definition of black holes is based on global
concepts (see textbooks like e.g. \cite{Wald84}).
Also, the choice of a local threshold criterion emphasizes the relation of these 
phenomena to other areas in nonlinear PDEs, where related phenomena occur, but the 
concept of black holes is absent.

Critical behavior of the kind originally found by Choptuik is usually referred
to as type II, because of its formal correspondence with type II phase
transitions of statistical physics.
A different type of critical solutions at the threshold of black hole
formation, corresponding to type I phase transitions, is provided by unstable
static configurations  -- like those found by Bartnik and McKinnon \cite{Bartnik88}.

Linear perturbation calculations of such critical solutions revealed exactly
one unstable mode, which confirmed their interpretation as intermediate attractors 
in the language of dynamical systems.
Critical phenomena in general relativity are reviewed in \cite{Gundlach97fd, Bizon96},
including discussions in terms of phase transitions and renormalization group
techniques familiar from statistical physics.

A massless scalar field in spherical symmetry exhibits type II critical collapse 
(there are no regular stationary or time-periodic solutions).
Type II critical solutions have been found to exhibit continuous or discrete
self-similarity in the past lightcone of the singularity.
In our case, the critical solution is known to be
discretely self-similar (DSS), and has been constructed directly as an
eigenvalue problem \cite{Gundlach95}.

A spacetime is said to be DSS 
\cite{Gundlach00a}
if it admits a discrete diffeomorphism $\Phi_\Delta$ which leaves the 
metric invariant up to a constant scale factor: 
\begin{equation}\label{intro:DSS-def}
\left(\Phi^*_\Delta\right)^n g = e^{2 n \Delta} g,
\end{equation}
where $\Delta$ is a dimensionless real constant and $n \in \mathbb{N}$.

We choose scalar field critical collapse in spherical
symmetry for several reasons: the model is very well studied and we can 
compare with a large amount of previous numerical and analytical results. 
Furthermore, the model is also very demanding:
The value of the echoing period in the DSS critical solution 
is $\Delta \simeq 3.44$, which is quite larger compared to
many other models. Note that larger values of $\Delta$ make it 
more difficult to resolve a large number of echos.

Our numerical method is based on a characteristic initial value problem, i.e.
we foliate spacetime by null cones. This allows for a very efficient evolution 
system and simplifies the study of the causal structure of the solutions. 
In spherical symmetry, caustics are restricted to the center of symmetry,
so we do not have to deal with the dynamical appearance of caustics, which causes 
potential problems for characteristic initial value problems in higher dimensions.

The numerical approach used in the compactified code\cite{Puerrer-Husa-PCA} is based on the ``DICE''
(Diamond Integral Characteristic Evolution) code, which has been documented in
\cite{Husa2000b}.
It mixes techniques from previous work of Garfinkle \cite{Garfinkle95}
and the Pittsburgh group \cite{Gomez92a,Winicour98}, in particular we follow Garfinkle
in moving along ingoing null geodesics to utilize gravitational focusing for increasing 
resolution in the region of large curvature. 
Furthermore, compactification methods are well studied and relatively 
straightforward to implement in characteristic codes. 

\begin{figure}
\centering
\includegraphics[width=.9\textwidth]{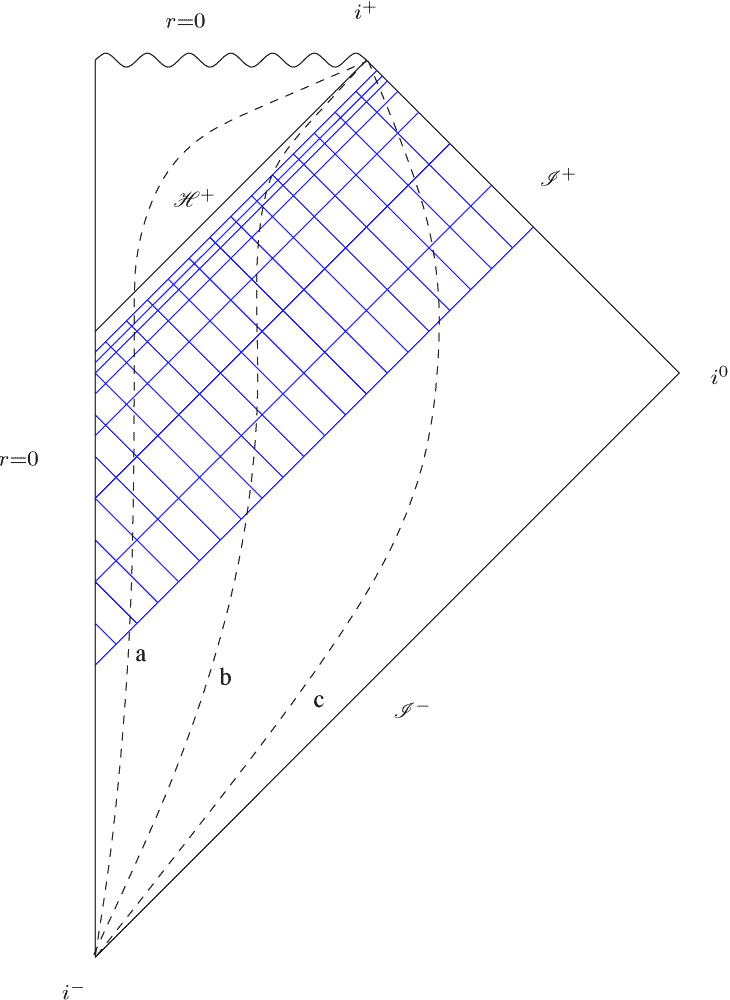}

\caption{\label{spacetime}A Penrose diagram of a typical collapse spacetime. 
Shown is our numerical null grid which extends to future null infinity $\Scri^+$.
The grid consists of the null slices $u=const$ and ingoing radial null geodesics $v=const$.
Evolution slows down in the vicinity of the future event horizon $\Scrh^+$.
We also indicate lines (a) $r = const < 2M_f$, (b) $r = const = 2M_f$
and (c) $r = const > 2M_f$, where $M_f$ is the final black hole mass.
}
\end{figure}

An important aspect of our compactified characteristic evolution scheme is that
at late times our null slices asymptotically 
approach the event horizon, see Fig. \ref{spacetime}.
Essentially this is because our coordinates can not penetrate a 
dynamical horizon \cite{Ashtekar02,Ashtekar03,Ashtekar04,Hayward94}
(they become singular at a marginally trapped surface, e.g.
at an apparent horizon),
which is spacelike if any matter or radiation falls through it and
null otherwise \cite{Hawking73a,Ashtekar04}.
Note that the dynamical horizon is contained inside of the event horizon,
and the outermost dynamical horizon approaches (or coincides with) the
event horizon at late times, assuming cosmic censorship holds.
This fact makes our approach in some sense complementary to previous critical
collapse studies, which were not adapted to the asymptotic regime. 

To further investigate physical quantities such as the mass function when a 
dynamical horizon forms, we employ an uncompactified double-null code
which can penetrate dynamical horizons. This code is based on the work of
Hamad\'e and Stewart \cite{Stewart96} and improvements by Harada and Carr
\cite{HaradaCarr}. The code has also been developed to analyze collapse
problems in $2+1$ dimensional gravity.

In this thesis we focus on those aspects of critical
collapse which are associated with global structure, and in particular the 
phenomenology seen by asymptotic observers. For the sake of completeness, We will 
also discuss other well-studied aspects such as mass scaling or local DSS behavior
in the past self-similarity horizon.

\section{Organization of this Thesis}

This thesis is organized as follows: in chapter \ref{ch:geometric-setup}
we discuss characteristic surfaces of hyperbolic PDE's, mention different horizon
concepts for black holes, and introduce our geometric setup, which is based on 
Bondi-type coordinates and double-null coordinates in spherical symmetry.
In this context we also state gauge and regularity conditions at the origin.

Chapter \ref{sec:ContinuumProblem} we present evolution systems for the
Einstein-massless scalar field problem in both Bondi and double-null coordinates.
We discuss a compactification scheme which introduces the Misner-Sharp mass-function
as an independent evolution variable, which renders our evolution system regular 
at null infinity. We also mention physical quantities of interest, such as the
Bondi-mass and Ricci scalar curvature.

Our numerical algorithms are presented in chapter \ref{ch:numerical-algorithms}.
In addition to the numerical schemes for the Bondi and double-null codes, we 
discuss issues of mesh-refinement, accuracy, convergence and the satisfaction of 
constraint equations in numerical evolutions.

Chapter \ref{ch:critical-phenomena} deals with critical phenomena in gravitational
collapse. We introduce the concept of continuous and discrete self-similarity,
the dynamical systems picture and present a standard derivation of the mass-scaling
law.

In chapter \ref{ch:qnm-tails}, we introduce quasinormal modes and tails and present
heuristic results for the presence of the least scalar damped QNM of a Schwarzschild
black hole in critical collapse evolutions. We also study power-law tail exponents
in different regions of numerical spacetime.

In chapter \ref{ch:crit-phen-results} we give a detailed discussion of our numerical
results pertaining to critical behavior, which include a comparison of the radiation 
signal at future null infinity with a heuristic estimate based on self-similar scaling and
the accumulation of matter exterior to the past self-similarity horizon via backscattering.

Our results and conclusions are summarized in chapter \ref{chp:discussion}.

In Appendix \ref{app:tensors-Bondi} and \ref{app:tensors-double-null} we list tensor components 
in Bondi and double-null coordinates.
Appendix \ref{app:numerics} discusses numerical methods, starting from discretizations 
of ordinary differential equations and presenting a heuristic error analysis for the 
NSWE-algorithm. Finally, appendix \ref{app:convergence-methodology} mentions the basic 
methodology which underlies convergence tests.

\section{Conventions}

We use the metric signature $(-+++)$ and work in ``geometrized units'' $G=c=1$.
Spacetime indices are Latin. 
Space indices are denoted by Greek letters. 
Angular indices or indices of two-dimensional tensors are denoted by capital Latin letters.


\chapter{Geometric Setup}
\label{ch:geometric-setup}

%
%
\section{Some Notes on Hyperbolic PDEs}

In this thesis we will essentially be dealing with the numerical solution
of nonlinear \emph{partial differential equations }(PDEs) on characteristic manifolds. 
Since this characteristic approach is entirely different from and not as common as 
the standard spacelike Cauchy problem, it is well worth the effort to investigate 
how characteristics and associated concepts feature in the theory of PDEs.

A quasilinear (hyperbolic in our case) partial differential equation
(PDE) of $2^{nd}$ order in n variables $x_1, ..., x_n$ for an
unknown function $u(x_i)$ can be written:
(See Refs. \cite{Garabedian}, \cite{Courant-Hilbert})
\begin{equation}
\label{eq:quasilinear-PDE-def}
L(u) =
\sum_{j,k = 1}^n a_{jk}(\partial u, u, x_i) \frac{\partial^2 u}{\partial x_j \partial x_k}
+ \sum_{j = 1}^n b_j (\partial u, u, x_i) \frac{\partial u}{\partial x_j} + c(\partial u, u,x_i) + d =0,
\end{equation}
and thus defines the differential operator $L(u)$.

A submanifold of dimension $n-1$ and equation $\phi(x_i) = 0$ is said
to be a {\em characteristic manifold (surface)} if its normals $\phi_{,i}$
fulfill the first order PDE\footnote{The eikonal equation of
geometric optics is of that form.}
\begin{equation}
\label{eq:characteristic-mf-def}
\sum_{j,k = 1}^n a_{jk} \pd{\phi}{x_j} \pd{\phi}{x_k} = 0.
\end{equation}

The Cauchy problem for a $2^{nd}$ order hyperbolic PDE can only be well
posed if the manifold on which the initial data is specified is not a
characteristic manifold. If it were posed on a characteristic manifold, the PDE would
only be first order in time, $L(u)$ being an ``interior'' (i.e. acting purely in the
characteristic surface) differential operator, and the $2^{nd}$ derivative
transversal to the initial data surface could, in this case, not be determined.
Thus, characteristic surfaces are exactly those, on which the Cauchy
problem is not well-posed.
Note that in the quasilinear case the characteristic condition depends on
the initial data, since the $a_{jk}$ are, in general, also functions
of $u(x_i)$.

Characteristic surfaces are, in turn, generated by {\em characteristic rays}
or {\em bicharacteristics}, which satisfy the following equation
\begin{equation}
\label{bicharacteristics-def}
\dot x_j = \sum_{k=1}^n a_{jk} \pd{\phi}{x_k}.
\end{equation}
It can be shown that the bicharacteristics of general relativity are
null geodesics which generate null hypersurfaces.
We will show in section \ref{sec:bicharacteristics} that the bicharacteristics of
a massless scalar field which is minimally coupled to gravity are null geodesics of
spacetime.
Therefore, since gravitational disturbances, and, in our case, disturbances in the
massless scalar field, are propagated along null geodesics, this makes a case
for introducing a foliation of spacetime based on a family of non-intersecting
null hypersurfaces.
It is also worth noting that these disturbances need not be smooth in general,
i.e. shock waves are possible in some matter models (e.g. perfect fluids).

As a simple example consider the flat space wave equation
\begin{equation}
u_{tt} = u_{xx} + u_{yy} + u_{zz}
\end{equation}
The characteristics then have to fulfill
\begin{equation}
\phi_t^2 = \phi_x^2 + \phi_y^2 + \phi_z^2.
\end{equation}
A special solution is the {\em characteristic cone}
\begin{equation}
\phi = (t-t_0)^2 - (x-x_0)^2 - (y-y_0)^2 - (z-z_0)^2 = 0,
\end{equation}
which represents spherical wavefronts propagating from a source located
at $(x_0,y_0,z_0)$.
The generators of the cone, which are bicharacteristics, may be represented
as light rays perpendicular to the wavefronts.

It is illuminating to compare the spacelike slices of the $3+1$ formalism
with a characteristic foliation based on outgoing null hypersurfaces
for the scalar wave equation $\square \phi = 0$ in Minkowski spacetime
with polar spherical spatial coordinates $(r,\theta,\phi)$.

In $3+1$ the result is well known:
\begin{equation}
\partial_{tt} \phi
        = \frac{1}{r^2} \partial_r \Bigl( r^2 \partial_r \phi \Bigr)
          + \frac{1}{r^2 \sin \theta}
            \partial_\theta \Bigl( \sin \theta \partial_\theta \phi \Bigr)
          + \frac{1}{r^2 \sin \theta} \partial_{\phi\phi} \phi.
\end{equation}
This equation is {\em second\/} order in time, \ie{} both the field
and its time derivative must be specified on an initial slice in order
to uniquely determine the future evolution.

In contrast, let us introduce a coordinate system $(u,r,\theta,\phi)$ based on
null cones $u = const$, which are generated by a two-dimensional set of null rays,
labeled by the angles $\theta,\phi$ and with $r$ a radial parameter along the rays.
These coordinates will be discussed in detail in section
\ref{sec:Bondi_coordinates}.
The line element is
\begin{equation}
ds^2 = - du^2 - 2du dr + r^2 d\Omega^2,
\end{equation}
so that the wave equation then becomes
\begin{equation}
2(r\phi)_{,ur}=(r\phi)_{,rr}-\frac{1}{r^2}L^2(r\phi)
\end{equation}
where $L^2$ is the angular momentum operator.
This equation is only {\em first\/} order in the retarded time $u$, so
only the value of the field needs to be specified on an initial slice
in order to uniquely determine the future evolution.  However, the
domain of dependence only extends to the inside of the cone;
\ie{} the initial null surface is not a Cauchy surface for the spacetime.

%
%

\subsection{The Bicharacteristics of a Scalar Field Coupled to Gravity}
\label{sec:bicharacteristics}

Let us consider the matter field equation for the massless scalar field
minimally coupled to gravity in spherical symmetry and 
assume that we can foliate spacetime by null hypersurfaces which are
labelled by a coordinate $u$.
We will show that the bicharacteristics of the scalar field are null 
geodesics of spacetime (see \cite{dinverno-book}).

The matter field equation is of the form of equation (\ref{eq:quasilinear-PDE-def})
\begin{equation}
  \square_g \phi = \nabla^a \nabla_a \phi
                 = g^{ab} \partial_a \partial_b \phi
                 - g^{ab} \Gamma_{ab}^c \partial_c \phi,
\end{equation}
and its principal part (the terms which contain the highest 
derivatives) is
\begin{equation}
  P\left[ \square_g \phi \right] = g^{ab} \partial_a \partial_b \phi.
\end{equation}
The coeffient matrix which appears here, the $a_{jk}$ of
(\ref{eq:quasilinear-PDE-def}), is just the inverse spacetime metric in
the (u,r)-submanifold.  According to equation
(\ref{eq:characteristic-mf-def}) the $u=const$ hypersurfaces are
characteristic manifolds since they fulfill
\begin{equation}
  g^{ab} \nabla_a u \nabla_b u = 0.
\end{equation}

By definition \eqref{bicharacteristics-def} the bicharacteristics which
generate the $u=const$ surfaces satisfy (with $\lambda$ being a parameter
along the bicharacteristics)
\begin{equation}\label{bichar2}
\frac{d x^a}{d\lambda} = g^{ab} \nabla_b u,
\end{equation}
i.e., they are the orbits of the vector field $\nabla^a u$.

Now we will prove that these bicharacteristics are, in fact, null geodesics.
We take the covariant derivative in the direction of the vector field $\nabla^a u$
of the last equation and find
\begin{equation}
\left(\nabla^c u\right) \nabla_c \left(\frac{d x^a}{d\lambda}\right) =
g^{ab} \left(\nabla^c u\right) \nabla_c \nabla_b u.
\end{equation}
Thanks to the symmetry of the connection this last expression is equal to
\begin{equation}
g^{ab} \nabla^c u \nabla_b \nabla_c u =
\frac{1}{2} g^{ab} \nabla_b\left( \nabla^c u \nabla_c u \right) = 0,
\end{equation}
where we have used that $\nabla^a u$ is null.  Thus, the
bicharacteristics of the massless scalar field turn out to be null
geodesics. This intuitively makes sense, since the scalar field
propagates at the speed of light.  We have also shown that $\nabla^a
u$ is an \emph{affine} null vector, that is, $\lambda$ (chosen
by equation \eqref{bichar2}), is an affine
parameter along these null rays.

\section{General Properties of Black Holes}\label{sec:bh-properties}

In the following we give some textbook definitions of a \emph{black hole}, both from 
a global and a local point of view \cite{Poisson04, Wald84}.

The \emph{causal past} of an event $p$, denoted by $J^-(p)$ is the set of all 
events that can be reached from $p$ by past-directed curves, either timelike or 
null. The \emph{causal past} of a set of events $S$, $J^-(S)$ is the union of 
the causal pasts of all events $p \in S$.

The \emph{black hole region} $B$ of the spacetime manifold $M$ is the set of all 
events that do not belong to the causal past of future null infinity:
\begin{equation}
  B = M - J^-(\mathscr{I}^+).
\end{equation}

The \emph{event horizon} $EH$ is defined to be the boundary of the black hole region:
\begin{equation}
  EH = \partial B = \partial \left(  J^-(\Scri^+) \right).
\end{equation}
Because the event horizon is a causal boundary, it must be a null hypersurface 
generated by null geodesics.
Note that this definition of a black hole depends crucially on the notion of 
future null infinity which can be defined for asymptotically flat spacetimes.
Moreover, to apply this definition one has to know the entire future development 
of the spacetime in question. In practice, it is often more convenient to use a 
quasi-local definition of a horizon.

A \emph{trapped surface} is a closed, spacelike two-dimensional surface $S$ such that for
both congruences (ingoing and outgoing) of future-directed null geodesics orthogonal to $S$,
the expansion $\theta$ is negative everywhere on $S$.

The \emph{trapping horizon}\cite{Hayward-bh-dyn} is the three-dimensional boundary of the region 
of spacetime that contains trapped surfaces.
\footnote{A \emph{trapping horizon} is the closure $\bar H$ of a three-surface $H$ foliated by \emph{marginal surfaces} on which $\theta_- |_H \ne 0$ and $\mathcal{L}_- \theta_+ |_H \ne 0$. 
A \emph{marginal surface} is a spatial two-surface $S$ on which one null expansion vanishes, fixed as
$\theta_+ |_S = 0$.
The trapping horizon
and marginal surfaces are said to be \emph{outer} if $\mathcal{L}_- \theta_+ |_H < 0$ and \emph{future}
if $\theta_- |_H < 0$.
The concept of \emph{dynamical horizons}\cite{Ashtekar04} is closely related to trapping horizons.
}

The two-dimensional intersection of the trapping horizon with a spacelike hypersurface $\Sigma$
(Cauchy surface) is called an \emph{apparent horizon}.
The apparent horizon is therefore a \emph{marginally trapped surface} -- a closed two-surface $S$
on which \emph{one} of the congruences of future-directed null geodesics orthogonal to $S$ 
has zero expansion.
\footnote{For technical details see \cite{Wald84} theorem 12.2.5.}
Note that the intersection of an outgoing null hypersurface with the trapping horizon can be 
an apparent horizon, or a three-dimensional subset of the trapping horizon.

The \emph{trapped region} $\mathscr{T}$ of a spacelike hypersurface $\Sigma$ is the portion of 
$\Sigma$ that contains trapped surfaces.

Under certain technical conditions the black hole region contains trapped surfaces. 
\footnote{For details see \cite{Wald84} proposition 12.2.3.}
While for stationary black holes the trapping horizon typically coincides with the event
horizon, it lies within the black hole region in dynamical situations (unless the null energy condition
is violated.)
We will not distinguish between the two-dimensional apparent horizon and the three-dimensional trapping
horizon and will refer to both as the apparent horizon.
We also use the term ``trapped region'' in a different sense as defined above, namely we refer to
a region of spacetime for which each point lies on a trapped surface \cite{Hayward-bh-dyn}.

In the following we will restrict ourselves to future outer trapping horizons (FOTH) and future 
marginally outer trapped surfaces (MOTS), which require $\theta_+ = 0$ and $\theta_- < 0$, where
$+$ and $-$ refers to the outgoing and ingoing null congruences, respectively.
A FOTH is either spacelike or null, being null if and only if the shear of the outer null normal
as well as the matter flux across the horizon vanishes \cite{Ashtekar04}.

The apparent horizon is the outermost (future) marginally outer trapped surface (MOTS)
in a spacelike hypersurface $\Sigma$.
If one looks for globally outermost MOTS in a Cauchy evolution, they can 
``jump'' \cite{AnderssonMarsSimon05}.
This does not happen for an outgoing null foliation.

%
%

\section{Bondi Coordinates in Spherical Symmetry}\label{sec:Bondi_coordinates_in_SS}

%
%
\subsection{Spherical Symmetry}
By definition, the metric of a spherically symmetric spacetime possesses
three spacelike Killing vector fields, which form the Lie-algebra of SO(3), the
rotation group in three dimensions.
The Killing vectors fulfill the following commutation relations
\begin{equation}
\left[V_i,V_j\right] = \epsilon^k_{ij} V_k,
\end{equation}
where the $\epsilon^k_{ij}$ denote the structure constants of SO(3).
The commutators close (that is, the commutator of any two fields in the set
is a linear combination of other fields in the set), so that by Frobenius'
theorem (see \cite{Wald84}) the integral curves of these vector fields
``mesh'' together to form submanifolds of the manifold on which they are defined.
The orbits of SO(3) are spacelike 2-spheres and the spacetime metric induces
a metric on each orbit 2-sphere.

The general form of a spherically symmetric metric is (see appendix B of \cite{Hawking73a})
\begin{equation}
\label{eq:generic-ss-metric}
  ds^2 = d\tau^2(t,r) + Y^2(t,r)d\Omega^2(\theta,\varphi),
\end{equation}
where $d\tau^2$ is a two-dimensional metric of indefinite signature $(-+)$ and
$d\Omega^2 = d\theta^2 + \sin^2\theta d\varphi^2$ is the canonical metric on
$S^2$. The coordinates $(t,r,\theta,\varphi)$ are chosen
so that the group orbits of SO(3) are the surfaces ${t,r = const}$
and the orthogonal surfaces are given by ${\theta,\varphi= const}$.

%
%

\subsection{Bondi Coordinates}\label{sec:Bondi_coordinates}

In the following, we define Bondi coordinates and derive the form of the spacetime metric.
(see \cite{dinverno-book} and \cite{Papapetrou}).
We start from the general form of a spherically symmetric metric given
in equation \eqref{eq:generic-ss-metric}.
We chose
\begin{equation}
x^0 = u = const
\end{equation}
to be a family of non-intersecting, outgoing null hypersurfaces, i.e.\\
$g^{-1}(du,du) = 0$.
We know from section \ref{sec:bicharacteristics} that these null hypersurfaces are generated
by null geodesics (the bicharacteristics of the surfaces). This gives us the
possibility to define a second coordinate in a purely geometrical way.
We choose
\begin{equation}
x^1 = r
\end{equation}
to be a radial parameter along the outgoing null geodesics that foliate the
$u=const$ surfaces, i.e. on each such null ray we have $u = const$ and $x^2$ and
$x^3 = const$.

The remaining coordinates are chosen to be polar angles, coinciding with the
usual flat space definitions for
\begin{equation}
x^2 = \theta, x^3 = \varphi
\end{equation}
near the center of spherical symmetry, as we will later choose $u$ to be central
proper time.
This completes our coordinate system to $(u,r,\theta,\varphi)$.

Now, we derive the form of the spacetime metric in these coordinates.
An outgoing null ray (light ray) is a coordinate curve
\begin{equation}
u = u_0, \quad \theta = \theta_0, \quad \phi = \phi_0 \qquad\text{all constant}
\end{equation}
where r is varying, or
\begin{equation}
du = d\theta=d\phi = 0
\end{equation}
The tangent vector to such a curve $x^a (r)$,
\begin{equation}
\left(\pd{}{r}\right)^a  = \frac{dx^a}{dr} = \delta^a_r,
\end{equation}
must be parallel to the normal vector to the $u = \text{const}$ hypersurfaces
\begin{equation}
\nabla^a u = g^{ab} \nabla_b u = g^{au},
\end{equation}
since, in the case of a null hypersurface, the normal vector is also the
tangent and the null rays generate the $u = \text{const}$ hypersurfaces.
Thus, we find
\begin{equation}
g^{au} \propto \delta^a_r,
\end{equation}
which entails
\begin{equation}
g^{uu} = 0,
\end{equation}
or equivalently
\begin{equation}
g_{rr} = 0.
\end{equation}

We still have freedom in the choice of parametrization along the null rays.
Following Bondi, we choose $r$ to be an areal radial coordinate or luminosity
distance, so that the surface area of $r=const$, $u=const$ 2-spheres
equals $4\pi r^2$.
\footnote{Other possible choices include an affine parametrization of the
  outgoing null rays, as used by Newman and Penrose in their work on
  gravitational radiation \cite{NewmanPenrose62}.
  An affine radial coordinate entails by definition
  $$\left(\pd{}{r}\right)^a \nabla_a \left(\pd{}{r}\right)^b = 0$$ which leads to
  $$0 = \partial_r \delta^a_r + \Gamma_{rb}^a \delta_r^b = \Gamma^a_{rr}.$$
  For null coordinates in spherical symmetry the only nonzero contribution comes
  from $$\Gamma^r_{rr} = g^{ur} g_{ur,r}.$$
  A vanishing of $g^{ur}$ and thus diverging of $g_{ur}$ would make the metric
  singular, therefore the condition to be satisfied is $$g_{ur,r} = 0.$$
  According to the general form of a spherically symmetric metric given in
  equation \eqref{eq:generic-ss-metric}, the line element must then be
  of the following form
  \begin{equation}
    ds^2 = -g_{uu}(u,r) du^2 - 2dudr + f(u,r) d\Omega^2.
  \end{equation}
}
This choice can be enforced by the condition
\begin{equation}
\begin{vmatrix}
g_{\theta\theta} & g_{\theta\varphi} \\
g_{\theta\varphi} & g_{\varphi\varphi}
\end{vmatrix}
= r^4 \sin^2\theta.
\end{equation}
Note that in spherical symmetry, the areal $r$-coordinate is defined in a
geometrically unique way (\cite{Hayward-grav-energy}).
It is also possible to compactify the radial coordinate; we discuss this
in detail in section~\ref{sec:compactification}.

Putting these requirements together, we find that the line element reduces to
the form
\begin{equation}
ds^2 = g_{uu}(u,r) du^2 + 2g_{ur}(u,r) du dr + r^2 d\Omega^2.
\end{equation}
Since
$\bigl( \begin{smallmatrix}
g_{uu} & g_{ur} \\
g_{ur} & g_{rr}
\end{smallmatrix}\bigr)$
is Lorentzian, we can choose $\pd{}{u}$ to be timelike, i.e.
\begin{equation}
g_{uu} = g\left(\pd{}{u},\pd{}{u}\right) < 0.
\end{equation}

We write the metric in the following way, adhering to Bondi's
conventions for the names of the metric functions
\begin{equation}\label{eq:def-Bondi-line-element}
ds^2 = -e^{2\beta(u,r)}\frac{V(u,r)}{r} du^2 - 2 e^{2\beta(u,r)} du dr + r^2 d\Omega^2.
\end{equation}
where $V > 0$ is assumed in order that $\pd{}{u}$ be timelike.

So the metric and inverse metric are
\begin{equation}
g_{ij}
        =
  \left(
  \begin{array}{cccc}
  - e^{2 \beta} \frac{V}{r}     & - e^{2 \beta} & 0     & 0     \\
  - e^{2 \beta}                 & 0             & 0     & 0     \\
  0                             & 0             & r^2   & 0     \\
  0                             & 0             & 0     &  r^2 \sin^2 \theta
  \end{array}
  \right)
\end{equation}
and
\begin{equation}
g^{ij}
        =
 \left(
 \begin{array}{cccc}
 0           & -e^{-2 \beta}            & 0                       & 0 \\
-e^{-2 \beta}&  e^{-2 \beta}\frac{V}{r} & 0                       & 0 \\
 0           &       0                  & \frac{1}{r^2}           & 0 \\
 0           &       0                  & 0      &  \frac{1}{r^2\sin^2\theta}
 \end{array}
 \right)
                                                \label{eq:BondiMetric_SS_upper}
\end{equation}
respectively.

The sign of $g_{ur}$ is due to our use of a retarded null-coordinate $u$.
In section \ref{sec:regularity} we will choose $u$ to be proper time at the
origin which implies that the metric reduces to the following flat metric
at the origin
\begin{equation}
ds_{flat}^2 = -du^2 - 2 du dr + r^2 d\Omega^2.
\end{equation}
One can also check that this is consistent with the signature $(-+++)$ by
introducing a time coordinate $t = u + r$ in flat space, so that
\begin{equation}
ds_{flat}^2 = -dt^2 + dr^2 + r^2 d\Omega^2.
\end{equation}
For purposes of comparison, we can also introduce an advanced null-coordinate
$v = t + r$ in flat space, and find
\begin{equation}
ds_{flat}^2 = -dv^2 + 2 dv dr + r^2 d\Omega^2.
\end{equation}

\begin{figure}[htbp]
  \centering
  \includegraphics[angle=0,height=9cm]{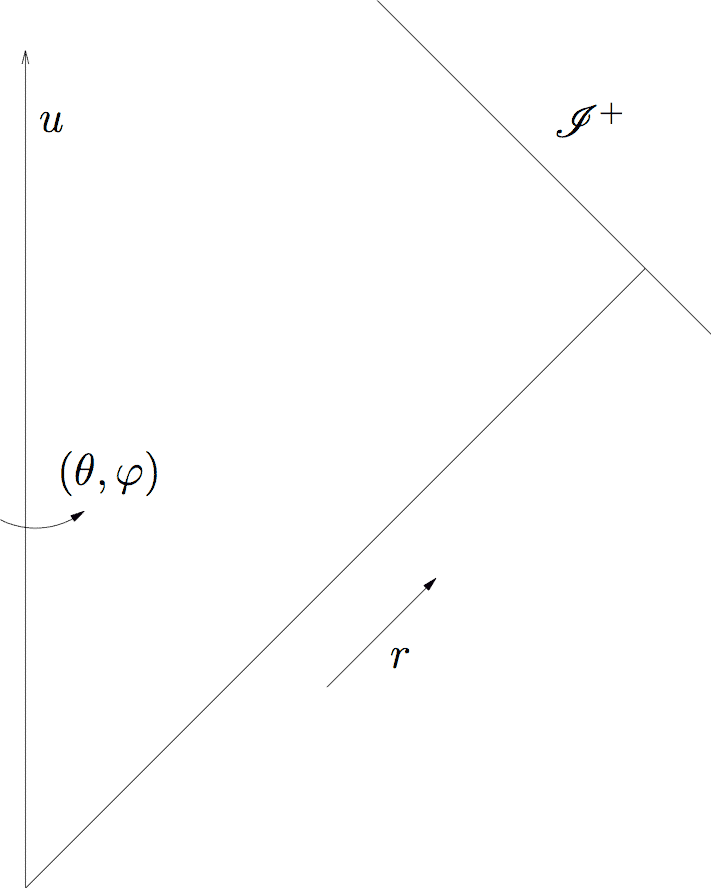}

  \caption{Bondi coordinates $(u,r,\theta,\varphi)$.
  $u$ is the proper time at the center of spherical symmetry, $r$ is an areal
  radial coordinate. A regular center is assumed at $r=0$.}\label{coordinates}
\end{figure}

In the context of Bondi coordinates,
we often write partial derivatives of some function $f(u,r)$ with respect to $u$
and $r$ as $\dot f$ and $f'$, repectively.

%
%

\subsection{Physical Interpretation of the Bondi Metric}\label{sec:Bondi-physical-interpretation}

To improve physical understanding, it is worthwhile to analyze the Bondi metric in 
some further detail.

First, let us take a look at the tangent and normal vectors as shown in table \ref{tab:coord-vectors}.
By construction, we have
\begin{equation}
\left(\pd{}{r}\right)^a = -e^{2\beta} \nabla^a u
\end{equation}
i.e., $\left(\pd{}{r}\right)^a$ lies in the u=const null surfaces. Or, to put
it differently, these null rays generate the $u=const$ null hypersurfaces.

Observe also that if $V$ is positive, then $r=const$ is a timelike surface and
$(\partial/\partial u)^a$ is also timelike. If $V=0$, both are null,
and if $V<0$ both are spacelike.
Correspondingly, $\nabla_a r$ is spacelike, null or timelike.

\begin{center}
\begin{table}[htbp]
  \setlength{\extrarowheight}{1.5mm}
  \begin{center}
    \begin{tabular}{|l|l|ll|}\hline
      $\nabla_a u = \left(1,0,0,0\right)$ &
      $\nabla^a u = \left(0,-e^{-2\beta},0,0\right)$ &
      $\nabla_a u \nabla^a u = 0$ & null\\

      $\nabla_a r = \left(0,1,0,0\right)$ &
      $\nabla^a r = \left(-e^{-2\beta},e^{-2\beta}\Vr,0,0\right)$ &
      $\nabla_a r \nabla^a r = e^{-2\beta}\Vr > 0$ & spacelike\\

      $\left(\pd{}{u}\right)^a = \left(1,0,0,0\right)$ &
      $\left(\pd{}{u}\right)_a = \left(-e^{2\beta}\Vr,-e^{2\beta},0,0\right)$ &
      $\left(\pd{}{u}\right)^a \left(\pd{}{u}\right)_a =
      - e^{2\beta}\Vr < 0$ & timelike\\
      $\left(\pd{}{r}\right)^a = (0,1,0,0)$ &
      $\left(\pd{}{r}\right)_a = (-e^{2\beta},0,0,0)$ &
      $\left(\pd{}{r}\right)^a \left(\pd{}{r}\right)_a = 0$ & null\\[1mm]
      
      \hline
    \end{tabular}
    \caption{Overview of tangent and normal vectors to coordinate
      hypersurfaces and coordinate lines. Here we rely on the assumption that
      $V > 0$.}
    \label{tab:coord-vectors}
  \end{center}
\end{table}
\end{center}

The particular choice of the algebraic form of the metric components serves
the simplicity of the resulting Einstein equations. The functions
$\beta$ and $V$ also have simple physical interpretations.
As will be shown in section \ref{sec:regularity} $\beta$ measures the 
redshift between an asymptotic coordinate frame of reference and the center 
of spherical symmetry. $V$ is the analog of the Newtonian potential and 
contains the Bondi-mass in its asymptotic expansion.

We define ingoing and outgoing null directions $l_{\pm}^a$ by
\begin{equation}
l_-^a = (1, -\frac{1}{2}\Vr,0,0),
\end{equation}
and
\begin{equation}
l_+^a = - \nabla^a u = (0,e^{-2\beta},0,0),
\end{equation}
which is an affine null vector, i.e. $l_+^a \nabla_a l_+^b = 0$.
For an affine null vector the null expansion $\Theta$ can be defined by
(see chapter 9 of \cite{Wald84})
\begin{equation}
\Theta = \nabla_a l^a.
\end{equation}
We find
\begin{equation}
\Theta_+ = \frac{2}{r}e^{-2\beta}.
\end{equation}

If $\Theta_+ > 0$ along an outgoing null ray, the surface area of
2-spheres increases and neighboring null rays diverge.
Since $\Theta_+$ is positive definite unless $\beta$ diverges,
the occurrence of marginally outer trapped surfaces,
$\Theta_{+}=0$ for some sphere, cannot happen in outgoing Bondi coordinates.
Thus, we cannot penetrate apparent horizons (see \ref{sec:bh-properties}).

\begin{figure}
  \centering
  \includegraphics[width=9cm]{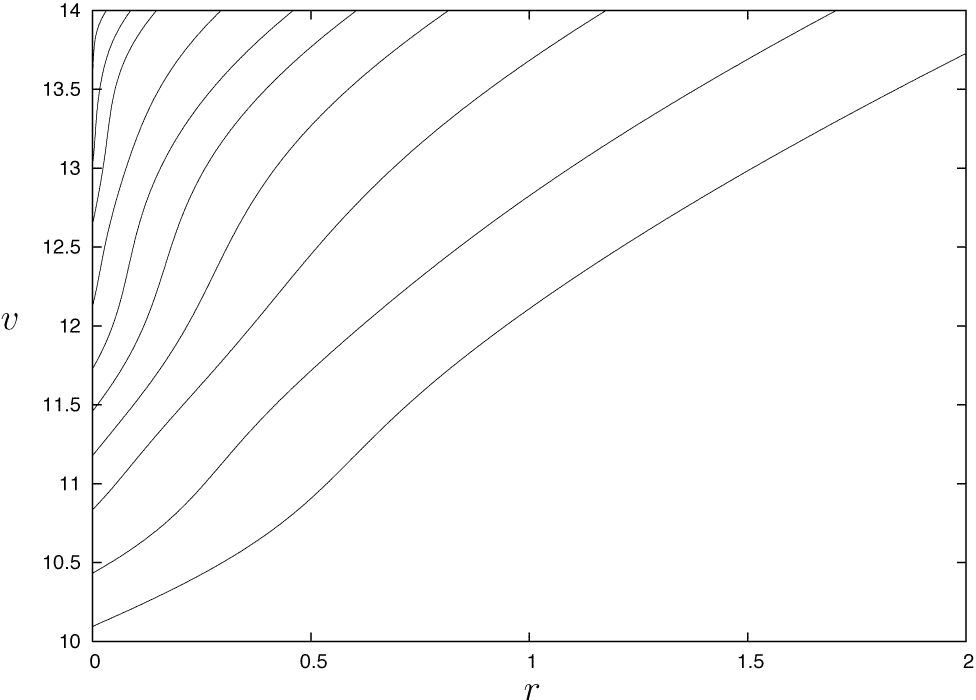}

  \caption{This figure shows how the $u=const$ null slices approach
    an apparent horizon in the late part of a near-critical time evolution.
    Observe how the first slices shown are still close
    to a flat space null-cone, whereas the last slices extend
    almost in a vertical direction, i.e. $\frac{dv}{dr} \to \infty$.
  }\label{fig-slices-lin}
\end{figure}
\begin{figure}
  \centering
  \includegraphics[width=9cm]{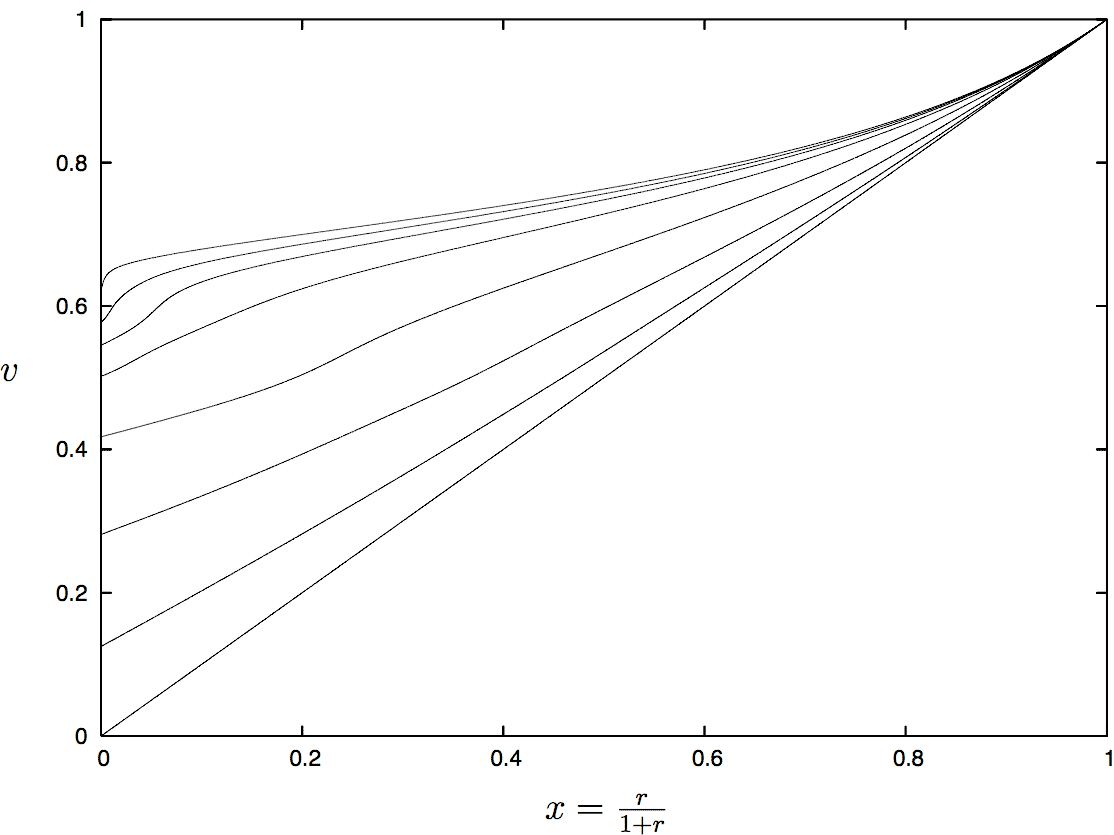}

  \caption{This figure shows the same situation as depicted in
    \ref{fig-slices-lin}, but for a near-critical compactified run.
    The initial slice is by definition a straight line. All slices
    extend to $\mathscr{I}^+$ at $x=1$. Late slices exhibit the formation
    of an apparent horizon close to the origin.
  }\label{fig-slices-cmp}
\end{figure}

Nevertheless, as we come close to an apparent horizon, our slices will
asymptote to it (see figures \ref{fig-slices-lin} and \ref{fig-slices-cmp})
and ultimately, as $\Theta_+ \to 0$, $r$ will cease to be a
good coordinate; it will no longer be a monotonically increasing function
of the advanced null coordinate $v$ which is defined by
$\pd{}{v} = l_{-}$, i.e. the curves $v=const$ are the ingoing null-geodesics
and $v = r$ on the initial slice. (In flat space $v=t+r$.)
Tiny numerical errors may lead to the
bending over of the slices and thus crash the code.

The radial ingoing null geodesics are given by integration of the equation
\begin{equation}\label{eq:ingoing-null-geodesics}
\frac{d}{du}r(u) = - \frac{V(u,r(u))}{2 r(u)}.
\end{equation}
Accordingly, the total derivative of any scalar quantity $f$
with respect to $u$ along an ingoing null geodesic is given by
\begin{equation}
\frac{d}{du}f\left(u, r(u)\right) = \left( \frac{\partial}{\partial u}-
                       \frac{V}{2 r} \frac{\partial}{\partial r} \right)f
= \mathscr{L}_{l_-} f.
\end{equation}
Since this is a Lie-derivative with respect to a non-coordinate vector field, 
it will in general not commute with partial derivatives like 
$\frac{\partial}{\partial u}$.

%
%

\subsection{Gauge and Regularity Conditions at the Origin}\label{sec:regularity}
Assume that there is a \emph{regular center}
at $r=0$ (no eternal black or white holes or naked singularities), so that we can
designate a finite geodesic line segment of a central observer at $r=0$ as
our spatial origin. 
We fix the gauge such that the family of outgoing null cones emanating from
the center is parametrized by the \emph{proper time} 
\footnote{
  With our choice of Lorentz signature $(-+++)$, the proper time of a 
  timelike curve $C$ with tangent $T^a$ and curve parameter $t$ is given by
  $$\tau = \int \left( -g_{ab}T^a T^b \right)^{1/2} dt.$$
},
$u$, of this central observer.
This choice is taken since we are interested in the
``echoing region'' near the center of spherical symmetry, where
oscillations in the  fields will pile up if we are near to the
``critical solution''.
Another common gauge is to choose $u$ as the proper time of an observer
at future null infinity, $\mathscr{I}^+$. This asymptotically flat 
time coordinate is the usual choice if one is interested in asymptotics.

Choosing $u$ as proper time at the origin implies the condition
\begin{equation}
  e^{2\beta}\Vr = 1,
\end{equation}
while a regular center, as defined by \cite{Hayward-grav-energy}, requires
$g^{-1}(dr,dr) - 1 = \bigO(r^2)$, which in turn entails
\begin{equation}
  e^{-2\beta}\Vr = 1 + \bigO(r^2).
\end{equation}
Together theses requirements imply local flatness and regularity of the metric, 
that is 
\begin{equation}
  \begin{split}
    e^{2\beta}(u_0,r) &= 1 + \bigO(r^2)\\
    \Vr(u_0,r)        &= 1 + \bigO(r^2),
  \end{split}
\end{equation}
at a fixed retarded time $u_0$.
These smoothness conditions are rederived in section
\ref{sec:Bondi-regularity-conditions}.

\subsubsection{Relation between Central and Bondi Time}
In the following, $u_\CS$ will always denote {\em central time} - proper time
at the center of spherical symmetry, whereas we denote the
asymptotically flat {\em Bondi time} by $u_\BS$.
The relation between the two time coordinates can easily be derived by
comparing with an asymptotically flat (standard Bondi) frame
(see section \ref{sec:asm-exp}).
For large $r$ the line elements behave as
\begin{align}
ds^2 &= -\left(e^{2H} + \bigO(r^{-2})\right) du_\CS
         \left[\left(e^{2H} + \bigO(r^{-1})\right) du_\CS
              + 2 dr\right] + r^2 d\Omega^2 \\
d\bar{s}^2 &= - d{u_\BS}^2 - 2 du_\BS dr  + r^2 d\Omega^2
\end{align}
which yields in the limit $r\to\infty$
\begin{equation}\label{eq:def_BondiTime}
du_\BS = e^{2H} du_\CS,
\end{equation}
where $H = \beta(u_\CS,\infty)$ and
$\lim_{r \to \infty} \frac{V}{r}(u_\CS,r) = e^{2H}$.

The following more geometrical argument yields the same result:
In general $u_\BS = u_\BS(u_\CS,r)$, but under the following
assumptions we can show that $u_\BS = u_\BS(u_\CS)$ only.
We demand that
\begin{equation}
  \left(\pd{}{u_\BS}\right)^a \left(\pd{}{u_\BS}\right)_a < 0,
\end{equation}
and
\begin{equation}
  \nabla^a u_\BS \nabla_a u_\BS = 0,
\end{equation}
i.e. $u_\BS = const$ are null hypersurfaces.

We find
\begin{equation}
  \begin{split}
    0 &= \nabla^a u_\BS(u_\CS,r) \nabla_a u_\BS(u_\CS,r)\\
      &= e^{-2\beta}\left(\pd{u_\BS}{r}\right)
         \left[ \Vr \left(\pd{u_\BS}{r}\right) - 2\left(\pd{u_\BS}{u_\CS}\right) \right],    
  \end{split}
\end{equation}
which yields
\begin{equation}
  \pd{u_\BS}{r} = 0.
\end{equation}
Thus $u_\BS = u_\BS(u\CS)$ only.
Now we have
\begin{equation}\label{eq:uCuB_vec}
  \left(\pd{}{u_\BS}\right)^a = f(u_\CS) \left(\pd{}{u_\CS}\right)^a.
\end{equation}
Since $u_\BS$ is proper time at infinity,
\begin{equation}
  g\left(\pd{}{u_\BS},\pd{}{u_\BS}\right)\bigg|_{r=\infty} = -1,
\end{equation}
plugging \eqref{eq:uCuB_vec} in to the metric yields
\begin{equation}
-1 = f^2 g\left(\pd{}{u_\CS},\pd{}{u_\CS}\right)\bigg|_{r=\infty}
= -f^2 e^{2\beta}\Vr \bigg|_{r=\infty} = f^2 e^{4H}.
\end{equation}
Therefore, $f = e^{-2H}$ and we have found the desired relation between central
and Bondi time: 
\begin{equation}
  du_\BS = e^{2H} du_\CS.
\end{equation}


\subsubsection{Redshift}

Consider a freely falling observer located at $r=0$ whose worldline is tangent
to the timelike geodesic parametrized by proper time at the origin $u_\CS$ with
tangent $\left(\partial/\partial u_\CS \right)^a$. He sends out a light signal which propagates
along the affinely parametrized outgoing radial null geodesic with tangent
$k^a = -\nabla^a u_\CS$ (As shown in section \ref{sec:bicharacteristics},
$k^a$ is an affine null vector). The signal is received
by an observer whose worldline is tangent to a timelike geodesic at $r=\infty$,
parametrized by proper time at infinity $u_\BS$ with tangent
$\left(\partial/\partial u_\BS\right)^a$.

The frequency (see Ref. \cite{Wald84}) of an emitted wave is given by minus the inner 
product of the wave vector (which is just the rate of change of the phase of the wave) 
with the 4-velocity of the observer.
The frequency of emission at the center is given by
\begin{equation}
  \omega_C = -k_a \left(\pd{}{u_\CS}\right)^a\bigg|_{r=0}
  = \nabla_a u_\CS \left(\pd{}{u_\CS}\right)^a\bigg|_{r=0} = 1.
\end{equation}
The frequency of reception at infinity is, using the relation between central
and Bondi time \eqref{eq:def_BondiTime},
\begin{equation}
  \omega_B = -k_a \left(\pd{}{u_\BS}\right)^a\bigg|_{r=\infty}
  = \nabla_a u_\CS \left(\pd{}{u_\BS}\right)^a\bigg|_{r=\infty} = e^{-2H}.
\end{equation}
The redshift-factor is defined as
\begin{equation}\label{eq:redshift-factor}
  z = \frac{\omega_C}{\omega_B} - 1 = e^{2H} - 1.
\end{equation}
For small $H$ we thus have $z = 2H + \bigO(H^2)$.

Now, because of the hypersurface equations (see section \ref{field_eq})
and gauge choice for $u$, the metric functions, $\beta$ and
$V$, are both monotonically increasing in the direction of increasing $r$ and
both are positive definite. Thus $H =\beta(u,\infty) > 0$ and
\begin{equation}
  \omega_B < \omega_C
\end{equation}
always, which is just the condition for signals to be redshifted.
In the case of horizon formation
$\theta_+ \to 0$, however, $H$ will tend to infinity, so that light rays
can no longer escape to infinity and the redshift will diverge
exponentially until our coordinates break down.

A crude argument for the redshift is also apparent just from the relation
\eqref{eq:def_BondiTime}.
If $H \to \infty$, a finite amount of central time $u_\CS$ corresponds to an
infinite amount of Bondi time $u_\BS$ and thus light signals originating from the
center are infinitely redshifted.

%
%

\subsubsection{Derivation of Regularity Conditions at the Origin of Spherical
Symmetry}\label{sec:Bondi-regularity-conditions}

In this section we apply the general procedure for deriving regularity conditions
set forth by Bardeen and Piran in \cite{Bardeen-Piran-regularity}. 
Although the latter deals with axisymmetric systems, it is nevertheless worthwhile 
to study this approach in the simpler setting of spherical symmetry.

A tensorial quantity is called \emph{regular} at $r=0$ if and only if its components
relative to Cartesian coordinates can be expanded in non-negative
integer powers
\footnote{
Otherwise the tensorial quantity or its derivatives will blow up at $r=0$.
}
of x, y and z; i.e. one demands the existence of a Taylor series expansion in
Cartesian coordinates.
We would like to derive what restrictions this assumption of regularity at the
origin implies for the metric functions $\beta$ and $V$ as functions of $r$.

We will work in a regular $(t,x,y,z)$ coordinate system with $t$ defined as
$t = u + r$.
\footnote{Note that the coordinates $(u,x,y,z)$ are not regular: The vector fields
$\pd{}{x}$, $\pd{}{y}$ and $\pd{}{z}$ all have a kink at the origin.}
Since we have taken $u$ to be proper time at the origin, $t$ also
measures proper time at $r=0$.
We also enforce \emph{local flatness} at the center, \ie{} the metric goes to
the Minkowski metric as $r$ tends to $0$,
\begin{equation}\label{flatness_gauge}
  \lim_{r\to 0} g_{ab} = \eta_{ab}.
\end{equation}
We will have to investigate which derivatives of the metric functions are allowed
to be non-zero and which are zero by regularity.

The relations between the two coordinate systems are:
\begin{xalignat*}{2}
t &= u + r                 & \qquad      u &= t - r \\
x &= r \sin\theta \cos\phi & \qquad      r &= \sqrt{x^2 + y^2 + z^2} \\
y &= r \sin\theta \sin\phi & \qquad \theta &= \arctan\sqrt{\frac{x^2+y^2}{z^2}} \\
z &= r \cos\theta          & \qquad \phi   &= \arctan\frac{y}{x} \\
\end{xalignat*}
The Jacobian for this coordinate transformation is
\begin{equation}
  \frac{\partial(u,r,\theta,\phi)}{\partial(t,x,y,z)} =
  \begin{pmatrix}
    1 & -\frac{x}{r} & -\frac{y}{r} & -\frac{z}{r} \\
    0 &  \frac{x}{r} &  \frac{y}{r} &  \frac{z}{r} \\
    0 & \frac{x}{r^2\sqrt{\frac{x^2+y^2}{z^2}}} &
     \frac{y}{r^2\sqrt{\frac{x^2+y^2}{z^2}}} & \frac{-\sqrt{x^2 + y^2}}{r^2} \\
    0 & -\frac{y}{x^2+y^2} & \frac{x}{x^2+y^2} & 0
  \end{pmatrix}.
\end{equation}

First we need to calculate the components of the metric in $(t,x,y,z)$
coordinates. Some interesting components as functions of the polar coordinates are
\begin{equation}
  g_{tt} = - e^{2\beta} \Vr
\end{equation}

\begin{equation}
  g_{tx} =   e^{2\beta} \Vr \sin\theta \cos\phi
           - e^{2\beta} \sin\theta \cos\phi
\end{equation}

\begin{equation}
  g_{xx} = - e^{2\beta} \Vr \sin^2\theta \cos^2\phi
           + 2 e^{2\beta}\sin^2\theta \cos^2\phi 
           + \cos^2\theta \cos^2\phi
           + \sin^2\phi
\end{equation}

\begin{equation}
  g_{yy} = - e^{2\beta} \Vr \sin^2\theta \sin^2\phi
           + 2 e^{2\beta}\sin^2\theta \sin^2\phi 
           + \cos^2\theta \sin^2\phi
           + \cos^2\phi
\end{equation}

\begin{equation}
  g_{zz} = - e^{2\beta}\Vr \cos^2\theta
           + 2e^{2\beta}\cos^2\theta
           + \sin^2\theta
\end{equation}

We will also need derivatives of the metric. Since the derivatives of $g_{xx}$ and
$g_{yy}$ are particularly messy and we do not need to consider them,
we leave them out.
\begin{equation}\label{gttx}
  g_{tt,x} = - 2 e^{2\beta}\beta_{,r}\Vr \sin\theta \cos\phi
             + e^{2\beta}\Vr \frac{\sin\theta\cos\phi}{r} 
             - e^{2\beta}V_{,r} \frac{\sin\theta\cos\phi}{r}
\end{equation}

\begin{multline}
  g_{tx,x} = e^{2\beta}\beta_{,r} \left(\Vr -1\right) \sin^2\theta \cos^2\phi
           + e^{2\beta}\left(\frac{V}{r^2} - \frac{1}{r}\right) \\
           + e^{2\beta}\left(-\frac{2V}{r^2} + \frac{1}{r} + \frac{V_{,r}}{r}\right)
             \sin^2\theta \cos^2\phi.
\end{multline}

\begin{multline}\label{gzzx}
  g_{zz,x} = \left(-2\beta_{,r}e^{2\beta}\Vr + 2e^{2\beta}\frac{V}{r^2}
                   + 4\beta_{,r}e^{2\beta} - \frac{4e^{2\beta}}{r}\right)
              \cos^2\theta \sin\theta \cos\phi \\
              + \left(\frac{2}{r} - e^{2\beta}\left(\frac{V_{,r}}{r}
                                                    - \frac{V}{r^2}\right)\right)
              \sin\theta\cos\phi
              -\frac{2}{r}\sin^3\theta \cos\phi.
\end{multline}

We make the following series ansatz for the metric functions $\beta$ and $\Vr$:
\begin{align}\label{series_first}
\beta &= a + b r + c r^2 + \bigO(r^3) \\
\Vr   &= d + e r + f r^2 + \bigO(r^3)
\end{align}
This yields
\begin{align}
\beta_{,r} &= b + 2c r + \bigO(r^2) \\
V_{,r}     &= d + 2e r + 3f r^2 + \bigO(r^3) \\
e^{2\beta} &= (1 + 2a + 2a^2) + (2b + 4ab)r 
              + (2c + 2b^2 + 4ac)r^2 + \bigO(r^3).
\end{align}

We now impose the local flatness gauge condition \eqref{flatness_gauge}.
\begin{equation}
  \lim_{r\to 0} g_{tt} \overset{!}{=} -1
\end{equation}
yields
\begin{equation}\label{gauge_c1}
  -d (1 + 2a + 2a^2) \overset{!}{=} -1.
\end{equation}
Similarly, we have
\begin{equation}
  \lim_{r\to 0} g_{tx}
  = \lim_{r\to 0} e^{2\beta} \sin\theta \cos\phi \left(\Vr - 1\right) \overset{!}{=} 0,
\end{equation}
from which we find
\begin{equation}
  d = 1,
\end{equation}
and thus, it follows from \eqref{gauge_c1} that
\begin{equation}
  a = 0.
\end{equation}
One does not need to consider any other equations, since the two constant terms
in \eqref{series_first} have now been determined. Indeed, one can also check that
$\lim_{r\to 0} g_{xx} = 1$ and $\lim_{r\to 0} g_{zz} = 1$ are fulfilled.

The series that respect this gauge choice are
\begin{align}\label{regularity_series}
\beta &= b r + c r^2 + \bigO(r^3) \notag \\
\Vr   &= 1 + e r + f r^2 + \bigO(r^3) \notag \\
\beta_{,r} &= b + 2c r + \bigO(r^2) \\
V_{,r}     &= 1 + 2e r + 3f r^2 + \bigO(r^3) \notag \\
e^{2\beta} &= 1 + 2b r + (2c + 2b^2)r^2 + \bigO(r^3). \notag
\end{align}

We are now in a position to investigate limits ($r\to 0$) of the derivatives
of the metric. The values should remain finite and be direction
independent, i.e. independent of the spherical angles $\theta$ and $\phi$,
since these are indefinite at the origin. This will then provide us with
conditions on the series expansions of $\beta$ and $\Vr$.
Consequently, we find from \eqref{gttx}, the equation for $g_{tt,x}$, that
\begin{equation}
  - 2\beta_{,r} e^{2\beta}\Vr + e^{2\beta}\left(\frac{V}{r^2}
    - \frac{V_{,r}}{r}\right) \overset{!}{=} \bigO(r).
\end{equation}
Applying the expansions \eqref{regularity_series}, we find
\begin{equation}
  \left(1+ 2br + \bigO(r^2)\right)
   \left[\frac{1}{r} + e + fr - \frac{1}{r} - 2e
    -3fr - 2b -4cr -2ber + \bigO(r^2)\right] \overset{!}{=} \bigO(r)
\end{equation}
and eventually
\begin{equation}\label{reg_cond_I}
 e + 2b = 0.
\end{equation}
Similarly, we find from \eqref{gzzx}, the equation for $g_{zz,x}$, that
\begin{equation}
  e^{2\beta}\left(-2\beta_{,r}\Vr + 4\beta_{,r} + \frac{2V}{r^2} - \frac{4}{r}\right)
  \cos^2\theta 
  - e^{2\beta}\left(\frac{V_{,r}}{r} - \frac{V}{r^2}\right)
  + \frac{2}{r}\left(1 - \sin^2\theta\right) \overset{!}{=} \bigO(r).
\end{equation}
Here, we can set all terms proportional to $\cos^2\theta$ to zero and
insert the series expansions \eqref{regularity_series}. We find
\begin{equation}
e + 2fr + 2ber + \bigO(r^2) \overset{!}{=} \bigO(r),
\end{equation}
which leads to the condition
\begin{equation}\label{reg_cond_II}
 e = 0.
\end{equation}
Combining the two conditions \eqref{reg_cond_I} and \eqref{reg_cond_II} we recover
\begin{align}
  b &= 0 \\
  e &= 0.
\end{align}

We did not consider $g_{tx,x}$, since it is already well behaved because
of the gauge condition alone. For $g_{xx,x}$ the calculation would have been more
involved but would have led to the same result.

Summarizing, we have found that the metric functions $\beta$ and $\Vr$ guarantee
that the metric is regular at the center of spherical symmetry if they satisfy,
at a fixed retarded time $u_0$,
\begin{equation}
\fbox{
  \begin{minipage}[c][1.7cm]{4cm} 
    \par\vspace{-0.4cm} 
    \begin{equation*}\label{eq:Bondi-regularity-cond}
      \begin{split}
        \beta(u_0,r) &= \bigO(r^2) \\
        \Vr(u_0,r)   &= 1 + \bigO(r^2)       
      \end{split}
    \end{equation*}
  \end{minipage}
}
\end{equation}
These regularity conditions are consistent
with the hypersurface equation \eqref{eq:V-prime-hypersurface}, $V_{,r} = e^{2\beta}$,
derived in chapter \ref{field_eq}.

The conditions we have derived are sufficiently accurate for our purposes,
though one could consider higher derivatives of the metric.
We will see from the following simple argument that, in fact, all odd
derivatives of the metric functions have to vanish at the origin.

Without loss of generality, define \emph{linear distance} $l$ by
\begin{equation}
l \equiv z =
  \begin{cases}
    -r & \text{if} \quad z < 0, \\
    +r & \text{if} \quad z > 0.
  \end{cases}
\end{equation}
We demand that the metric does not have a kink as the origin is crossed along a
ray parametrized by linear distance $l$.
If a metric function $f(r)$ contains a linear term in its Taylor expansion
around the origin, then
\begin{equation}
\frac{d}{dr} f(r)\Big\lvert_{r=0} = C
\end{equation}
where $C$ is a constant.
\begin{equation}
\frac{d}{dl} f(r)\Big\lvert =
  \begin{cases}
    -C + \bigO(r) & \text{if} \quad r < 0 \\
    +C + \bigO(r) & \text{if} \quad r > 0
  \end{cases}
\end{equation}
has a jump discontinuity and is thus not regular.
It will be regular provided that the constant $C$ vanishes.
This is equivalent to the condition that
\begin{equation}
\frac{d}{dr} f(r)\Big\lvert_{r=0} = 0
\end{equation}
Note that this argument can be extended to higher derivatives. Roughly speaking,
each time one applies $\frac{d}{dl}$ to a function $f(r)$, $f(r)$ reverses its sign
for $r < 0$. Therefore, all derivatives of an odd order have to vanish at the
origin, so that a regular function of $r$ will have a Taylor series which contains
only even terms in $r$:
\begin{equation}
  f(r) = c + c_2 r^2 + c_4 r^4 + \dots.
\end{equation}

In addition, we can also derive the Taylor series expansion for the scalar
field by making use of the hypersurface equation \eqref{eq:beta-prime-hypersurface}
$\beta_{,r} = 2\pi r \left(\phi_{,r}\right)^2$ and find that there is no
restriction on $\phi$:
\begin{equation}\label{eq:Bondi-phi-regularity}
  \phi(r) = \phi_0 + \phi_1 r + \bigO(r^2).
\end{equation}
In this respect it is also interesting to view this last fact from a different
perspective: When working in polar coordinate systems, the d'Alembertian
usually contains terms similar to $\frac{2}{r}\partial_r$. Regularity of
the field $\phi$ at the origin then amounts to requiring that
$\frac{\partial\phi}{\partial r}\lvert_{r=0}$
vanishes. But using a null coordinate $u$ makes a difference. There is now a
rivaling $-\frac{2}{r}\partial_u$ term in the d'Alembertian \eqref{wave-eq} which
happens to cancel the other term at the origin.

%
%

\subsection{Historical Notes on the Bondi Coordinate System}

The original motivation for Bondi and his co-workers 
(see \cite{Bondi-axisymmetric-vacuum}) in developing this
coordinate system was to analyze radiation from an isolated system.
To do so, one needed to deal with expansions in negative
powers of a radial distance for various quantities.
Such a treatment is, however, made impossible by the appearance of
logarithmic terms in $r$, as is the case, in Schwarzschild coordinates.

Later, the conformal compactification of null infinity $\mathscr{I}$
allowed a geometrized description of the asymptotic physical properties of
radiative spacetimes. This novel technique also proved very beneficial
for the study of gravitational radiation, see e.g. \cite{Frauendiener04}.

One demands that in a region from $\mathscr{I}^+$ inwards the
coordinate system should be non-singular. Farther in, null rays will,
in general, focus and cross, forming caustics and causing the
coordinates to break down and become singular.  \footnote{Except for
  the case where there exists a second regular center, where rays may
  caustic, this cannot happen in spherical symmetry, because the
  lightrays are 'running' after each other but never touch. A second
  regular center is however ruled out by the assumed existence of
  future null infinity $\mathscr{I}^+$.
  The only caustic that is present in spherical symmetry is a point
  caustic at the center itself, which has to be handled by regularity
  conditions on the metric and the matter fields.  Of course, for
  nontrivial topologies the situation is entirely different: e.g. on
  the Einstein cylinder lightrays may form caustics.}

The Bondi-like coordinates defined in section
\ref{sec:Bondi_coordinates} cover all of spacetime provided there are
no caustics or singularities and $\mathscr{I}^+$ exists.
\footnote{For the existence of $\mathscr{I}^+$, which is a part of
  conformal infinity, the spacetime metric has to satisfy
  certain falloff conditions, i.e. it must tend to a flat metric
  sufficiently fast as one goes to infinity in an outgoing null
  direction. In a coordinate independent formulation a spacetime is
  said to be asymptotically flat, if the physical spacetime can be
  mapped into a new, ``unphysical'' spacetime via a conformal isometry
  which has to satisfy certain, very technical conditions. Details can
  be found in chapter 11 of Ref. \cite{Wald84}.}

\section{Double-Null Coordinates}\label{sec:double-null-coords}

As we have seen in section \ref{sec:Bondi-physical-interpretation}, 
outgoing Bondi coordinates cannot penetrate apparent horizons, because the areal radial coordinate $r$ fails to be a good coordinate when an AH forms: $\nabla^a r \to 0$.
A coordinate system which does not suffer from this deficiency
and is suitable for spherically symmetric critical collapse evolutions are \emph{double-null coordinates}.

In addition to the retarded null-coordinate $u$ which is constant on outgoing light rays,
we now introduce an advanced null-coordinate $v$, which is constant on ingoing light rays 
and replaces $r$ as a coordinate. Along with the standard polar coordinates on the 
group orbits of SO(3), $S^2$, the coordinate chart becomes $(u,v,\theta,\phi)$.
The line element is then of the following form
\begin{equation}\label{eq:dn-line-element}
ds^2 = -a^2(u,v) du dv + r^2(u,v) d\Omega^2.
\end{equation} 
The area radius $r$ is now a metric function that depends on the null coordinates.
The double-null form is natural in the sense that each two-sphere possesses two 
preferred normal directions, the null directions $\partial / \partial_u$ and 
$\partial / \partial_v$.

The components of the metric and inverse metric are given by
\begin{equation}
g_{ab} =
\begin{pmatrix}
0 & -a^2/2 & 0 & 0\\
-a^2/2 & 0 & 0 & 0\\
0 & 0 & r^2 & 0\\
0 & 0 & 0 & r^2\sin^2\theta,
\end{pmatrix}
\end{equation}
and
\begin{equation}
g^{ab} =
\begin{pmatrix}
0 & -2/a^2 & 0 & 0\\
-2/a^2 & 0 & 0 & 0\\
0 & 0 & 1/r^2 & 0\\
0 & 0 & 0 & 1/(r^2\sin^2\theta),
\end{pmatrix}
\end{equation}
and the metric determinant is
\begin{equation}
det g = -\frac{a^4 r^4 \sin^2\theta}{4}.
\end{equation}

A relation between Bondi coordinates and double-null coordinates can be established 
by a coordinate transformation (see Ref. \cite{PretoriusLehner04}) of the form
\begin{equation}
  dv = e^{2\alpha} \left( \Vr du + 2 dr \right),
\end{equation}
with an integrating factor $\alpha(u,r)$.

The line-element is
\begin{equation}
  \begin{split}
ds^2 &= -e^{2\beta}\left( \Vr du + 2 dr \right)du + r^2 d\Omega^2\\
     &= -e^{2\beta - \alpha} du dv + r^2 d\Omega^2.
  \end{split}
\end{equation}
In the new coordinates $r=r(u,v)$ and we can introduce a metric function $a(u,v)$, 
so that $a^2(u,v) = e^{2\beta(u,r(u,v)) - \alpha(u,r(u,v))}$ and the line-element 
becomes equation \eqref{eq:dn-line-element}.

Since $dr = \npd{r}{u}|_v du + \npd{r}{v}|_u dv$ we have the relations
\begin{align}
  \pd{r}{u}\Big|_v &= -\frac{V}{2r}\\
  \pd{r}{v}\Big|_u &= \frac{1}{2} e^{-2\alpha}  
\end{align}

In the context of double-null coordinates we denote partial derivatives of some 
function $f(u,v)$ with respect to $u$ and $v$ by $\dot f$ and $f'$, respectively.

\subsection{Null Expansions}\label{sec:dn-null-expansions}
To compute marginally trapped surfaces we compute the null expansions 
(see Ref. \cite{Wald84})
\begin{equation}
\theta_\pm = \nabla_a l^a_\pm,
\end{equation}
where the affine ingoing and outgoing null vectors $l_\pm^a$ are defined through 
their covectors as
\begin{align}
\left(l_+\right)_a &:= -\nabla_a u\\
\left(l_-\right)_a &:= -\nabla_a v.
\end{align}
This yields
\begin{align}
l^a_+ &= -g^{au} = (0,2/a^2,0,0)^a\\
l^a_- &= -g^{av} = (2/a^2,0,0,0)^a
\end{align}
It is easy to check that the $l_\pm^a$ are indeed affine null vectors, that is $l_\pm^b\nabla_b l^a_\pm = 0$ and that the null expansions $\theta_\pm$ are given by
\begin{align}\label{eq:theta_dn_p}
\theta_+ = \frac{4 r'}{a^2 r}\\
\theta_- = \frac{4 \dot r}{a^2 r}.\label{eq:theta_dn_m}
\end{align}
The signs of the null expansions are geometrical invariants, while their actual values 
are not \cite{Hayward-grav-energy}.
For future marginally outer trapped surfaces (MOTS) (see section \ref{sec:bh-properties}) 
we need to have $\theta_+ = 0 $ while $\theta_- < 0$.
Since the denominators in equations \eqref{eq:theta_dn_p} and \eqref{eq:theta_dn_m} are 
non-negative, the sign of the null expansions is determined by the quantities $r'$ and 
$\dot r$ and it suffices to check the conditions $r' = 0$ while $\dot r < 0$.

The possible presence of multiple MOTS on an evolution hypersurface depends on the slicing. 
For outgoing null slices, there cannot be more than one MOTS on a slice, whereas for spacelike slices there can be.
The geometric property of future outer trapping horizons being spacelike or null (the latter 
only if there is no more infalling matter when following the null geodesic generators further
outwards), but never timelike, applied to the intersection of an MOTS with an outgoing 
null-slice shows that we cannot have multiple MOTS on such a null-slice, which is equivalent 
to the function $r'(u=const,v)$ having at most one zero.

\subsection{Gauge Choice}\label{sec:dn-gauge}

The form of the metric is preserved under diffeomorphisms $u \to U(u)$ and $v \to V(v)$.
We fix this residual gauge freedom present in the null coordinates $u$ and $v$ by choosing 
\begin{itemize}
\item $r=0$ to be at $u=v$
\item $r(u=0,v) = v/2$.
\end{itemize}

\forceemptypage

\newpage
\chapter{The Continuum Problem}\label{sec:ContinuumProblem}

We consider the coupled spherically symmetric Einstein-massless scalar field system,
both in Bondi and in double-null coordinates.
After deriving the Einstein equations and the curved space wave equation for the scalar 
field, we analyze the evolution systems and formulate the characteristic initial value
problem to be solved. We also discuss compactification for Bondi coordinates and
highlight some quantities of physical interest which we will use in the numerical
evolution codes described in the following chapter.

\section{Einstein's Equations}\label{sec:Einstein-equations}
To derive Einstein's equations we start with the action functional 
which consists of the Einstein-Hilbert action plus the matter action:
\begin{equation}\label{eq:total-action}
  S[g;\phi] = \int_V \left( \frac{R}{16\pi} + \mathscr{L} \right) \sqrt{-g}\, d^4 x,
\end{equation}
where $R$ denotes the Ricci scalar and the Lagrangian density of the massless 
scalar (Klein-Gordon) field is given by
\begin{equation}
  \mathscr{L} = -\frac{1}{2} g^{ab} \nabla_a\phi \nabla_b\phi.
\end{equation}

Variation of the total action \eqref{eq:total-action} with respect to the spacetime 
metric yields the Einstein equations, 
\begin{equation}\label{eq:Einstein}
  G_{ab} = 8\pi T_{ab}
\end{equation}
where
\begin{equation}\label{eq:stress-energy-tensor}
  T_{ab} = \nabla_a \phi \nabla_b \phi - \half g_{ab} \nabla_c \phi \nabla^c \phi
\end{equation}
is the energy momentum tensor of the massless scalar field obtained via
variation of the matter action with respect to the spacetime metric.

To complete the system, variation of the matter action with respect to 
the scalar field $\phi$ yields the matter field equation, the curved wave 
equation for the massless scalar field
\begin{equation}
  \square_g\phi \equiv \nabla^a \nabla_a\phi= 0.
\end{equation}

The Einstein and Ricci tensors in outgoing Bondi coordinates $(u,r,\theta,\varphi)$ 
are given in appendix \ref{app:tensors-Bondi} along with the Christoffel symbols,
while the tensor components for double-null coordinates are given in appendix
\ref{app:tensors-double-null}.

%
%

\section{Hierarchy of Einstein Equations in Bondi Coordinates}

In spherical symmetry, the four algebraically independent Einstein equations,
the $(u,u)$, $(u,r)$, $(r,r)$, and $(\theta, \theta)$ 
components of equation \eqref{eq:Einstein}, are not differentially independent.

Essentially, we have four equations for two unknown metric functions, $\beta$ and $V$.
In principle, one could freely choose two of these equations and complement them
with the matter field equation to complete the evolution system.
Numerical considerations dictate, however, a preference for the constraint
equations (which here act in the outgoing null hypersurfaces, i.e. 
they are ODEs in the radial coordinate $r$). 
Choosing these two \emph{hypersurface} equations instead of the other wave-type 
Einstein equations naturally ensures enforcement of the constraints. 
Unconstrained (or free) evolution systems let the solutions drift
off the constraint surface of the Einstein equations and may excite numerical 
unstable modes, while constrained evolution cuts back on numerical constraint 
violations and offers greater stability.

There is an analysis of the Einstein equations for the original
\emph{axissymmetric} vacuum Bondi problem according to which the equations
decompose into the following sets 
(see Refs. \cite{Bondi-axisymmetric-vacuum} and \cite{dinverno-book}):

\begin{itemize}
\item four \emph{main equations}:
  \begin{itemize}
  \item \emph{dynamical equation}:
    \begin{equation}\label{eqn-axis-dyn}
      R_{\theta\theta} = 0
    \end{equation}
  \item \emph{hypersurface equations}:
    \begin{equation}\label{eqn-axis-hyp}
    R_{rr} = R_{r\theta} = g^{AB}R_{AB} = 0,
    \end{equation}
    where $A,B=\theta,\phi$.
    They contain only derivatives that act in the $u=const$ hypersurfaces
  \end{itemize}
\item \emph{symmetry conditions}:
  \begin{equation}\label{eqn-axis-symm}
  R_{u\phi} = R_{r\phi} = R_{\theta\phi} = 0
  \end{equation}
  These components vanish identically because of axisymmetry.
\item \emph{trivial equation}:
  \begin{equation}\label{eqn-axis-triv}
  R_{ur} = 0
  \end{equation}
\item \emph{supplementary (conservation) conditions} (for energy and angular momentum)
  \begin{equation}\label{eqn-axis-supp}
  R_{uu} = R_{u\theta} = 0
  \end{equation}
  If they are imposed on a world line, the conservation conditions reduce to
  regularity conditions on the vertices of the null cones.
  They are the analog with respect to an r-foliation of the 3+1 momentum
  constraints.
\end{itemize}
It follows by the Bianchi identities that, if the main equations are satisfied,
the trivial equation is automatically fulfilled and the conservation
conditions are satisfied on a complete outgoing null-cone if they hold on a
single spherical cross-section (e.g. at infinity).

Our situation is, of course, entirely different, we work in spherical symmetry
and have a matter field coupled to gravity.
Still, one might expect to obtain some useful insights into the hierarchy
of the equations by attempting a similar analysis.
We now show how Einstein's equations split into hypersurface equations, 
and equations which are automatically satisfied if certain conditions are met.

In the context of Bondi coordinates, we often abbreviate partial derivatives
of a function $f(u,r)$ with respect to $u$ and $r$ by $\dot f$ and $f'$, 
respectively.

The hypersurface equations follow indeed from straightforward analogues of
equations \eqref{eqn-axis-hyp}:
\begin{equation}
G_{rr} = 8\pi T_{rr}
\end{equation}
yields
\begin{equation}\label{eqn-hyp-beta}
\beta' =2\pi r (\phi')^2,
\end{equation}
while from
\footnote{We may rewrite Einstein's equations as 
$R_{ab} = 8\pi \left(T_{ab} - \half g_{ab} T\right)$, 
where $T$ is the trace of the energy momentum tensor of the scalar field.}
\begin{equation}
g^{AB}R_{AB} = 8\pi g^{AB}\left( T_{AB} - \frac{1}{2}g_{AB}T\right),
\end{equation}
we have, using equation \eqref{eqn-trace-T},
\begin{equation}
\frac{2}{r^2} R_{\theta\theta}
          = 8\pi \left( T - \frac{1}{2} g^{AB} g_{AB} T\right) = 0,
\end{equation}
and finally
\footnote{Equation \eqref{eqn-hyp-V} can also be derived from the $(u,r)$ 
component of Einstein's equations when plugging in equation \eqref{eqn-hyp-beta}.}
\begin{equation}\label{eqn-hyp-V}
V' = e^{2\beta}.
\end{equation}

The dynamical equation \eqref{eqn-axis-dyn} is satisfied if the hypersurface
equations \eqref{eqn-hyp-beta} and \eqref{eqn-hyp-V} hold everywhere
and the energy momentum tensor is covariantly divergence free:
Inserting the hypersurface equations in equation \eqref{eqn-axis-dyn} leads to
\begin{equation}
  V r \phi' \phi'' + r \left(\phi'\right)^2 V' - 2r^2\dot\phi' \phi'
+ V \left(\phi'\right)^2 - 2r\dot\phi \phi' = 0.
\end{equation}
This equation is satisfied if $\nabla^a T_{ar} = 0$.

Clearly, this equation cannot contribute to the dynamics in our
setting, since in spherical symmetry there are no gravitational
degrees of freedom and the dynamics resides entirely in the matter
field.
\footnote{
  Birhoff's theorem shows that a spherically symmetric vacuum solution
  is necessarily static. The conservation of mass and angular momentum
  in general relativity prohibits the existence of spherically symmetric
  and of dipole waves in the gravitational field. The lowest possible
  symmetry is that of a quadrupole which already requires at least
  axisymmetry.
}

The symmetry conditions, equations \eqref{eqn-axis-symm}, vanish
identically because of spherical symmetry.
What remains to be discussed is the analogue of the trivial equation, equation
\eqref{eqn-axis-triv}, $G_{ur} = 8\pi T_{ur}$
and one of the conservation conditions \eqref{eqn-axis-supp},
$G_{uu} = 8\pi T_{uu}$.

The trivial equation is satisfied if the hypersurface equations
\eqref{eqn-hyp-beta} and \eqref{eqn-hyp-V} hold
everywhere:
\begin{equation}
\begin{split}
G_{ur} = 8\pi T_{ur} \\
\frac{1}{r^2}\left(e^{2\beta} + 2 V \beta' - V'\right)= 8\pi\half\Vr(\phi')^2\\
4\pi r V (\phi')^2 = 4\pi r V (\phi')^2
\end{split}
\end{equation}

Assuming that the hypersurface equations \eqref{eqn-hyp-beta} and
\eqref{eqn-hyp-V} hold everywhere, the conservation condition
$G_{uu} = 8\pi T_{uu}$ can be rewritten as
\begin{equation}
-2 V \dot\beta + \dot V = 8\pi r^2 \left((\dot\phi)^2 - \Vr\dot\phi\phi'\right).
\end{equation}
At the world line of the central observer, $r=0$, this equation is
satisfied if the regularity conditions, equations \eqref{eq:Bondi-regularity-cond},
imposed by spherical symmetry hold.
The $u$- component of the contracted Bianchi identities $\nabla^a G_{ab} = 0$,
which must also hold for the energy momentum tensor as a
consequence of the Einstein equations, yields the following equation
\begin{equation}
  \nabla^u E_{uu} = -\nabla^r E_{ur},
\end{equation}
where we use the shorthand $E_{ab} = G_{ab} - 8\pi T_{ab}$.
Since we have already shown that $E_{ur} = 0$ everywhere, we have
\begin{equation}
  0 = \nabla^u E_{uu} = -e^{-2\beta} \nabla_r E_{uu},
\end{equation}
and thus
\begin{equation}
  \nabla_r E_{uu} = 0.
\end{equation}
Combined with $E_{uu} = 0$ at $r=0$, we have shown that $E_{uu} = 0$
everywhere.

If one considers the set of hypersurface equations plus the matter field
wave equation, it is apparent that three integration constants will be
necessary to carry out an actual time evolution.
These constants have been fixed by our choice of proper time at the
origin.
Otherwise, for asymptotically flat coordinates, as in Bondi's
original work, one chooses the integration constants at infinity.
In section \ref{sec:asm-exp} will make an acquaintance with these 
integration constants (which in our gauge are functions of $u$) and we 
will see that the conservation condition $E_{uu}=0$ imposed at 
future null infinity, $\mathscr{I}^+$, yields a differential relation 
between two of them, the Bondi mass-loss equation.

%
%
\section{Field Equations for the Gravitating Massless Scalar Field}
\label{field_eq}
Given the massless scalar (Klein-Gordon) field, all gravitational quantities 
can be determined by integration along the characteristics of the null 
foliation, since the matter field is the only dynamical field in spherical
symmetry and needs to  be included to make the system non-Schwarzschild.
This is a coupled problem since the scalar wave equation involves the curved
space metric.

The field equations consist of the wave equation for the scalar field
\begin{equation}
\square_g\phi=0
\end{equation}
and two hypersurface equations for the metric functions:
\begin{equation} \label{eq:beta-prime-hypersurface}
\beta_{,r}=2\pi r(\phi_{,r})^2
\end{equation}
\begin{equation} \label{eq:V-prime-hypersurface}
V_{,r}=e^{2\beta}.
\end{equation}
Note that both $\beta$ and $V$ are monotonic in $r$.
Combined with the gauge conditions imposed in section \ref{sec:regularity} 
monotonicity implies
\begin{align}
  \beta & \ge 0 \\
   V/r    & \ge 1.
\end{align}

In spherical symmetry the curved space d'Alembertian
\footnote{This can easily be obtained by using the formula
$\nabla_a T^a = \frac{1}{\sqrt{-g}}\partial_a (\sqrt{-g}T^a)$ and setting
$T^a = g^{ab}\nabla_b \phi$.}
is
\begin{equation}
\square_g = e^{-2\beta}\left[\left(\frac{2V}{r^2} +
\left(\frac{V}{r}\right)_{,r} \right)\partial_r
- \frac{2}{r}\partial_u - 2\partial_u\partial_r +
\frac{V}{r}\partial_{rr}\right].
\end{equation}
We will now derive a useful relation between the four-dimensional wave operator
and the wave operator in the two-dimensional (u,r) submanifold.
The intrinsic metric in the (u,r) submanifold is
\begin{equation}
(ds^2)_h = -e^{2\beta}\left(\Vr du^2 + 2dudr\right)
\end{equation}
and thus the two-dimensional d'Alembertian follows as
\begin{equation}
\square_{h} = e^{-2\beta} \left[\left(\Vr\right)_{,r}\partial_r
              -2\partial_u\partial_r + \Vr\partial_{rr}\right]
\end{equation}

Now we introduce a \emph{rescaled field} which factors out the known
falloff of $\phi$ at large $r$
\begin{equation}\label{eq:def-psi}
  \psi = r\phi.   
\end{equation}
It is then straightforward to derive the following identity:
\begin{equation}
\square_g\phi = \frac{1}{r}\square_h\psi -
\left(\Vr\right)_{,r}\frac{e^{-2\beta}}{r^2}\psi
\end{equation}
and use this to write the wave equation $\square_g\phi = 0$ as
\begin{equation}\label{wave-eq-h}
\square_h\psi - \left(\Vr\right)_{,r} \frac{e^{-2\beta}\psi}{r} = 0.
\end{equation}
The motivation for introducing the rescaled field $\psi$ is the
following: The amplitude of an outgoing spherical wave packet
decreases with $1/r$, a rescaled field $\psi=r \phi$ thus behaves
similarly to a plane wave. Indeed the following identity holds: In the
flat background case, the field $\psi$ satisfies the usual
$1+1$-dimensional wave equation $\Box_{h}\psi = 0$. 
Thanks to the plane wave behavior of $\psi$, numerical accuracy
is expected to benefit for large $r$ since the amplitude of the wave 
does not change as fast.

There exist two standard algorithms for solving the wave equation based on
this identity, which will henceforth be called the Piran-Goldwirth-Garfinkle
(PGG) and the NSWE (North-South-West-East) ``diamond'' algorithm which are
described in sections \ref{sec:PGG} and \ref{sec:diamond-algorithm}, respectively.

%
%
\section{The PGG Evolution Algorithm}\label{sec:PGG}

Here we briefly discuss a characteristic evolution scheme due to
Piran, Goldwirth and Garfinkle \cite{Goldwirth-Piran-mKG}, \cite{Garfinkle95}.
We essentially follow the original algorithm although the terminology is
slightly different.

It is useful to define the quantity
\begin{equation}\label{def_f}
  f(r,u) := (r \phi(r,u))_{,r} \quad (=\psi_{,r}),
\end{equation}
so that
\begin{equation}
  \phi(r,u)_{,r}=\frac{f(r,u)-\phi(r,u)}{r}, \quad 
  \phi(r,u)=\frac{1}{r}\int_{r_0}^r f(r',u)\,\,dr' .
\end{equation}
The wave equation (\ref{wave-eq-h}) in $u-r$ coordinates then becomes
\begin{equation}
  f_{,u}- \frac{1}{2}\Vr f_{,r} = \frac{1}{2}\left(\Vr\right)_{,r} (f-\phi)
\end{equation}
According to equation~(\ref{eq:ingoing-null-geodesics}), the left hand
side can be written as
\begin{equation}\label{Field_Evolution_ODE}
 \frac{ df(r(u),r)}{du}=\frac{1}{2}\left(\Vr\right)_{,r}(f-\phi),
\end{equation}
if the radial positions follow null geodesics.

The evolution system is thus formed by the ODE's
(\ref{def_f}, \ref{Field_Evolution_ODE}, \ref{eq:ingoing-null-geodesics}).

%
%
\section{The Gomez-Winicour ``Diamond'' Algorithm}
\label{sec:diamond-algorithm}

\begin{figure}[htbp]
\begin{center}
\begin{psfrags}
 \psfrag{u0}[][r][1.5]{$u_0$}
 \psfrag{u1}[][r][1.5]{$u_0 + \Delta u$}
 \psfrag{r0}[][r][1.5]{$r=0$}
 \psfrag{N}[][r][2]{$N$}
 \psfrag{S}[][r][2]{$S$}
 \psfrag{W}[][r][2]{$W$}
 \psfrag{E}[][r][2]{$E$}
 \includegraphics[angle=0,totalheight=8cm]{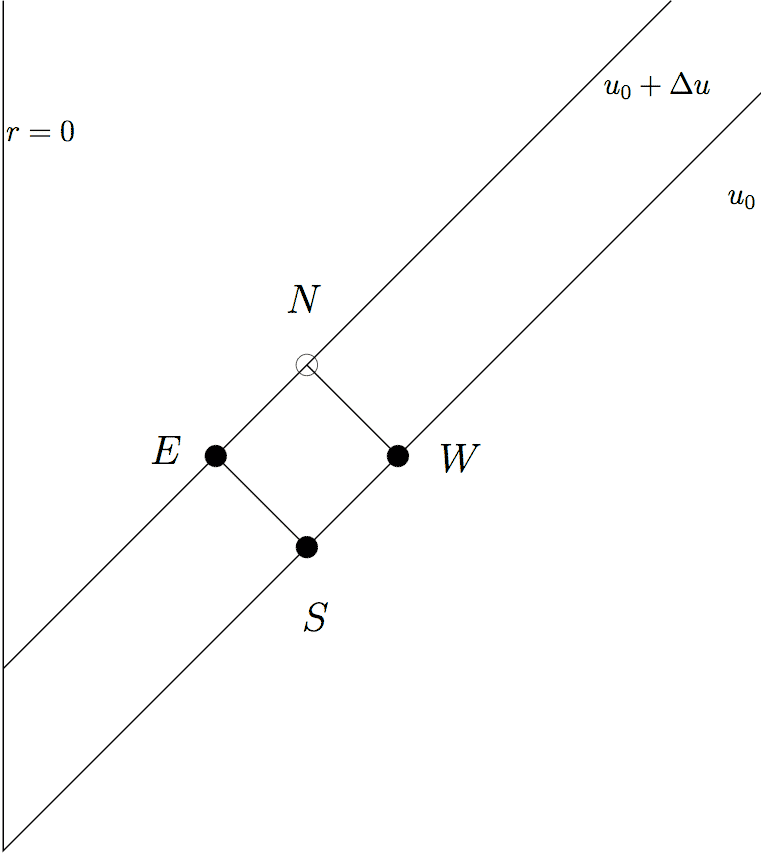}
\end{psfrags}
\caption{The computational molecule for the NSWE scheme is made up by
two $u=const$ hypersurfaces and two ingoing $v=const$ null geodesics.
}\label{fig-diamond}
\end{center}
\end{figure}

The basic idea of this algorithm (see Refs.\cite{Gomez92a},
\cite{Gomez92b},\cite{Winicour98}) is to integrate the wave
equation over the null parallelogram $\Sigma$ spanned by the points
N, S, W, and E (see figure~\ref{fig-diamond} on page~\pageref{fig-diamond}).
Using the identity (\ref{wave-eq-h}) and the volume element
$\sqrt{-h} ~ du \wedge dr = e^{2\beta} ~ du \wedge dr$
We have
\begin{equation}\label{eq:phi-evo-continuum}
\int\limits_\Sigma\square_h\psi = \int\limits_\Sigma du dr
\left(\Vr\right)_{,r}\frac{\psi}{r}
\end{equation}

The crucial trick of the whole undertaking is the following: Since h 
is a two-dimensional metric, it is conformally flat.
As we will shortly see, $\square_h$ has conformal weight
\footnote{See \cite{Wald84} Appendix D for a definition.}
$-2$ whereas the surface area element $d^2x\sqrt{-h}$ has conformal weight $+2$.
Thus, the surface integral over $\square_h\psi$ is identical to the flat space
result which can easily be obtained.

Let $\tilde{g}_{ab} = \Omega^2 g_{ab}$ be a conformal rescaling of the
metric $g_{ab}$.
First we need to know how the determinant of the metric transforms under
conformal
transformations. For a two-dimensional metric ${}^{(2)}g$ we have
$$\det{}^\two\!\tilde{g} \; \propto \; ({}^\two\tilde{g}_{AB})^2
\; \propto \; \Omega^4 ({}^\two g_{AB})^2$$
and therefore $$\sqrt{-{}^\two\tilde{g}} = \Omega^2 \sqrt{-{}^\two g},$$
\ie{} the volume element in two dimensions has conformal weight $+2$.
The formula for the d'Alembertian is $$\tilde{\square}_g\psi =
\frac{1}{\sqrt{-\tilde{g}}} \partial_a (\sqrt{-\tilde{g}} \tilde{g}^{ab}
\partial_b \psi)$$
Applying the previous result yields
$$\quabtwoT \psi = \Omega^{-2}\frac{1}{\sdetgtwo}\partial_A
\left( \Omega^2 \sdetgtwo \, \Omega^{-2} \,
{}^\two\tilde{g}^{AB} \partial_B \psi \right) = \Omega^{-2} \quabtwo \psi,$$
\ie{} the two-dimensional d'Alembertian has conformal weight -2.
So, all in all we can rewrite the original surface integral in the following
manner:
$$\int\limits_\Sigma d^2x\sqrt{-h}\,\square_h\psi = \int\limits_\Sigma du dv
\sqrt{-h_{flat}}
\,\square_{flat}\psi$$

The flat two-dimensional (Minkowsi-)metric $ds^2=-dt^2+dr^2$ can be rewritten in
double-null coordinates as
$ds^2 = -du dv$. The metric is
$(m_{AB}) = \left( \begin{smallmatrix} 0    & -1/2 \\
                                              -1/2 & 0
                          \end{smallmatrix} \right).$
So that the d'Alembertian becomes $\square_{flat} = -4 \partial_u \partial_v$.
Now our integral can easily be evaluated:
\begin{eqnarray*}
\int\limits_\Sigma du dv \sqrt{-m}\,\square_{flat}\psi
&=& -2 \int\limits_\Sigma du dv \,\psi_{,uv} \\
&=& -2 \int_{v0}^{v1}dv (\psi_{,v}\vert_{u1} - \psi_{,v}\vert_{u0}) \\
&=& -2 (\psi\vert_{u1,v1} - \psi\vert_{u1,v0} - \psi\vert_{u0,v1} +
\psi\vert_{u0,v0}) \\
&=& -2 (\psi(N) - \psi(W) - \psi(E) + \psi(S))
\end{eqnarray*}
where we have introduced the grid points $N$, $S$, $E$, and $W$ as
shown in figure~\ref{fig-diamond}

Using (\ref{wave-eq-h}) and the volume element $\sqrt{-h} ~ du \wedge dr$,
the wave equation can thus be written as
\begin{equation}
\fbox{
$\displaystyle \psi(N) = \psi(W) + \psi(E) - \psi(S) -
\frac{1}{2}\int\limits_\Sigma
du dr \left(\Vr\right)_{,r}\frac{\psi}{r}.$
}   \label{eq:diamond-scheme}
\end{equation}

\section{Compactification of the Radial Coordinate}\label{sec:compactification}

Compactification is rather simple for characteristic codes, and has
been used extensively in the  characteristic approach to numerical
relativity \cite{Gomez92a,Gomez92,Gomez92b,Winicour98}.

The type of compactification we pursue here is not \emph{conformal compactification}
\emph{per se} (see Ref. \cite{Frauendiener04} for a review),
where the spacetime manifold is completed by ``attaching'' a boundary at ``infinity''.
The method of conformal compactification maps physical spacetime onto a bounded open 
region of unphysical spacetime, introduces an \emph{unphysical} metric via a conformal
rescaling $\tilde g_{ab} = \Omega^{-2} g_{ab}$ ($\Omega$ approaches zero (at infinity) 
at an appropriate rate), and factors out known asymptotic behavior of geometrical 
quantities by a conformal transformation.

Rather, we use an ad-hoc regularization adapted to the simple geometry which could be 
related to conformal rescalings \cite{Sascha-pm}.
We employ a simple coordinate transformation with respect to the radial 
coordinate, which maps a half-infinite domain $r \in [0,\infty)$ to a compact interval
$x \in [0,1]$.
In this ad-hoc approach the metric is not altered (no unphysical metric is introduced)
and the evolution equations ``notice where infinity is'' because they degenerate there
\cite{Frauendiener04}.
The advantage of this compactification method are its simplicity and that we can evolve till 
future null infinity, $\Scri^+$ and thus study global and radiative properties of 
collapse spacetimes in detail.

We would like to emphasize that the use of conformal rescalings does not imply that one 
has to employ the rather difficult framework of Friedrich's regular conformal Einstein 
equations \cite{Friedrich-conf-null-equations}.
In fact such an approach has also been suggested in \cite{Andersson02,Husa05}.

Before introducing compactification we want to say a few words about
asymptotic series expansions.

\subsection{Asymptotic Series Expansions for the Massless Scalar Field}\label{sec:asm-exp}

Assuming initial data that are smooth at $\Scri^+$, one can
expand the massless scalar field $\phi$ in powers of $1/r$ near $\Scri^+$
\footnote{A constant term in the expansion of $\phi$ has been
  omitted since one can trivially rescale the coupled
  Einstein massless scalar field system: A transformation
  $\phi \rightarrow \phi \,+\, \text{const}$ leaves $\square\phi = 0$ invariant.
} 
\begin{equation}\label{eq:phi-asm-exp}
\phi(u,r) = \frac{c(u)}{r} + \frac{\NP}{r^2} + \bigO(r^{-3}).
\end{equation}
The coefficient $\NP$ of the $1/r^2$- term in the expansion is a 
Newman-Penrose constant \cite{Newman68} of the scalar field.

Inserting the expansion \eqref{eq:phi-asm-exp} into the hypersurface equations 
\eqref{eq:beta-prime-hypersurface} and \eqref{eq:V-prime-hypersurface} yields
\begin{equation}\label{eq:beta-asm-exp}
\beta(u,r) = H(u) - \frac{\pi c^2(u)}{r^2} + \bigO(r^{-3}),
\end{equation}
and
\begin{equation}\label{eq:V-asm-exp}
  V(u,r) = e^{2H(u)} \left(r - 2M(u) + \frac{\pi c^2(u)}{r}\right) + \bigO(r^{-2}),
\end{equation}
where integration constants $H(u)$ and $M(u)$ have been introduced.

$H(u) = \lim_{r\to\infty}\beta(u,r)|_{u=const}$ indicates redshift since Bondi time $\uB$ 
is related to proper time at the center via the relation \eqref{eq:def_BondiTime}.
\begin{equation}
  M(u) = \lim_{r\to\infty}
         \left(\frac{1}{2}r^2 e^{-2H(u)}\left(\frac{V}{r}\right)_{,r}\right).
\end{equation}
is the Bondi mass which is in general not conserved.
In terms of the Misner-Sharp mass-function defined in section \ref{sec:Misner-Sharp-mass} we have 
the relation
\begin{equation}
  M(u) = \lim_{r\to\infty} m(u,r)|_{u=const},
\end{equation}
and the asymptotic expansion
\begin{equation}
  m(u,r) = M(u) - \frac{\pi c(u)^2}{r} + \bigO(r^{-2}).
\end{equation}

%
%

\subsection{Compactified Evolution Scheme}\label{sec:compactified-evolution-scheme}
First, we introduce a compactified radial coordinate, then we recalculate our evolution
equations and aim to regularize them at null infinity.

Starting with our standard Bondi-like coordinates (u,r,$\theta$,$\phi$) we
introduce a compactified radial coordinate
\begin{equation}
x := \frac{r}{1+r}
\end{equation}
which maps $r\in [0,\infty] \mapsto x \in [0,1]$, so that points at 
$\Scri^+$ are automatically included in the grid at $x=1$.
Consequently, we have the relations
\begin{equation}
r = \frac{x}{1-x}, \qquad  dr = (1-x)^{-2} dx, \qquad \text{and} \qquad
\frac{d}{dr} = (1-x)^2 \frac{d}{dx}.
\end{equation}
The line element, equation \eqref{eq:def-Bondi-line-element}, then becomes
\begin{equation}
ds^2 = -e^{2\beta} du \left( V\left(\frac{1-x}{x}\right)du
       + \frac{2}{(1-x)^2}dx\right) + \left(\frac{x}{1-x}\right)^2 d\Omega^2,
\end{equation}
which evidently contains singular terms. Regularity, however, only requires that
the field equation for the scalar field and the hypersurface
equations for the metric functions be well behaved.

A naive approach of rewriting the hypersurface equations in terms 
of the $x$-coordinate leads to a singular equation for the quantity $V$
\begin{align}\label{eq:beta-cmp-naive}
  \beta_{,x} &= 2\pi x (1-x) (\phi_{,x})^2,\\
  V_{,x}     &= \frac{e^{2\beta}}{(1-x)^2},\label{eq:V-cmp-naive}
\end{align}
while the equation for $\beta$ is fine.

The introduction of a renormalized quantity $\bar{V} := V(1-x)$ 
(chosen such that $\frac{\bar{V}}{x} = \frac{V}{r}$) lets us rewrite the
hypersurface equation \eqref{eq:V-cmp-naive} in the following form
\begin{equation}\label{eq:Vbar-hyper-cmp-naive}
  \left(\frac{\bar{V}}{x}\right)_{,x} = \frac{e^{2\beta} - \bar{V}/x}{x(1-x)},
\end{equation}
which remedies the situation somewhat.
Still, it is not obvious that $(\frac{\bar{V}}{x})_{,x}$ is
well behaved at $\mathscr I^+$, since the denominator of equation 
\eqref{eq:Vbar-hyper-cmp-naive} tends to zero. 
An asymptotic expansion shows that $(\frac{\bar{V}}{x})_{,x}$ is nonsingular,
\begin{equation}
  \left(\frac{\bar{V}}{x}\right)_{,x} = 2 M(u) e^{2H(u)} + \bigO(r^{-1}),
\end{equation}
but the cancellation in the numerator of \eqref{eq:Vbar-hyper-cmp-naive} that
brings about this result is very delicate in numerical terms.

As described in \cite{Puerrer-Husa-PCA}, to obtain a fully regular system of 
evolution equations, we eliminate $V$ 
(or $\bar V$) by the Misner-Sharp mass-function 
\begin{equation}\label{eq:m-cmp}
  m(u,x) = \frac{x}{2(1-x)}\left(1 - \frac{\bar{V}}{x} e^{-2\beta}\right),
\end{equation}
which satisfies
\begin{equation}\label{eq:dm-cmp}
m_{,x} = 2\pi x^2 \frac{\bar{V}}{x}e^{-2\beta} \left(\phi_{,x}\right)^2.
\end{equation}
The set of hypersurface equations then becomes:
\begin{equation}
\fbox{
  \begin{minipage}[c][2.1cm]{6.5cm} 
    \par\vspace{-0.4cm} 
    \begin{equation*}\label{eq:m-beta-cmp-hyper}
      \begin{split}
        \beta_{,x} &= 2\pi x (1-x) (\phi_{,x})^2,\\
        m_{,x} &= 2\pi x^2 \left[ 1 - \frac{2(1-x)}{x} m \right] \left( \phi_{,x} \right)^2.
      \end{split}
    \end{equation*}
  \end{minipage}
}
\end{equation}
These equations are now completely regular.
Note that the term $m \frac{1-x}{x} = m/r$ does not cause problems because of 
the smoothness of the metric at the regular center \eqref{eq:Bondi-regularity-cond}.

In section \ref{sec:Bondi-evolution-scheme},
we will choose our gridpoints to freely fall along ingoing radial null geodesics 
$x(u)$ which fulfill
\begin{equation}
\frac{dx}{du} \bigg|_v = -\frac{1}{2} (1-x)^2 e^{2\beta} \left(1 - 2m \frac{1-x}{x} \right).
\end{equation}
In section \ref{sec:dice-amr} we will argue that this choice is crucial to resolve DSS
phenomena.

For $\psi$ the matter field equation takes the form
\begin{multline} 
  0 = \square_g \phi = e^{-2\beta} \Biggl\{
                 (1-x)^3 \left(\frac{2\bar V}{x^2}
                               + (1-x)\left(\frac{\bar V}{x}\right)_{,x} \right)
                   \left[\psi_{,x}\left(\frac{1-x}{x}\right) - \frac{\psi}{x^2}\right] \\
                 -2\left(\frac{1-x}{x}\right)^2 \dot\psi
       -2\left[\dot\psi_{,x}\left(\frac{1-x}{x}\right) - \frac{\dot\psi}{x^2}\right] \\
                 + \frac{\bar V}{x}\left(1-x\right)^2
                         \left[-\psi_{,x}\left(\frac{1-x}{x}\right)^2
                               - \frac{2\psi}{x^2}\left(\frac{1-x}{x}\right)\right]
                             \Biggr\}.
\end{multline}
At $\mathscr{I}^+$ this reduces to the ODE
\begin{equation}
  e^{-2H} \dot\psi = 0.
\end{equation}

The diamond scheme can be derived by applying the transformation
to the x-coordinate to equation \eqref{eq:diamond-scheme}, which yields
\begin{equation}\label{eq:diamond-m-cmp}
\fbox{
  $ \displaystyle
    \psi_N = \psi_W + \psi_E - \psi_S - \frac{1}{2} \int du dx
              \left(\frac{1-x}{x^3}\right) 2m \, e^{2\beta} \psi.    
  $
}
\end{equation}

Equations \eqref{eq:m-beta-cmp-hyper} and \eqref{eq:diamond-m-cmp}
now form a manifestly nonsingular set of evolution
equations for the massless scalar field coupled to gravity.

The gauge and regularity conditions \eqref{eq:Bondi-regularity-cond} outlined in section  \ref{sec:Bondi-regularity-conditions} and the regularity of the scalar field $\phi$ \eqref{eq:Bondi-phi-regularity} along with the definition of the rescaled field
$\psi$, equation \eqref{eq:def-psi} now become
\begin{equation}
\begin{split}
  \beta(u,x) &= \mathcal O (x^2),\\
  m(u,x)     &= \mathcal O (x^3),\\
  \psi(u,x)  &= \mathcal O (x). 
\end{split}
\end{equation}

In section \ref{sec:Taylor-expansions-compactified} we discuss Taylor expansions
in the vicinity of the origin.

\section{Diagnostics}\label{sec:diagnostics}

\subsection{The Misner-Sharp Mass}\label{sec:Misner-Sharp-mass}

In spherical symmetry, there exists a well defined notion of quasilocal energy,
the Misner-Sharp mass-function (see Refs. \cite{Misner-Sharp-1964} and
\cite{Hayward-grav-energy}):
\begin{equation}\label{m_MS}
  m = \frac{r}{2} 
  \left[ 1 - g^{rr}\right],
\end{equation}
which in outgoing Bondi coordinates yields
\begin{equation}\label{eq:m_MS}
  m(u,r) = \frac{r}{2} \left[ 1 - \Vr e^{-2\beta} \right].
\end{equation}
Note that $m/r$ is a smooth function. The 
Misner-Sharp mass measures the energy content of a sphere of radius $r$ and 
reduces the ADM and Bondi masses in the appropriate limits.
We also refer to the Misner-Sharp mass-function defined in equation 
\eqref{eq:m_MS} as $m_{MS}$, in order to distinguish it from an integral 
expression for the local mass, $m_\rho$ which we define below.

We write equation~\eqref{eq:m_MS} as an integral along outgoing
radial null geodesics $(u= const, \theta = const, \phi = const)$:
\begin{equation}
\label{mass-function-as-integral}
m_\rho (u,r) = \int\limits_0^r d{\tilde r} \  m'(u,{\tilde r})
\end{equation}
and then use Einstein's equations to simplify the integrand:
\begin{equation}
\begin{split}
m'(u,r)
        &= \frac{1}{2} \left(1 - \frac{V}{r} e^{-2 \beta} \right)
                + \frac{r}{2} e^{-2 \beta}
                  \left(
                  \frac{V}{r^2} - \frac{V'}{r} + 2\beta' \frac{V}{r}
                  \right)                                             \\
        &=      2 \pi r^2
                  \frac{V}{r} e^{-2 \beta} (\phi')^2.
\end{split}
\end{equation}
Since $m'$ is a form of local energy density,
we refer to its integral $m_\rho$
(defined by equation~\eqref{mass-function-as-integral}) as the
``integrated-scalar-field mass-function''.

On the continuum level, the mass-functions $m_\MS$ and $m_\rho$ are
clearly identical, but numerically they will in general differ
by a small amount due to having different finite differencing errors;
their difference is a good ``diagnostic'' of the code's accuracy.

To this end, we define
\begin{equation}
\delta m(u,r) = \frac{m_\MS - m_\rho}{m_{\total,\,\init}}
\label{eqn-delta-m-defn}
\end{equation}
where $m_{\total,\,\init} = m_\MS(u{=}0, r_\text{max})$ is the total mass of
our initial slice, \ie{} the mass-function at the outer grid boundary
on the initial slice.
$\delta m$ is a dimensionless measure of how well our field variables
satisfy the Einstein equations; we must have
$|\delta m| \ll 1$ everywhere in the grid at all times for our results
to be trustworthy.

%
%
\subsection{The Bondi Mass and the News Function}

Taking the limit of the Misner-Sharp mass-function, $m(u,r)$, 
as $r \to \infty $ at constant $u$, we obtain the Bondi mass:
\begin{equation}\label{Bondimass}
M(u) = \lim\limits_{r \to \infty} m(u,r).
\end{equation}

In an isolated system outgoing waves can radiate physical energy to
future null infinity, $\mathscr I^+$.
Therefore, the Bondi mass is in general not conserved
in retarded time.  (In contrast, the ADM mass, which is defined
by taking the limit of $m(u,r)$ as $r \to \infty$ at constant $t$, that
is, on a spatial slice, {\em is\/} conserved in $t$.)

Moreover, one can show (see Ref. \cite{Wald84}) that there exists a flux $f$ such that
\begin{equation}\label{eq:deltaMBondi}
\Delta M = - \int\limits_{{\mathscr S} \times I} f
\end{equation}
where $\mathscr S$ is a cross-section of $\mathscr{I}^+$ with a $u = const$
null hypersurface and $I$ is a real interval.

We will derive a relation between the outgoing radiation flux, which
is described in terms of the scalar news function, and the change of the
Bondi mass in time. This relation is known as the \emph{Bondi mass-loss equation}.
(As always, $u$ denotes central time and not Bondi time.
The two time coordinates are related by equation \eqref{eq:def_BondiTime}.)

We insert the asymptotic series expansions of the fields
$\phi, \beta$ and $V$, given in equations \eqref{eq:phi-asm-exp}, 
\eqref{eq:beta-asm-exp} and \eqref{eq:V-asm-exp}, respectively, 
into the $(u,u)$-component of the Einstein equations
\begin{multline}
0 = r^2 E_{uu} = r^2 \left( G_{uu} - 8\pi T_{uu} \right) \\
  = \frac{2 V^2 \beta'}{r} + \Vr\left(\ebeta - V' - 2r\dot\beta\right)
    + \dot V - 8\pi\left(r\dot\phi\right)^2
    + 4\pi V r \left(\Vr (\phi')^2 - 2\dot\phi \phi' \right)
\end{multline}
and obtain
\begin{equation}
\begin{split}
0 &= \left(O(r^{-2})\right) \\
  &+ \left( e^{2H}\left(1-\frac{2M}{r}\right)\left(e^{2H} - e^{2H}
                       - 2r\dot H\right)
           + 2\dot H e^{2H}\left(r - 2M\right) + \bigO(r^{-1}) \right) \\
  &+ \left(e^{2H}\left(-2\dot M + \bigO(r^{-1})\right)\right) \\
  &- \left(8\pi \left(\dot c\right)^2 + \bigO(r^{-2})\right) \\
  &+ \left(O(r^2)\left(O(r^{-4}) - \bigO(r^{-3})\right)\right)
\end{split}
\end{equation}
As it turns out, the third and the fourth terms are crucial, all the other
terms are at least $O(r^{-1})$ or cancel out.
We find
\begin{equation}
0 = \lim_{r\to\infty} \left(r^2 E_{uu}\right)
  = -2 e^{2H(u)} \dot M(u) - 8\pi \left(\dot c(u)\right)^2
\end{equation}
Now, we introduce the scalar news function as
\begin{equation}
N(u_\CS) = e^{-2H(u_\CS)} \dot c(u_\CS).
\end{equation}
Combining the last two equations yields the Bondi mass-loss equation
\begin{equation}\label{eq:mass-loss-uC}
e^{-2H(u_\CS)} \frac{d M(u_\CS)}{d u_\CS} = -4\pi N^2(u_\CS).
\end{equation}
In Bondi time, it becomes, using relation \eqref{eq:def_BondiTime},
\begin{equation}\label{eq:mass-loss-uB}
\frac{d M(u_\BS)}{d u_\BS} = -4\pi N^2(u_\BS).
\end{equation}
From equation \eqref{eq:deltaMBondi} we find
\begin{equation}
\begin{split}
\Delta M &= -4\pi\int_{\Delta u_\CS} e^{2H(u_\CS)} N^2(u_\CS) du_\CS \\
         &= -4\pi \int_{\Delta u_\BS} N^2(u_\BS) du_\BS
         = - \int d\Omega \int_{\Delta u_\BS} du_\BS N^2(u_\BS)
\end{split}
\end{equation}
Thus, the square of the news function is just the flux that appears in the
integral \eqref{eq:deltaMBondi}:
\begin{equation}
N^2(u_\BS) \equiv f.
\end{equation}
The positivity of the flux-function $f$ entails that
\begin{equation}
\Delta M \le 0
\end{equation}
always.
``News'', that is radiation, can only decrease the Bondi mass contained
on a collection of null-slices that extend to $\mathscr{I}^+$.
But the Bondi mass can never increase to the future.

%
%
%
\subsection{The Bondi Mass as a Linkage Integral}

Let $\xi^a$ be a generator of an asymptotic time-translation and let
$\mathscr S_\alpha$ be a one-parameter family of spheres which approach the
cross-section $\mathscr S$ of $\mathscr{I}^+$.
Then the Bondi mass is defined as (see Refs. \cite{Geroch-Winicour-Linkages}
and \cite{Wald84})
\begin{equation}\label{eq:def-MBondi-linkage}
M_{Bondi} = - \lim_{\mathscr S_\alpha \to {\mathscr S}}
           \frac{1}{8\pi} \int_{\mathscr S_\alpha} \epsilon_{abcd} \nabla^c \xi^d,
\end{equation}
where the gauge condition $\nabla_a \xi^a = 0$ must be fulfilled
in a neighborhood of $\mathscr{I}^+$.
For the Bondi metric, equation \eqref{eq:def-Bondi-line-element},
the condition $\nabla_a \xi^a = 0$ gives
\begin{equation}
\xi^u_{,u} + \xi^r_{,r} + 2 \beta_{,u} \xi^u + 2 \beta_{,r} \xi^r = 0
\end{equation}
if $\xi^a = (\xi^u, \xi^r, 0, 0)$.
The generator of (asymptotic) time translations
\begin{equation}
\xi^a = (e^{-2\beta},0,0,0)
\end{equation}
satisfies the vanishing divergence condition as can be verified by making
use of the asymptotic expansions for $\beta$ and $V$ given in section
\ref{sec:asm-exp}.
Now the integrand of equation \eqref{eq:def-MBondi-linkage} is
\begin{equation}
 \epsilon_{abcd} \nabla^c \xi^d \, dx^a dx^b = \sqrt{-g} (\nabla^u\xi^r -
\nabla^r\xi^u) \, d\theta \, d\phi = -2M(u) \sin\theta \, d\theta \, d\phi
\end{equation}
so that one obtains
\begin{equation}
M_{Bondi} = M(u),
\end{equation}
where $M(u)$ is the $1/r$ coefficient in the asymptotic expansion
of $\frac{V(u,r)}{r}$.
%

\subsection{The Ricci Scalar Curvature}\label{ricci-scalar}

From the Einstein equations \eqref{eq:Einstein} for a minimally coupled massless
scalar field $\phi$ one finds by taking the trace 
\begin{equation}
  tr \, G_{ab} = -R = -8\pi \, tr \, T_{ab} = -8\pi \phi^{,a}\phi_{,a}
\end{equation}
the following expression for the Ricci scalar
\begin{equation}
R = 8\pi g^{ab} \phi_{,a} \phi_{,b}.
\end{equation}

For the Bondi metric \eqref{eq:def-Bondi-line-element} this yields
\begin{equation}
R = 8\pi e^{-2\beta} \phi' (-2\dot\phi + \frac{V}{r}\phi').
\end{equation}

\section{Double Null Equations}\label{sec:dn-equations}
To complement the DICE code which is based on Bondi coordinates, we introduce 
a first order formulation of the Einstein-scalar field system in double null 
coordinates which offers the advantage that it can penetrate apparent horizons.

We consider the Einstein massless scalar field equations 
$R_{ab} = 8\pi \phi_a \phi_b$
for the spherically symmetric metric introduced in section \ref{sec:double-null-coords}
\begin{equation}
ds^2 = -a^2(u,v) du dv + r^2(u,v) d\Omega^2
\end{equation}
in double-null coordinates $(u,v)$.
The tensor components are given in appendix \ref{app:tensors-double-null}.
We often use shorthands for partial derivatives:
\begin{equation}
  \dot{} = \pd{}{u} \qquad \prime{} = \pd{}{v}.
\end{equation}
We set $s:= \sqrt{4\pi} \phi$ to simplify the appearance of the matter field terms
and define additional variables in order to write the equations in first order form
following Ref. \cite{Stewart96}.

First, we define the following evolution variables
\begin{align}
&\text{\bf D1:} &  p &:=\dot s\\
&\text{\bf D2:} &  q &:= s'\\
&\text{\bf D3:} &  f &:= \dot r\\
&\text{\bf D4:} &  g &:= r'\\
&\text{\bf D5:} &  d &:= \frac{a'}{a}
\end{align}
The Einstein and matter field equations then become
\begin{align}
&\text{\bf E1:} & f' &= - \frac{fg + a^2/4}{r}\\
&\text{\bf E2:} & \dot d &= \frac{fg + a^2/4}{r^2} - pq\\
&\text{\bf C1:} & \dot f &= 2 \frac{\dot a}{a} f - r p^2\\
&\text{\bf C2:} & g' &= 2dg - r q^2\\
&\text{\bf S1:} & \dot q &= - \frac{fq + gp}{r}\\
&\text{\bf S2:} & p' &= - \frac{fq + gp}{r},
\end{align}
where we denote Einstein equations of wave-type by the letter \textbf{E},
constraint equations by the letter \textbf{C}, and the scalar field
wave equation by the letter \textbf{S}.

Boundary conditions at the center of spherical symmetry are dictated by
regularity and gauge choices as detailed in sections \ref{sec:dn-gauge},
\ref{sec:dn-regularity-BCs} and \ref{sec:dn-regularity-field-eqns}
and 
\begin{align}
g &= -f = a/2\\
p &= q\\
\pdrt{s} &= 0\\
\pdrt{a} &= 0
\end{align}

To evolve the coupled first order system, we choose a constrained evolution using equations \textbf{E2}, \textbf{S1} and \textbf{D2}, \textbf{D5}, \textbf{C2}, \textbf{D4}.
This leaves equations \textbf{C1}, \textbf{D1} and \textbf{D3} for checking the
numerical solution.

We need to specify the variables $a, r, s, p, f, g$ at the axis, 
while $q$ and $d$ are calculated from evolution equations. Obviously, $r=0$ at the axis.

\subsection{Regularity and Boundary Conditions}\label{sec:dn-regularity-BCs}

\subsubsection{Regular Variables at the Center}
We demand that our evolution variables (metric functions, scalar field) be well 
defined at the center of spherical symmetry.
See \cite{Alcubierre05} for a detailed analysis for Cauchy problems.
As we have seen in section \ref{sec:Bondi-regularity-conditions},
this is most easily obtained by demanding that the variables must not have a kink as the 
origin is crossed along a ray parametrized by linear distance $l$. Effectively, this 
forces all odd derivatives of the metric with respect to $r$ at constant $t$ to vanish 
at the origin.
\footnote{We may use flat space coordinates $(t,r)$ since we impose local flatness later.}

Therefore, $a^2(u,v)$ must be an even function of $r$, i.e.
\begin{equation}
a^2 = a_0^2 + \mathcal{O}(r^2) 
\end{equation}
so that we arrive at the  boundary condition
\begin{equation}
\pdrt{a}=0.
\end{equation}

Similarly, we have for the scalar field
\begin{equation}
\pdrt{s}=0.
\end{equation}
We can also investigate the behavior of  $g_{rr} = -\frac{4fg}{a^2}$
\begin{equation}
g_{rr} = g_{rr}^0 + \mathcal{O}(r^2)
\end{equation}

\subsubsection{Local Flatness at the Center}
As it turns out, one has to also impose local flatness to ensure a regular behavior 
in the evolution equations. Again, see \cite{Alcubierre05} for details.
Local flatness means that the spatial metric locally looks like the flat metric
\begin{equation}
dl^2 = dr^2 + r^2 d\Omega^2.
\end{equation}
In double-null coordinates $(u,v)$ the flat spacetime metric is given by
\begin{equation}
ds^2 = -a^2 du dv + r^2 d\Omega^2
\end{equation}
where $a = a(u,u) = const$ and $r$ is given by
\begin{equation}
r = \frac{a(v-u)}{2}.
\end{equation}
It then follows that $r' = - \dot r$ or, in the notation of section 
\ref{sec:dn-equations}),
\begin{equation}
g = -f = \frac{a}{2}
\end{equation}
and also that $f g + a^2/4 = 0$.

Due to spherical symmetry and local flatness, $r$ has to decrease as fast 
(with u) on an ingoing (radial) null ray as it has to increase (with v) on an outgoing 
null ray close to the origin.
Also, a null ray through the origin only appears to change direction in spherical 
coordinates, while it obviously passes straight through in Cartesian coordinates.

As we impose local flatness at the center, we may use flat space 
$(t,r)$ coordinates to express boundary conditions, such as $\pd{}{r}|_t = 0$. 
It is straight forward to implement such boundary conditions on an equispaced double 
null grid.

Issues of regularity of the evolution equations are discussed in section \ref{sec:dn-regularity-field-eqns}.
\subsection{Regularity of Evolution Equations}\label{sec:dn-regularity-field-eqns}
At $r=0$, the formally singular right hand sides of equations {\bf E1}, {\bf E2}, 
and {\bf S1}, {\bf S2} yield the following conditions:
\begin{gather}
f g + a^2/4 = 0\\
\partial_r \left( f g + a^2/4 \right) = 0\\
f q + g p = 0
\end{gather}
The last condition leads to the final boundary condition needed, which is simply
\begin{equation}
p = q.
\end{equation}

\subsection{Diagnostics for the Double-Null System}\label{sec:dn-diagnostics}
We calculate the density function $2m/r$,
\begin{equation}
  2m/r = 1 + \frac{4 f g}{a^2},
\end{equation}
where $m$ is the Misner-Sharp mass, defined in section
\ref{sec:Misner-Sharp-mass},
\begin{equation}
  m = 
      \frac{r}{2} \left[ 1 - g^{ab}  \nabla_a r \nabla_b r \right]
    = \frac{r}{2} \left[ 1 + \frac{4 f g}{a^2} \right].
\end{equation}
The quantity $\max_v 2m/r$ is used as an indicator for critical behavior and 
for the closeness of a slice to the formation of an apparent horizon.
To detect and measure an apparent horizon we search for a zero in 
the function $g = r_v$ (see \ref{sec:dn-null-expansions}).

The scalar curvature can be expressed in terms of the scalar field or the 
geometry:
\begin{equation}
\begin{split}
R &= -8 p q / a^2\\
   & = \frac{2 a^4 + 8 r^2 \left( a {\dot a}'  -\dot a a' \right) + 8 a^2 \left(2 r {\dot r}' + \dot r r' \right)}{a^4 r^2}.
\end{split}
\end{equation}
The scalar field energy density is given by
\begin{equation}
  \rho = \frac{p^2 + q^2}{8\pi a^2},  
\end{equation}

It is instructive to introduce observers at constant radius $r(u,v)$. 
We would then like to compute the proper time $t(u,r)$ along each of these 
worldlines of constant  $r$. 
Following \cite{Stewart96} we invert $r=r(u,v)$ to obtain $v=v(u,r)$. 
At $r=const$, it follows that $dv=\pd{v}{u}|_r du$. 
The line element \eqref{eq:dn-line-element} then becomes
\begin{equation}
ds^2\big|_{r=const} = -a^2(u,v) \pd{v}{u}\Big|_r du^2 + r^2(u,v) d\Omega^2.
\end{equation}
Then, we have
\begin{equation}\label{eq:dn-proper-time}
t(u,r=const) = \int_0^u \sqrt{-g_{\bar u \bar u}} d\bar u = \int_0^u a(\bar u, v(\bar u, r)) \sqrt{\pd{v}{u}\Big|_r} d\bar u
\end{equation}
At $r=0$, local flatness implies $v(u,r) = \frac{2r}{a(\bar u, \bar u)} + u$, and thus
\begin{equation}
t(u,0) =  \int_0^u a(\bar u, \bar u)  d\bar u
\end{equation}

\forceemptypage


\chapter{Numerical Algorithms}
\label{ch:numerical-algorithms}

\section{The DICE-code}\label{sec:algorithm}

The compactified code used in \cite{Puerrer-Husa-PCA} is based on the ``DICE''
(Diamond Integral Characteristic Evolution) code, which has been documented in
\cite{Husa2000b} (there particular emphasis is given to detailed
convergence tests).

\subsection{Our Overall Computational Scheme}\label{sec:compuScheme}
While this section discusses the uncompactified Bondi evolution algorithm,
it can be applied to the compactified system, by a few simple replacements.
Most notably, the metric function $V$ has been replaced by the Misner-Sharp
mass $m$. See section \ref{sec:compactification} for further details.

%
%
\subsubsection{Summary}\label{sec:Bondi-evolution-scheme}

We first construct initial data:
\begin{enumerate}
\item   Choose the field $\psi(u_0,r)$ on a null slice $u = u_0$.
\item   Also choose the positions, $r_i(u_0)$, of our grid points on this
        same initial slice.
\item   Radially integrate the hypersurface geometry
        equations~\eqref{eq:beta-prime-hypersurface}
        and~\eqref{eq:V-prime-hypersurface}
        outwards from the origin to compute $\beta$ and $V/r$ at
        all the grid points on the initial slice.
\end{enumerate}
Our integration scheme starts by first integrating the geodesics back in time
one step $\Delta u$, since our geodesic integrator needs two slices to work.
Now suppose we know all the gravitation and matter fields for time levels
$\leq k$.  To determine them for time level $k+1$, we use the following
algorithm (we use the usual notation where superscripts denote time levels):
\begin{enumerate}
\item   Determine the Taylor expansion of $\psi^k(r)$ near the origin
        (\cf{}~section~\ref{sec:Bondi-Taylor-expansions}).
\item   For each grid point in some small (typically 3~grid points)
        \defn{Taylor series region} starting at the origin and working
        outwards,
        \begin{enumerate}
        \item   Integrate the geodesic
                equation~\eqref{eqn-ingoing-null-geodesic} ahead
                one time level to determine the position of this
                grid point at time level $k+1$.
        \item   Using the $\psi^k$ Taylor coefficients and the Einstein
                equations (\cf{}~section~\ref{sec:Bondi-Taylor-expansions}),
                compute first $\psi^{k+1}$, then $\beta^{k+1}$ and
                $(V/r)^{k+1}$ at this grid point
        \end{enumerate}
\item   Now for each grid point in the rest of the grid, starting
        just outside the Taylor series region and working outwards,
        \begin{enumerate}
        \item   Integrate the ingoing null-geodesic
                equation~\eqref{eq:ingoing-null-geodesics} ahead
                one time level to determine the position of this
                grid point at time level $k+1$.
        \item   Use the Winicour diamond scheme~\eqref{eq:diamond-scheme}
                to compute $\psi^{k+1}$ at this grid point.
        \item   Radially integrate the hypersurface geometry
                equations~\eqref{eq:beta-prime-hypersurface}
                and~\eqref{eq:V-prime-hypersurface}
                outwards one spatial grid point to compute $\beta^{k+1}$
                and $(V/r)^{k+1}$ at this grid point.
        \end{enumerate}
\end{enumerate}

%
%
\subsubsection{Discretization of the Evolution System}
\label{sec:Bondi-discretization-evolution-system}

We evaluate the integral in equation \eqref{eq:diamond-scheme}
by treating the integrand as constant over the (small) null parallelogram $\Sigma$,
as shown in figure \ref{fig-diamond}.
Thus we can compute the integrand at the center of $\Sigma$ which in turn equals 
its average between the points $W$ and $E$ to second order accuracy:  
\begin{equation}
  \text{Integrand at center} 
  = \frac{\left[ (V/r)' \psi/r \right]_W + \left[ (V/r)' \psi/r \right]_E }{2} + \bigO(h^2),
\end{equation}
where $h = \frac{1}{2}\sqrt{\left(\Delta u\right)^2 + \left(\Delta r\right)^2}$.
Then we multiply the integrand by the area of the null parallelogram $\Sigma$,
which can be approximated by
\begin{equation}
  A_\Sigma = \Delta u \frac{(r_E - r_S) + (r_N - r_W)}{2}
\end{equation}

As will be discussed in section \ref{app:errors-NSWE}, this algorithm is globally 
2nd~order accurate; for the special case of a flat background
($\left(\Vr\right)_{,r}=0$) the algorithm is exact.

We use the explicit trapezoidal method (see equation \eqref{eq:explicit-trapezoidal} 
in appendix \ref{app:numerics}) to discretize the hypersurface  and ingoing null 
geodesic equations.
In principle the discretization is straightforward, with the exception of the
null geodesic equation. The corrector requires evaluating the right hand side at
the $n+1$ time-level, but this has not been computed yet at the time we do the
geodesic integration. To solve this problem, we radially extrapolate the needed
$V/r$ value from $V/r$ and $(V/r)'$ one spatial gridpoint inwards at the same
$n+1$ time-level. At the origin, we know $V/r = 1$ thanks to local flatness, anyway.

%
%
\subsubsection{Freely Falling Gridpoints}
\label{sect-freely-falling-grid-points}

The original Winicour algorithm uses a fixed grid (\ie{} the radial
positions of the grid points
do not move in the course of evolution. As the coordinate velocity of
light is in general not constant in a curved space-time, the vertices
$N$, $E$, $S$ and $W$ of the null parallelogram $\Sigma$ cannot be chosen
to lie exactly on the grid.  Therefore one has to use interpolation to
compute the matter field and metric functions at the vertices.

In the DICE code we used a different approach (which is, in fact, borrowed
from a different characteristic evolution scheme due to Piran, Goldwirth
and Garfinkle \cite{Garfinkle95} as described in section \ref{sec:PGG}):  
Our gridpoints free-fall along ingoing
null geodesics.  To compute the locations of the gridpoints on a new slice
we integrate the ingoing null geodesic equation
\begin{equation}
\left. \frac{dr}{du} \right|_v
        = - \frac{1}{2}\frac{V}{r}
                                \quad \hbox{.}
                                        \label{eqn-ingoing-null-geodesic}
\end{equation}

With this approach, the vertices of the null parallelogram $\Sigma$
in the diamond algorithm (\cf{} section~\ref{sec:diamond-algorithm})
{\em are} gridpoints, so we don't have to interpolate there.

As well as avoiding interpolation, this scheme has the additional
advantage that it provides a sort of adaptive grid refinement for free,
since, in regions of strong gravity, the null geodesics are being focussed
and thus the density of gridpoints is increased.

Since our grid points move along ingoing null geodesics,
each grid point eventually reaches the origin.
At this point it disappears, \ie{} we remove it from the grid.
%
%
\subsubsection{Stability of the Winicour algorithm and adaptive choice of timesteps}
\label{sect-stability}

Assuming that the gridpoints follow ingoing null geodesics, we want to 
impose a restriction on the ``drift'' of individual gridpoints from one slice 
to the next. It follows from the null geodesic equation 
\eqref{eqn-ingoing-null-geodesic} that, if we want the change in r
from a slice $u^n$ to $u^{n+1}$ to be smaller than a factor $D$
('drift\_limit') times $\Delta r$, we have to demand
\begin{equation}
\Delta u < D \frac{2r}{V} \Delta r
\end{equation}
for all gridpoints.
(The minus sign in the null geodesic equation is absorbed by the $\Delta r$
which is a directed
quantity.)
The code assumes that $D <= 1$, \ie{} we drop at most one grid point per time step.

%
%

\subsection{Taylor Expansions}\label{sec:Bondi-Taylor-expansions}

To start off the integration of the evolution system on a new null slice
at retarded time $u = u_0 + \Delta u$ and time-level $k+1$, 
we Taylor expand the scalar field on previous slices, and employ
the matter field wave equation to propagate the information to the new slice.
\subsubsection{Calculation of $\psi$}
First, we compute a Taylor expansion of the rescaled matter field $\psi$ on the 
slice at $u = u_0$ and time-level $k$ in a region that encompasses the origin of 
spherical symmetry, $r=0$:
\begin{equation}
  \psi^k(u_0,r) = t_0 r + t_1 r^2 + \bigO(r^3).  
\end{equation}
We have omitted the constant term due to the regularity of the matter field 
$\phi$ and the definition of the rescaled field $\psi = \phi r $.
For the sake of simplicity, we determine the coefficients $t_0$ and $t_1$ by 
fitting a parabola through two sample points (and the origin).
A fancier approach used in the code is to use a general linear least squares 
fit over a number of gridpoints adjacent to the origin, say 5 points. This 
way of handling the origin has in practice proven valuable in smoothing out
discontinuities in the fields near the center.

Next, we establish a relation between the field values on the new slice, 
at time-level $k+1$, and the field on the old slice, at time-level $k$.
To achieve this we make use of the
field equation
\begin{equation}
\square^\two\psi + \left(\frac{V}{r}\right)_{,r}\frac{e^{-2\beta}\psi}{r} = 0,
\end{equation}
where
\begin{equation}
\square^\two\psi = e^{-2\beta}\left[ \left(\frac{V}{r}\right)_{,r}
\partial_r - 2\partial_u\partial_r + \frac{V}{r}\partial_{rr}\right]\psi.
\end{equation}
Solving for the $\partial_u$-term we arrive at
\begin{equation}
\partial_u \psi^\prime = \frac{1}{2}\left[ \left(\frac{V}{r}\right)^\prime
\left(\frac{\psi}{r}+\psi^\prime\right) +
\frac{V}{r}\psi^{\prime\prime}\right].
\end{equation}
Replacing the u derivative by a finite difference approximation
\begin{equation}
  \left( \partial_u \psi^\prime \right)^k = \frac{\psi^{\prime k+1} 
  - \psi^{\prime k}}{\Delta u} + \bigO(\Delta u)  
\end{equation}
yields
\begin{equation}\label{eq:Taylor-psip-relation}
\psi^{\prime k+1}(r) = \psi^{\prime k} +\frac{\Delta u}{2}
\left[ \left(\frac{V^k}{r}\right)^\prime \left(\frac{\psi^k}{r}+\psi^{\prime k}\right) +
\frac{V^k}{r}\psi^{\prime\prime k} \right].
\end{equation}
Using the behavior of $V$ near the origin
\begin{equation}
\frac{V}{r}=1 + \bigO(r^2) \qquad \text{and} \qquad \left(\frac{V}{r}\right)^\prime
= \bigO(r),
\end{equation}
and integrating the above relation, equation \eqref{eq:Taylor-psip-relation}, 
with respect to $r$ we recover
\begin{equation}\label{eq:Taylor-psi-relation}
\psi^{k+1}(r) = \psi^k(r) + \frac{\Delta u}{2}
             \int_0^r d\bar r \, \left[\psi^{\prime\prime k}(\bar r) + \bigO(\bar r) \right].
\end{equation}
Finally, we insert the Taylor expansion for $\psi$ into equation 
\eqref{eq:Taylor-psi-relation} and find the desired result
\begin{equation}
\psi^{k+1}(r) = t_0 r + t_1 r^2 + t_1 \Delta u \, r  + \bigO(\Delta^3),
\end{equation}
where $\Delta$ denotes a small quantity (either $r$ or $\Delta u$).

The calculation works the same way for higher orders.
To third-order accuracy we find
\begin{equation}\label{eq:psi-Taylor}
\psi^{k+1}(r) = t_0 r + t_1 r^2 + t_1 r \Delta u + \frac{3}{2} t_2 r^2 \Delta u
                + \frac{3}{4}t_2 r (\Delta u)^2 + t_2 r^3 + \bigO(\Delta^4).
\end{equation}

%
%
\subsubsection{Calculation of the Metric Functions}
We can determine the metric functions on the new slice by inserting the expansion
of the rescaled field, equation \eqref{eq:psi-Taylor}, into the hypersurface equations
\begin{align}
  \beta_{,r} &= 2\pi r \left( \phi_{,r} \right)^2 \\
  V_{,r}     &= e^{2\beta}. 
\end{align}
We first determine
\begin{equation}
\phi(r) =\frac{\psi}{r} = t_0 + t_1 r + t_1 \Delta u + \bigO(\Delta^2)
\end{equation}
and thus
\begin{equation}
\beta^\prime(r) = 2\pi r \left( t_1 \right)^2 + \bigO(\Delta^2)
\end{equation}
and 
\begin{equation}
\beta(r) = \pi \left( t_1 \right)^2 r^2 + \bigO(\Delta^3).
\end{equation}
Along the same lines we have
\begin{equation}
V^\prime(r) = 1 + 2\beta + \bigO(\beta^2) = 1 + 2\pi \left( t_1\right)^2 r^2 + \bigO(\Delta^3)
\end{equation}
and eventually
\begin{align}
\frac{V}{r}(r) &= 1 + \frac{2}{3}\pi \left(t_1\right)^2 r^2 + \bigO(\Delta^3)\\
\left(\frac{V}{r}\right)^\prime(r) &= \frac{4}{3}\pi\left(t_1\right)^2 r + \bigO(\Delta^2).
\end{align}
The third-order accurate expansions are
\begin{equation}
\beta^{k+1}(r) = 8\pi \left(\frac{1}{8} t_1^2 r^2
                      + \frac{3}{8}t_1 t_2 r^2 du + \frac{1}{3} t_1 t_2 r^3 \right)
                      + \bigO(\Delta^4),
\end{equation}
and
\begin{equation}
V^{k+1}(r) = r - \frac{\Lambda}{3} r^3 + 8\pi
                    \left(\frac{1}{12} t_1^2 r^3 \right) + \bigO(\Delta^4).
\end{equation}

The Ricci Scalar curvature (see section \ref{ricci-scalar}) has the following
expansion at the center:
\begin{equation}
R(u,r=0) = 16 \pi t_1^2.
\end{equation}

\subsubsection{Taylor Series for the Compactified Code}
\label{sec:Taylor-expansions-compactified}

For small $r$ the functional behavior of the x-coordinate is similar to r, so that 
we can reuse the Taylor series expansions of the fields as functions of $(u,r)$ 
as given in section \ref{sec:Bondi-Taylor-expansions} and transform them to the 
$x$-coordinate by applying the expansion
\begin{equation}
r = \frac{x}{1-x} = x + x^2 + x^3 + \bigO(x^4).
\end{equation}

To third order accuracy we obtain ($\Delta$ denotes a small quantity $\Delta u$ or $x$)
\begin{multline}
\psi^{k+1} (u,x) = \left(t_0 + t_1 \Delta u + \frac{3}{4}t_2\, \Delta u^2\right) x
                 + \left(t_0 + t_1 + \left(t_1 + \frac{3}{2}t_2\right)\, \Delta u\right) x^2 \\
                 + \left(t_0 + 2t_1 + t_2\right) x^3 + \bigO(\Delta^4),
\end{multline}

\begin{equation}
\beta^{k+1} (u,x) =  \left(\pi t_1^2 + 3\pi t_1 t_2\, \Delta u\right)x^2
                        + \left(2\pi t_1^2 + \frac{8\pi}{3}t_1 t_2\right)x^3 +O(\Delta^4),
\end{equation}
We include one more order for $m$, since it appears as $m/x$ in the field equations.
\begin{multline}
  m^{k+1} (u,x) = \left( \frac{2\pi}{3} t_1^2 + 2\pi t_1 t_2\, \Delta u
                            - \frac{3\pi}{4} t_2^2\, \Delta u^2 \right) x^3 \\
               + \left( \frac{1}{2} + 2\pi t_1^2 + 2\pi t_1 t_2 - \pi t_2^2\, \Delta u
                      - \frac{9\pi}{4} t_2^2\, \Delta u^2 \right) x^4 + \bigO(\Delta^5),
\end{multline}
and
\begin{multline}
  \left(\frac{\partial m}{\partial x}\right)^{k+1} (u,x)
            = \left(2\pi t_1^2 + 6\pi t_1 t_2\, \Delta u
                    - \frac{9\pi}{4} t_2^2\, \Delta u^2\right) x^2\\
              + \left(2 + 8\pi t_1^2 + 8\pi t_1 t_2 - 4\pi t_2^2\, \Delta u
                      - 9\pi t_2^2\, \Delta u^2 \right) x^3 + \bigO(\Delta^4).
\end{multline}

\subsection{Mesh Refinement}\label{sec:dice-amr}
\begin{figure}
  \centering
  \includegraphics[width=.9\textwidth]{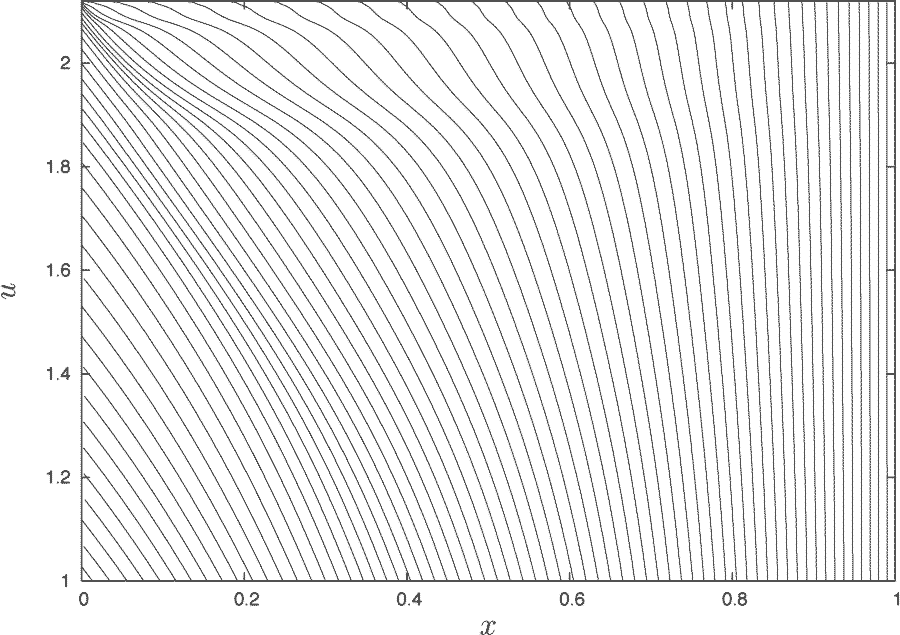}

  \caption{\label{fig:nullgeo}This figure shows the focussing of ingoing 
  null-geodesics by gravity in the late stages of a slightly supercritical evolution.
  The discretely self-similar dynamics causes the density of the geodesics to
  increase in a periodic manner.
  }
\end{figure}

Hamad\'e and Stewart \cite{Stewart96} have implemented full Berger-Oliger 
mesh refinement in double-null coordinates (without compactification) to
achieve sufficient resolution to study critical collapse. Garfinkle 
\cite{Garfinkle95} has shown, that this is not really necessary -- here
we follow his approach to increase resolution:
Most importantly, we choose our gridpoints to follow ingoing radial
nullgeodesics. This leads to a rapid loss of gridpoints in the early
phase of collapse, but to an accumulation of gridpoints in the region of strong
curvature for the late stages of critical collapse (see figure \ref{fig:nullgeo}). 
Furthermore, when half of the gridpoints have reached the origin we refine the grid
and thus obtain a very simple but effective form of mesh refinement
which is a crucial ingredient in the calculation of critical collapse spacetimes.
In previous work \cite{Husa2000b} we have also tuned our outermost gridpoint to
be located just outside of the past self-similarity horizon (SSH), 
the backwards lightcone of the accumulation point of the self-similar solution.
Here we choose to go out all the way to null infinity.
The most effective approach in this situation would be to just refine the region inside
the SSH when half of the gridpoints in this region have reached the origin.
While this is straightforward to implement, we found the penalty on the 
resolution that the original condition (the loss of half of \emph{all} gridpoints) 
causes to be acceptable for the results presented here.

\subsection{Accuracy and Convergence}\label{sec:dice-accuracy-convergence}
\begin{figure}
  \centering
  \includegraphics[width=.9\textwidth]{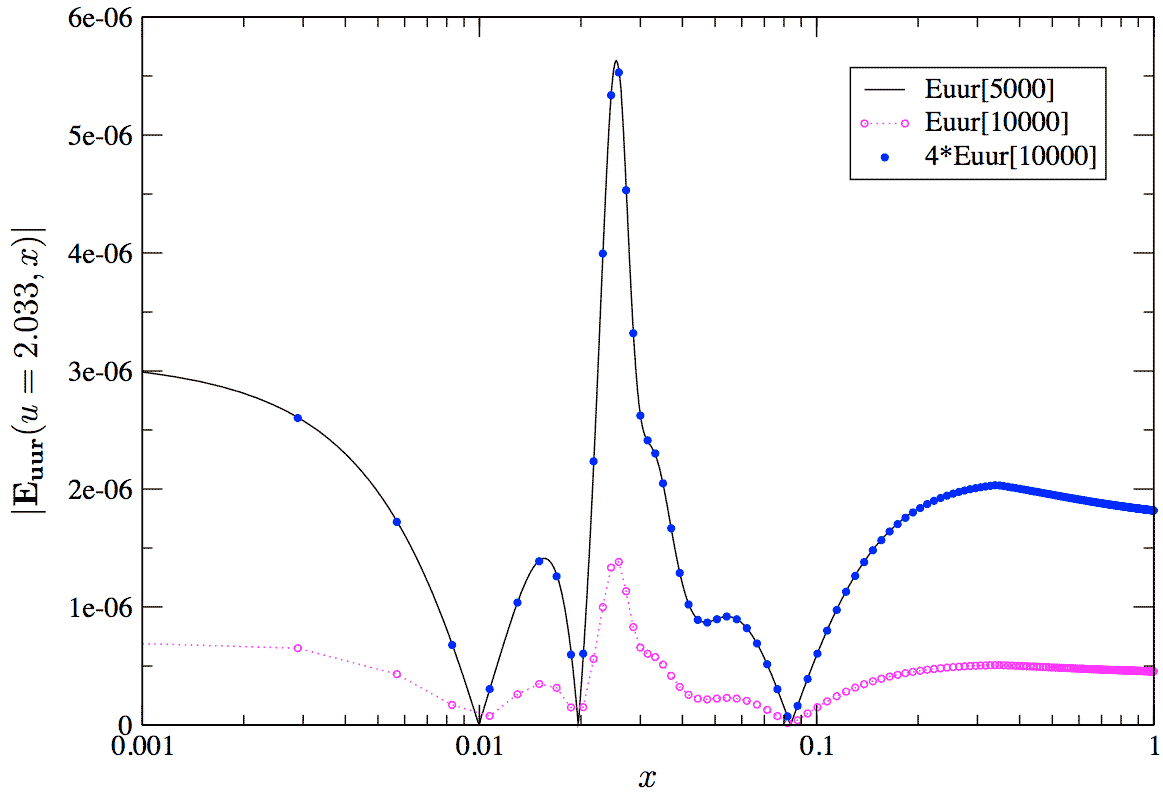}

  \caption{\label{fig:Euur-convergence}The convergence of the error diagnostic 
  $\mathbf{E_{uur}}$ with increasing grid resolution for two near-critical evolutions. 
  Evolution (1) uses 5000 gridpoints and $p=p^*[5000] + 10^{-10}$
  and evolution (2) uses 10000 gridpoints and $p=p^*[10000] + 10^{-10}$. 
  The data displayed as dots and circles have been sampled.}
\end{figure}

The code is globally second order accurate. 
Figure \ref{fig:Euur-convergence} shows a convergence test for near-critical
evolutions. 
We want to emphasize that the critical
value $p^*$ of the initial data parameter depends on the grid resolution.
This fact is essential when doing convergence tests for near critical
evolutions, as has been discussed in our previous paper \cite{Husa2000b}.

To monitor the accuracy of the code during runs we use components of the 
Einstein equations which are automatically satisfied if the evolution 
equations hold, such as the following linear combination of the $(u,u)$ 
and $(u,r)$ components 
\begin{equation}
- E_{uur} \equiv r^2 \Bigl( G_{uu} - 8 \pi T_{uu} \Bigr) 
                        - r^2 (V/r) \Bigl( G_{ur} - 8 \pi T_{ur} \Bigr).
\end{equation}
This can be rewritten in the following form
\begin{equation}
E_{uur} = 2 e^{2\beta} \dot m + 8\pi \left[ \dot\psi^2 - e^{2\beta}\left(\frac{1}{1-x} - \frac{2m}{x}\right)
                                                         (1-x)^2 \dot\psi
                                                         \left(\psi_{,x} (1-x) - \frac{\psi}{x} \right)
                                     \right],
\end{equation}
where $\dot f = \pd{f}{u}|_x$.
Since this expression is a linear combination of tensor components,
we use a suitably normalized quantity ${\mathbf E_{uur}} = \frac{E_{uur}}{1 + E_{|uur|}}$, 
where $E_{|uur|}$ is the sum of the running maxima ($\max_{i\le j} f_i$ over a $u=const$-slice) 
of the absolute values of the individual terms of $E_{uur}$.
We must have $\mathbf{E_{uur}} \ll 1$ for our finite difference solution to be a good
approximation to the continuum solution.
A convergence test for this quantity is shown in figure \ref{fig:Euur-convergence}.

\begin{figure}
  \centering
  \includegraphics[width=12cm]{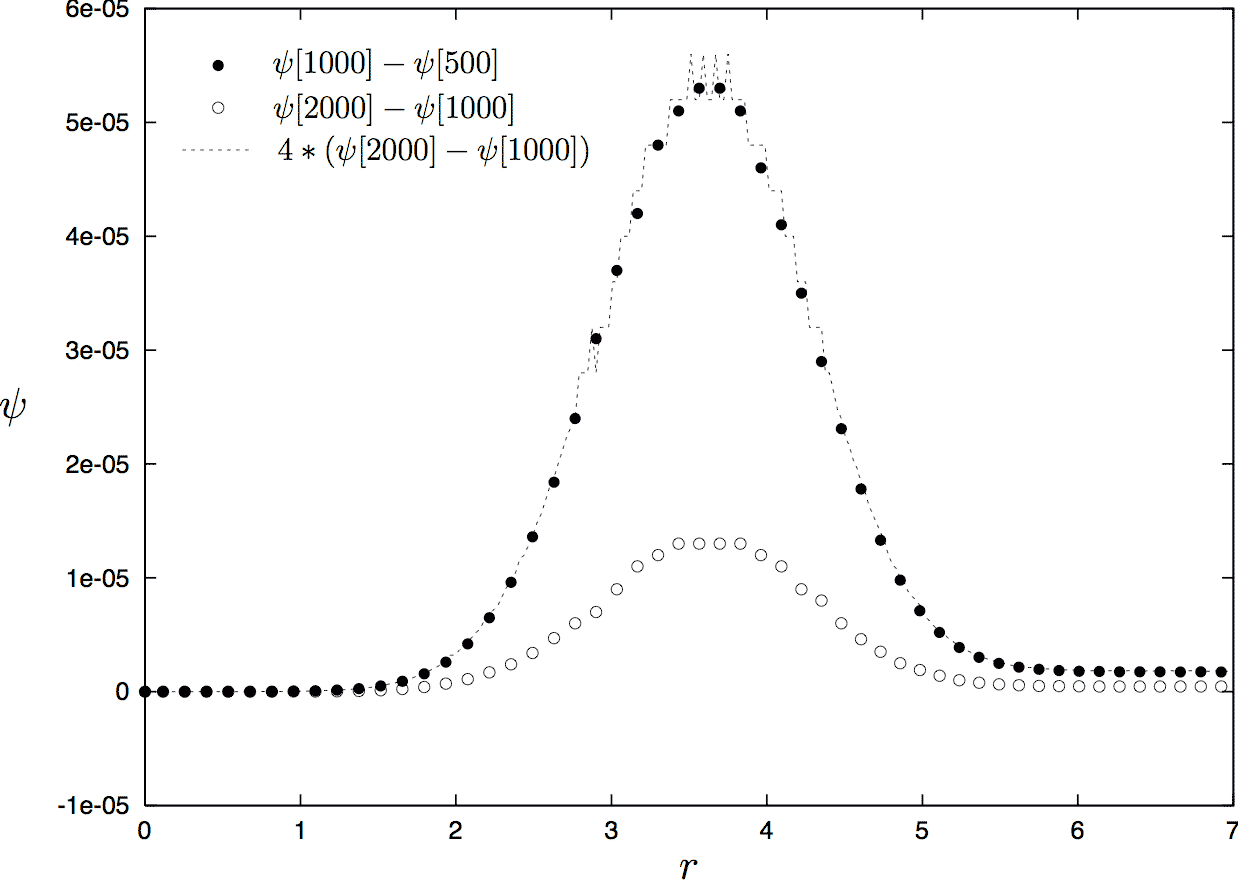}

  \caption{\label{fig:psi-convergence}
  A 3-level convergence test for $\psi(r)$}.
  \label{fig-psi-convergence}
\end{figure}

\begin{figure}
  \centering
  \includegraphics[width=12cm]{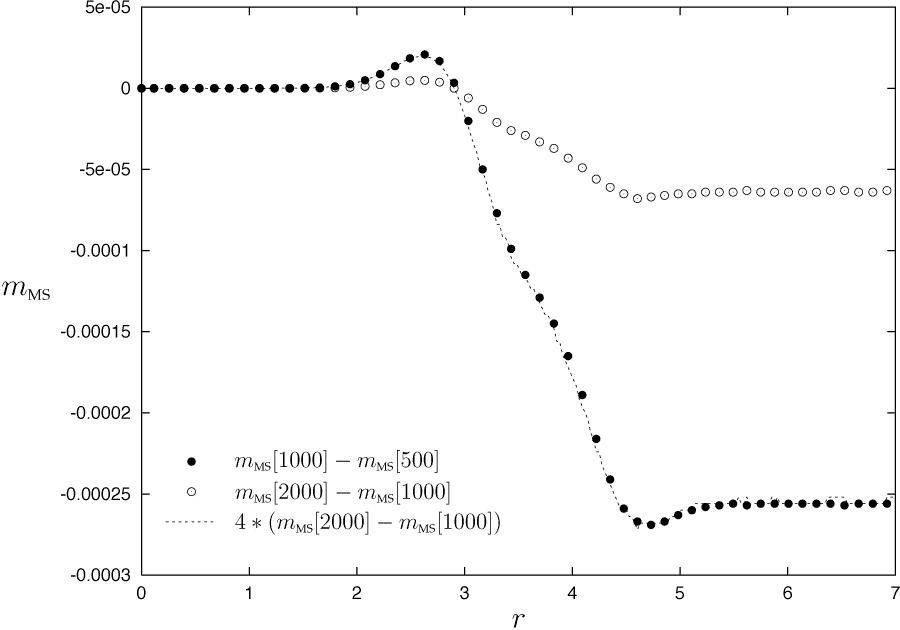}

  \caption{\label{fig:m-ms-convergence}
  A 3-level convergence test for $m_\MS(r)$.
  }\label{fig-m_MS-convergence}
\end{figure}

\begin{figure}
  \centering
  \includegraphics[width=12cm]{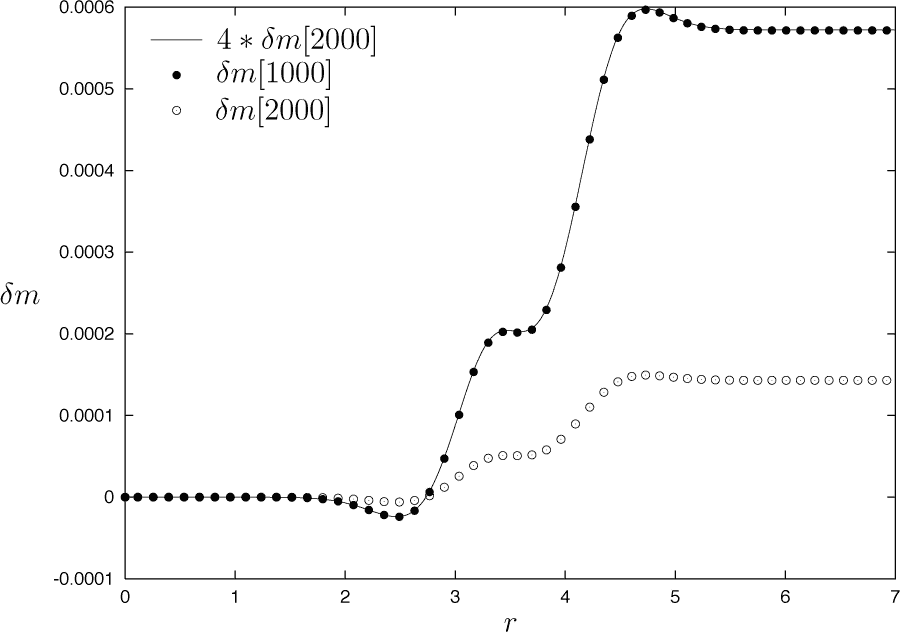}

  \caption{\label{fig:delta-m-convergence}
  A 2-level convergence test for $\delta m$.
  }\label{fig-delta_m-convergence}
\end{figure}

We also display numerical convergence tests (see appendix \ref{app:convergence-methodology})
for $\psi$ and $m$ (3-level) and $\delta m$ (2-level) in figures \ref{fig:psi-convergence},
\ref{fig:m-ms-convergence}, and \ref{fig:delta-m-convergence} which show that
our code computes these gridfunctions with the accuracy expected from the theoretical 
considerations of appendix \ref{app:errors-NSWE}.


\clearpage

\section{The Double-Null Code}\label{sec:dn-code}
The numerical solution scheme for the double-null code follows the work of
Hamad\'e and Stewart \cite{Stewart96} and further improvements made by Harada 
and Carr \cite{HaradaCarr} which makes the scheme second order accurate.

We choose to refer to the evolution equations by the mnemonic abbreviations 
\textbf{Dx}, \textbf{Ex}, \textbf{Cx} and  \textbf{Sx}, (with \textbf{x}, a number),
given in section \ref{sec:dn-equations}, rather than usual equation numbers.

\subsection{Regularization}\label{sec:dn-eq-regularization}

\begin{figure}[htbp]
  \centering
  \includegraphics[width=0.7\textwidth]{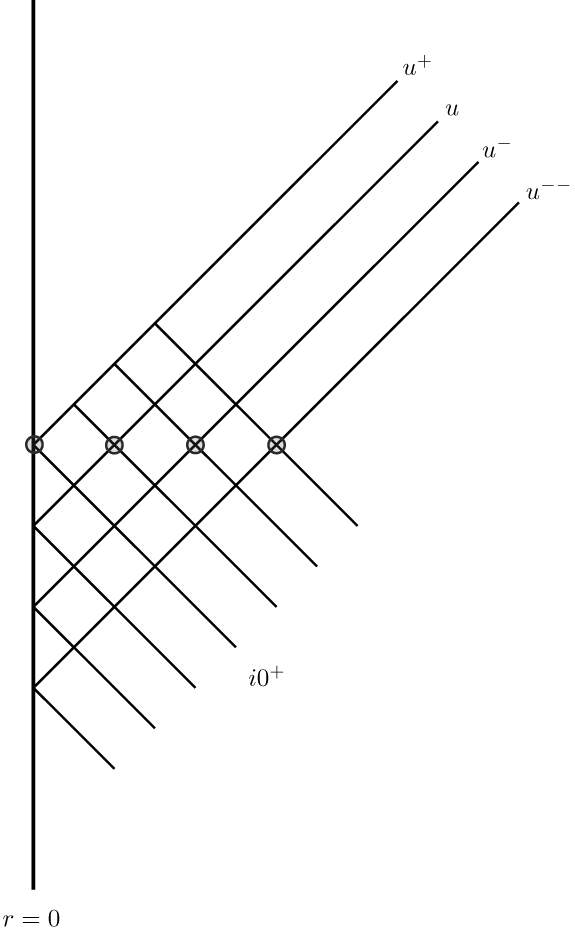}
  
  \caption{
  This figure shows the double-null grid around the origin. We have indicated 
  the new slice being computed by $u^+$ (with time index $+$) and the v-index
  that hits the origin on that slice, by $i0^+$.
  Four gridpoints which lie at constant $t$ (flat space time coordinate)
  are indicated by filled circles. Using these gridpoints, one can approximate
  derivatives of type $\pd{}{r}\vert_t$.
  }\label{fig:dn-origin-grid}
\end{figure}

At the center of spherical symmetry we need to regularize equations 
{\bf E1, E2, S1} and {\bf S2} using l'Hospital's rule.
To simplify matters we use derivatives $\pd{}{r}\vert_t$ evaluated at $r=0$. 
Derivatives of the type $\pd{}{r}\vert_t$ are straightforward to write down on an
equidistant grid in $u$ and $v$. See figure \ref{fig:dn-origin-grid} for an illustration.
However, we need finite difference approximations for unevenly spaced grids, 
as our grid is only evenly spaced in $u$ and $v$, but not in $r$. 
Such finite difference approximations can be obtained by determining a low order
polynomial through the required number of points and using symbolic differentiation.
This is most easily done in a computer algebra system, such as Mathematica \cite{mma52}:

For example, a second order accurate approximation of the first derivative of
a function $f$, given the function values $(f_0, f_1, f_2)$ at three points 
$(r_0, r_1, r_2)$, evaluated at the point $r_0$, given by
\begin{equation}\label{eq:D1G2}
  f'(r_0) = \frac{f(\text{r2}) (\text{r0}-\text{r1})^2+(2
     \text{r0}-\text{r1}-\text{r2}) (\text{r1}-\text{r2})
     f(\text{r0})-(\text{r0}-\text{r2})^2
     f(\text{r1})}{(\text{r0}-\text{r1})
     (\text{r0}-\text{r2}) (\text{r1}-\text{r2})},
\end{equation}
can be obtained by the compact expression
\begin{verbatim}
  rgrid = {r0, r1, r2};
  FullSimplify[NDSolve`FiniteDifferenceDerivative[ \
  Derivative[1], rgrid, Map[f, rgrid], \
  DifferenceOrder -> 2]][[1]]
\end{verbatim}
Similar expressions can be derived for higher order accuracy and second derivatives.

Depending on the number of previous slices that we have already computed, we employ
different general finite difference approximations for these radial derivatives at 
constant $t$. As it is not always possible to use second order accurate approximations,
we have to resort to lower orders in the first two timesteps of an evolution.
This is not a big problem, though, if we stick to initial data of compact support,
such as Gaussian data, where the field initially almost vanishes at the origin.

To regularize the \textbf{E2} equation $\dot d = \frac{fg + a^2/4}{r^2} - pq$ 
according to l'Hospital's rule, we take the second derivative of the numerator
of the singular part of the right hand side, $fg+a^2/4$, using a second order accurate
general finite difference stencil that requires four points.
All other singular equations require only the computation of first derivatives
as the denominator of the singular right hand sides are $r$.
To regularize equation \textbf{S1} we compute an approximation to 
$-\pd{}{r}\vert_t \left( fg + a^2/4  \right)$ using formula \eqref{eq:D1G2}, given above.

The right hand side of equation \textbf{S2} is identical to that of equation 
\textbf{S1}, so that we can reuse the right hand side approximation computed 
for \textbf{S1}.
As it turns out, the right hand side of equation {\bf E1} is in fact zero analytically.
This is due to regularity of equation {\bf E2} which, in addition to
$fg + a^2/2 = 0$, also forces 
\begin{equation}
  \pdrt{(fg + a^2/2)} = 0.  
\end{equation}

\subsection{Initial Data}
We may freely specify unconstrained data for the scalar field $s(u=const,v)$ on an 
initial nullcone, usually chosen to lie at $u=0$. 
For the sake of simplicity, we restrict considerations to \emph{Gaussian initial data}:
\begin{equation}
s(0,v) = A \exp\left[-\left(\frac{v - v_c}{\sigma}\right)^2\right].
\end{equation}
Together with the specification of the gauge for the null coordinate $v$, such as
\begin{equation}\label{eq:dn-v-gauge}
  r(0,v) = v/2,   
\end{equation}
we can construct initial data for all remaining evolution variables,
$a$,$d$,$f$,$g$,$p$, and $q$.
\footnote{Note that Hamad\'e and Stewart \cite{Stewart96} fix the gauge by choosing 
$d(0,v) = 0, instead.$}
It is then trivial to reconstruct $q(0,v)$ and $g(0,v)$ by analytic differentiation
and also easy to compute $d(0,v)$:
\begin{align}
&\text{\bf D2:} & q(0,v) &= s'(0,v) = - 2 \frac{v-v_c}{\sigma^2} s(0,v)\\
&\text{\bf D4:} & g(0,v) &= r' = 1/2\\
&\text{\bf C2:} & d(0,v) &= \frac{r q^2}{2g}.
\end{align}

The remaining quantities $p(0,v), f(0,v)$ and $a(0,v)$ have to be reconstructed by
numerical integration:
\begin{align}
&\text{\bf D5:} & (\ln a(0,v))' &=  \frac{a'}{a} = d\\
&\text{\bf E1:} & f'(0,v) &= - \frac{fg+a^2/4}{r}\\
&\text{\bf S2:} & p'(0,v) &= - \frac{fq+gp}{r}.
\end{align}
We use Simpson's rule \eqref{eq:Simpsons-rule-local} (and the Trapezoidal rule
\eqref{eq:trapezoidal-rule} for the first gridpoint outwards from the origin) 
to perform the integration.

We also apply l'Hospital's rule for formally singular right-hand sides at the center
as described in section \ref{sec:dn-eq-regularization}.
Only the equation for p is formally singular at the origin.
Since we do not know any previous slices, we cannot use $\frac{\partial}{\partial r}|_t$
derivatives and instead use $\frac{\partial}{\partial v}|_u$ and the gauge choice
\eqref{eq:dn-v-gauge}:
\begin{equation}
\begin{split}
p'|_0 &= - \frac{fq+gp}{r}\Big\vert_0\\
&= - 2 \frac{\partial}{\partial v}\Big\vert_{u,0} \left( fq + gp \right)\\
&= - 2 \frac{\partial}{\partial v}\Big\vert_{u,0} \left( fq + p/2 \right)\\
&= - 2 \left( fq \right)'|_0  - g p'|_0
\end{split}
\end{equation}
Subscripts of a gridfunction $f_i$ denote $v$ indices on the numerical grid, 
starting at $i=0$ at the origin.
Using this expression we can derive the following approximation for $p_1$ by applying 
the Trapezoidal rule
\begin{equation}
p_1 = \frac{p_0 + \frac{\Delta v}{2}\left( -2 \frac{f_1 q_1 - f_0 q_0}{\Delta v} + 2\frac{g_0 p_0}{\Delta v} 
                                     - \frac{f_1 q_1}{r_1} \right) }
           { 1 + \frac{\Delta v}{2}\left( \frac{2 g_0}{\Delta v} + \frac{g_1}{r_1} \right) }.
\end{equation}
Similarly, we can derive an expression for $p_2$ by applying the Simpson rule
\begin{equation}
p_2 = \frac{ p_0 + \frac{\Delta v}{3} \left( - 2 \frac{-3 f_0 q_0 + 4 f_1 q_1 - f_2 q_2}{2 \Delta v} - 2 \frac{g_0 ( -3p_0 + 4p_1 ) }{2 \Delta v}
                                       - 4 \frac{f_1 q_1 + g_1 p_1}{r_1} 
                                       -   \frac{f_2 q_2}{r_2}
															  \right) }
           { 1 + \frac{\Delta v}{3} \left( - \frac{g_0}{\Delta v}  + \frac{g_2}{r_2} \right) }.
\end{equation}
For indices $i>2$ we have
\begin{equation}
p_i = \frac{ p_{i-2} + \frac{\Delta v}{3} \left( -   \frac{fq + gp}{r}\Big\vert_{i-2}
                                           - 4 \frac{fq + gp}{r}\Big\vert_{i-1}
                                           -   \frac{fq}{r}\Big\vert_i
															      \right) }
           { 1 + \frac{\Delta v}{3} \frac{g_i}{r_i} }.
\end{equation}

Due to the regularity of equation {\bf E2} we have 
\begin{equation}
f'|_0 = 0.
\end{equation}
Using Trapezoidal and Simpson's rules we arrive at the following expressions for $f$
on the initial slice:
\begin{equation}
f_1 = \frac{f_0 + \frac{1}{2} \Delta v \left( 0 - \frac{a_1^2}{4 r_1} \right)}
           {1 + \frac{1}{2} \Delta v \frac{g_1}{r_1}},
\end{equation}
\begin{equation}
f_2 = \frac{f_0 + \frac{1}{3} \Delta v \left(   0 
																				- 4 \frac{f_1 g_1 + a_1^2/4}{r_1}
																				- \frac{a_2^2}{4 r_2} 																				
																 \right)}
           {1 + \frac{1}{3} \Delta v \frac{g_2}{r_2}},
\end{equation}
\begin{equation}
f_i = \frac{f_{i-2} + \frac{1}{3} \Delta v \left( -   \frac{f g + a^2/4}{r}\Big\vert_{i-2}
																				    - 4 \frac{f g + a^2/4}{r}\Big\vert_{i-1}
																				    - \frac{a^2}{4 r}\Big\vert_i 																				
																     \right)}
           {1 + \frac{1}{3} \Delta v \frac{g_i}{r_i}}.
\end{equation}

\subsection{Evolution Scheme}
The evolution scheme evolves the massless scalar field coupled to Einstein's equations 
for one time-step. In section \ref{sec:dn-equations} we have chosen to use a constrained
first order system consisting of the equations 
\textbf{E2}, \textbf{S1} and \textbf{D2}, \textbf{D5}, \textbf{C2}, \textbf{D4}.

Basically, since the equations to be solved are coupled ODEs, we need to alternate prediction/
correction steps in the $u$- and $v$-directions.
We have implemented the integration of the u-equations via the explicit Trapezoidal
Runge-Kutta method, equation \eqref{eq:explicit-trapezoidal}, while we use the implicit 
Trapezoidal method, equation \eqref{eq:implicit-trapezoidal}, for the v-equations to 
render the scheme stable.
The set of discretized equations forms a nonlinear coupled algebraic system where some of the 
predicted function values need to be used in the correctors of other equations.
The predicted value for a gridfunction $f$ is denoted by the hatted quantitiy $\hat f$.

First, we tackle the boundary conditions at the center of spherical symmetry $r=0$.
They can be categorized as follows:
\begin{itemize}
 \item algebraic boundary conditions for r, g, and f
 \item homogeneous differential boundary conditions for a and s involving first derivatives $\pdrts{}$
 \item nonhomogeneous differential boundary conditions for p=q involving first derivatives $\pdrts{}$
\end{itemize}

Depending on the number of past slices that have already been calculated, we could 
discretize the differential boundary conditions in a number of ways.
For first derivatives at the origin we could use the following stencils:
\begin{align}
0 &= \frac{\partial a}{\partial r}\Big\vert_{t,r=0} = \frac{-3a_0^+ + 4 a_1 - a_2^{-}}{2 \Delta r} + \mathcal{O}(\Delta r^2)\\
p_i^+ &= \frac{\partial s}{\partial u}\Big\vert_v = \frac{3s^+_i -4 s_i + s^-_i}{2 \Delta u} + \mathcal{O}(\Delta u^2)
\end{align}
It would, however, be wrong to use those approximations, since the grid will not, in general, be 
equidistant in $r$. Therefore we must resort to more general finite difference formulae
as has already been discussed in section \ref{sec:dn-eq-regularization}.

We assume that indefinite expressions (at $r=0$) of the form $0/0$ which occur in 
the right-hand sides of some of the ODE's have already been regularized. (Again, see section
\ref{sec:dn-eq-regularization}.)

The core evolution scheme uses RK2 (explicit Trapezoidal) for the evolution equations in the
$u$-directions (the $d$- and $q$-equations), and implicit Trapezoidal integration for the $v$-equations.
Luckily, almost all of the resulting coupled non-linear algebraic equations in $v$ can trivially 
be made explicit by solving them in a fixed order. Only the $g$- and $r$-equations need to be 
decoupled by solving a linear system (See Ref. \cite{HaradaCarr}).

Effectively, we first calculate predictor values $\hat d$ and $\hat q$ 
(via an explicit Euler step) for the 
$u$-equations. Then, we proceed to calculate $\hat s$, $\hat a$, $\hat g$ and $\hat r$,
and finally $\hat f$ and $\hat p$, which are already the final values for those quantities.
Last, we correct d and q according to the RK2 explicit trapezoidal step and
set the final values equal to the hatted values for $s, a, g, r, f$ and $p$.
We give a detailed description of the algorithm in pseudocode:

\paragraph{Evolution Algorithm:}
\texttt{Assuming that the boundary conditions have already been
enforced at the origin gridpoint with index $i0^+$, we loop over all 
remaining gridpoints in a $u=const > 0$ slice:}
\begin{enumerate}
  \item \texttt{Predictors}
\begin{itemize}
  \item \texttt{$u$-equations: Euler forward}
\begin{align}
\hat d &= d_i + \Delta u \left( \frac{f_i g_i + a_i^2/4}{r_i^2}  - p_i q_i \right)\\
\hat q &= q_i - \Delta u \left( \frac{f_i q_i + g_i p_i}{r_i}  \right)
\end{align}
\item \texttt{$v$-equations: implicit Trapezoidal}
\begin{align}
\tilde s &= s^+_{i-1} + \frac{1}{2}\Delta v \mspace{2mu} q^+_{i-1}\\
\hat s &= \tilde s + \frac{1}{2} \Delta v \mspace{2mu} \hat q\\
\tilde a &= a^+_{i-1} + \frac{1}{2} \Delta v \mspace{2mu} a^+_{i-1} d^+_{i-1}\\
\hat a &= \frac{\tilde a}{1 - \frac{1}{2} \Delta v \mspace{2mu} \hat d}\\
%
\tilde g &= g^+_{i-1} + \frac{1}{2} \Delta v \left( 2 d^+_{i-1} g^+_{i-1} - r^+_{i-1} (q^+_{i-1})^2 \right)\\
\tilde r &= r^+_{i-1} + \frac{1}{2} \Delta v \mspace{2mu} g^+_{i-1}\\
\hat g &= \frac{\tilde g - \frac{1}{2} \Delta v \mspace{2mu} \hat q^2 \tilde r}{1 - \Delta v \mspace{2mu} \hat d + \left(\frac{\Delta v \mspace{2mu} \hat q}{2}\right)^2}\\
\hat r &= \frac{\frac{1}{2} \Delta v \mspace{2mu} \tilde g + (1 - \Delta v \mspace{2mu} \hat d) \tilde r}{1 - \Delta v \mspace{2mu} \hat d + \left(\frac{\Delta v \mspace{2mu} \hat q}{2}\right)^2}\\
\intertext{\texttt{If we are at the first gridpoint out from the origin}}
\tilde f &= f^+_{i-1} + \frac{1}{2} \Delta v \hat f_{\text{regularized RHS}}\\
\intertext{\texttt{else}}
\tilde f &= f^+_{i-1} - \frac{1}{2} \Delta v \frac{f^+_{i-1} g^+_{i-1} + \frac{1}{4}(a^+_{i-1})^2 }{r^+_{i-1}}\\
\hat f &= \frac{\tilde f - \frac{1}{8} \Delta v \mspace{2mu} \hat a^2 / \hat r}{1 + \frac{1}{2} \Delta v \mspace{2mu} \hat g / \hat r}\\
\intertext{\texttt{If we are at the first gridpoint out from the origin}}
\tilde p &= p^+_{i-1} + \frac{1}{2} \Delta v \mspace{2mu} \hat p_{\text{regularized RHS}}\\
\intertext{\texttt{else}}
\tilde p &= p^+_{i-1} - \frac{1}{2} \Delta v \frac{f^+_{i-1} q^+_{i-1} + g^+_{i-1} p^+_{i-1}}{r^+_{i-1}}\\
\hat p &= \frac{\tilde p - \frac{1}{2} \Delta v \mspace{2mu} \hat f \hat q / \hat r}{1 + \frac{1}{2} \Delta v \mspace{2mu} \hat g/\hat r}\\
\end{align}
\end{itemize}
\item \texttt{Correctors}
\begin{itemize}
  \item \texttt{$u$-equations: explicit Trapezoidal}
\begin{align}
d^+_i &= \frac{1}{2} \left[ \hat d + d_i + \Delta u \left( \frac{\hat f \hat g + \frac{1}{4} \hat a^2}{\hat r^2} - \hat p \hat q \right) \right]\\
q^+_i &= \frac{1}{2} \left[ \hat q + q_i - \Delta u  \frac{\hat f \hat q + \hat g \hat p}{\hat r} \right]\\
\end{align}
  \item \texttt{$v$-equations: trivial}
\begin{align}
s^+_i &= \hat s\\
a^+_i &= \hat a\\
g^+_i &= \hat g\\
r^+_i &= \hat r\\
f^+_i &= \hat f\\
p^+_i &= \hat p
\end{align}
\end{itemize}

\end{enumerate}

\subsection{Horizon Detection and Excision}\label{sec:horizon-detection-excision}

To detect apparent horizons we employ a utility function that 
reliably finds a possible zero in the gridfunction \texttt{g[i]} that represents
$g(v) = r_v(v)$ for a given null slice $u=const$.
This is accomplished by looking for sign changes in the discrete gridfunction 
via a linear search. Such an algorithm is straightforward to implement and safe 
to use, as it cannot overlook zeroes.
As we have discussed in section \ref{sec:dn-null-expansions}, due to our use
of an outgoing null foliation and the causal properties of the outer trapping horizon,
the function $g(v)$ can, on a given null slice, at most have one zero, 
so that we can safely ignore any further zeroes as numerical artifacts.

In the code, we compute a quantity \verb|r_AH| defined as the average 
of \verb|r[i_left]| and \verb|r[i_right]|, the interval bracket of the root of 
\verb|g[i]|. I.e. the actual zero $r_\text{AH}$ is contained in the interval \verb|[r[i_left],r[i_right]]|.

\begin{figure}[htbp]
  \centering
  \includegraphics[width=\textwidth]{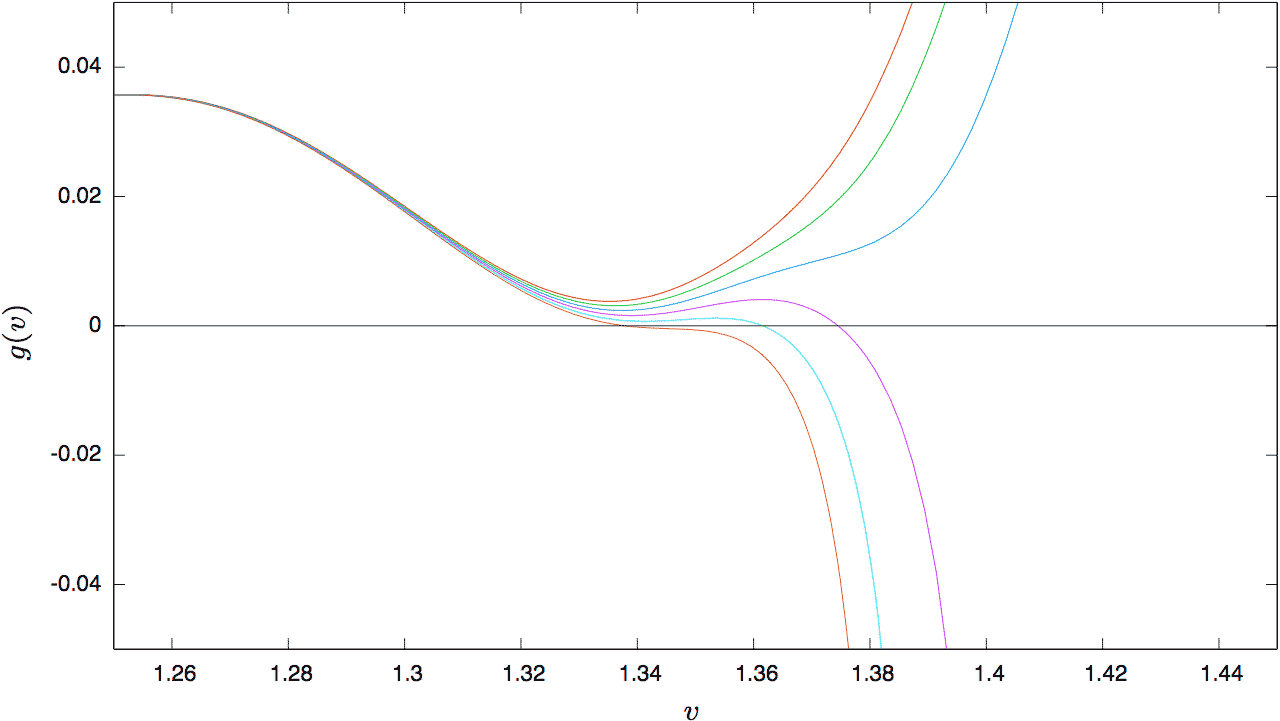}

  \caption{\label{fig:dn-g-AH-detection}
  This figure shows $g(v) = r'$ for a series of $u=const$ null-slices.
  The first three slices are untrapped, while the last three contain an apparent horizon,
  the radius of which decreases with increasing retarded time $u$.}
\end{figure}

We illustrate the behavior of the function $g$ by some results from numerical 
evolutions. Figure \ref{fig:dn-g-AH-detection} shows $g(v)$ for some null slices

\begin{figure}[htbp]
  \centering
  \includegraphics[width=\textwidth]{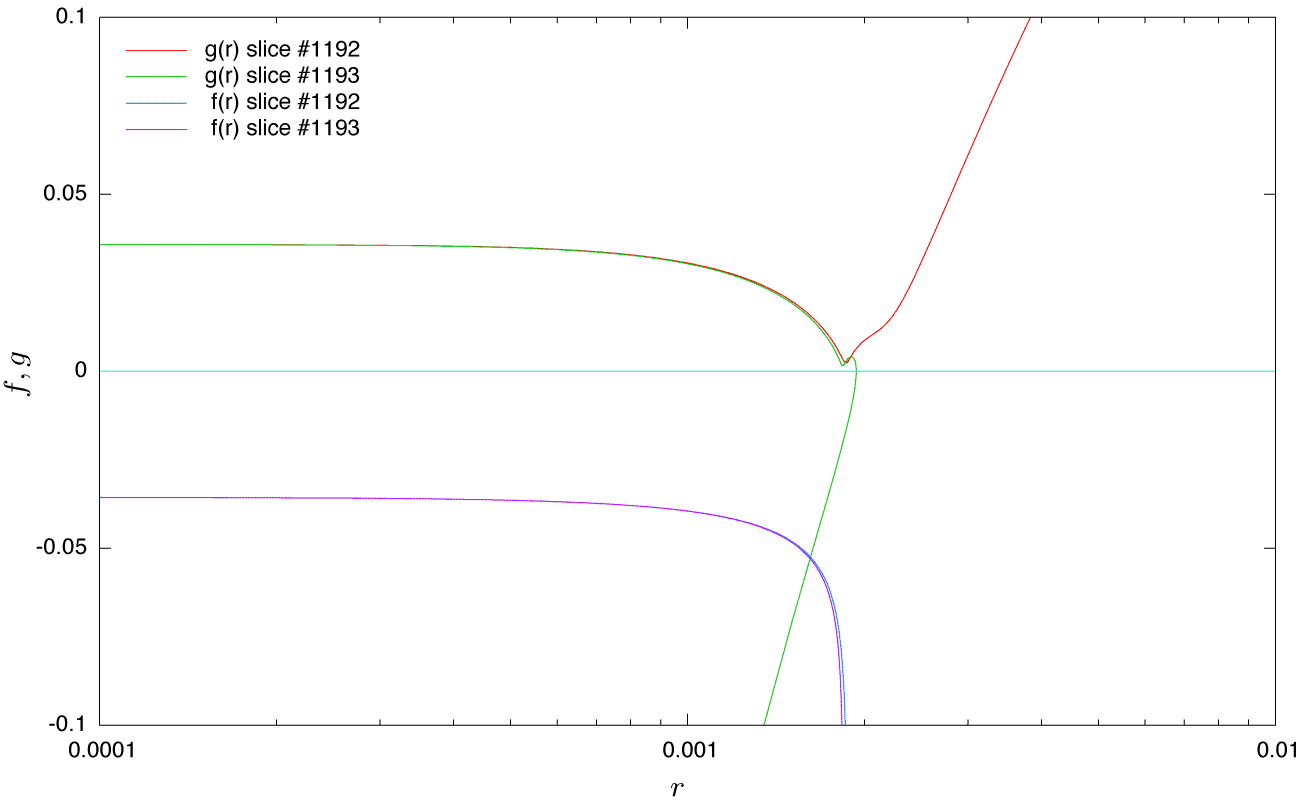}

  \caption{\label{fig:dn-f-g-theta-AH-detection}
  Instead of showing the null expansions $\theta_\pm$ which diverge at the origin
  due to their $1/r$ dependence, we depict $f=\dot r$ and $g=r'$ which decide the
  sign of the expansions as functions of $r$. 
  We show one slice shortly before apparent horizon formation
  (red and blue curves) and a second slice containing a marginally trapped surface
  (green, violet). Note that the function $f$ barely changes between the two slices,
  while $g$ abruptly bends down into negative values.
  In contrast to figure \ref{fig:dn-g-AH-detection}, where we plotted $g$ as a function 
  of advanced time $v$, $g(r)$ first grows with $r$, and after reaching the horizon radius,
  it decreases again; the null slice ``bends'' back towards the curvature singularity in the
  black hole.
  The condition for an apparent horizon, $\theta_+ =0$ while $\theta_-<0$ is fulfilled.}
\end{figure}

\begin{figure}[htbp]
  \includegraphics[width=\textwidth]{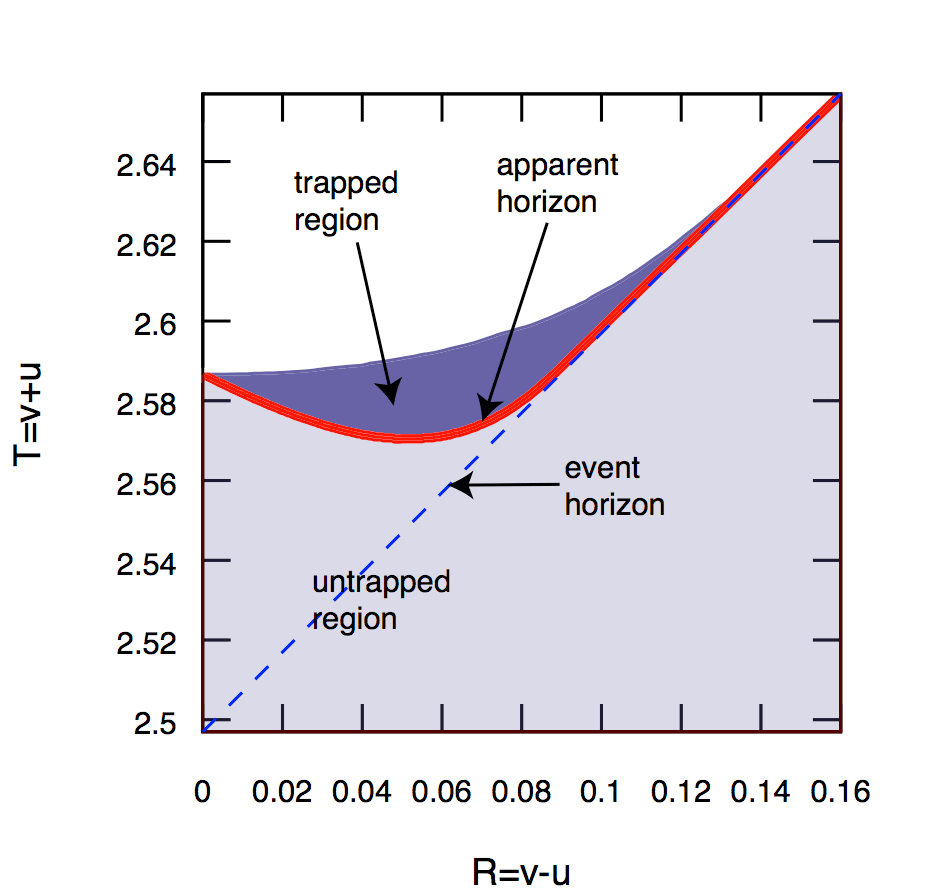}

  \caption{\label{fig:spacetime-diag-AH-T-R}
  This diagram shows the location of the trapping horizon in a numerically
  generated spacetime by the double-null code.
  The coordinates are linear combinations of the null coordinates $(u,v)$. 
  This choice allows us to map null slices into diagonal lines.
  The trappped and untrapped regions have been shaded light and dark blue.
  The trapped region is bounded by the trapping horizon and the outer grid boundary.
  In addition, we indicate an approximation of the event horizon. }
\end{figure}

Owing to the properties of the outer trapping horizon (see section \ref{sec:bh-properties}), 
one might assume that in a numerical collapse evolution using an outgoing null 
slicing the first AH is always detected at the outermost gridpoint of the slice.
In practice, this is simply not the case. Due to the finiteness of the timestep, 
$\Delta u$, it is highly unlikely to exactly ``hit''  the null slice $u_\text{EH}$ 
which coincides with the event horizon. Instead, one would expect to ``jump'' over the 
event horizon and detect an AH for a $u > u_\text{EH}$ and thus, since $v_\text{AH}$ decreases 
with increasing $u$ (if the AH is spacelike, which it has to be if matter falls in), 
we detect the AH somewhere inside the grid, but not at the outer grid boundary. 
On a slice that penetrates the trapping horizon, all gridpoints for which 
$v > v_\text{AH}$ are \emph{trapped} and the slice bends back towards the curvature 
singularity at $r=0$, i.e. $r$ decreases until it reaches $r=0$ for some $v > v_\text{AH}$.
Effectively, this forces us to cut off a sizeable chunk of the grid on each slice that
contains trapped surfaces: We look for trapped gridpoints, $g[i] < 0$, that have 
fallen into the curvature singularity, that is they fulfill $r[i] < 0$, and 
\emph{excise} them and all other gridpoints that we deem too close to the singularity.
This is simply done, by setting the outer grid boundary index N to a new value 
(smaller than the old one) such that all offending gridpoints are removed from the grid.
In contrast to Cauchy evolution codes, excision is very natural and straightforward to
implement in a double-null code.

\subsection{Mesh Refinement}\label{sec:dn-amr}

Although we closely follow the work of Hamad\'e and Stewart \cite{Stewart96} 
in our implementation of the double-null code, we employ Garfinkle's \cite{Garfinkle95}
approach to increase resolution which we already described for the DICE code in
section \ref{sec:dice-amr}, rather than the more involved Berger Oliger type
adaptive mesh refinement \cite{BergerOliger84}, which was also used in the 
seminal work by Choptuik \cite{Choptuik92,Choptuik93}. The application of 
Berger Oliger type AMR in a characteristic framework has been discussed by Pretorius
and Lehner \cite{PretoriusLehner04}.

\begin{figure}[htbp]
  \centering
  \includegraphics[totalheight=8cm]{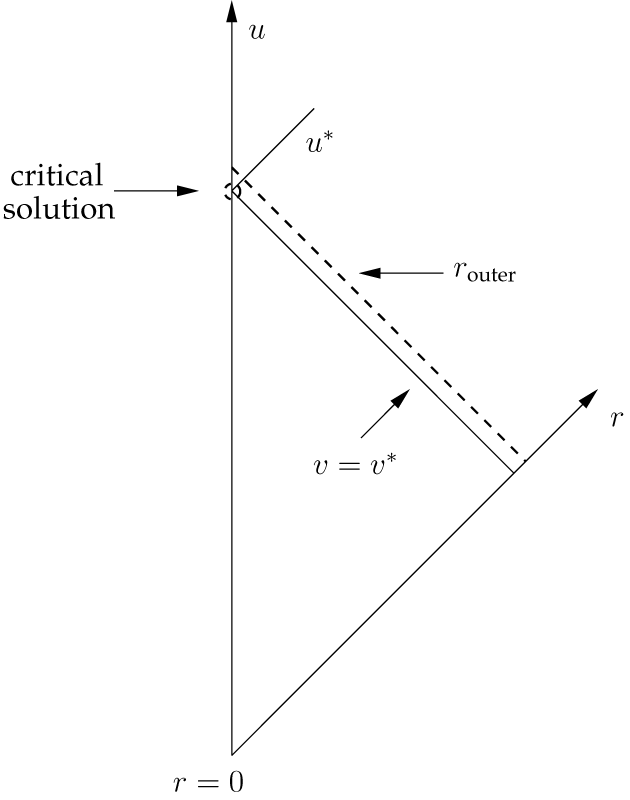}

  \caption{
  This figure shows how to adjust the outer grid boundary $r_\text{outer}$
  (or equivalently $v_\text{outer}$) in order to accurately resolve 
  the self-similar features of critical collapse close to the origin.
  }\label{fig:adjust-outer-grid-boundary-vstar}
\end{figure}

Concerning the mesh refinement strategy, the only difference to the DICE code, 
is that the double-null code is uncompactified and we can adjust the outer 
boundary of the grid to be slightly larger than the null ray $v=v^*$ which hits the 
accumulation point at $(u^*,v^*)$ of the self-similar solution as described in figure \ref{fig:adjust-outer-grid-boundary-vstar}. Self-similar solutions and critical 
collapse phenomena are discussed in detail in chapter \ref{ch:critical-phenomena}.

\subsection{Constraints and Convergence}\label{sub:constraints_and_convergence} 
\begin{figure}[htbp]
  \centering
  \includegraphics[width=\textwidth]{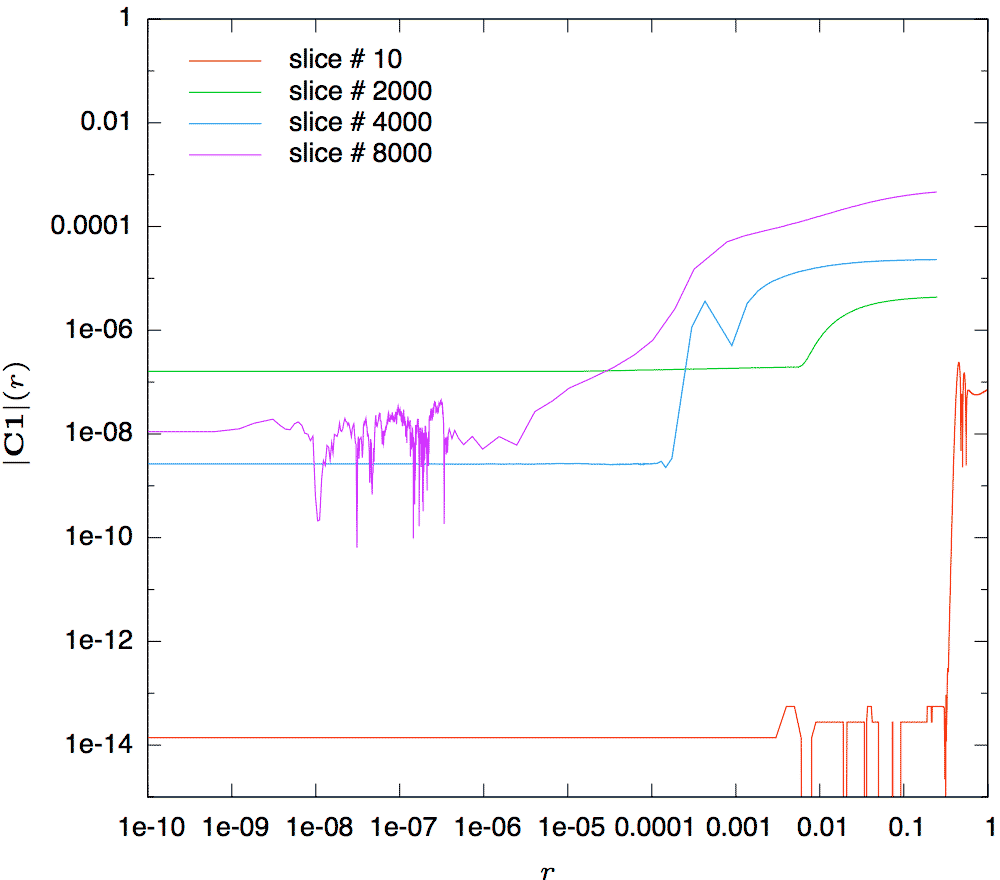}	
  \caption{\label{fig:dn-C1-convergence-sub}
  Here we show the spatial variation of the absolute value of the constraint \textbf{C1}
  for some slices of a subcritical evolution.
  The maximum of $|\textbf{C1}|$ over the course of the evolution is below $10^{-3}$.
  Figure \ref{fig:dn-C1-convergence-super} shows the constraint for the supercritical case.
  }
\end{figure}

\begin{figure}[htbp]
  \centering	
  \includegraphics[width=\textwidth, trim=20 0 0 0]{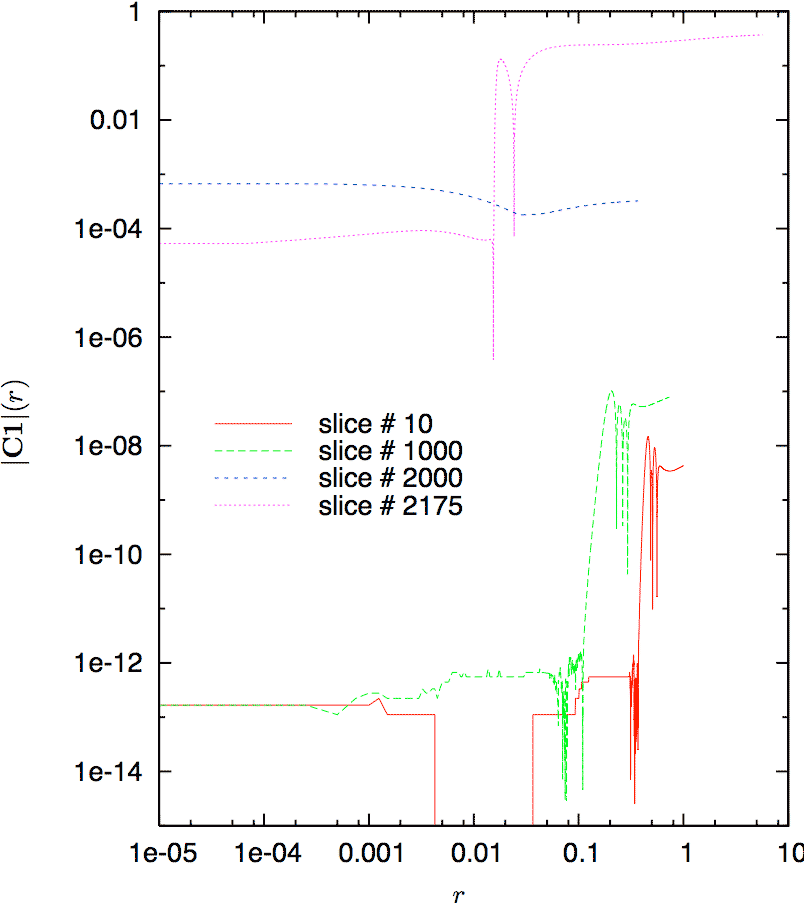}	

  \caption{\label{fig:dn-C1-convergence-super}
  This figure shows the absolute value of the constraint \textbf{C1}
  for a few slices of a supercritical evolution.
  On the null slice \# 2175, an apparent horizon has formed at $r \approx 0.015$
  which causes the constraint to grow about three orders of magnitude.
  }
\end{figure}

To check that our numerical solution fulfills the Einstein equations we calculate 
the constraint 
\textbf{C1}:
\begin{equation}
  \dot f = 2 \frac{\dot a}{a} f - r p^2
\end{equation}
and suitably normalize it with the 
absolute value of the running maximum of its individual terms over a slice, as described
in section \ref{sec:dice-accuracy-convergence}.
We also compute the discrete $\ell_2$-norm of the normalized constraint.
Figures \ref{fig:dn-C1-convergence-sub} and \ref{fig:dn-C1-convergence-super}
show the absolute value of the constraint \textbf{C1} as a function of $r$
for a couple of null slices in a subcritical and a supercritical evolution, respectively.


Along with the monitoring of constraint equations, convergence testing is an 
important strategy to ensure the correctness of numerical results, namely that,
as the gridspacing tends to zero, the numerical approximation converges to the 
analytical evolution system according to the accuracy of the discretization.
The basic methodology is explained in appendix \ref{app:convergence-methodology}.

In the following, we present results from a convergence test using 1001 gridpoints 
as the base resolution and 2001 and 4001 gridpoints for the medium and finest 
resolution, respectively. As the initial data are not as close to criticality as
those used in the convergence test shown in figure \ref{fig:Euur-convergence} 
for the DICE code, it was not necessary to adjust the initial data parameter 
$p$ for each resolution.
Due to the size of the gridspacing and second order accuracy of the code, the finite 
difference error is approximately $10^{-6}$ for the coarsest,  $2\times 10^{-7}$ 
for the medium and $6 \times 10^{-8}$ for the finest resolution.
The set of evolutions under consideration forms an apparent horizon at approximately
$u=1.088$ with a mass of $0.0076 \pm 0.0001$. For details see table \ref{tab:dn-m_AH-convergence}.
\begin{table}[htbp]
  \begin{center} 
      \begin{tabular}[b]{|l|ll|}
      \hline
      N & $u_{AH}$  & $m_{AH}$\\
      \hline
      1001 & 1.088 & 0.00775767\\
      2001 & 1.088 & 0.00765987\\
      4001 & 1.088 & 0.00763467\\
      \hline
      \end{tabular}
      \caption{\label{tab:dn-m_AH-convergence}
      The initial AH masses (i.e. when we first detect an AH),  
      for the three evolutions of the convergence test.
      }
\end{center} 
\end{table}

\begin{figure}[htbp]
  \centering
  \includegraphics[width=\textwidth,trim=80 0 0 0]{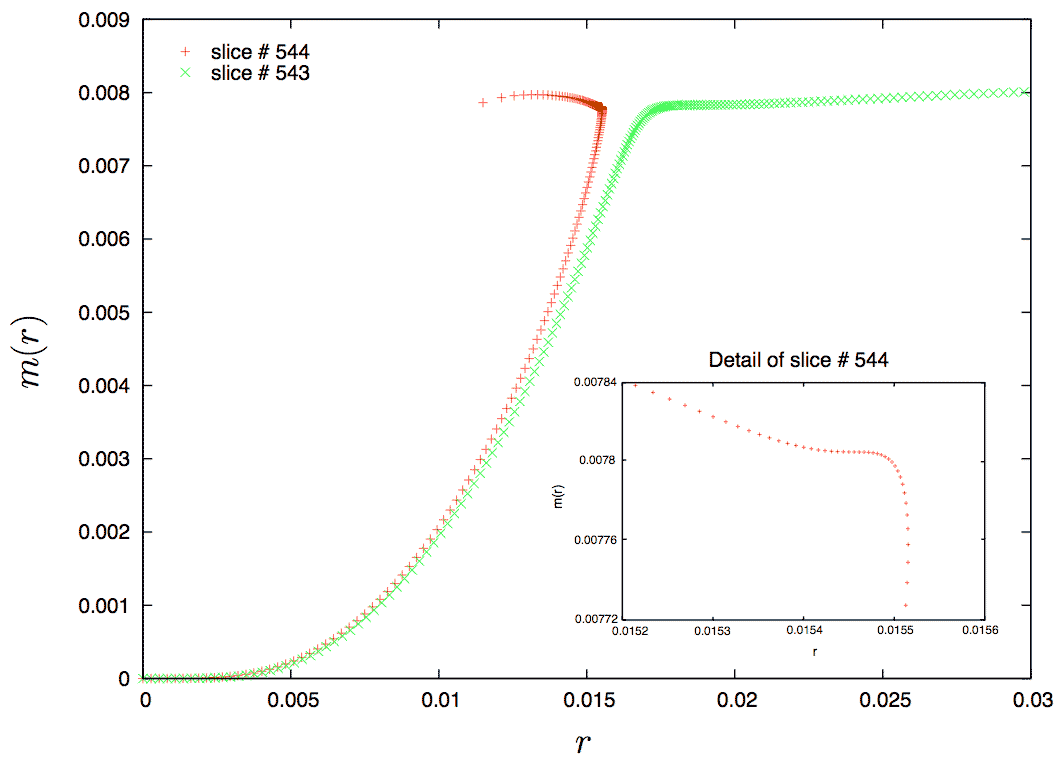}

  \caption{\label{fig:dn-AH-resolution-check}
  This figure shows the mass-function $m(r)$ for two successive null slices
  in the lowest resolution (1001 gridpoints) run of the convergence test.
  An apparent horizon is detected on slice 
  \# 544 
  at $u=1.088$.
  The detail shows that the AH is well resolved.}
\end{figure}

\begin{figure}[tbp]
  \includegraphics[width=\textwidth,trim=20 0 0 0]{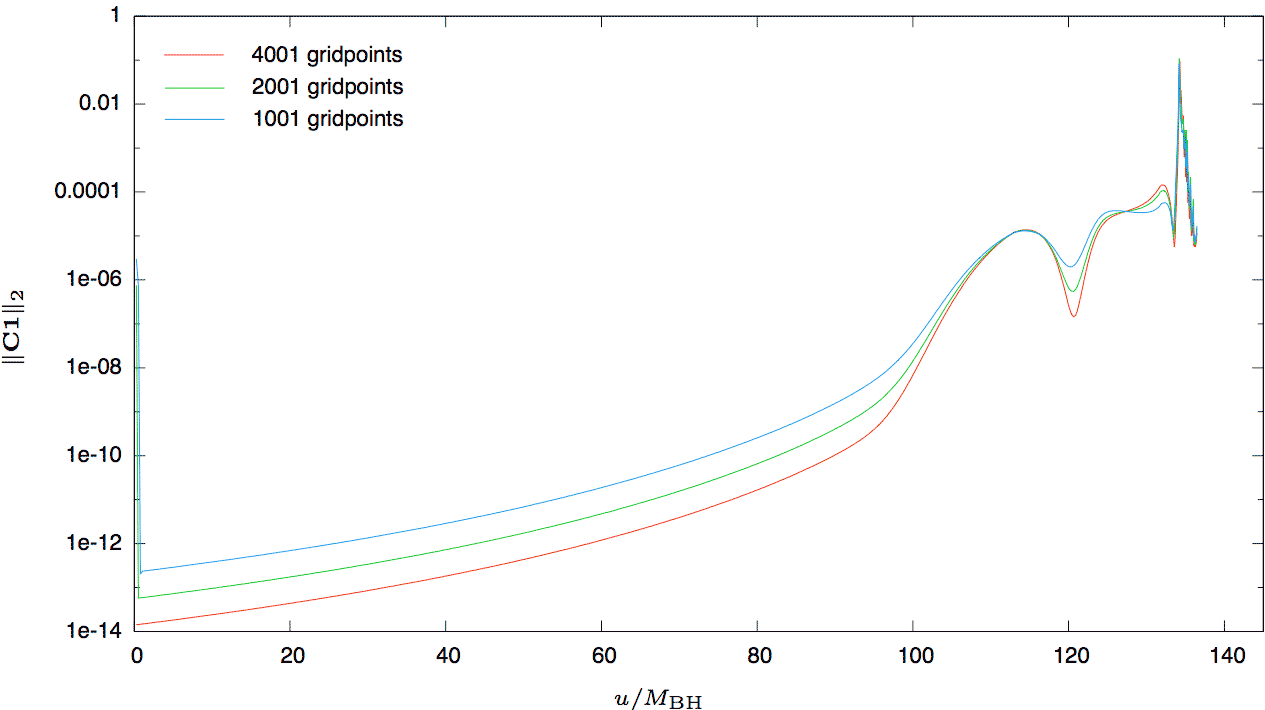}	

  \caption{\label{fig:dn-C1-convergence}
  This figure shows the convergence of the $\ell_2$-norm of the constraint \textbf{C1}. 
  Initially, the convergence is fourth order, but note that the finite 
  difference error for the lowest resolution run is about $10^{-6}$. For $t/M_{BH} > 100$
  convergence is roughly second order.
  An apparent horizon formes at $u=1.088$ with a mass of $0.0076 \pm 0.0001$.
  The constraint violations reach 0.1 at the formation of the apparent horizon, 
  but otherwise are fine.
  }
\end{figure}

\begin{figure}[htbp]
  \includegraphics[width=\textwidth,trim= 120 0 0 0]{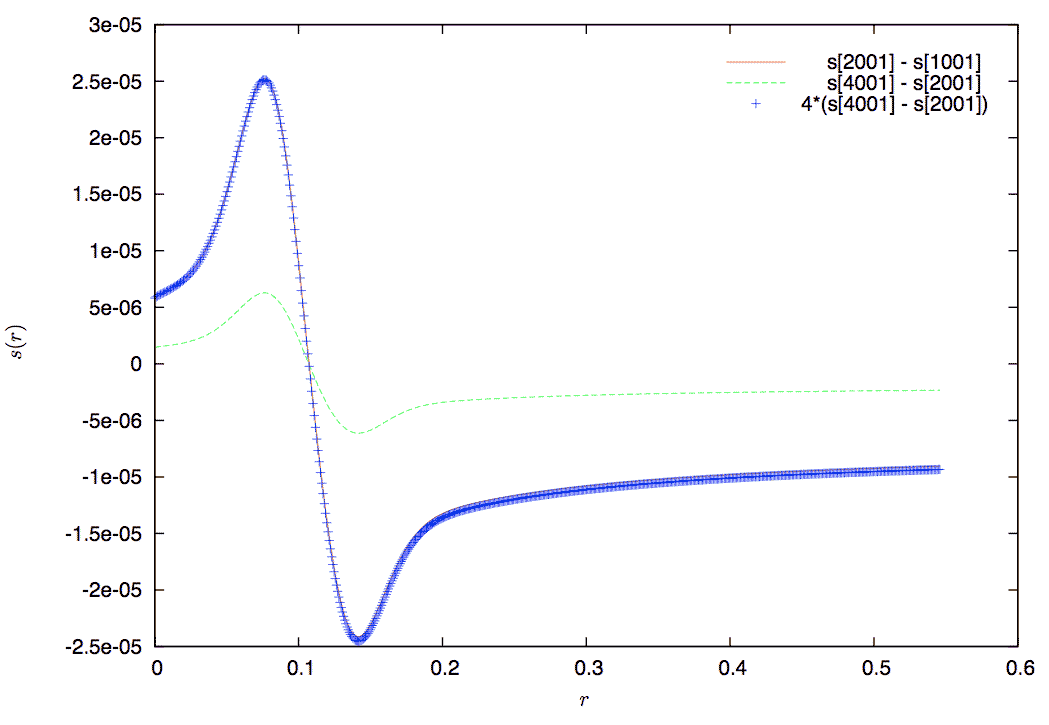}	

  \caption{\label{fig:dn-s-convergence}
  This figure shows a 3-level convergence test for the  scalar field $s$.
  The field is clearly second order convergent.}
\end{figure}

As can be seen from figure \ref{fig:dn-AH-resolution-check}
the AH is well resolved even for the coarsest resolution of 1001 gridpoints.
A convergence test for the $\ell_2$-norm of the constraint \textbf{C1} 
is shown in figure \ref{fig:dn-C1-convergence}.
Interestingly, the first part of the evolution exhibits fourth order
convergence. Later on we have solid second order convergence. The constraint
violations reach $0.1$ at the formation of the AH, which is the most demanding
part of a collapse evolution, but otherwise are fine.
In figure \ref{fig:dn-s-convergence} we show that the scalar field $s$ 
is second order convergent.

\subsection{Timelike Observers}\label{sec:numalg-timelike-observers}

In the following, we describe how to calculate the proper time of an $r=const$ observer 
from the expression \eqref{eq:dn-proper-time} derived in section \ref{sec:dn-diagnostics}.
As the $r$-values of our gridpoints on the double-null grid keep changing and we need to 
integrate over time, we fix a set of predetermined radii \verb|r[i]| and use interpolation 
to obtain the corresponding set of $v$ coordinates for each $u=const$ slice.
We store the running integrals in a vector \verb|tp[i]|.
The vector is initialized at the initial slice by setting \verb|tp[i] = r[i]|, since we assume a flat geometry, initially.
On each null slice, and for each observer at $r = \verb|r[i]|$ we
\begin{enumerate}
\item Determine $\bar v$ and $\dbar{v}$ via the relations 
\begin{equation}
  r(u,\bar v)=\texttt{r[i]}  
\end{equation}
and 
\begin{equation}
  r(u-\Delta u,\dbar{v})=\texttt{r[i]},   
\end{equation}
respectively, by using cubic spline interpolation.
\item Evaluate 
\begin{equation}
  \pd{v}{u}\Big|_{r=\texttt{r[i]}} \approx \frac{\bar v - \dbar{v}}{\Delta u}  
\end{equation}
to first order accuracy.
\item Evaluate $a(u,\bar v)$ using spline interpolation.
\item Update the running integral by adding the term 
$\Delta u \, a(u,\bar v) \, \frac{\bar v - \dbar{v}}{\Delta u}$.
\end{enumerate}

\begin{figure}[htbp]
  \centering
  \includegraphics[width=0.6\textwidth]{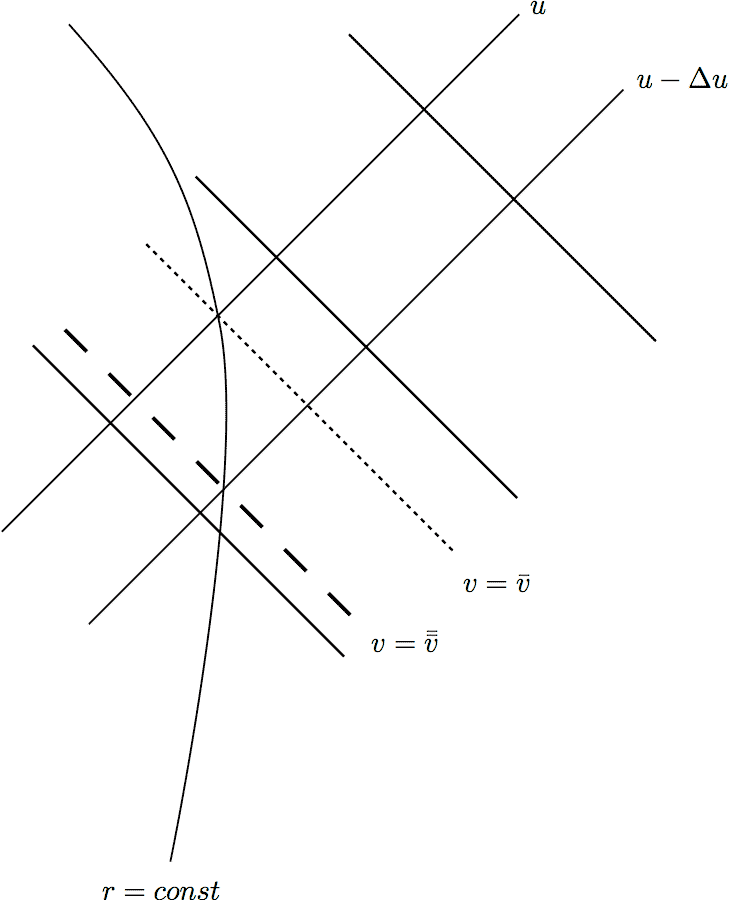}
  \caption{Calculation of $\pd{v}{u}\big|_r$}\label{default}
\end{figure}

If the slice contains an apparent horizon, things are more complicated, as $t(u,r)$ is no longer a single-valued function of $r$.
We first compute the index of the gridpoint with the maximum $r$-value in the grid.
We can then use spline interpolation in each of the two halves of the grid, where 
$r(u=const,v)$ is either an increasing (untrapped region) or decreasing (trapped region)
monotonic function of $v$ or gridpoint index. In the trapped part of a slice 
$\pd{v}{u}\big|_r <0$, so we take its absolute value in \eqref{eq:dn-proper-time}.
We now have to keep track of one running integral for proper time for each half of the slice.

For near-critical evolutions, the trapped part of the slices is very small and hardly 
warrants the difficulties incurred. Moreover, due to excision of gridpoints that come 
close to the singularity at $r=0$, we are missing data to compute \verb|tp[i]| in the 
outer half of the slice. 

\forceemptypage


\chapter{Critical Phenomena}\label{ch:critical-phenomena}

In general relativity, critical phenomena can be subsumed
by three essential features: scaling, self-similarity and universality.
They arise in the behavior of near-critical evolutions, in the course of
which a universal self-similar solution is approached for some time
and the idiosyncrasies of the initial data are mostly ``forgotten'' and 
replaced by symmetries inherited from the attractor. Ultimately,
the self-similar solution causes the evolution to approach an
endstate of the system. While self-similarity can be analyzed in single
evolutions, scaling only becomes apparent when considering the final
parameters of a one-parameter family of evolutions. Dimensionful quantities,
such as the black hole mass, exhibit a power-law behavior with an exponent
that is related to the unstable mode of the self-similar solution.

We first define the special symmetries, discrete and continuous self-similarity,
that lie at the core of critical phenomena. Next, we give a qualitative description
of near-critical evolutions in the dynamical systems picture.
Finally, we derive the scaling law for dimensionful quantities.

\section{Self-Similarity in General Relativity}\label{sec:self-similarity-in-GR}

First, we introduce the concepts of discrete (DSS) and continuous (CSS)
self-similarity and then go on to talk about coordinates adapted to
these symmetries.
Although we are not interested in CSS per se, it is often much easier to 
handle and useful for modelling situations where the periodicity of DSS
does not play a decisive role.


A spacetime $(M,g)$ is said to be \emph{discretely self-similar} (DSS) \cite{Gundlach00a}
if it admits a discrete diffeomorphism $\Phi_\Delta : M \to M$ which leaves the 
metric invariant up to a constant scale factor: 
\begin{equation}\label{eq:DSS-def}
\left(\Phi^*_\Delta\right)^n g \big|_p = e^{2 n \Delta} g \big|_p \qquad {\forall p \in M},
\end{equation}
where $\Delta$ is a dimensionless real constant, $n \in \mathbb{N}$ and
\begin{equation}
\left( \Phi^*_\Delta g \right)_{ab} \big|_p 
= \pd{\Phi_\Delta^i}{x^a}\pd{\Phi_\Delta^j}{x^b} g_{ij} \bigg|_{\Phi_\Delta (p)}  
\end{equation}
is the \emph{pullback} of the spacetime metric under the diffeomorphism $\Phi_\Delta$
and we denote the Jacobian of this map by $\pd{\Phi_\Delta^a}{x^b}$.

Similarly, a spacetime that admits a one-parameter family of such diffeomorphisms,
parametrized by $\Delta$, with $\Phi_0$ being the identity, is called 
\emph{continuously self-similar} (CSS).
The generating vector field $\xi = \frac{d}{d\Delta} \Phi_\Delta \big|_{\Delta = 0}$
is \emph{homothetic}, that is, it obeys the conformal Killing equation with a 
constant coefficient on the right hand side:
\begin{equation}\label{eq:def-homotheticity}
\mathscr{L}_\xi \, g = 2 g.
\end{equation}
The choice of the constant is pure convention.

Following Gundlach \cite{Gundlach97f} we introduce coordinates $(\tau, x^\alpha)$
such that if a point $p$ has coordinates $(\tau, x^\alpha)$, its image 
$\Phi_\Delta (p)$ has coordinates $(\tau - \Delta, x^\alpha)$.
One can then verify that DSS in these coordinates is equivalent to the condition
\begin{align}\label{eq:DSS-periodicity-condition}
  g_{ab}(\tau, x^\alpha) &= e^{2\tau} \tilde g_{ab}(\tau, x^\alpha),
  \intertext{where}\\
  \tilde g_{ab}(\tau, x^\alpha) &= \tilde g_{ab}(\tau + \Delta, x^\alpha).
\end{align}

To explain the connection between CSS and DSS one may 
introduce a vector field $\xi = \partial/ \partial \tau$.
The discrete diffeomorphism $\Phi_\Delta$ is then realized as Lie-dragging 
along $\xi$ by a distance $\Delta$.
The discrete diffeomorphism cannot uniquely determine $\xi$, as, loosely speaking,
the vector field $\xi$ is free to do whatever it wants between DSS ``echoing'' periods.

\begin{figure}[htbp]
  \centering
  \includegraphics[width=\textwidth]{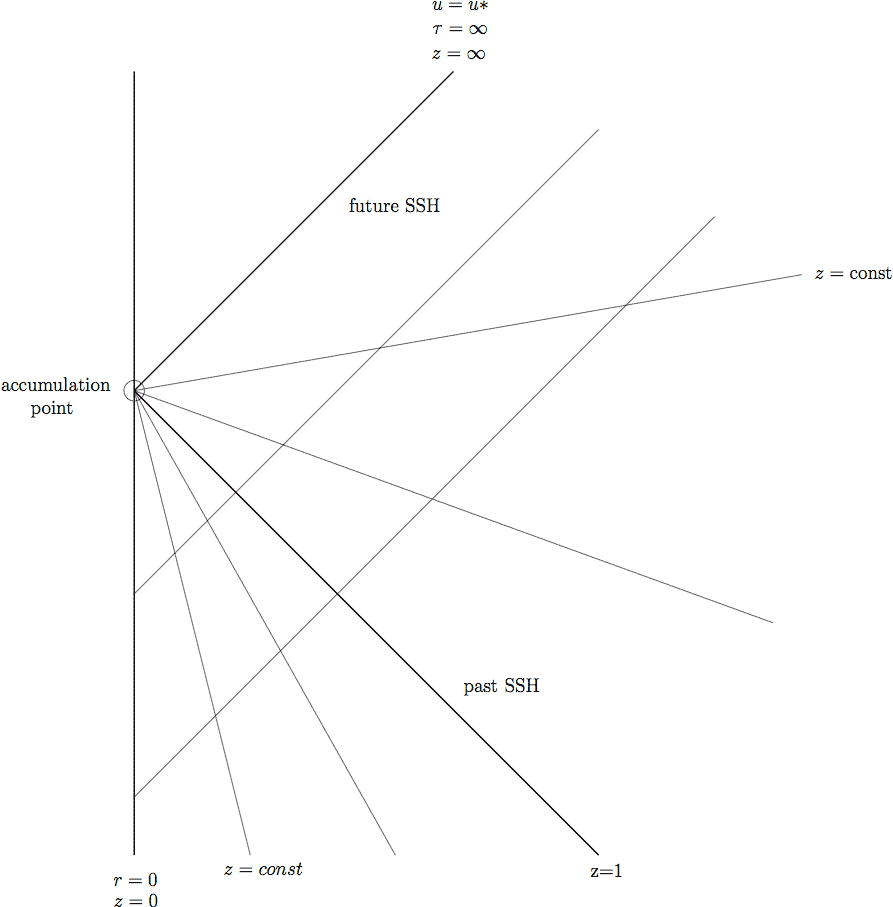}

  \caption{
    This diagram shows a prototypical self-similar spacetime.
    The adapted coordinates $(\tau, z)$ defined in equations
    \eqref{eq:tau-z-def} and \eqref{eq:z-def} cover only the region $u<u^*$.
    The metric functions $\beta$, $\Vr$ and $m$ are constant along lines 
    of constant $z$. Thus, their features shrink to zero as one approaches the
    accumulation point at $(u=u^*,r=0)$. Self-similarity is confined to the
    backwards lightcone of the accumulation point, which extends from the origin
    to the past self similarity horizon at $z=1$.
  }\label{fig:tau_z_diagram}
\end{figure}

We introduce \emph{adapted coordinates} $(\tau, z, \theta, \phi)$ 
based on the Bondi coordinates defined in section \ref{sec:Bondi_coordinates}:
\begin{eqnarray}\label{eq:tau-z-def}
\tau &=& -\ln \frac{u^* - u}{u^*},\\
z    &=& \frac{r}{(u^* - u)\zeta(\tau)} = \frac{r e^\tau}{\zeta(\tau) u^*},\label{eq:z-def} 
\end{eqnarray}
where $u^*$ is a real number which denotes the \emph{accumulation time} of DSS 
and $\zeta(\tau + \Delta) = \zeta(\tau)$.
A detailed derivation of these adapted coordinates is given in \cite{LechnerPhD}.
We also require that $\zeta(\tau) > \dot \zeta(\tau)$ for all $\tau$ such that 
$r_\tau (\tau,z) < 0$ for all $\tau$.
The remaining gauge freedom in $\zeta(\tau)$ is chosen such that $\partial_\tau$
(which is timelike at the origin) becomes null at $z=1$.
This null hypersurface, the backwards lightcone of the accumulation point,
is known as the past \emph{self-similarity horizon} (SSH).
See figure \ref{fig:tau_z_diagram} for a spacetime diagram that illustrates the
behavior of these coordinates.

Next, we transform the Bondi line-element \eqref{eq:def-Bondi-line-element} to 
these new coordinates using the inverse transformation
\begin{align}
  u &= u^* \left( 1 - e^{-\tau} \right)\\
  r &= u^* z e^{-\tau} \zeta(\tau),
\end{align}
which yields
\begin{multline}
  ds^2 = e^{-2\tau} \biggl\lbrace   -e^{2\beta(\tau, z)} {u^*}^2
          \left[ \Vr(\tau,z) + 2z\left( \dot\zeta(\tau) - \zeta(\tau) \right) \right] d\tau^2\\
          - 2e^{2\beta(\tau,z)}{u^*}^2 \zeta(\tau) d\tau dz
          + \left[ u^* z \zeta(\tau) \right]^2 d\Omega^2    \biggr\rbrace.
\end{multline}
This line-element fulfills the condition \eqref{eq:DSS-periodicity-condition} if the
metric functions $\beta$ and $V/r$ are periodic in $\tau$ with period $\Delta$:
\begin{align}
  \beta(\tau,z) &= \beta(\tau + \Delta,z)\\
  \Vr(\tau,z) &= \Vr(\tau + \Delta,z).
\end{align}

Which conditions does the symmetry \eqref{eq:DSS-def} imply for the scalar field?
Clearly, matter has to share at least some of the symmetry of the geometry, as it
is coupled via Einstein's equations.
First, we need to know how the Einstein tensor behaves under the action of the 
discrete diffeomorphism. Thanks to general covariance, the pullback of the
Einstein tensor is equal to the Einstein tensor formed from the pulled-back metric:
\begin{equation}
  \Phi_\Delta^* G[g] = G\left[ \Phi_\Delta^* g \right].
\end{equation}
It follows from the definition of DSS, equation \eqref{eq:DSS-def}, that
$ \Phi_\Delta^* G[g] = G\left[ e^{2\Delta} g \right]$.
Because the structure of the Einstein tensor is such that in every term that
contains the metric, the inverse metric also appears, and since
$\Phi_\Delta^* g^{-1}|_p = e^{-2\Delta} g|_p$ we find 
$G\left[ \Phi_\Delta^* g  \right] = G[g]$ and thus
\begin{equation}
  \Phi_\Delta^* G[g] = G[g].
\end{equation}
Via Einstein's equations this symmetry also holds for the stress energy tensor:
\begin{equation}
  \left( \Phi_\Delta^* T \right)_{ab} = T_{ab}.
\end{equation}
Taking the pullback of the scalar field stress energy tensor, equation
\eqref{eq:stress-energy-tensor}, yields
\begin{equation}
  \begin{split}
     \left(\Phi_\Delta^* T \right)_{ab} \big|_p &= 
     \pd{\Phi^i_\Delta}{x^a}\pd{\Phi^j_\Delta}{x^b} 
     \left[ \nabla_i\phi \nabla_j\phi 
                 - \frac{1}{2}g_{ij}g^{cd} \nabla_c\phi \nabla_d\phi    
     \right] \Big|_{\Phi_\Delta (p)}\\
     &= \pd{\Phi^i_\Delta}{x^a}\pd{\Phi^j_\Delta}{x^b} 
     \nabla_i\phi \nabla_j\phi \Big|_{\Phi_\Delta (p)}
     - \frac{1}{2} g_{ab}(p)  \pd{\Phi^c_\Delta}{x^k}\pd{\Phi^d_\Delta}{x^l}  g^{kl}(p)
      \nabla_c\phi \nabla_d\phi \Big|_{\Phi_\Delta (p)}
 \end{split}
\end{equation}
where we have used the pushforward of the inverse metric,
\begin{equation}
  \left(\left(\Phi_\Delta\right)_* g^{-1}\right)^{ab}\bigg|_{\Phi_\Delta (p)} = 
  \pd{\Phi^a_\Delta}{x^c}\pd{\Phi^b_\Delta}{x^d} g^{cd} \big|_p =
  e^{2\Delta} g\big|_{\Phi_\Delta (p)}
\end{equation}
and the definition of DSS in terms of the metric, equation \eqref{eq:DSS-def}.
To simplify the expression we use the adapted coordinates \eqref{eq:tau-z-def} and
\eqref{eq:z-def}, so that the point $p$ has coordinates $(\tau,z)$ and 
$\Phi_\Delta (p)$ has coordinates $(\tau - \Delta,z)$ and drop the angular dependence:
\begin{multline}
    \nabla_a\phi(\tau,z)  \nabla_b\phi(\tau,z) 
    - \frac{1}{2}g_{ab}(\tau,z) g^{cd}(\tau,z)   \nabla_c\phi(\tau,z)  \nabla_d\phi(\tau,z) \\
    =   \nabla_a\phi(\tau - \Delta,z)   \nabla_b\phi(\tau - \Delta,z)
    - \frac{1}{2}g_{ab}(\tau,z) g^{cd}(\tau,z)   
    \nabla_c\phi(\tau-\Delta,z)  \nabla_d\phi(\tau-\Delta,z).    
\end{multline}
Therefore, we find the condition
\begin{equation}
  \nabla_a\phi(\tau,z) = \pm \nabla_a\phi(\tau-\Delta,z).
\end{equation}
For the scalar field, the most general ansatz compatible with DSS 
is (see Ref. \cite{Gundlach97f})
\begin{equation}\label{eq:DSS-general-anasatz-phi}
  \phi(x^i) = \tilde\phi(\tau,z) + \kappa \tau
  \quad \text{with} \quad
  \tilde\phi(\tau+\Delta,z) = \pm \tilde\phi(\tau,z).
\end{equation}
The constant $\kappa$ is determined by the $(uu)$ component of Einstein's equations 
written in terms of the adapted coordinates $(\tau,z)$.
As discovered numerically by Choptuik, $\kappa = 0$, for unknown reasons.

The vanishing of $\kappa$ together with the choice of the minus sign in
equation \eqref{eq:DSS-general-anasatz-phi} for $\tilde\phi$ gives rise to a 
further symmetry (see \cite{Gundlach97f}, \cite{LechnerPhD})
that has been numerically observed:
If the metric functions are periodic with a period $\tilde\Delta$ 
and the scalar field is ``anti-periodic'' with respect to this period, then,
the field is obviously periodic with respect to the period $2\tilde\Delta$.
We say that the solution is DSS with period $\Delta=2\tilde\Delta$ and has
the additional symmetry
\begin{equation}\label{eq:DSS-metric-functions-additional-symmetry}
f^*(\tau + n\Delta/2, z) = f^*(\tau, z),
\end{equation}
where $f^*$ denotes the metric functions $\beta$, $V/r$, $m/r$ and the 
function $\zeta(\tau)$, while the scalar field $\phi$ satisfies
\begin{equation}\label{eq:DSS-scalar-field-additional-symmetry}
\left(\Phi^*_{\Delta/2}\right)^n \phi = \left(-1\right)^n \phi.
\end{equation}
This is consistent with the fact that $\beta, V/r, m/r$ are even in $\phi$.


\section{General Relativity as an Infinite Dimensional Dynamical System}

%
%

Following \cite{Gundlach00a}, we give a brief qualitative introduction 
to the dynamical systems picture. We pretend that general relativity 
can be treated as an infinite-dimensional dynamical system and ignore
problematic issues such as convergence to an attractor.

We restrict ourselves to isolated self-gravitating systems, such as
a ball of radiation or a star.
The phase space consists of the space of initial data for an isolated
\footnote{
Note that DSS or CSS do not lie in this phase space since they are
not asymptotically flat. Solutions which are only self-similar inside 
a finite radius and asymptotically flat can be included in the phase space.
}
system, which in our case is just the space of $\phi(u_0,r)$.
Possible end-states of the system correspond to \emph{basins of attraction} in phase
space. The boundary between two basins of attraction is called a
\emph{critical surface}.
We restrict attention to the existence of two distinct endstates;
otherwise, more complex critical points would arise at the intersection
of more than two basins of attraction.
Minkowski space is an attractive fixed point in the dispersion basin, while
Schwarzschild black holes form a half-line of attracting fixed points, 
parametrized by mass.

\begin{figure}
  \centering
  \includegraphics[width=\textwidth]{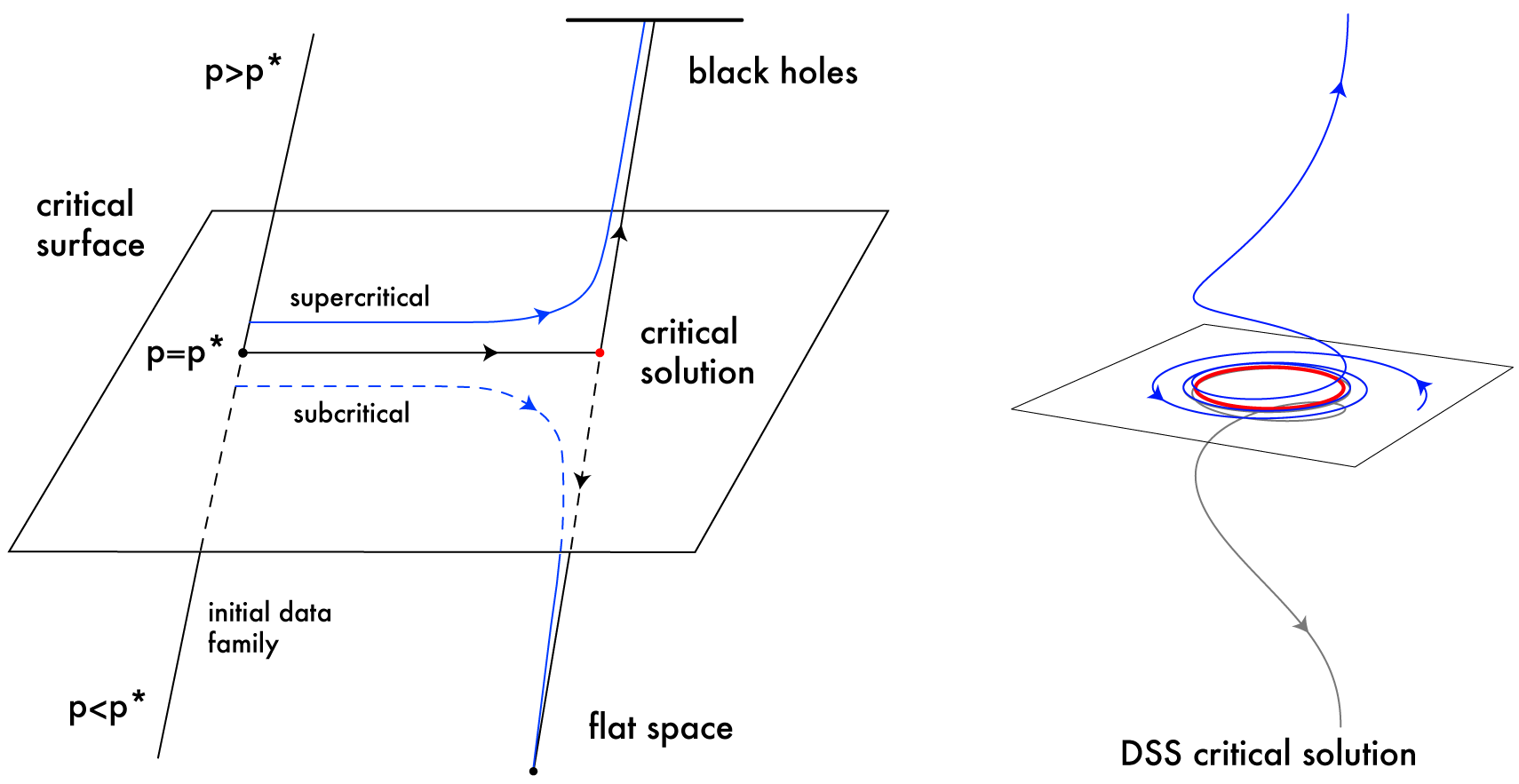}

  \caption{A qualitative picture of the phase-space of near critical evolutions.
    Every point corresponds to one configuration $\phi(u=\text{const}, r)$.
    The stable manifold contains a codimension one discretely self-similar 
    solution, depicted as a limit cycle, as it is periodic in $\tau$.
    We have indicated a one-parameter family of initial data by a line that
    intersects the critical surface on the left. At the ``critical'' parameter
    value $p=p^*$, the trajectory is attracted by the stable modes of the 
    critical solution and never leaves the stable manifold, ultimately arriving
    at the DSS solution in the limit $\tau \to \infty$.
    Trajectories with $p>p^*$ may approach the critical solution for some time,
    as long as the admixture of the unstable mode is small, but ultimately 
    leave the vicinity of the critical solution and move on to form a black hole.
    In contrast, configurations with $p<p^*$ eventually disperse and reach flat space.
  }\label{fig-attractor}
\end{figure}

Consider an attracting fixed point or limit cycle, the \emph{critical solution},
that lies in this critical surface, as shown in figure \ref{fig-attractor}.
We assume that this critical solution, in turn, is an attractor of codimension 
one in a neighborhood of phase space, which means that it possesses a single 
growing perturbation mode transverse to the critical surface and an infinite 
number of decaying modes tangential to the surface.
\footnote{
  Attractors which possess more than one unstable mode would be
  exceedingly hard to ``run into'' in the course of numerical evolutions,
  as they would require the concurrent fine-tuning to ``critical values''
  of more than one parameter.
}
Trajectories starting out in the vicinity of the critical surface at first 
move close to the surface towards the critical solution. 
In the so-called \emph{intermediate asymptotics} (a term coined by P. Bizon)
they spend some time in the vicinity of the \emph{universal} critical solution, 
shedding all the details of initial data from which they originated.
After a while the growing mode becomes dominant and the trajectories depart 
towards the flat space or black hole fixed point depending on whether the data 
were sub- or supercritical, respectively.

In the context of this work the terms ``critical solution'' and ``DSS solution''
(or ``self-similar solution'') are often used interchangeably. 
Due to self-similarity the DSS solution is evidently not asymptotically flat. 
It is however not so clear whether the term critical solution entails asymptotic 
flatness or not. Here, we want to make this sloppy usage more precise:
We mean by critical solution an asymptotically flat 
spacetime which agrees in a finite region near the centre with the DSS solution, 
but falls off at null infinity. E.g. it would be consistent to have the exact DSS 
spacetime in the region defined within the backward light cone of the DSS
singularity (the past self similarity horizon) and outside to fall off smoothly to infinity. 
Similarly, we call a solution near-critical, if it comes close to the critical solution 
(in the above sense) near the centre of collapse for some time during the evolution. 
The spacetime region in which an evolution spends in the neighbourhood of the critical 
solution depends on the fine tuning and the class of initial data.
Therefore, the failure of the DSS solution to be asymptotically flat is perfectly consistent
with its role in the dynamics of a localized object emitting radiation to null infinity.

\section{The Black Hole Mass-Scaling Law}\label{sec:scaling-law-derivation}
In this section we present an argument for the scaling of the black hole mass.
We follow Gundlach (\cite{Gundlach00a}, \cite{Gundlach97f}) and Lechner \cite{LechnerPhD}.

For the sake of simplicity, we first consider the spherically symmetric CSS case.
Assume that the self-similar solution is an attractor of codimension one in phase
space, that is, it has exactly one unstable mode with eigenvalue $\lambda_0$.
Furthermore, assume that the system has two possible endstates, dispersion to flat
space or black hole formation, such that the stable manifold of the self-similar
solution divides phase space into subcritical and supercritical data.
(See figure \ref{fig-attractor} for an illustration.)
Let $Z$ denote any scaling variable $(\beta, \Vr, \phi, \frac{m}{r})$.
We use adapted coordinates to CSS: 
\begin{equation}
  \tau = -\ln \frac{u^*-u}{u^*} \quad \text{and} \quad
  z = \frac{r}{u^*-u}.  
\end{equation}

General near-critical initial data are attracted via the stable modes of the CSS
solution, which we denote by $Z_*(z)$. 
In contrast to a DSS solution, which would be periodic in the adapted coordinate 
$\tau$, a CSS solution is independent of $\tau$.
In the echoing region, where $Z_*$ dominates we write the solution as a small
perturbation of the CSS solution:
\begin{equation}\label{eq:Z-perturb-CSS}
  Z(\tau,z) = Z_*(z) + C_0(p) \, e^{\lambda_0 \tau} \, Z_0(z) + \delta Z_\text{stable} (\tau,z).
\end{equation}
We assume that the single unstable eigenvalue $\lambda_0$ is real and positive.
As $\tau \to \infty$ all other perturbations, which are contained in 
$\delta Z_\text{stable} (\tau,z)$, vanish.
The amplitude of the unstable mode $Z_0(z)$ contains information on the initial data,
in particular, the parameter $p$ that characterizes a one-parameter family of initial data.
For so-called \emph{critical initial data}, $p=p^*$, the unstable mode
is completely tuned out, so that $C_0(p^*) = 0$.
For near-critical data, we may then linearize around $p^*$:
\begin{equation}
  C_0(p) = \frac{d C_0}{dp}\bigg|_{p^*} \left( p - p^* \right) 
         + \bigO\left( \left( p - p^* \right)^2 \right),
\end{equation}
so that for suitable $\tau$
\begin{equation}\label{eq:Z-CSS-vicinity}
  Z(\tau,z) \simeq  Z_*(z) 
  + \frac{d C_0}{dp}\bigg|_{p^*} \left( p - p^* \right) \, e^{\lambda_0 \tau} \, Z_0(z).
\end{equation}
Clearly, as $p \to p^*$, we see more and more of the critical solution. Equation
\eqref{eq:Z-CSS-vicinity} also sheds light on the \emph{universality} of the 
self-similar solution: The critical phenomena are independent of the family
of initial data.

Let $\tau_p$ be a value of the adapted time chosen such that the stable modes are 
negligible compared to the unstable mode and that the amplitude of the
unstable mode equals a small number $\epsilon \ll 1$:
\begin{equation}
  \frac{d C_0}{dp}\bigg|_{p^*} \left( p - p^* \right) \, e^{\lambda_0 \tau_p} = \epsilon.
\end{equation}
Note that $\tau_p$ depends on the initial data parameter $p$.
We rewrite this relation as
\begin{equation}
  e^{-\tau_p} = \text{const} \left( p - p^* \right)^{1/\lambda_0},
\end{equation}
where the constant is independent of $p$.

For large $\tau$ the linear approximation eventually breaks down when the unstable 
mode has grown beyond a certain bound. If, as we assume in the following, 
the data are supercritical, a black hole forms at late times.
We emphasize that we need not follow the evolution into the nonlinear regime;
it is sufficient to note that intermediate data which we can extract at $\tau=\tau_p$
depend on $r$ only through the coordinate $z$:
\begin{equation}
  Z(\tau_p,z) \simeq Z_*(z) + \epsilon Z_0(z).
\end{equation}
In Bondi coordinates we have
\begin{equation}
  Z(u_p,r)  \simeq Z_*\left(\frac{r}{u^* - u_p}\right) 
           + \epsilon Z_0\left(\frac{r}{u^* - u_p}\right).
\end{equation}

The crucial point of this argument is that these intermediate data at $u=u_p$ depend
on the initial data, say at time $u=0$, only through the overall \emph{scale}
\begin{equation}
  L = u^* - u_p = u^* e^{-\tau_p}.
\end{equation}
The field equations do not have an intrinsic scale so that in the absence of an 
external scale, such as a cosmological constant, the field equations are invariant
under rescalings $u \to \eta u$, $r \to \eta r$.
Therefore, the solution based on the data at $u=u_p$ must be \emph{universal} up to
the overall scale.

Because the black hole mass has dimension of length in geometrized units ($G=c=1$),
it must be proportional to $L$:
\begin{equation}\label{eq:mass-scaling-def}
  \boxed { M_\text{BH} \propto L \propto \left( p - p^* \right)^{1/\lambda_0} }
\end{equation}
and we define the \emph{critical exponent}
\begin{equation}
  \boxed {\gamma := \frac{1}{\real \lambda_0} }.
\end{equation} 
Relation \eqref{eq:mass-scaling-def} was derived independently by 
Koike et al. \cite{Koike95} and Maison \cite{Maison96}.

As pointed out by Garfinkle and Duncan \cite{GarfinkleDuncan98}, 
it is also possible to analyze scaling in subcritical evolutions via the
scaling behavior of the maximum over a whole evolution of the Ricci scalar 
curvature at the axis. Due to its dimension being $1/L^2$, the Ricci scalar scales as
\begin{equation}\label{eq:curvature-scaling}
  \max_u R(u,0) \sim \left(p-p^*\right)^{-2/\lambda_0}.
\end{equation}
Numerically, it is easier to measure the critical exponent from the subcritical 
scaling of the Ricci scalar than from the supercritical scaling of the black
hole mass, since one avoids numerical difficulties incurred in apparent horizon formation. 

The mass-scaling relation \eqref{eq:mass-scaling-def} is universal in the sense that
it holds for all one-parameter families of initial data, since its derivation relies 
on perturbations of the universal (for each model) critical solution.
Thus, all solutions which approach the attractor for a while will exhibit this same
scaling law.

Although the critical solution itself is not asymptotically flat due to its
self-similarity, one can restrict the analysis to a region of a sufficiently large,
but finite radius, so that the black hole mass is not affected.

\subsection{DSS Fine Structure}
In the case of a discretely self-similar critical solution, the scaling law
\eqref{eq:mass-scaling-def} is modified by an oscilliatory ``fine structure'' 
overlaid on the linear relation between $\ln M_\text{BH}$ and $\ln (p - p^*)$.
This phenomenon was first predicted by Gundlach \cite{Gundlach97f} and
Hod and Piran \cite{Hod97}. 

Similar to the CSS case, where the perturbation
modes of the background solution would retain its CSS symmetry 
(independence of $\tau$), the modes are now periodic in $\tau$ and equation
\eqref{eq:Z-perturb-CSS} becomes
\begin{equation}
  Z(\tau,z) = Z_*(\tau, z) + C_0(p) \, e^{\lambda_0 \tau} \, Z_0(\tau, z) 
  + \delta Z_\text{stable} (\tau,z).
\end{equation}
Note that $\delta Z_\text{stable} (\tau,z) = \sum_{n=1}^\infty 
C_n e^{\lambda_n \tau} Z_n(\tau,z)$
is not periodic in $\tau$, whereas the individual modes $Z_n$ are, since
the perturbation must break the DSS symmetry or the critical solution would
not be isolated \cite{Gundlach97f}.
Again, for suitably large $\tau$ so that the perturbation is still small and 
we can neglect the stable modes, we obtain for near-critical data
\begin{equation}\label{eq:Z-DSS-vicinity}
 Z(\tau,z) \simeq  Z_*(\tau,z) 
  + \frac{d C_0}{dp}\bigg|_{p^*} \left( p - p^* \right) \, e^{\lambda_0 \tau} \, Z_0(\tau, z).
\end{equation}
As before, we extract intermediate data at a time $\tau_p$ defined as
\begin{equation}
  \tau_p = - \frac{1}{\lambda_0} \ln \left[ (p-p^*) \frac{1}{\epsilon} 
  \frac{d C_0}{dp}\bigg|_{p^*}  \right],
\end{equation}
where $\epsilon$ is a small constant.
These data depend on $r$ only through the dimensionless coordinate $z$:
\begin{equation}
  Z(\tau_p,z) \simeq Z_*(\tau_p, z) + \epsilon Z_0(\tau_p,z),
\end{equation}
so that in Bondi coordinates we have
\begin{equation}
  Z(u_p,r)  \simeq Z_0\left(u_p, \frac{r}{\left(u^* - u_p\right) \zeta(u_p)}\right) 
           + \epsilon Z_0\left(u_p, \frac{r}{\left(u^* - u_p\right) \zeta(u_p)}\right),
\end{equation}
where $u_p = u^* \left( 1 - e^{-\tau_p} \right)$.
The entire solution evolved from these data scales with
\begin{equation}
  \left(u^* - u_p\right) \zeta(u_p) f(u_p) 
  = u^* e^{-\tau_p} \zeta(\tau_p) f(\tau_p)
  = e^{-\tau_p} \tilde{f}(\tau_p),
\end{equation}
where $f$, $\tilde f$ (and $\zeta$) are periodic functions of their argument.
Therefore, the black hole mass obeys the following scaling law
\begin{equation}
  m_\text{BH} = c_1 \left( p - p^* \right)^{1/\lambda_0} 
  \tilde{f} \left[ -\ln c_1 - 1/\lambda_0\ln\left( p - p^* \right) \right]
\end{equation}
where the constant 
$c_1 = \left[ \frac{1}{\epsilon} \frac{d C_0}{dp}\big|_{p^*} \right]^{1/\lambda_0}$ 
depends on the family of initial data. Equivalently,
\begin{equation}\label{eq:scaling-fine-structure}
  \boxed {
    \ln m_\text{BH} = \ln c_1 + \gamma \ln\left( p - p^* \right)
     + \hat{f} \left[ -\ln c_1 - \gamma \ln\left( p - p^* \right) \right],
  }
\end{equation}
where $\hat f = \ln \tilde f$ and we have introduced the scaling exponent
$\gamma = 1/\lambda_0$.
As a function of $\ln (p-p^*)$, $\hat f$ is periodic with period 
$\Delta/(2\gamma) \approx 4.61$ (for the massless scalar field).
This is due to the additional symmetry of the metric functions 
mentioned in equation \eqref{eq:DSS-metric-functions-additional-symmetry}
(see Ref. \cite{Gundlach97f}).

\clearpage

\forceemptypage

\chapter{Quasinormal Modes and Tails}
\label{ch:qnm-tails}

\section{Introduction}

In general relativity, \emph{quasinormal modes} (QNM) (see \cite{Kokkotas99a, NoviFro98, Chandrasekhar83} 
for a review) arise as perturbations of stellar or black hole
spacetimes. Due to energy loss caused by radiation, these infinite systems do not exhibit normal mode
oscillations which are characteristic of compact linear oscillating systems without damping. Instead,
the frequencies become ``quasi-normal'', i.e. complex, with the real part representing the frequency of
the oscillation and the imaginary part representing the damping.
Moreover, in contrast to normal modes, quasinormal modes do not form a complete set of basis functions.

In addition, for late times, the weak decay of the Schwarzschild potential causes the quasinormal 
oscillations to be swamped by the radiative tail\cite{Price72}, which only decays polynomially in time.

In the following we give a brief review of the evolution of weak fields on a Schwarzschild background 
(or similar) \cite{NoviFro89}. 
The general procedure is to expand the field in spherical harmonics (according to its spin $s$).
For each radiative multipole $l \ge s$ there is a scalar field $\phi^{(s)}_l$ which depends on
the Schwarzschild coordinates $t$ and $r$.
Each such scalar function $\phi_l^{(s)}(t,r)$ satisfies an equation
\begin{equation}
\pd{^2\phi_l^{(s)}}{r_*^2} - \pd{^2\phi_l^{(s)}}{t^2} = V_l^{(s)}(r) \phi_l^{(s)}
\end{equation}
where $V_l^{(s)}$ is the effective potential and $r_*$ is the usual Schwarzschild ``tortoise'' 
coordinate, which is related to $r$ by $dr/dr_* = 1 - 2M/r$.

With the ansatz $\phi_l^{(s)}(r,t) = \psi_l^{(s)}(r_*) e^{-i\omega t}$ we arrive at the following
second order ODE that is similar to the one-dimensional Schr\"odinger equation for a particle
encountering a potential barrier on the inifinite line\cite{Iyer87a}:
\begin{equation}\label{eq:SS-perturb}
\frac{d^2 \psi_l^{(s)}}{d r_*^2} + \left[ \omega^2 - V_l^{(s)}(r_*) \right] \psi_l^{(s)} = 0.
\end{equation}
For scalar fields the potential is
\begin{equation}
V_l^{(0)} = \left( 1 - \frac{2M}{r} \right) \left[ \frac{2M}{r^3} + \frac{l(l+1)}{r^2} \right].
\end{equation}
The effective potential behaves as a potential barrier, as it is appreciably distinct from zero
only in the neighborhood of $r_* \approx 0$ ($r \approx 3M$). As $r_* \to \pm \infty$ 
the potential falls off very rapidly.

The so-called quasinormal modes of the black hole are solutions to the perturbation equation 
\eqref{eq:SS-perturb} which satisfy radiation boundary conditions for purely outgoing waves at 
(spatial) infinity and purely ingoing waves at the horizon. The real part of the QNM represents
the oscillation frequency, while the imaginary part represents the decay.

The radiation produced in response to a perturbation of the field around a black hole can be divided
into three stages:
\begin{itemize}
\item radiation emitted directly by the source of perturbation
\item radiation due to the damped oscillations of quasinormal modes excited by the perturbation 
source (``ringing radiation'')
\item power law tails of radiation, caused by scattering of waves by the effective potential
\end{itemize} 
A distant observer first records the radiation from the perturbation source, then the quasinormal ringing, 
which decay exponentially, and finally the radiation tails which decay much slower, by a power law, and 
have a very small amplitude in comparison to the first two components.

As an example, the dimensionless frequency $M\omega$ for the $l=2, n=0$ mode of gravitational 
perturbations is $M\omega \approx 0.37 - 0.089i$, while the $l=0, n=0$ mode for scalar perturbations is 
$M\omega \approx 0.11 - 0.11 i$ \cite{Iyer87a}. Both are fundamental modes.

To convert from the dimensionless frequency $M\omega$ in geometrized units ($G=c=1$) to
non-geometrized units, we note that the dimension of $M\omega$ is mass/time which yields the
conversion factor $c^3/G$. The QNM frequency for a 10 solar mass black hole is then
\begin{equation}
f = \frac{1}{2\pi} \frac{c^3}{G}  \frac{\real M\omega}{10 M_\odot} \text{Hz} 
\approx 3.2  \real(M\omega) \text{kHz}
\end{equation}
and the damping time is
\begin{equation}
t_\text{damp} = \frac{G}{c^3} \frac{10 M_\odot}{|\imag M\omega|} \text{s} 
\approx \frac{4.97 \times 10^{-2}}{|\imag M\omega|} \text{ms}.
\end{equation}
For the $l=0, n=0$ scalar mode, this corresponds to a frequency of $357 \text{Hz}$ and a damping
time of $0.45 \text{ms}$, while for the $l=2, n=0$ gravitational mode, this corresponds to a frequency of $1.2 \text{kHz}$ and a damping time of $0.55 \text{ms}$.
Compared to the Schwarzschild radius for the 10 solar mass black hole, which is
$r_S = \frac{2 G M_\odot}{c^2} \approx 30 km$, the wavelengths of the fundamental QNM are much bigger,
$841 \text{km}$ and $250 \text{km}$ for the $l=0, n=0$ and the $l=2, n=0$ mode, respectively.

\section{Quasinormal Modes In Critical Collapse}\label{sec:QNMs}
As we have mentioned, QNM excitations are, in general, obtained from linear perturbations 
off a fixed background, together with their associated (complex) eigenvalues. 
Thus, in a highly dynamical setting, such as in critical collapse evolutions, 
one would not expect to see (identify) quasinormal modes.
However, it turns out that for our setting, the least damped spherically symmetric mode 
for scalar perturbations of a Schwarzschild black hole plays a relevant role.

Perturbation theory\cite{Iyer87a} gives the following value for the half-period
\begin{equation}
\frac{T_0}{2} = \frac{\pi}{\real \omega} \approx 28.43 M_{bg},
\end{equation}
where $\real \omega = 0.11$ is the real part of the $n=0, l=0$ QNM.
This mode has previously been detected in supercritical evolutions (far away from criticality) 
for a self-gravitating massless scalar field by Gundlach et al. \cite{Gundlach94b}. 

\begin{figure}[htbp]
  \centering
  \includegraphics[width=.9\textwidth,trim=30 0 0 0]{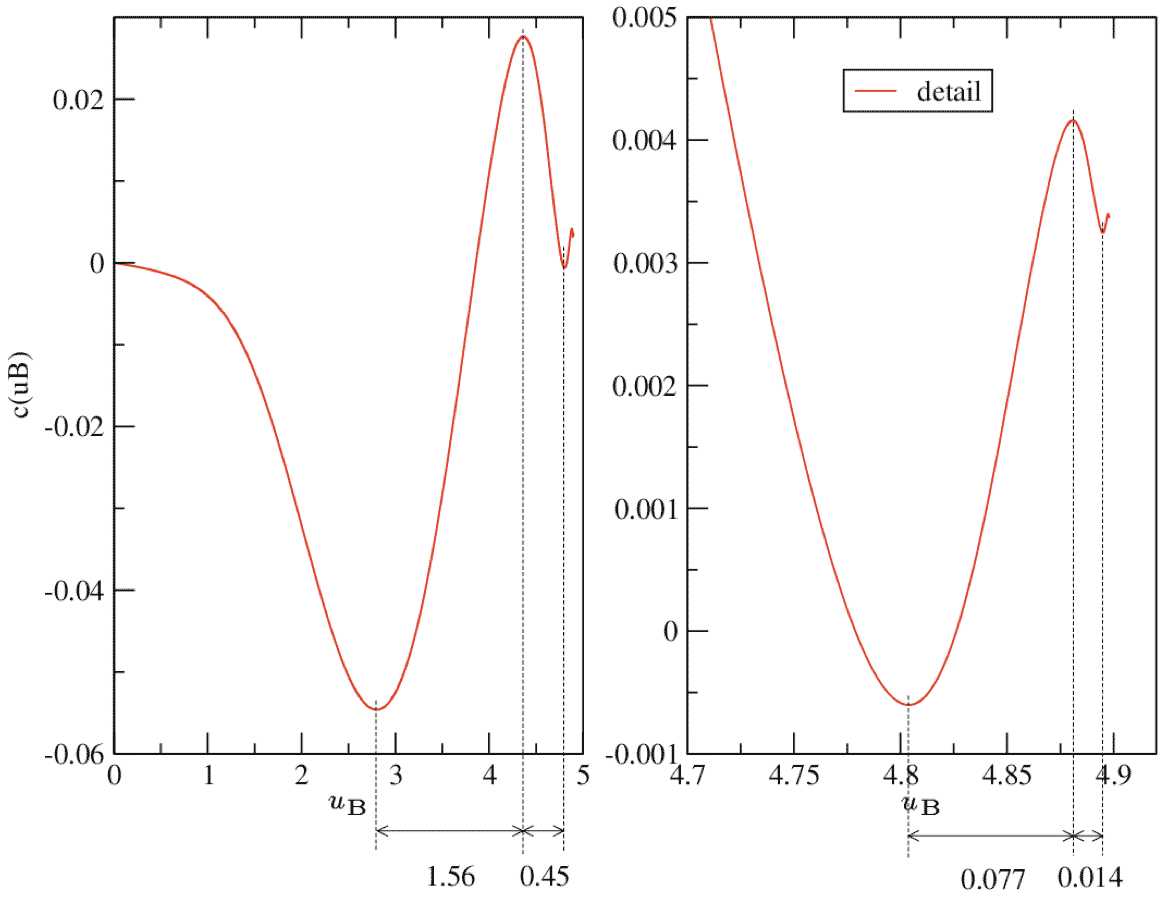}
  
  \caption{\label{c_QNM}This figure shows the scalar field monopole moment $c(\uB)$ for a 
  near-critical (but barely supercritical) evolution. The half-periods
  measured from one extremum to the next roughly agree with the prediction
  of perturbation theory shown in figure \ref{T0_QNM}.
  }
\end{figure}

\begin{figure}[htbp]
  \centering
  \includegraphics[clip,width=.9\textwidth]{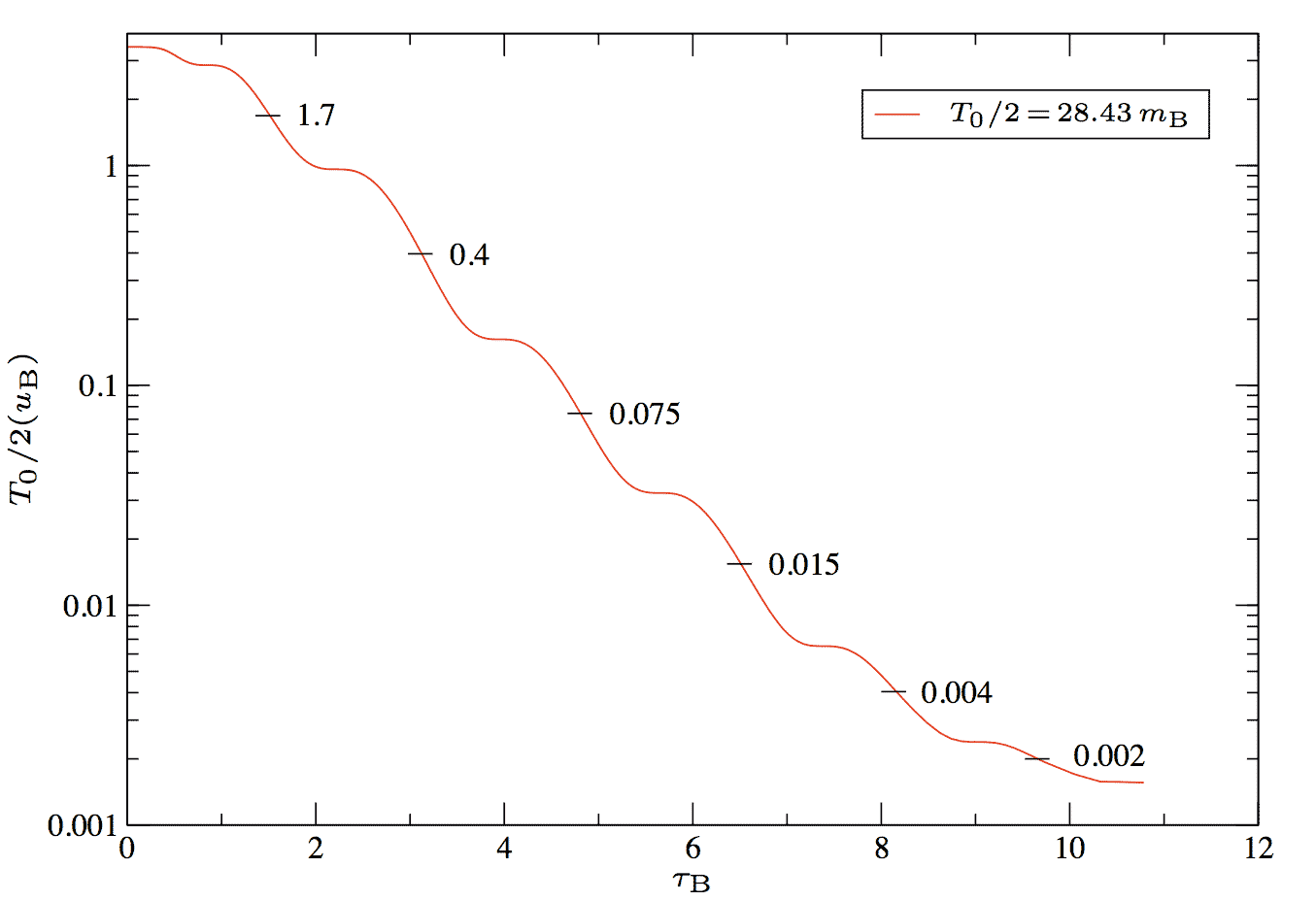}

  \caption{\label{T0_QNM}The exponential decay of the QNM half-periods predicted by 
  perturbation theory is shown with annotated values at the midpoints
  between the points of inflection for the same near--critical evolution as
  in figure \ref{c_QNM}. 
  }
\end{figure}

In the following, we analyze radiation signals for near-critical evolutions,
where the notion of a fixed background mass does no longer apply.
We find that the monopole moment of the scalar field $c(\uB)$ shows a damped 
oscillation with exponentially increasing frequency (see figure \ref{c_QNM}).
Moreover, the sizes of the half-periods measured from one extremum to the next in
$c(\uB)$ roughly agree with the half-periods obtained from
the least damped quasi-normal mode (QNM) of a Schwarzschild black hole with a 
strongly changing ``background'' mass $M_{bg}(\uB)$ as shown in figure
\ref{T0_QNM}.
$M_{bg}(\uB)$ is obtained by evaluating $\mB(\uB)$  at the mean value 
between the extrema of $c(\uB)$ (which are inflection points of $\mB(\uB)$).

\section{Quasinormal Modes for the Vaidya Metric}\label{Vaidya-QNM}
\subsection{The Vaidya Metric}
The \emph{Vaidya} metric\cite{Poisson04, GirottoSaa04,AbdallaChirentiSaa06}
is a generalization of the Schwarzschild solution with a changing mass.

We express the Schwarzschild metric in the hybrid \emph{outgoing Eddington-Finkelstein} coordinates
$(\bar u, r)$
\begin{equation}
  ds^2 = - \left( 1 - \frac{2M}{r} \right) d{\bar u}^2 - 2 d\bar u dr + r^2 d\Omega^2,
\end{equation}
where the null coordinate $\bar u = t - r_*$ is defined in terms of the \emph{tortoise coordinate}  
$r_* = r + 2M \ln(r/2M - 1)$.
Then we allow the mass parameter $M$ to become a function of retarded time: $M \to m(\bar u)$. 
The resulting metric is the \emph{outgoing Vaidya metric}
\begin{equation}\label{eq:Vaidya-outgoing}
  ds^2 = - \left( 1 - \frac{2m(\bar u)}{r} \right) d{\bar u}^2 - 2 d\bar u dr + r^2 d\Omega^2.
\end{equation}
The only nonvanishing component of the Einstein tensor is 
$G_{{\bar u}{\bar u}} = - (2/r^2)$ $(dm/d{\bar u})$.
For the metric \eqref{eq:Vaidya-outgoing} to be a solution of the Einstein equations the
stress-energy tensor must be of the form 
\begin{equation}
  T_{ab} = - \frac{1}{4\pi r^2} \frac{dm}{d\bar u} l_a l_b,  
\end{equation}
where $l_a = - \nabla_a \bar u$ is a radial null vector.
This stress-energy tensor describes \emph{null dust}, a pressureless fluid with energy density
$\rho = - 1/(4\pi r^2) (dm/d \bar u)$ moving with a four-velocity $l^a$.
All the standard energy conditions \cite{Poisson04} are satisfied by $T_{ab}$ if $dm/d \bar u \le 0$.
This solution of Einstein's equations describes a unidirectional radial flow of unpolarized 
radiation in the geometric optics (high frequency) approximation.

In contrast, the \emph{ingoing Vaidya metric} is given by
\begin{equation}
  ds^2 = - \left(1 - \frac{2m(\bar v)}{r}\right) d{\bar v}^2 + 2d{\bar v}dr + r^2 d\Omega^2.
\end{equation}
and $m$ must be a monotonically increasing function of advanced time $\bar v = t + r_*$
in order to satisfy the energy conditions.

\begin{figure}[htbp]
  \centering
  \includegraphics[height=3in,trim=0 0 50 0]{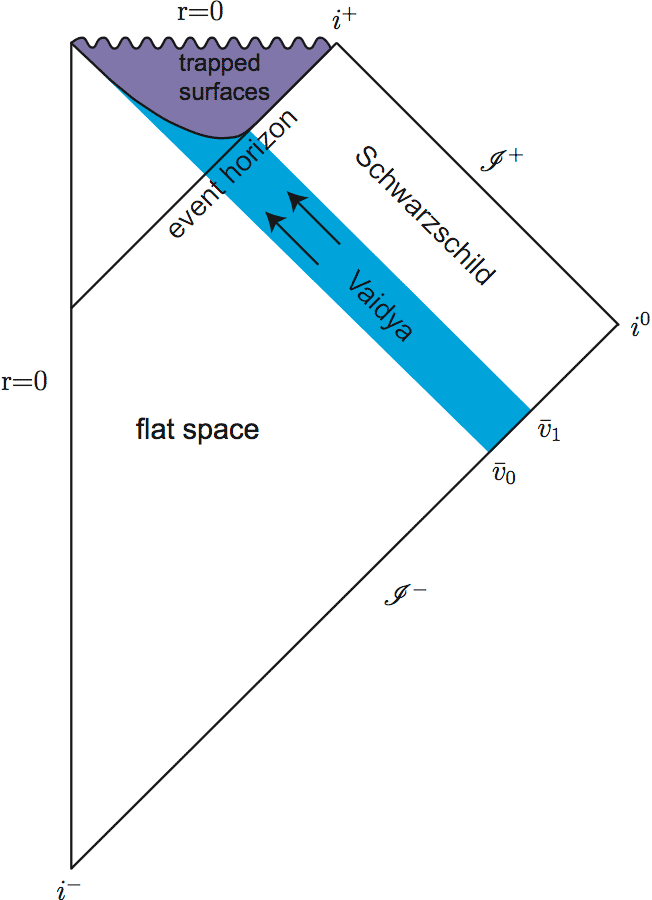}
    
  \caption{This figure shows a Penrose diagram of null dust collapse. For $\bar v \le \bar v_0$
  the spacetime is flat. In the interval $\bar v_0 \le \bar v \le \bar v_1$ the ingoing massless
  radiation described by the Vaidya metric forms a black hole. For $\bar v \ge \bar v_1$ the inflow of
  matter has stopped and this part of spacetime is isometric to the exterior region of Schwarzschild.
  } \label{fig:figures_Vaidya_collapse_diagram}
\end{figure}

Figure \ref{fig:figures_Vaidya_collapse_diagram} shows an imploding shell of massless radiation (modelled by
the ingoing Vaidya solution) which forms a black hole. For a given value of retarded time, $\bar u$, let the
shell be contained in the interval $[\bar v_0,\bar v_1]$ of advanced time, i.e. let $m(\bar v) = 0$ for
$\bar v \le \bar v_0$ and $m(\bar v) = M$ for $\bar v \ge \bar v_1$. Assume that in the interval 
$[\bar v_0,\bar v_1]$ the mass-function is chosen such that no shell focussing singularity appears \cite{Kuroda84}.
\footnote{According to Kuroda\cite{Kuroda84}, possible choices of mass-function that do not generate 
a shell focussing singularity are $m(\bar v) \sim \mu \bar v^n$, where $n<1$, or $n=1$ and $\mu > 1/16$.} 
For $\bar v < \bar v_0$ observers are unaware of the infalling shell of radiation. For late times 
$\bar u$ observers close to $r=0$ will experience tidal stresses and will begin to fall inwards. 
By that time, they are already trapped in the event horizon. The apparent horizon of the Vaidya 
spacetime is always located at $r = 2m(\bar v)$ \cite{Poisson04}.

\subsection{QNM Modes in a Vaidya Background}
The analysis of QNM for time-dependent situations, such as accretion processes or Hawking radiation
is discussed for the Vaidya metric by Abdalla, Chirenti, and Saa in \cite{AbdallaChirentiSaa06}. 
Building on earlier work by Girotto and Saa\cite{GirottoSaa04}, they
employ a semi-analytical approach in which one considers the Vaidya metric in 
double-null coordinates $(u,v)$ \emph{ab initio}.
After choosing a smooth mass-function $m(v)$, it is possible to numerically reconstruct the metric
functions. Then, scalar (or electromagnetic) perturbations on the constructed Vaidya background 
are evolved using a characteristic algorithm. Analysis of the perturbation field close
to the event horizon then yields estimates for the real and imaginary parts of the QNM.

The authors point out the existence of a \emph{stationary adiabatic regime} where the real part 
of the QNM varies inversely with the mass-function, just as one would obtain when modelling the 
Vaidya solution as a series of Schwarzschild slices with changing mass as we have done in
section \ref{sec:QNMs}.
In addition, they formulate a heuristic criterion that indicates
when to expect the appearance of non-stationary behavior:
\begin{equation}\label{eq:QNM-nonstationarity}
|m''(v)| \gtrapprox |\imag\omega(v)|  
\end{equation}
The non-stationary behavior manifests itself as an \emph{inertial effect} in the real part of 
the QNM: $\real\omega(v)$ no longer behaves as $1/m(v)$, as one would expect for a stationary
adiabatic regime.

In the following, we apply their criterion for non-stationary behavior to our numerical results.
Note that we measure the QNM at $\Scri^+$ and we are dealing with a monotonically decreasing
function, the Bondi mass $M_B(u_B)$.
\footnote{The DICE code from which these results have been calculated cannot penetrate apparent
horizons, and, at least for critical collapse, we have not been able to read off QNM close 
to the event horizon.}
In section \ref{sec:QNMs} we have observed that the real parts of the fundamental scalar field QNM 
roughly agrees with its Schwarzschild value (for a decreasing mass) during critical collapse.
We have verified this (see figures \ref{c_QNM}, \ref{T0_QNM}) by comparing the half-periods measured 
from one extremum to the next in $c(u_B)$ with the half-periods predicted by perturbation theory 
with a strongly changing background mass $M_{bg}(u_B)$.

\begin{figure}[htbp]
  \centering
  \includegraphics[width=\textwidth,trim=30 0 0 0]{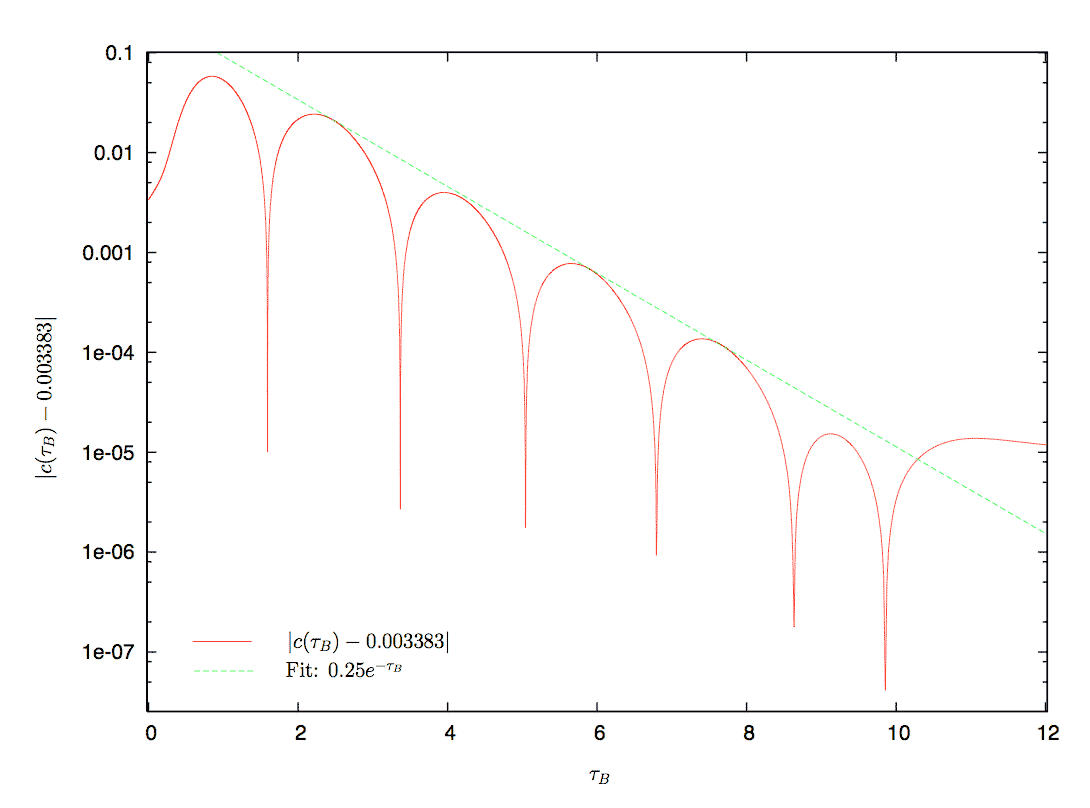}
  
  \caption{The monopole moment of the scalar field decays in accordance with self-similar scaling.
  We have corrected a small offset in $c$ and plot $|c(\tau) - 0.003383|$.}
  \label{fig:figures_c_uB_corr_tau_fit}
\end{figure}

We have also tried to read off the imaginary parts of the QNM from $c(u_B)$, but to no avail:
One the one hand, this is very difficult, because we do not have many oscillations to accurately 
measure the decay.
But, more importantly, as shown in figure \ref{fig:figures_c_uB_corr_tau_fit}, $c(\tau)$ behaves 
as $e^{-\tau}$. This is due to self-similarity, since the monopole moment $c$ has dimension of length. 
We note that we have corrected a slight offset in $c(u_B)$.

\begin{figure}[htbp]
  \centering
  \includegraphics[height=3in]{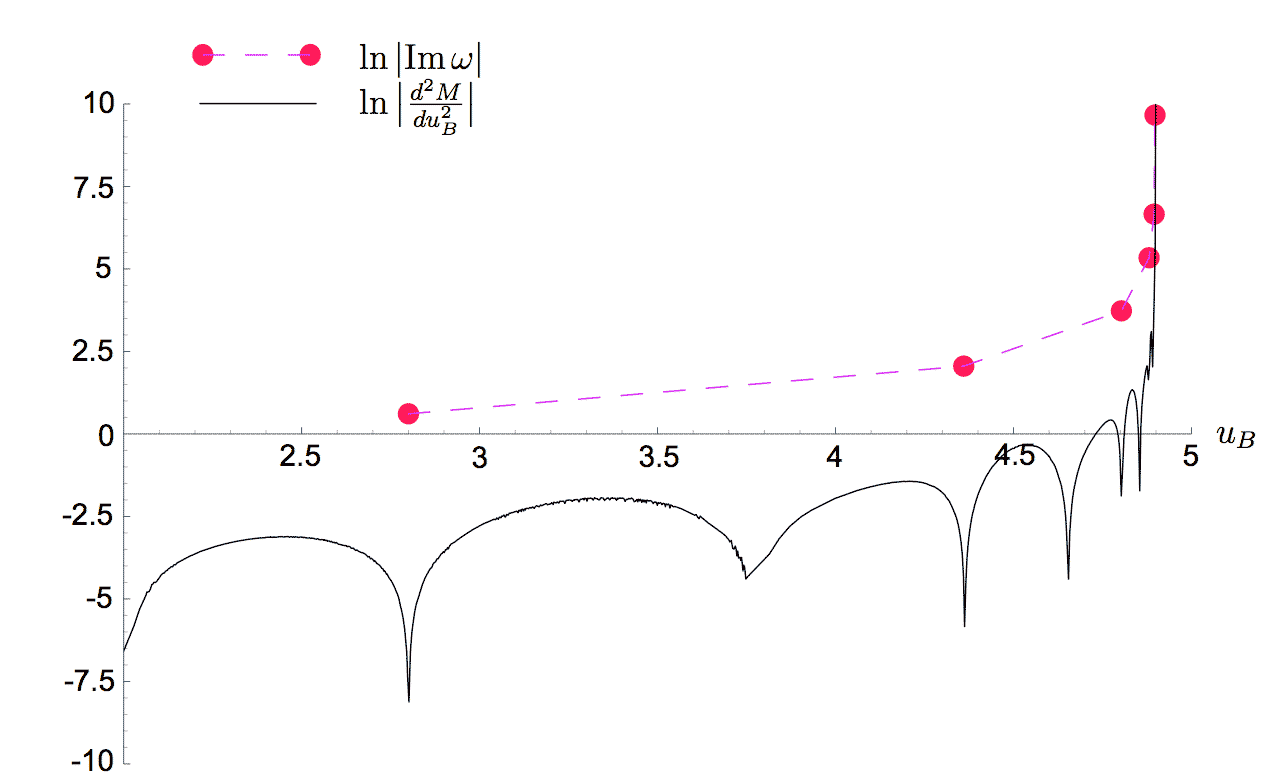}
  
  \caption{This figure checks that the non-stationarity criterion for QNM for critical collapse 
  simulated by the DICE code. The connected filled circles represent $\ln |\omega_{\bf I}|$ 
  at the extrema of $c(u_B)$, while the lower curve shows $\ln\left|d^2 M/du_B^2\right|$.
  With the exception of the very late stages of collapse we are in the stationary regime.
  Figure \ref{fig:figures_ddm_omega_zoomed} shows a detail of this plot.}
  \label{fig:figures_ddm_omega}
\end{figure}

\begin{figure}[htbp]
  \centering
  \includegraphics[height=3in]{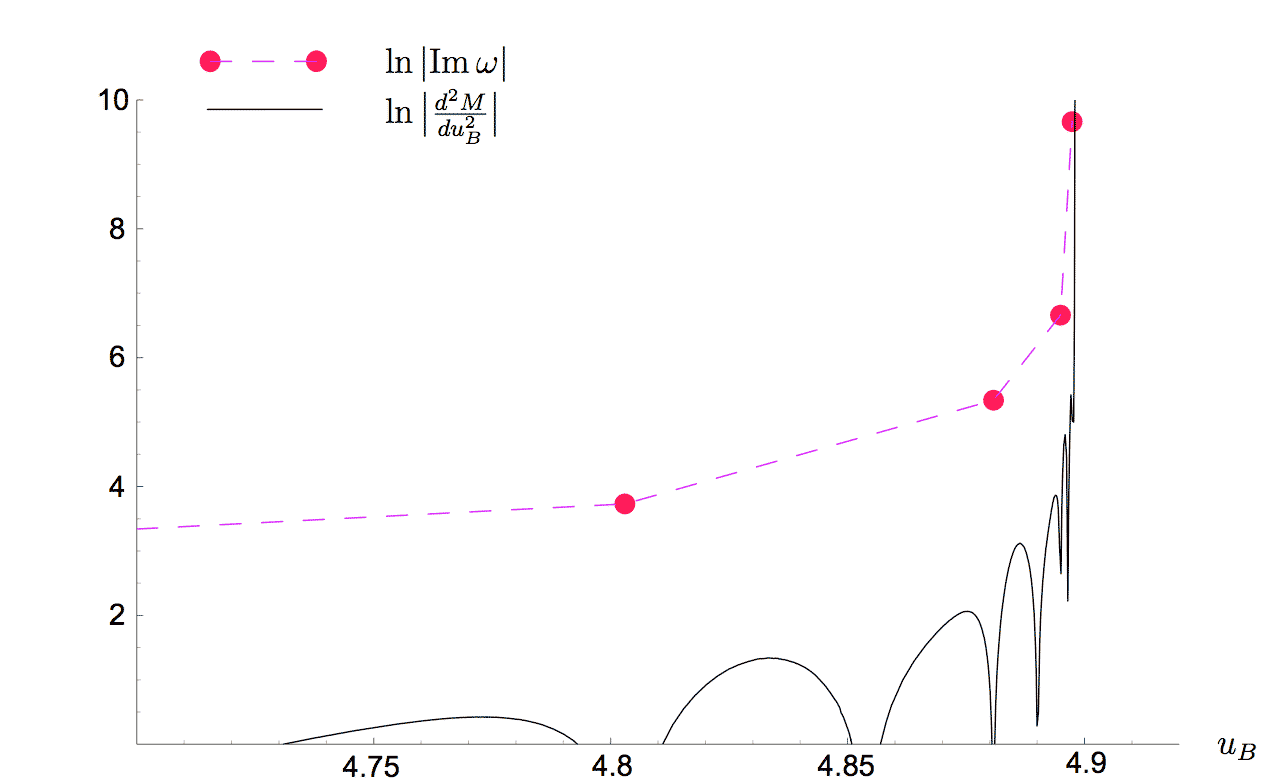}
  
  \caption{This figure shows a detail of \ref{fig:figures_ddm_omega}}.
  \label{fig:figures_ddm_omega_zoomed}
\end{figure}

Instead of directly reading off the imaginary part of the QNM, we can still estimate the 
decay from perturbation theory for a few background masses and compare the result to 
$d^2 M/du_B^2$ in order to check that the non-stationarity criterion mentioned above, 
\eqref{eq:QNM-nonstationarity} is not fulfilled.
As can be seen in the figures \ref{fig:figures_ddm_omega} and \ref{fig:figures_ddm_omega_zoomed}, 
stationarity is well satisfied with the exception of the very late stages of collapse, where we 
leave the self-similar regime and approach black hole formation.

\section{Power-law Tails}\label{sec:tails}

As has been established by Price \cite{Price72}, perturbation fields outside
of Schwarzschild black holes die off with an inverse power-law tail at late times.
In contrast to quasinormal modes this behavior does not depend on the details of the
collapse process, but only on the asymptotic falloff of the effective potential;
(i.e. in a curved spacetime, wave propagation is not confined to the light cones,
rather waves spread inside the light cones, due to scattering off spacetime curvature.)
Therefore, tail phenomena can be observed independently of the endstate of the evolution.
For Gaussian initial data the Newman-Penrose constant \cite{Gomez94b}
vanishes and for late times perturbation theory \cite{Gundlach94a,Gundlach94b}
predicts that the field falls off as 
$\phi \propto \uB^{-2}$ near $\Scri^+$ and $\phi \propto \uB^{-3}$
near timelike infinity $i^+$.

To be more precise \cite{Barack99a,Barack99b}, consider a distant static 
observer at $r=const$ and Bondi time $\Delta \uB$ elapsed since the ``main pulse'' 
of radiation has reached the observer (the duration of the main pulse
has been assumed negligible in \cite{Barack99a,Barack99b}). 
Null infinity is then found to be approximated by the 
region where $\Delta \uB \ll r$ within the context
of a perturbative analysis of tail behavior \cite{LeaverJMP,LeaverPRD,Barack99a}.
This regime has been termed the ``astrophysical zone'' by
Leaver \cite{LeaverJMP,LeaverPRD,Barack99a}.

In addition, the convergence of the perturbation expansion in \cite{Barack99a,Barack99b} 
requires $\Delta \uB \gg M$, where $M$ is the mass of the background. 
In our case, the mass which gives rise to the effective potential is bounded 
from above by the mass of the initial (ingoing) pulse, $M_i$. Therefore, we 
demand that $\Delta \uB \gg M_i$.
Numerically, we observe tails only for $\Delta \uB > 10^3 M_i$.

\begin{figure}
  \centering
  \includegraphics[width=\textwidth]{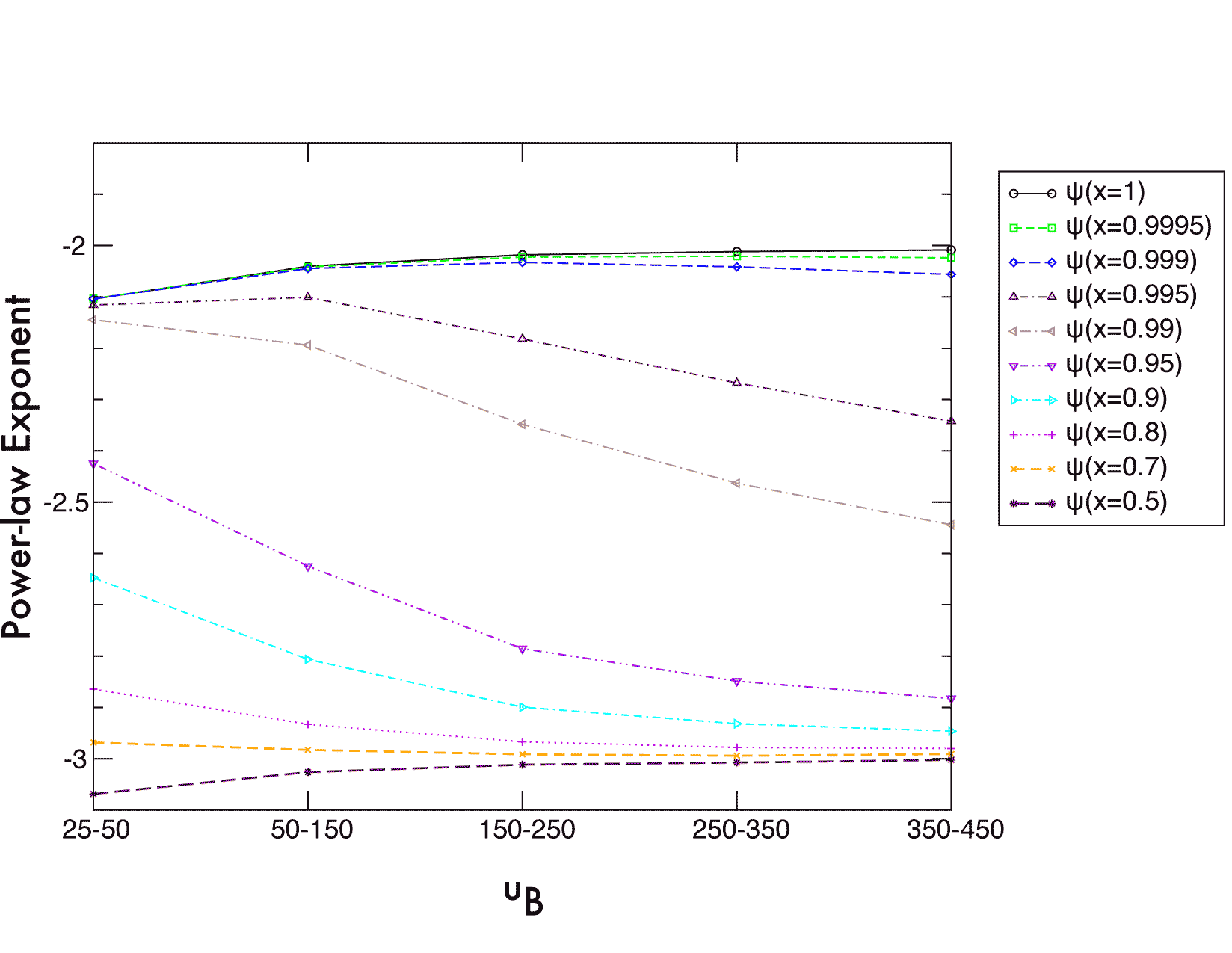}

  \caption{\label{fig:power-law}This figure shows power-law exponents for a 
  subcritical evolution and illustrates the domains of validity of the 
  predictions of perturbation theory for the two zones: $-2$ near 
  $\Scri^+$ and $-3$ near $i^+$. 
  }
\end{figure}

\begin{figure}
  \centering
  \includegraphics[clip,width=.8\textwidth]{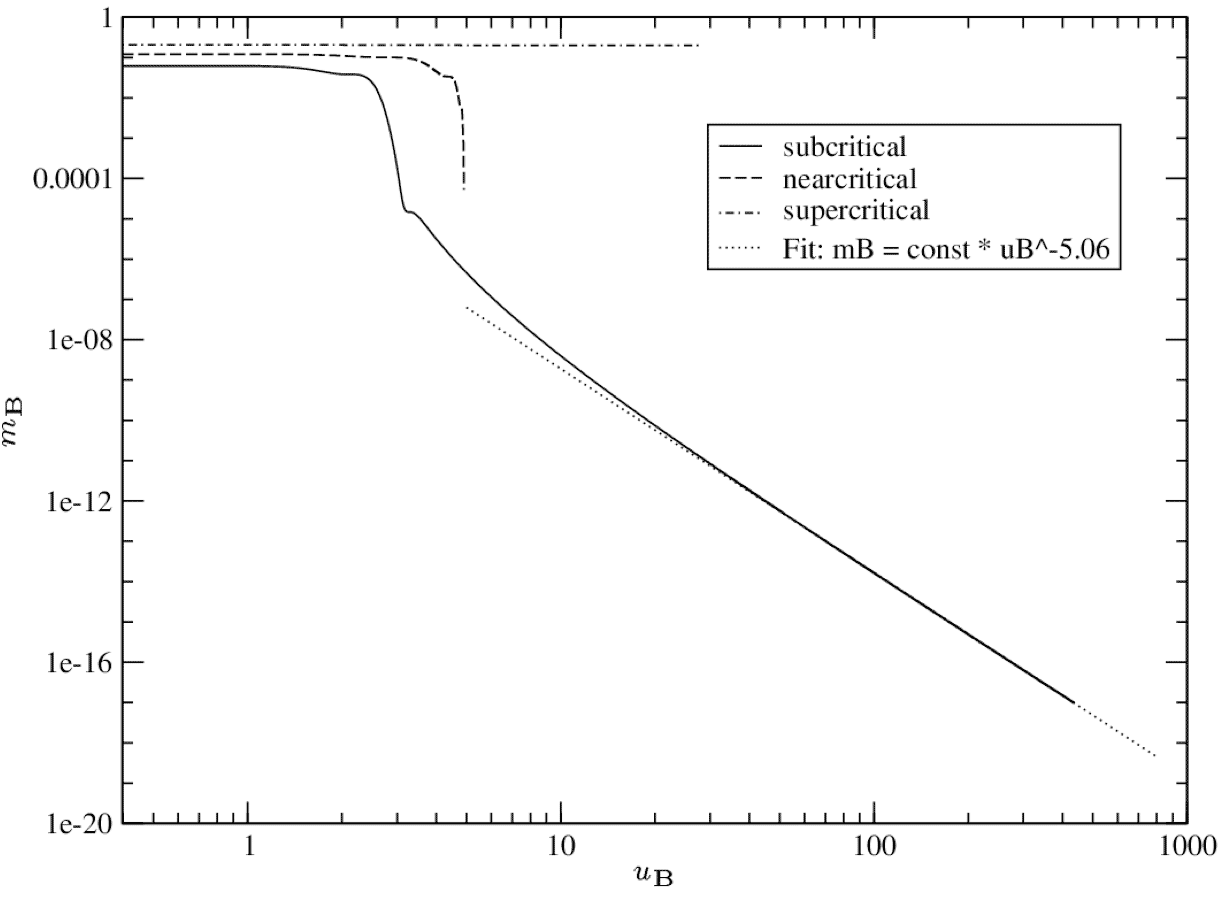}

  \caption{\label{fig:m_tail}We compare the decay of the Bondi mass in supercritical, 
  near--critical and subcritical evolutions. In the subcritical case, the Bondi mass 
  is found to decay for late times with a power-law exponent of approximately $-5$.}
\end{figure}

Closeness to timelike infinity, on the other hand, demands $\Delta \uB \gg |r_*|$,
where $r*$ is the usual ``tortoise'' coordinate $r_* = r + 2M \ln \bigl( \frac{r-2M}{2M} \bigr)$.
In figure \ref{fig:power-law} we show power-law exponents determined
by fits of $\psi$ at different $r=const$ curves over a series of time intervals for a 
subcritical evolution.
As described in section \ref{sec:algorithm} we use spline interpolation to 
calculate $\psi(x=const)$.
The exponents have been determined by fitting the field $\psi$ at $x=constant$ 
against a power of $\uB$ in 5 distinct time intervals of the evolution.
The domains of validity of the exponents predicted by perturbation theory, $-2$ near 
$\Scri^+$ and $-3$ near $i^+$, can be observed here. 
It is clear that the outermost gridpoints in this evolution (using 10000 gridpoints) 
are indeed located in the ``astrophysical zone'' since $r(x=0.9995) \approx 2000 \gg \Delta \uB$,
where $\Delta \uB$ is the Bondi time elapsed since the main pulse of radiation has reached
the observer at about $\uB \approx 3$ (see figure \ref{fig:m_tail}). 
On the other hand, closeness to timelike infinity $i^+$ demands that
$\Delta \uB \gg r_*$. Note that $r_* \approx r$ for $r \gg 2M_i$, where $M_i \approx 0.06$ 
is the initial Bondi mass. 
It is also apparent that for $\Delta \uB \approx r$ the observers are in between the two zones
and the power-law exponents seem to change smoothly.

Figure \ref{fig:m_tail} displays the power-law decay, with exponent $-5$, of the Bondi mass
for the same subcritical evolution.
This behavior can be explained by integrating the Bondi mass-loss equation \eqref{eq:mass-loss-uB} 
with $c \propto \uB^{-2}$ in the regime of power-law tails.

Note that in situations involving realistic sources and realistic detectors, 
power-law tails only play a minor (if any) role as an actual signal, but the power-law
tail results still can provide hints on how to resolve the very important question
of whether null infinity is a useful idealisation for gravitational wave detectors.
Accordingly, we suggest to generalize the term ``astrophysical zone'' to
more general, non-perturbative situations, by re-enterpreting -- in a very loose sense --
$\Delta \uB$ as a suitably chosen large time scale characteristic of the source.
The physical idea is that the distance from observers of astrophysical phenomena, 
e.g. gravitational wave detectors, to the radiation sources is very large compared 
to the time during which substantial radiation from the source can be observed.
This is at least expected for sources where general
relativity is important, as opposed to problems for which a (Post-) Newtonian approach
and the quadrupole formula are sufficient. An example would be a binary black hole  
merger in another galaxy, which might
have a characteristic dynamical time scale of a fraction of a second, and which 
might be observed during several thousand cycles, including a portion of the QNM 
ringdown.

\newpage

\chapter{Numerical Evidence for Critical Phenomena}\label{ch:crit-phen-results}

In this chapter we present numerical results of critical behavior in the
Einstein-massless scalar field system computed using the DICE code and 
the double-null code.

Section \ref{sec:DICE-results} focusses on results from the compactified dice 
evolution code which stress the global observability of critical phenomena at 
future null infinity. We also show results from uncompactified evolutions for cases
in which we did not have compactified results available. 
We would like to emphasize that for the local detection of discrete self-similarity, 
inside the past self-similarity horizon, as discussed in section \ref{sec:detection-of-DSS},
both DICE codes are basically equivalent, the uncompactified DICE code having a slight 
edge over the compactified code as to resolution.

The results of the double-null code shown in section \ref{sec:dn-results} 
complement the local picture given by the DICE code. The double-null code is
uncompactified but, in contrast to the DICE code, it can penetrate apparent 
horizons and thus enables us to probe supercritical spacetimes a little longer. 
The reason for introducing this code was mostly to analyze the fate of the 
exterior mass that accumulates outside of the past self-similarity horizon 
in near-critical evolutions, as will be discussed in section \ref{sec:mass_news}.
Moreover, it is always preferable to have more than one way of computing a 
numerical result, which, by comparison, allows us to rule out numerical 
artifacts specific to a particular numerical scheme.

\section{Results from the DICE Code}\label{sec:DICE-results}

\subsection{Identification of Critical Behavior}\label{sec:crit_behavior}

We consider one-parameter families of initial data $\phi = \phi_p(u_0,x)$,
such that for small values of $p$ we have dispersion, while for large
values of $p$ we have black hole formation. 
It has been found numerically \cite{Choptuik92,Choptuik93} that, for any initial 
data family considered, the evolution of near-critical data approach a 
universal DSS critical solution.
In this thesis, we assume universality and restrict ourselves
to Gaussian-like initial data
\begin{equation}\label{Gaussian-ID}
\phi(u_0,x) = A \, r(x)^2 \exp \Bigl[ - \bigl(\frac{r(x) - r_0}{\sigma}\bigr)^2 \Bigr],
\end{equation}
where $r(x) = \frac{x}{1-x}$. This choice makes it easy to compare 
compactified to uncompactified evolutions using the same initial data.
All results from the compactified DICE code presented here use 
$r_0 = 0.7$ and $\sigma = 0.3$ and a radial resolution of $10000$ gridpoints,
unless stated otherwise.
The criticality parameter $p$ is identified with the amplitude $A$. 

As has become common practice, we find near-critical data through
a \emph{bisection search} in $p$. This procedure yields in particular
a numerical approximation to the critical value $p=p*$
which defines the threshold of singularity formation.

In the bisection search, a number of criteria are possible to distinguish
dispersion from collapse. Their equivalence can be checked a posteriori 
when comparing the final result, i.e. subcritical and supercritical solutions
close to criticality.
A typical criterion is to monitor the ratio $2 m/r$, where $2 m/r = 1$
signifies the presence of an apparent horizon. This criterion has been
used successfully also in combination with slicing conditions, that do not
penetrate apparent horizons -- as is the case in our approach.
Remarkably, in practice it turns out, that  $2 m/r \geq 0.6$ is a
sufficient criterion to mark a scalar field evolution as supercritical.
For practical and historical reasons, this is the approach we have adopted 
for our code.

A number of other options come to mind, in particular in our context of 
evolving out to null infinity, one could e.g. monitor the redshift 
\eqref{eq:redshift-factor} or Bondi mass.
In a dispersion evolution, the redshift will decay to zero, while it will 
approach infinity when a black hole forms.
Similarly, the Bondi mass will decay to zero with a characteristic 
tail behavior, as is shown in Fig. \ref{fig:m_tail}, in
a dispersion evolution, and will asymptote to a (positive) constant 
when the field does not disperse.

\begin{figure}
  \centering
  \includegraphics[width=.7\textwidth]{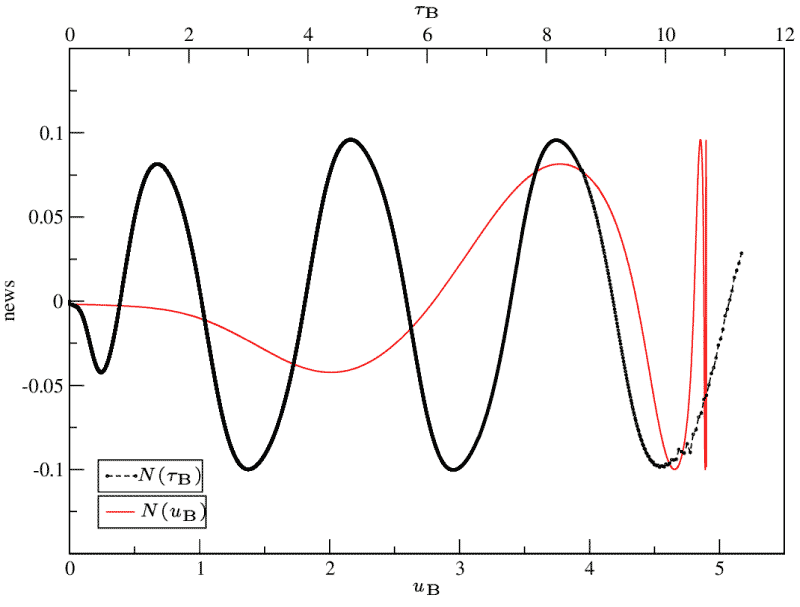}
  \caption{\label{fig:news}We show the news function $N(u)$, first as a function of
  the natural time coordinate $\uB$ of an asymptotic observer 
  and also as a function of a suitably adapted time
  $\tauB = -\ln \frac{\uB^* - \uB}{\uB}$ where $N(\tauB)$ is periodic 
  with period $\Delta \simeq 3.44$ after the spacetime has come close to the
  critical solution.
  Even if the constant $\uB^*$ is not known, it can be determined by
  a fit to periodicity in $\tauB$. Thus, it is possible to observe
  DSS at $\Scri^+$ and to extract the critical exponent $\Delta$.
  }
\end{figure}

The local detection of discrete self-similarity and extraction of the critical
echoing parameter $\Delta \simeq 3.44$ from a near-critical numerical evolution is
discussed in section \ref{sec:detection-of-DSS}. 
Here we are concerned with the global picture:
In the course of a near-critical evolution, remnants of the self-similar dynamics 
which occur locally, inside the SSH, are radiated to future null infinity. 
Remarkably, we observe that the imprints of DSS behavior are still present in 
asymptotic quantities such as the news function and the Bondi mass
(see figures \ref{fig:news} and \ref{fig:cnews-mass-compare-mext}).

\begin{figure}
  \begin{center}
    \includegraphics[angle=0,width=\textwidth]{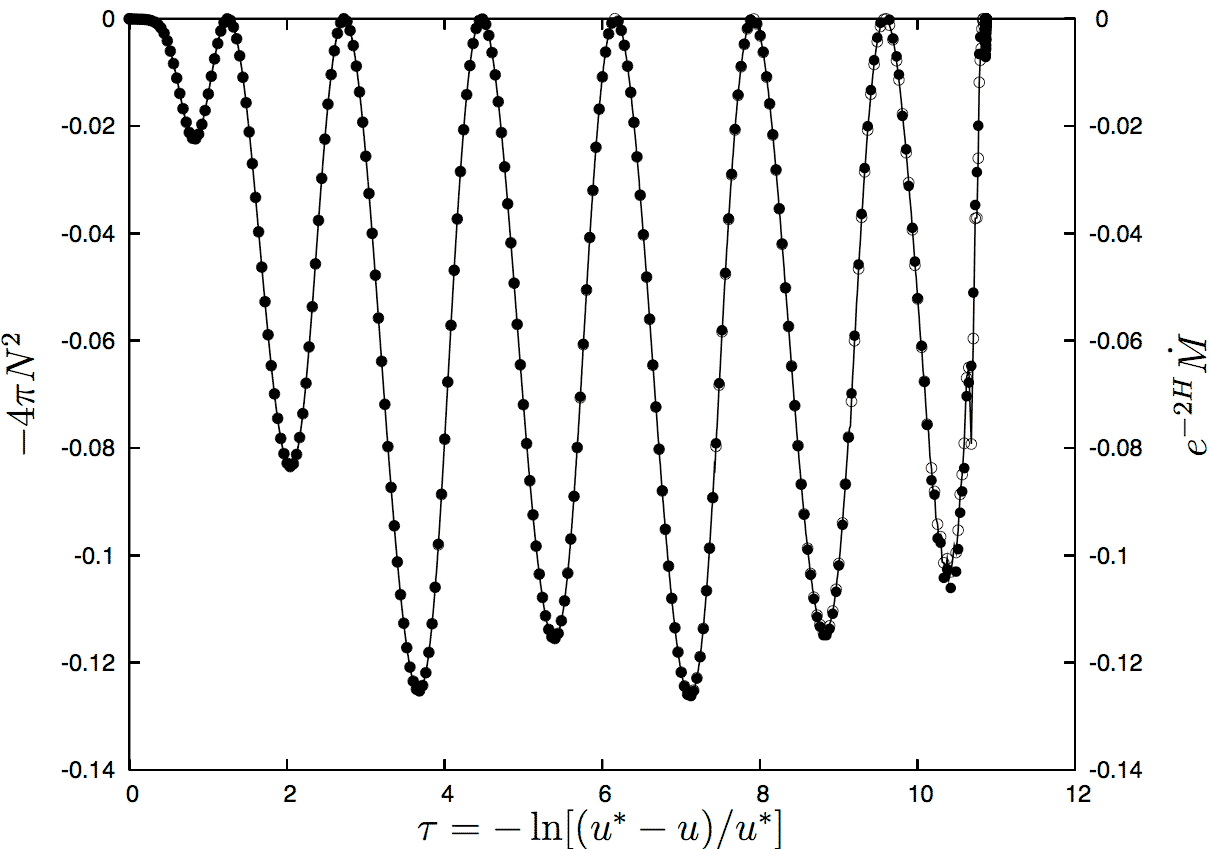}

    \caption{This figure superimposes the left and right hand sides
      (empty and filled circles, respectively) of equation
      \eqref{eq:mass-loss-uC}, the Bondi mass-loss equation, and plots them
      against adapted time $\tau$ for a very near critical evolution.
      The data have been sampled in order that individual data points
      are visible. Until $\tau = 9.0$ the empty and filled circles are
      indistinguishable in the plot, i.e. the mass-loss equation holds
      with a high degree of accuracy. Around $\tau = 11.0$ a small
      black hole is starting to form, the resolution has been exhausted 
      and numerical results are unreliable.
      It is apparent that the bursts of scalar field radiation which
      escape to $\mathscr I^+$ are periodic in $\tau$ with the same
      period $\Delta/2$ which is found in the DSS behavior of fields
      near the origin. 
    }\label{fig:Bondi-mass-loss-DSS-Scri}
  \end{center}
\end{figure}

\begin{figure}
  \begin{center}
    \includegraphics[angle=0,width=0.7\textwidth]{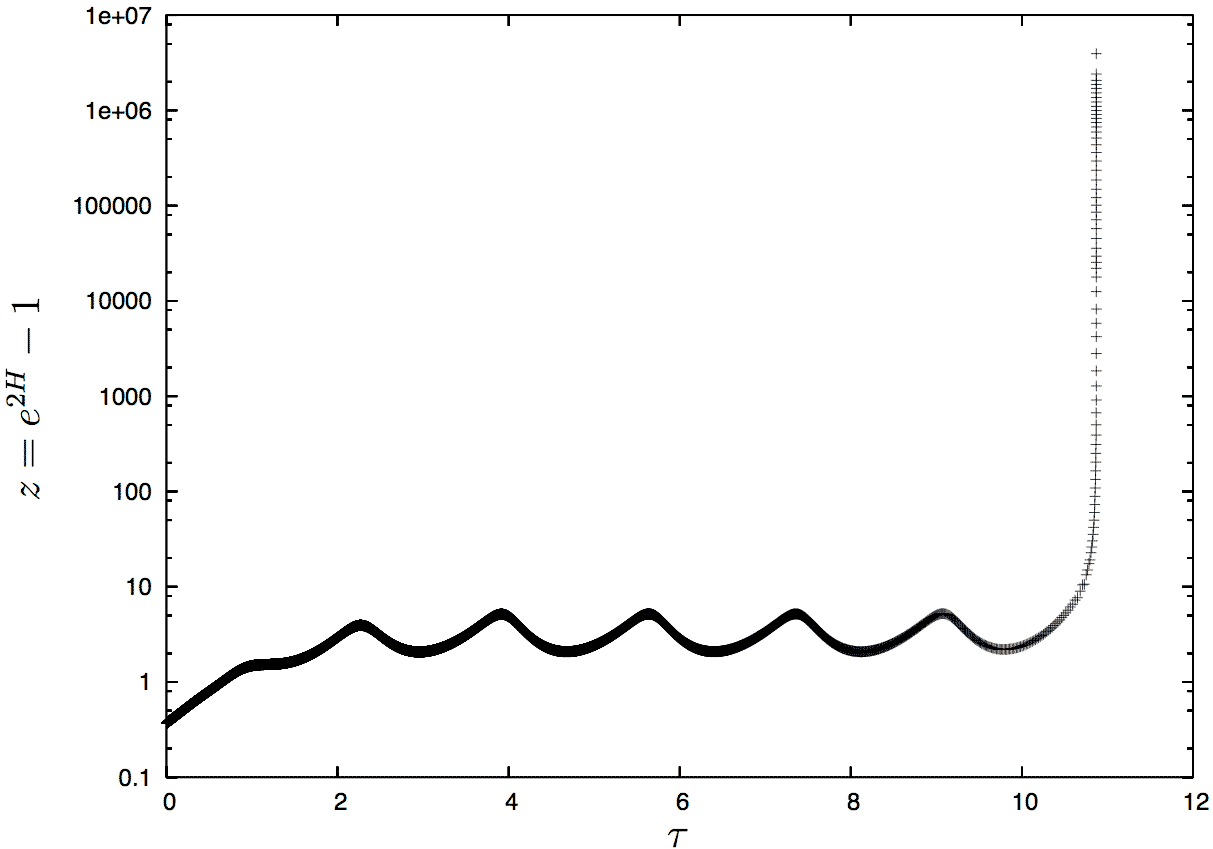}

    \caption{This figure shows the redshift factor $z=e^{2H}-1$ for
      the compactified evolution depicted in figure \ref{fig:Bondi-mass-loss-DSS-Scri}.
      Variations in the density ($2m/r$) of the discretely self-similar scalar 
      field subject the outgoing light rays to a varying amount of focussing.  
      When a black hole starts to form around $\tau = 11.0$, the redshift factor 
      diverges exponentially. 
      }\label{fig:dice-cmp-DSS-redshift-factor}
  \end{center}
\end{figure}

To further test the accuracy of the compactified DICE code in the asymptotic
regime, we have plotted the left and right hand sides of the Bondi mass-loss
equation, \eqref{eq:mass-loss-uC}, in figure \ref{fig:Bondi-mass-loss-DSS-Scri}.
Figure \ref{fig:dice-cmp-DSS-redshift-factor} shows the redshift factor, equation
\eqref{eq:redshift-factor}, for the same evolution.

\begin{figure}
  \centering
  \includegraphics[angle=0,clip,width=0.7\textwidth]{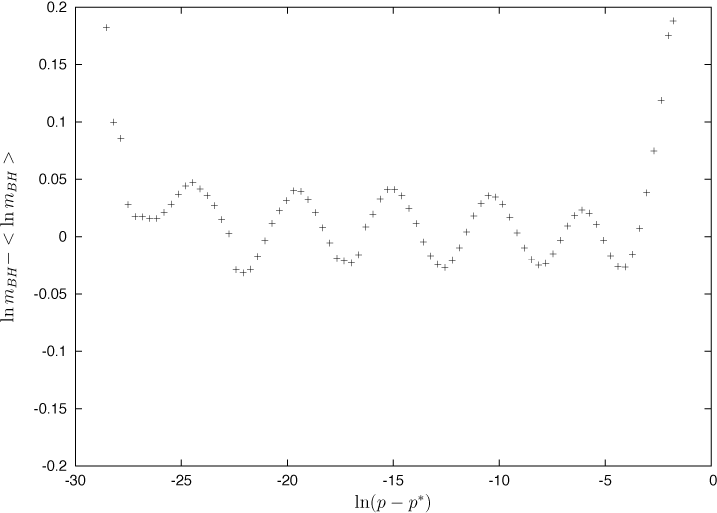}

  \caption{
    \label{fig:fine-structure}
    The fine-structure in $\mBH$ after subtracting a linear fit. The measured period $4.6$
    is close to the value predicted by perturbation theory 
    $\frac{1}{2}\Delta/\gamma \approx 4.61$.
  }
\end{figure}

\begin{figure}
  \begin{center}
    \includegraphics[angle=0,width=0.7\textwidth]{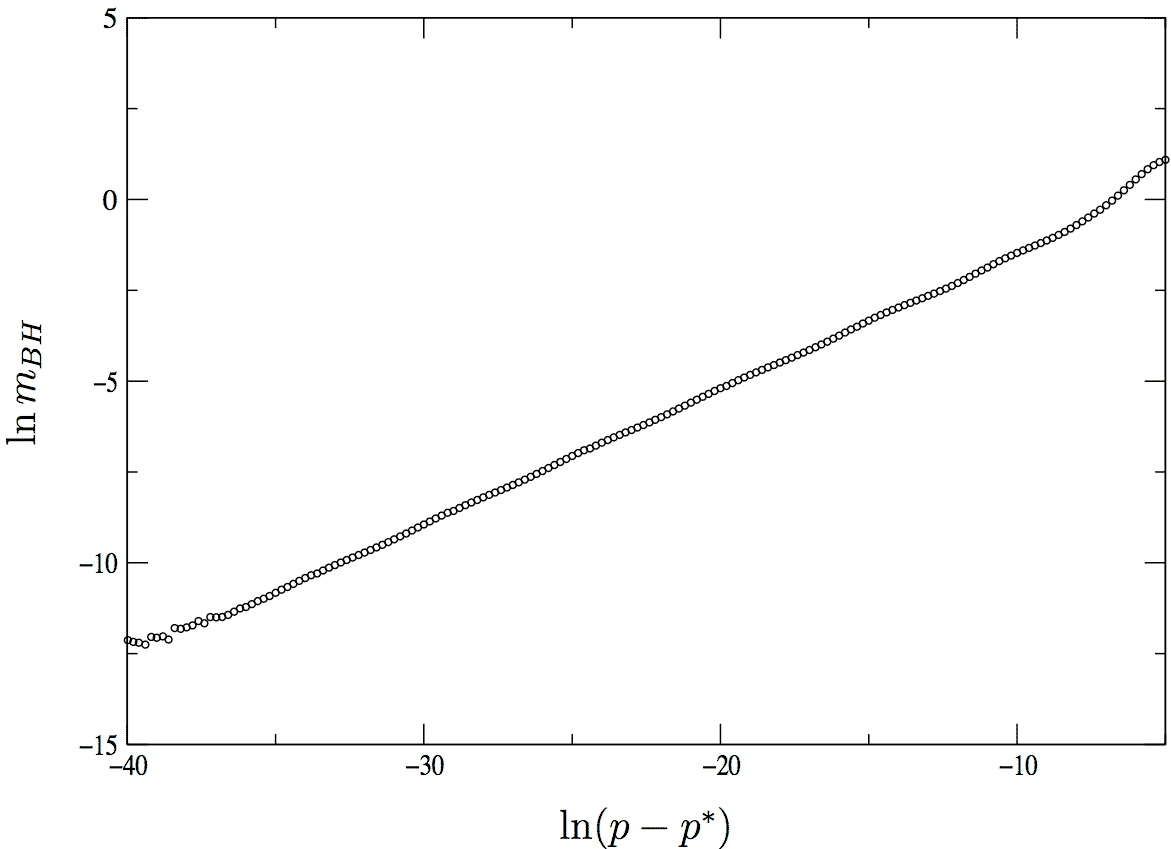}

    \caption{
    This figure shows the mass-scaling law for a series of uncompactified
    evolutions at 5000 gridpoints radial resolution. $\ln m_{BH}$ is
    well fit by a straight line with slope $\gamma = 0.373$.
    For $\ln(p-p^*) \lesssim -35$ which corresponds
    to $p-p^* \lesssim 10^{-15}$ our code reaches the precision limit for
    double precision floating point numbers. On the other hand, for
    $\ln(p-p^*) \gtrsim -10$ we are leaving the range of validity of linear
    perturbation theory which predicts the simple power-law behavior of the mass.}
    \label{fig:dice-uncompactified-mass-scaling}
  \end{center}
\end{figure}

In addition, in numerical evolutions of supercritical data, the black hole mass
has been found to exhibit a universal scaling law \eqref{eq:scaling-fine-structure},
in accordance with the discussion in section \ref{sec:scaling-law-derivation},
(see \cite{Hod97,Koike95,Gundlach97f}):
\begin{equation}
\ln \mBH = \gamma \ln(p-p^*) + \Psi(\ln(p-p^*)) + const,
\end{equation}
where $\gamma \approx 0.373$ and the function $\Psi$ is periodic with 
period $\frac{1}{2}\Delta/\gamma$
in $\ln(p-p^*)$.
Figure \ref{fig:dice-uncompactified-mass-scaling} shows this scaling
behavior of the mass.
To analyze the fine-structure of the mass-scaling law, one
conveniently runs a series of evolutions with masses spaced linearly
in $\ln(p-p^*)$.
One also needs to ensure to measure the mass of the outermost component 
of the horizon in order to obtain the fine-structure of the scaling law 
shown in figure \ref{fig:fine-structure}.

This scaling law has been derived by arguments based on linear
perturbation theory around the  critical solution and dimensional analysis
however without using a precise definition of the black hole mass 
\cite{Hod97,Gundlach97f}.
A typical approach in numerical simulations based on coordinates that do not
penetrate apparent horizons, as far as we are aware, is
to follow a peak in $2m/r$ until this quantity almost reaches unity, at which
point the simulation is usually slowed down by a CFL-type condition, and then
read off the approximate horizon mass at this point. We follow this heuristic
approach, and are able to reproduce the mass scaling and fine structure.

At least conceptually, this approach is problematic without quantifying how
much mass-energy still remains outside of the peak in $2m/r$ where
the horizon-mass is read off. From the perturbation theory argument,
one expects scaling behavior only for quantities within the self-similar
region of spacetime, which does not necessarily extend beyond the SSH. 
We will return to this point in the discussion section \ref{chp:discussion}.

\begin{figure}
  \begin{center}
    \includegraphics[angle=0,width=0.6\textwidth]{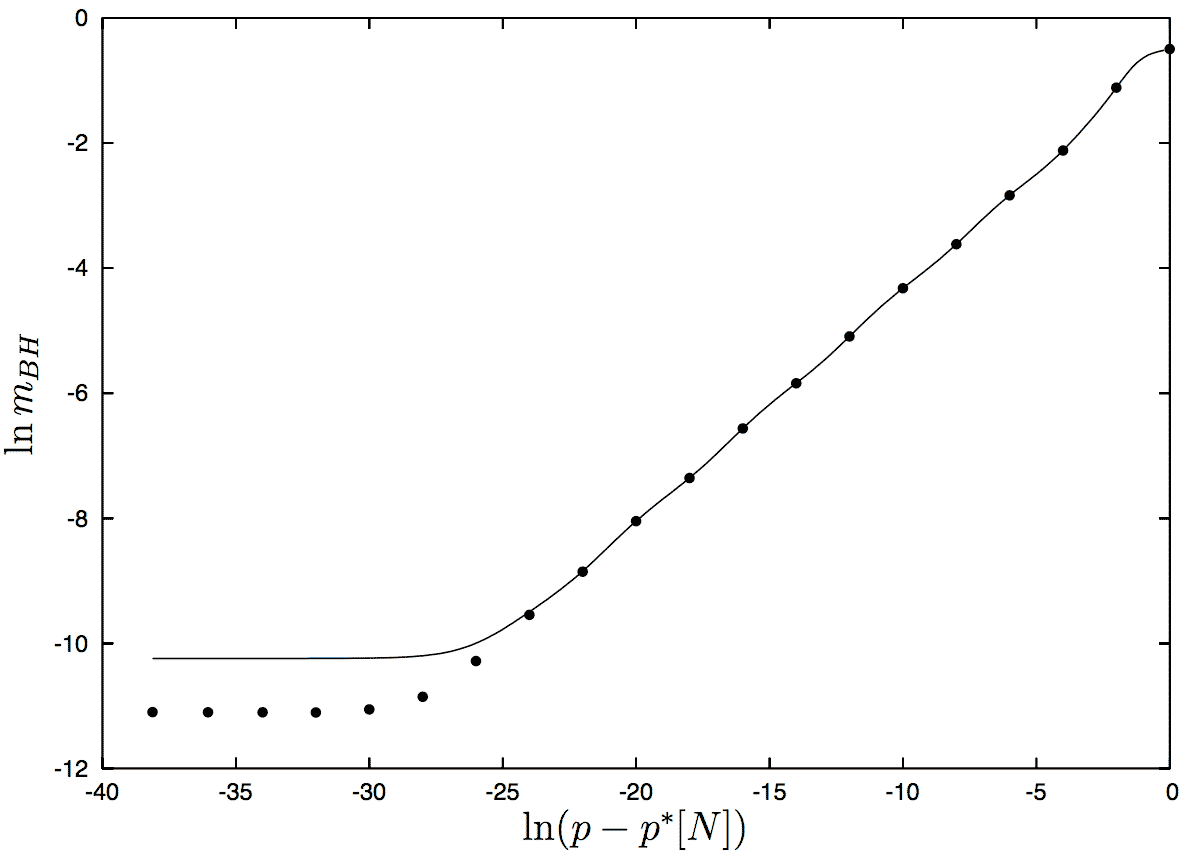}

    \caption{
      This figure shows the mass-scaling law for a series of 200
      compactified evolutions using 8000 gridpoints 
      (depicted as a continuous line) and 20 evolutions using 16000
      gridpoints (depicted as black dots).  Note that as $p^*$ depends on
      the gridresolution $N$, one has to use $p^*[8000]$ and $p^*[16000]$,
      respectively for the two datasets.  
      For $\ln(p-p^*) \lesssim -30$ the
      black hole masses asymptote to a different constant value for each
      resolution.
      For $\ln(p-p^*) \gtrsim -5$ we are leaving
      the range of validity of linear perturbation theory which predicts
      the simple power-law behavior of the mass.  In the intermediate
      range $-25 \lesssim \ln(p-p^*) \lesssim -5$, $\ln m_{BH}$ is well
      fit by a straight line with slope $\gamma = 0.373$.
      }\label{fig-dice-cmp-mass-scaling-8000-16000-level-off}
  \end{center}
\end{figure}

We have found that the mass scaling law for compactified evolutions exhibits 
a "leveling-off" effect for strongly fine-tuned evolutions. 
Figure \ref{fig-dice-cmp-mass-scaling-8000-16000-level-off} shows that
for $\ln \left(p-p^*\right) \lessapprox -30$ the logarithm of the detected
black hole mass is constant, while further away from criticality, the scaling
agrees with the expected power law behavior given in equation 
\eqref{eq:scaling-fine-structure}.
The fact that this constant is resolution dependent, i.e. it decreases with 
increasing resolution, indicates that it is a numerical artifact of the compactified code.

In contrast, uncompactified evolutions generated by the DICE code where the outer grid boundary 
has been adjusted to be close to the ingoing null ray, $v=v^*$, that hits the 
accumulation point, are completely unaffected by this leveling off in the scaling law
(see figure \ref{fig:dice-uncompactified-mass-scaling}).
Nor are codes affected that penetrate the apparent horizon and base their horizon 
detection on the first actual apparent horizon detected on a slice, such as our double 
null code (see figure \ref{fig:dn-mass-scaling} in section \ref{sec:dn-scaling}).
%

Note that for a near-critical solution, the approximate value of the accumulation
time naturally defines an approximate location (i.e. advanced time) of the SSH.

For observers at null infinity, a natural time coordinate is Bondi time
$\uB$ as defined in eq. (\ref{eq:def_BondiTime}) -- Bondi time can be
identified with the proper time of timelike observers
at large distances -- see the discussion in Sec. \ref{chp:discussion}.
In order to analyze critical phenomena, we define
a time coordinate which is suitably adapted to self-similar critical collapse
and set 
\begin{equation}\label{eq:tauB-def}
\tauB = -\ln \frac{\uB^* - \uB}{\uB^*}.
\end{equation}
The adapted time coordinate $\tauB$ can be defined for spacetimes which are
close to the critical solution inside of the SSH, so that 
the value of the accumulation time $\uB^*$ can be determined by
a fit to periodicity in $\tauB$. We have used a fit to periodicity
of the news function -- $N(\tauB)$ is periodic 
with period $\Delta \simeq 3.44$ after the spacetime has come close to the
critical solution (see figure \ref{fig:news}).

Note that $\tauB$ is only an approximate adapted coordinate since it depends on the relation
between Bondi and central time \eqref{eq:def_BondiTime}.
In order to gain some insight into the behavior of $\tauB(\tau)$ consider
the simpler case of a CSS collapse. We assume that $\beta$ changes only little
outside the past SSH, i.e. $H(u) \approx \beta_\SSH(u)$.
Then, since $\beta_\SSH(u) = constant$ in the CSS regime, we find by integrating
\eqref{eq:def_BondiTime} that
\begin{equation}
\uB \approx C u,
\end{equation}
where $C = e^{2\beta_\SSH}$ and we have chosen the initial condition $\uB(0) = 0$. 
Furthermore, it follows from the definition of the adapted time coordinates, 
equations \eqref{eq:tau-z-def} and \eqref{eq:tauB-def}, that
\begin{equation}\label{eq:tauB-tau-rel}
\tauB \approx \tau,
\end{equation}
in the CSS regime.
Numerically, the deviations from this relation for the DSS case turn out to be
quite small.
We cannot expect this relation to hold, if the mass outside the past SSH is large
compared to the inside, which we observe in the very late stages of strongly 
fine-tuned evolutions. Nevertheless, we have still been able to resolve
DSS phenomena using the coordinate $\tauB$ (see fig. \ref{fig:cnews-mass-compare-mext}).

\begin{figure}
  \centering
  \includegraphics[clip,width=.7\textwidth]{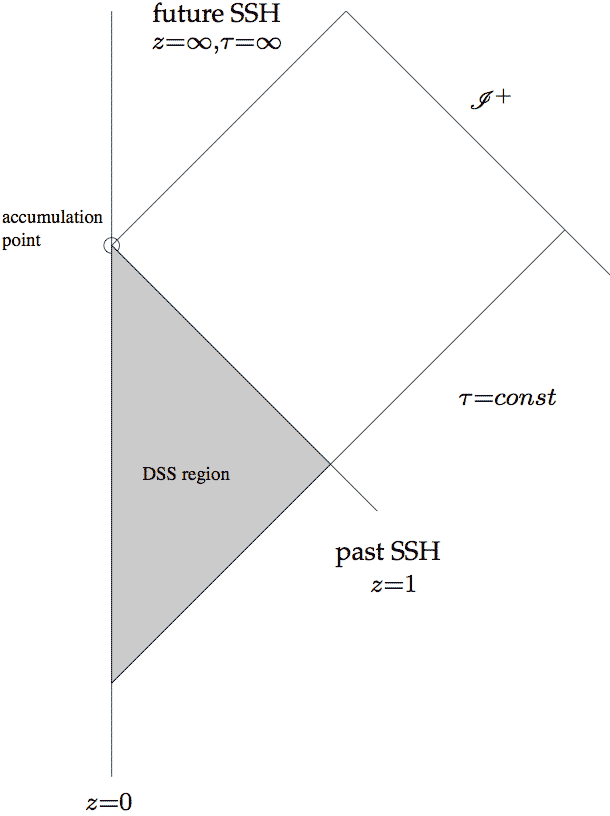}

  \caption{\label{fig:SSH}
  A conformal diagram of a critical collapse spacetime.
  In the backwards light cone of the accumulation point the dynamics are close
  to the DSS critical solution. The lightlike boundary is also called the
  past self-similarity horizon (SSH). Depending on whether the initial data are sub- or
  supercritical, the spacetime will for late times be Minkowski or Schwarzschild.}
\end{figure}

\subsection{DSS Behavior in the Bondi Mass and the News Function}\label{sec:mass_news}
The following argument suggests that the news function is approximately 
periodic in $\tauB$, as shown in figure \ref{fig:news}.
Assume effective DSS data for the scalar field and the metric functions on 
the past self-similarity horizon (SSH) (see figure \ref{fig:SSH}). 
Moreover, we assume that the contribution of the right hand side integral in the wave 
equation for the scalar field \eqref{eq:phi-evo-continuum} can be neglected outside of 
the SSH, such that the DSS data on the SSH are linearly propagated to $\Scri^+$ 
without backscattering, i.e. 
\begin{equation}
\lim_{r\to\infty} \psi(u,r) \approx \psi_\SSH (u).
\end{equation}
Furthermore, assume that changes in $\beta$ outside of the SSH are small, so that
$\beta_\SSH(u) \approx H(u)$.
It then follows that
\begin{equation}
\begin{split}
N(\uB) &= \frac{d c(\uB)}{d\uB}  \approx \frac{d\psi_\SSH(u)}{d\uB}\\
       &= \frac{d}{d\tau}\Bigl[ \phi_\SSH(\tau) \zeta(\tau) e^{-\tau} u^* \Bigr] \frac{e^\tau}{u^*} e^{-2H} \\
       &= e^{-2\beta_\SSH} \Bigl[ \frac{d}{d\tau} \left(\phi_\SSH(\tau) \zeta(\tau) \right)
                                - \phi_\SSH(\tau) \zeta(\tau) \Bigr],
\end{split}
\end{equation}
which shows that the news function $N(\uB)$ is approximately
periodic in $\tau$ (and in $\tauB$ if equation \eqref{eq:tauB-tau-rel} holds) 
with period $\Delta$ and satisfies $N(\tau + n\Delta/2) = (-1)^n N(\tau)$.

In order to determine the behavior of the Bondi mass,
we can then rewrite the Bondi mass-loss equation \eqref{eq:mass-loss-uB} 
\begin{equation}
\frac{d\mB}{d\tauB} = -4\pi N^2(\tauB) \frac{d\uB}{d\tauB} = - 4\pi \uB^* e^{-\tauB} N^2(\tauB).
\end{equation}
Since $N^2(\tauB)$ is $\Delta/2$-periodic in $\tauB$, $\mB$ then takes the following form 
\begin{equation}\label{eq:mBondi-DSS}
\mB(\tauB) \approx e^{-\tauB} f_\SSH (\tauB),
\end{equation}
where $f_\SSH (\tauB)$ is $\Delta / 2$-periodic in $\tauB$.

This behavior mimics the behavior of the mass-function:
We rewrite the mass-function in adapted coordinates $(\tau,z)$, using \eqref{eq:z-def},
\begin{equation}
m(\tau,z) = \frac{1}{2} z e^{-\tau} \zeta(\tau) u^* 
            \Bigl[1 - \Vr(z,\tau) e^{-2\beta(z,\tau)} \Bigr],
\end{equation}
and evaluate it at the SSH, which, by a judicious choice of the periodic function $\zeta(\tau)$ 
can be chosen to be at $z=1$ (since the past SSH is a null surface, one needs to ensure that
$\nabla_a z$ becomes null at $z=1$).
We obtain
\begin{equation}
\mSSH (\tau) = e^{-\tau} f^*_{\scriptscriptstyle\text{SSH}}(\tau),
\end{equation}
where $f^*_{\scriptscriptstyle\text{SSH}}(\tau)$ is periodic with period $\Delta/2$.

\begin{figure}
  \centering
  \includegraphics[clip,width=\textwidth]{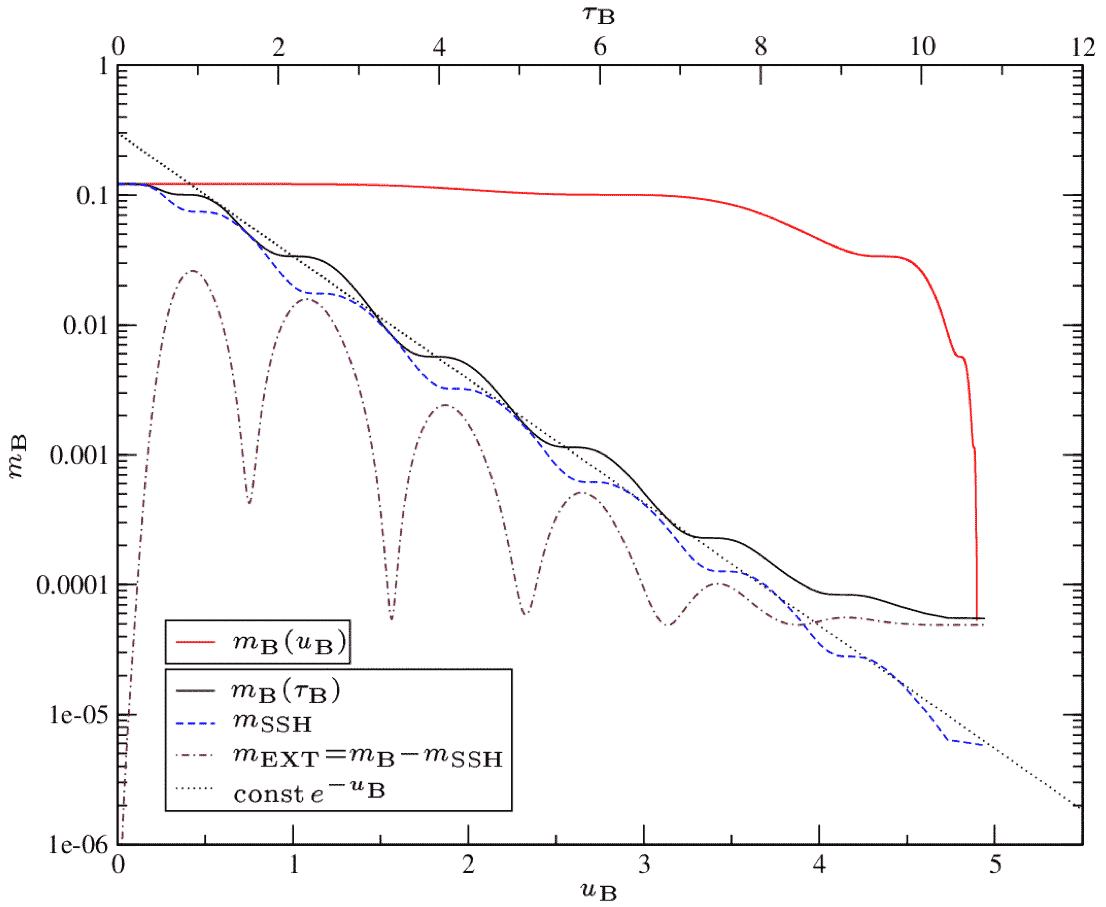}

  \caption{\label{fig:cnews-mass-compare-mext}
  This figure plots the Bondi mass $\mB$ against both $\uB$
  and the adapted time $\tauB$ for a barely supercritical evolution with final
  black hole mass $M_f \approx 5 \times 10^{-6}$. The Bondi mass $\mB$ and the mass at
  the past SSH, $\mSSH$, are found to decrease exponentially in $\tauB$ 
  (with an overlayed $\tauB$-periodic oscillation with period 
  $\Delta /2$), once the evolution has sufficiently approached the critical solution
  near the center of spherical symmetry. 
  We also show $\mEXT$, the energy present outside of the SSH.
  }
\end{figure}

\begin{figure}
  \centering
  \includegraphics[width=\textwidth]{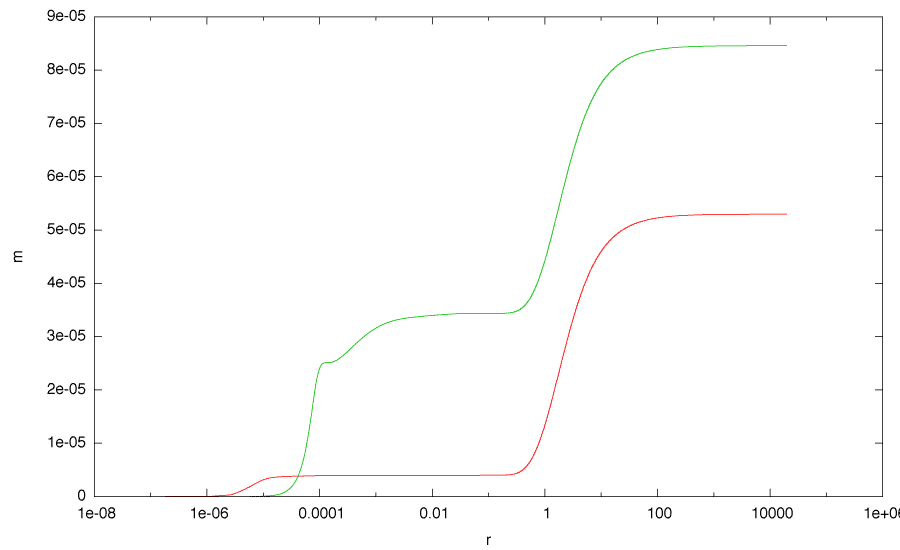}

  \caption{
  Shown is the mass-function on two null-slices very late in the nearcritical evolution
  depicted in fig. \ref{fig:cnews-mass-compare-mext}. Excluding the gridpoint at $\Scri^+$,
  the compactified grid extends to large values of the areal radial coordinate $r$.
  For last slice shown, the location of the past SSH is approximately at $r \approx 0.0001$. 
  The matter exterior to the SSH, $\mEXT$ is mostly concentrated in the radial region 
  $0.5 < r < 100$.}
  \label{fig:m_cmp_vs_logr_mext}
\end{figure}

The periodicity and exponential decay of $\mB$ and $\mSSH$ are confirmed by our 
numerical calculations (see fig. \ref{fig:cnews-mass-compare-mext}).
Note that the Bondi mass levels off at roughly
$10^{-4}$ of the mass in the initial configuration, whereas the mass contained
within the backwards light cone $\mSSH$ continues to scale according to the
prediction of critical collapse evolution, equation \eqref{eq:mBondi-DSS}. 
Since the difference, i.e. $\mEXT = \mB - \mSSH$, 
is almost zero at the initial slice, we conjecture that, at late times, $\mEXT$ 
is the energy due to backscattering in a critical evolution.
We do not have a physical explanation for the oscillations seen in 
$\mEXT$ for earlier times. This behavior may, however, be due to errors in the 
determination of the location of the past SSH.
In figure \ref{fig:m_cmp_vs_logr_mext} it is apparent that $\mEXT$ is localized in a 
region outside the past SSH.

We observe that $\mEXT$ contained in a slice close to horizon formation
is almost constant for different near critical evolutions (down to the
numerical limit of fine tuning).
In the very late stages of the evolution, the growing redshift, $\beta \to \infty$ 
effectively halts the numerical evolution, while the error norm 
$\max |\mathbf{E_{uur}}(\tauB > 10)|$ approaches $10^{-1}$.

\begin{figure}
  \centering
  \includegraphics[clip,width=.6\textwidth]{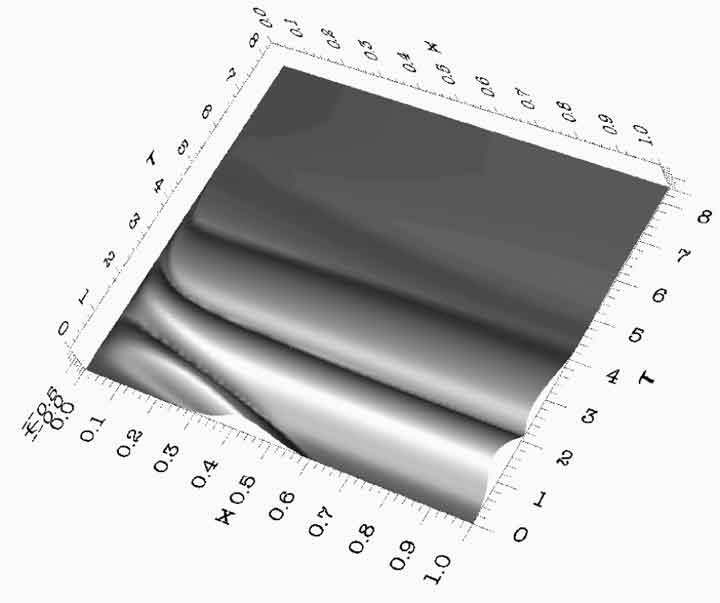}

  \caption{\label{fig:psi-perspective}This figure shows a surface plot 
  of $|\psi(\tau,x)|$ for the same nearcritical  evolution as in figures 
  \ref{fig:cnews-mass-compare-mext}, \ref{c_QNM} and \ref{T0_QNM}.
  When the initial Gaussian reaches the origin, it is  ``instantly'' (in retarded time $u$)
  radiated to future null infinity $\Scri^+$ (located at $x=1$) by interfering nonlinearly with
  the field that has not yet reached the origin. Once the evolution has come close to
  the critical solution, the matter field $\psi(u,x) = \phi r$ decays exponentially;
  further self-similar features are thus not visible in this plot.
  }
\end{figure}


\subsection{Local Detection of Discrete Self-Similarity}\label{sec:detection-of-DSS}

We discuss some methods to extract the echoing exponent of discrete
self-similarity from a near-critical numerical evolution while restricting
attention to the past self-similarity horizon.

\subsubsection{max(2m/r)-Plots}
The black-hole formation diagnostic $2m/r$ furnishes a necessary condition
for discrete self-similarity, namely periodicity in the function
\begin{equation}
  \max_r \frac{2m}{r}(\tau),
\end{equation}
where the maximum is taken over $u=const$ null slices and
$m(u,r)$ denotes the mass-function in spherical symmetry, as defined in
equation \eqref{m_MS}. 

The viability of this approach depends on the choice of initial data.
For the usual Gaussian initial data family, experience shows that
the maximum is well-defined. For other choices, such as the double-Gaussian
initial data of section \ref{sec:dn-double-Gaussian}, two (or more) peaks 
in $2m/r$ may compete with each other and thus the maximum can jump 
from one slice to another.

\begin{figure}[htb]
  \centering
  \includegraphics[angle=0,width=0.8\textwidth]{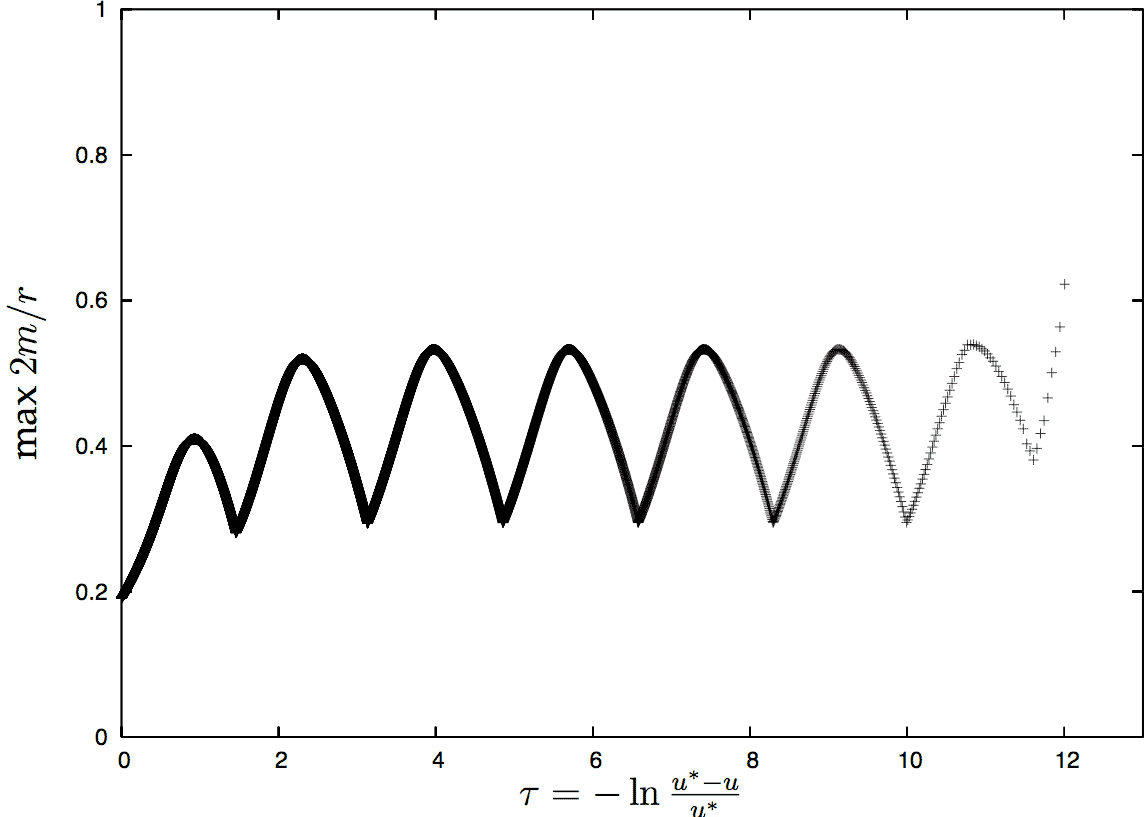}

  \caption{\label{fig:max-two-m-r-DSS}
    This is a typical plot of the black hole diagnostic $\max 2m/r$ vs. $\tau$
    computed using the compactified DICE code with 8000 gridpoints.  
    The parameter $\Delta$ can be readily extracted from the picture. The
    accumulation time $u^*$ was determined as the time of the formation
    of the very small black hole with mass $m_\BH = 3.5 \times 10^{-5}$.
    For $\tau \gtrsim 12$, $\max 2m/r$ rapidly approaches unity, which 
    heralds the formation of an apparent horizon.
    }
\end{figure}

The echoing exponent $\Delta$ can be determined by analyzing
plots of $\max(2m/r)$ vs. $\tau$ and adjusting $u^*$ such that
successive echoes obey a constant spacing, which equals $\Delta/2$.
In practice, a crude estimate of $u^*$ can be obtained in
slightly supercritical evolutions as the coordinate time of the
formation of the (small) black hole.
This process can be carried out by hand or automatized by a fitting procedure.
The initial transient in which the system first approaches the vicinity of 
the attractor has to be ignored.
The result of such a fit is shown in figure \ref{fig:max-two-m-r-DSS}.

\subsubsection{DSS-Snapshots}

Although plots of the diagnostic $\max(2m/r)$  are handy for determining 
$\Delta$ and $u^*$, since they are easy to generate, they do not encompass 
the full scope of DSS, since, by taking the maximum, the information available 
is collapsed down to just one data point per null-slice. Given a function on 
spacetime which is DSS, it shows this behavior at any given 
radial position. This much stronger criterion can be investigated with
the following method (adapted from \cite{Choptuik-pune}).

In coordinates $(t,r)$ DSS corresponds to a logarithmic rescaling in both 
time and space. 
As has been discussed in section \ref{sec:self-similarity-in-GR},
in adapted coordinates $(\tau, z)$ DSS amounts to
\begin{equation}
  \phi(\tau +n\Delta/2, z) = \left(-1\right)^n \phi(\tau, z).
\end{equation}
Here, we prefer to use $\ln r$ instead of $z$. From the definition of $z$, 
\begin{equation}
  z = \frac{r e^\tau}{\zeta(\tau) u^*},
\end{equation}
observe that
\begin{equation}
  \ln z = \ln r + \tau - \ln(\zeta(\tau) u^*).
\end{equation}
Thus, the following equality holds
\begin{multline}
  \phi(\tau +n\Delta/2, \ln r + \tau + n\Delta/2 -\ln(\zeta(\tau + n\Delta/2) u^*) ) \\
  = \left(-1\right)^n \phi(\tau, \ln r + \tau -\ln(\zeta(\tau) u^*)),
\end{multline}
and, for a given $\tau=const$, and using $\zeta(\tau + n\Delta/2) = \zeta(\tau)$, we have
\begin{equation}\label{eq:shifts-in-lnr}
  \phi(\tau +n\Delta/2, \ln r + n\Delta/2) = \left(-1\right)^n \phi(\tau, \ln r).
\end{equation}

\begin{figure}[H]
  \centering
  \includegraphics[width=\textwidth]{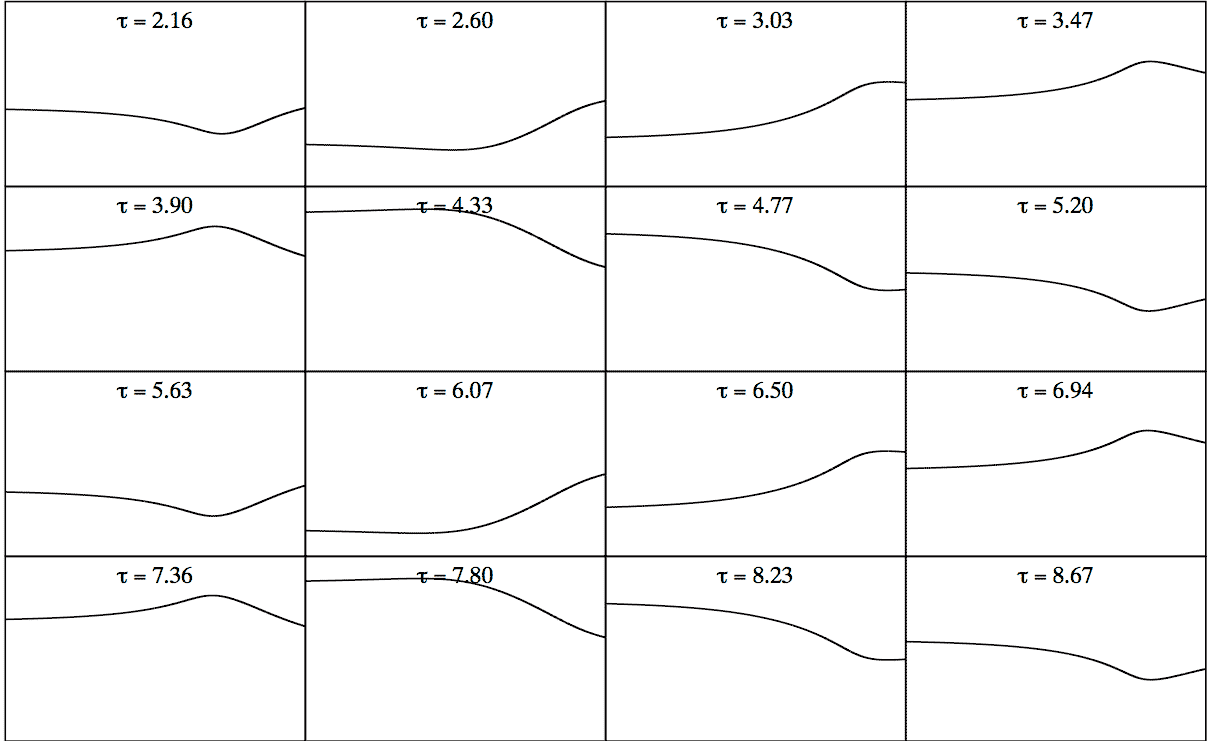}

  \caption{\label{fig:phi-DSS-snapshots}
    This figure displays snapshots of a near-critical uncompactified
    evolution with $5000$ gridpoints of the massless scalar field
    $\phi$ as a function of $\ln r$.
    The frames are evenly spaced in the adapted time coordinate 
    $\tau = -\ln \frac{u^*-u}{u^*}$. Observe that the shape of $\phi$ is 
    identical in every 8th frame; thus, $\phi$ is periodic with
    period $\Delta \simeq 3.44$. Moreover, $\phi$ is antisymmetric with respect to
    every 4th frame, i.e. it satisfies the half-period self-similarity condition
    \eqref{eq:DSS-scalar-field-additional-symmetry}. Figures \ref{2mr-DSS-snapshots}
    and \ref{fig:Psi-overlay-DSS} show other discretely self-similar
    quantities from the same numerical evolution.}
\end{figure}

\begin{figure}[H]
  \centering
  \includegraphics[angle=0,width=\textwidth]{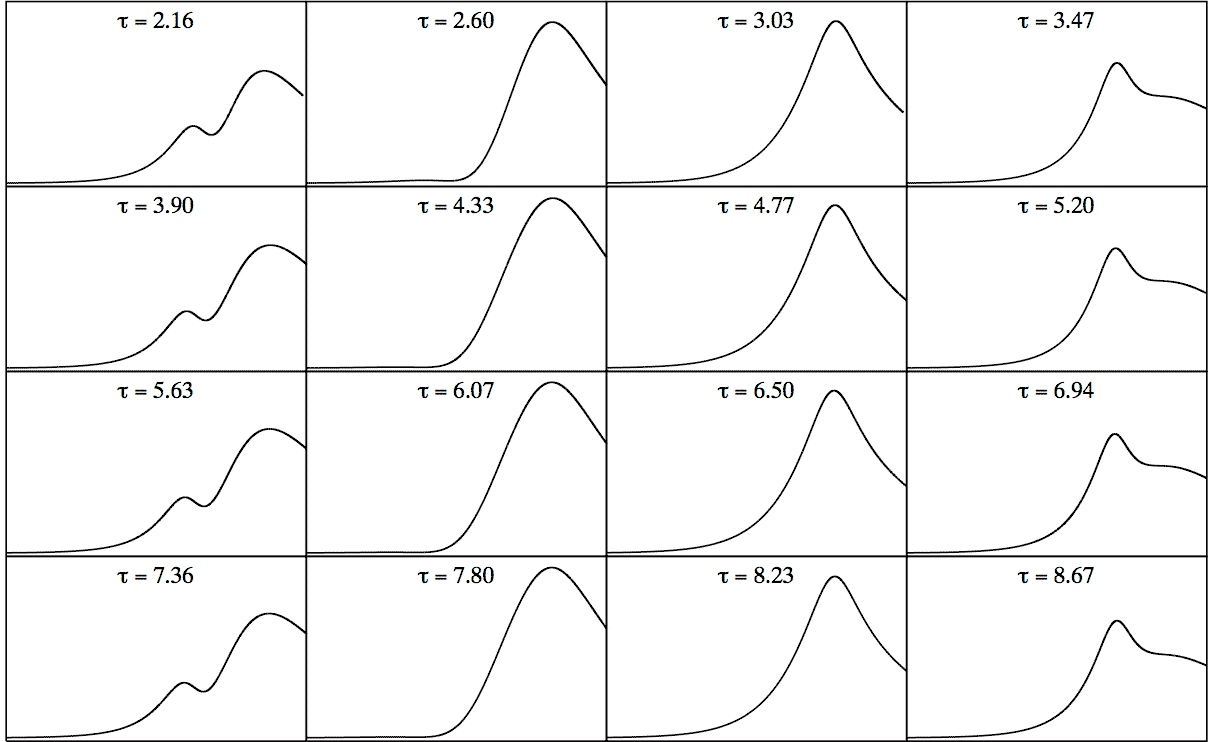}

  \caption{Snapshots of the dimensionless black hole formation diagnostic $2m/r$ for the
   same near-critical evolution as shown in figure \ref{fig:phi-DSS-snapshots}.}
  \label{2mr-DSS-snapshots}
\end{figure}

Successive frames in the plots shown in figures \ref{fig:phi-DSS-snapshots}
and \ref{2mr-DSS-snapshots} are equally spaced in the adapted
time coordinate $\tau$. The quantities are plotted against $\ln r$ with 
appropriate ``shifts'' in the null slices according to equation \eqref{eq:shifts-in-lnr}.

\begin{figure}[H]
  \centering
  \includegraphics[angle=0,width=\textwidth]{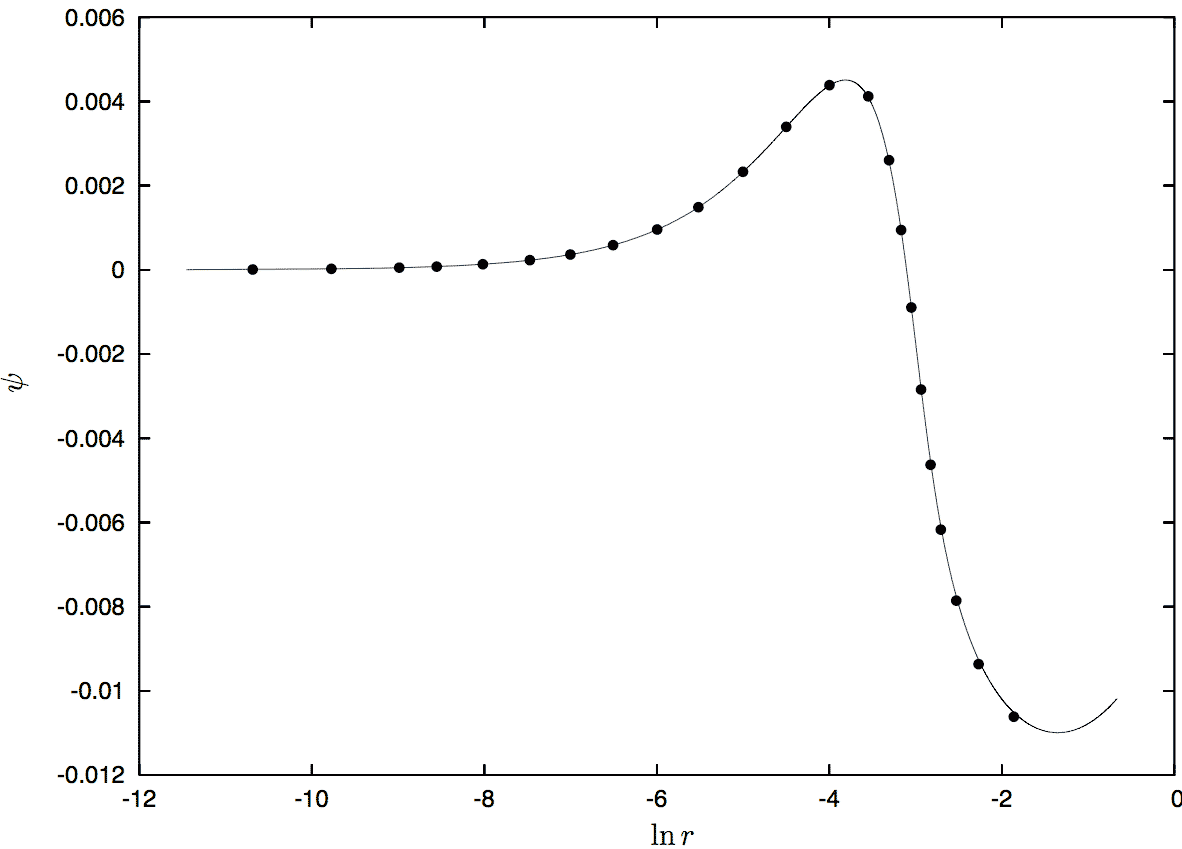}

  \caption{This figure shows $\psi(\tau_0,\ln r)$ (depicted by a few sampled dots)
  at a time $\tau_0=4.77$, when the evolution was in the echoing region
  overlaid with its image under the DSS diffeomorphism $\Phi_\Delta$,
  $e^\Delta \psi(\tau_0+\Delta,\ln r + \Delta)$ (represented by the continuous line).
  Since $\psi$ is not dimensionless, its amplitude had to be rescaled
  by $e^{\Delta}$.}\label{fig:Psi-overlay-DSS}
\end{figure}

While dimensionless quantities ($\phi$, $2m/r$, $\beta$, $\Vr$) are mapped
according to \eqref{eq:shifts-in-lnr} without a change of scale under the
diffeomorphism associated with DSS, dimensionful quantities such as
$m$ or $\Psi$ also scale according to their dimensions, e.g.
$[\Psi] = L$ and thus
\begin{equation}
  \psi(\tau + \Delta/2, \ln r + \Delta/2) = - e^{\Delta/2} \psi(\tau, \ln r).
\end{equation}
As an example, figure \ref{fig:Psi-overlay-DSS} shows the scalar field and
its image under the DSS diffeomorphism.

\begin{figure}
  \centering
  \includegraphics[width=\textwidth]{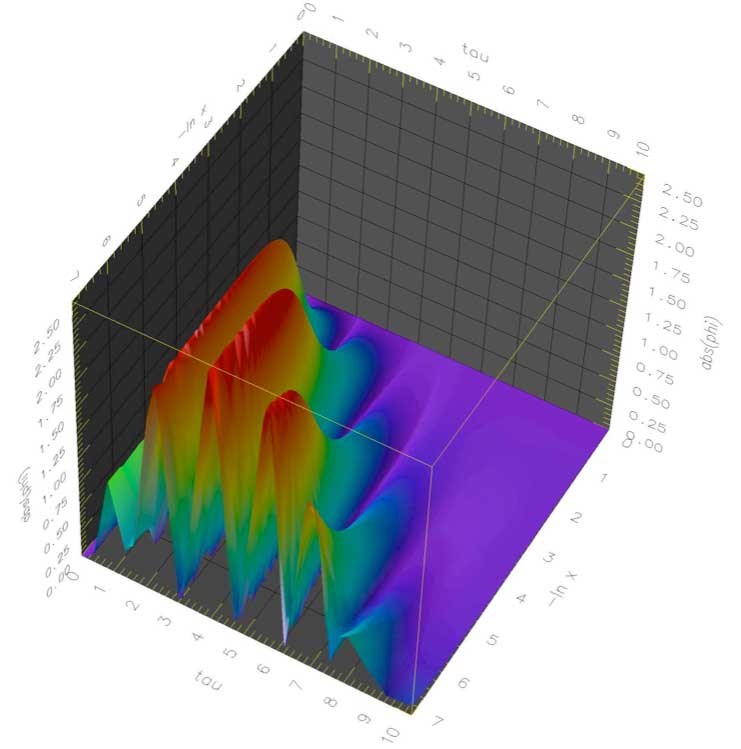}

  \caption{This figure shows $\lvert \phi(\tau, -\ln x) \rvert$ for a near-
  critical compactified evolution. In the direction of increasing $\tau$
  the features of the scalar field repeat themselves with a period
  $\Delta/2$ since $\phi$ fulfills the half-period self-similarity condition
  \eqref{eq:DSS-scalar-field-additional-symmetry}. 
  Concurrently, these features are shifted towards
  the origin by $\Delta/2$ in $-\ln r$ ($\ln x \approx \ln r$ near the origin).
  Figure \ref{fig:DSS-3d-flat} shows a planar view of the same surface.}
  \label{fig:DSS-3d-perspective}
\end{figure}

\begin{figure}
  \centering
  \includegraphics[width=\textwidth]{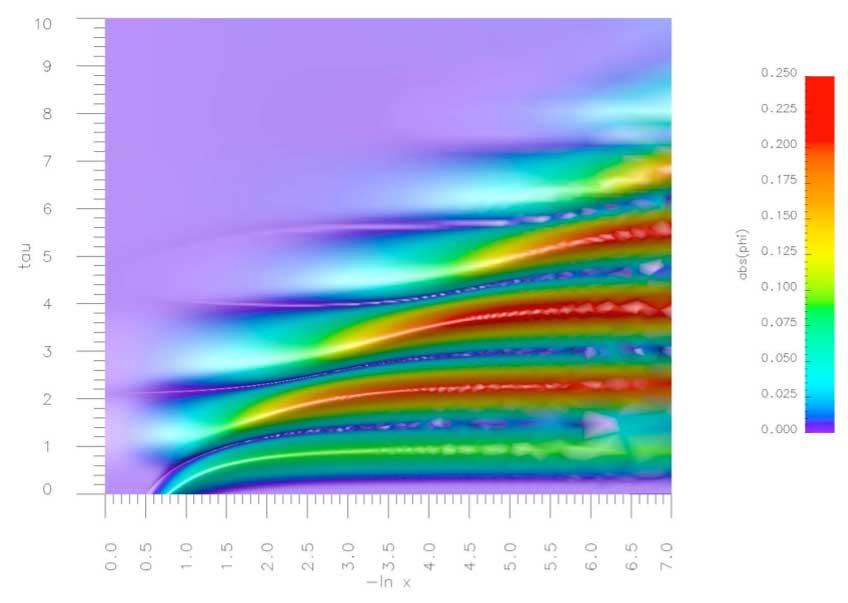}

  \caption{This figure shows a planar view of $\lvert \phi(\tau, -\ln x) \rvert$
  for the same evolution depicted in figure \ref{fig:DSS-3d-perspective}.
  The ingoing wave packet is clearly visible on the initial slice ($\tau=0$) of
  the evolution.
  The the tail ends of the bumps are enveloped by the null geodesic that
  barely hits the accumulation point which is shown separately
  in figure \ref{fig:accumulation-point-null-geodesic}.}\label{fig:DSS-3d-flat}
\end{figure}

\begin{figure}
  \centering
  \includegraphics[width=\textwidth]{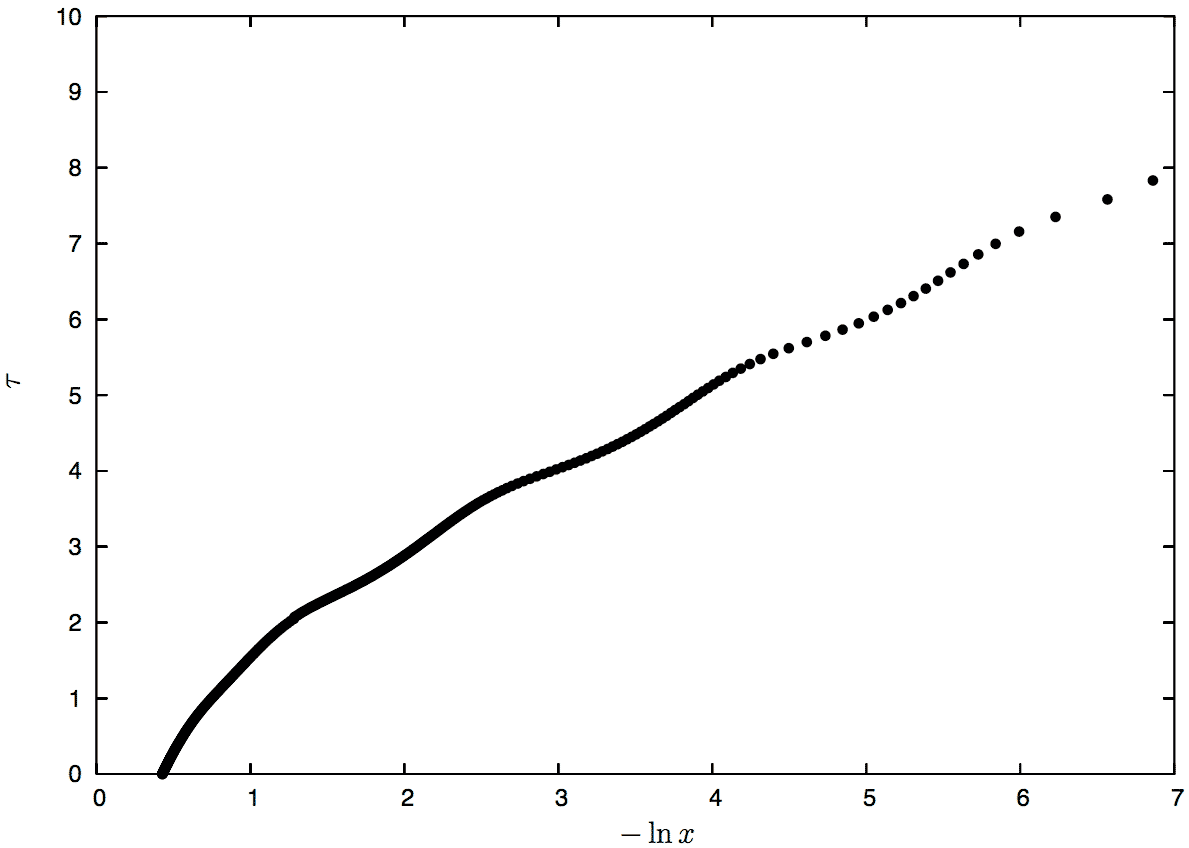}

  \caption{This figure shows the null geodesic which almost hits the
  accumulation point in the evolution shown in figures \ref{fig:DSS-3d-flat}
  and \ref{fig:DSS-3d-perspective}  }\label{fig:accumulation-point-null-geodesic}
\end{figure}

To be able to more clearly visualize discrete self-similarity, we show 
the scalar field as a function of $\tau$ and $-\ln x$ for a near-critical
compactified evolution in figures \ref{fig:DSS-3d-perspective} and
\ref{fig:DSS-3d-flat}. Figure \ref{fig:accumulation-point-null-geodesic}
depicts the ingoing null geodesic that hits the accumulation point
of the DSS-solution.

\enlargethispage*{-1cm}



\clearpage

\section{Results from the Double-Null Code}\label{sec:dn-results}
The results shown here are based on fine-tuning of Gaussian initial data
\begin{equation}
  s(u=0,v) = A \exp\left[-\left(\frac{v - v_c}{\sigma}\right)^2\right].
\end{equation}
In section \ref{sec:dn-double-Gaussian} we introduce \emph{double Gaussian data} that 
are merely a superposition of two such Gaussians with parameters adjustable
for each pulse.

\subsection{Scaling}\label{sec:dn-scaling}

\begin{figure}[hbt]
  \centering
  \includegraphics[width=0.7\textwidth]{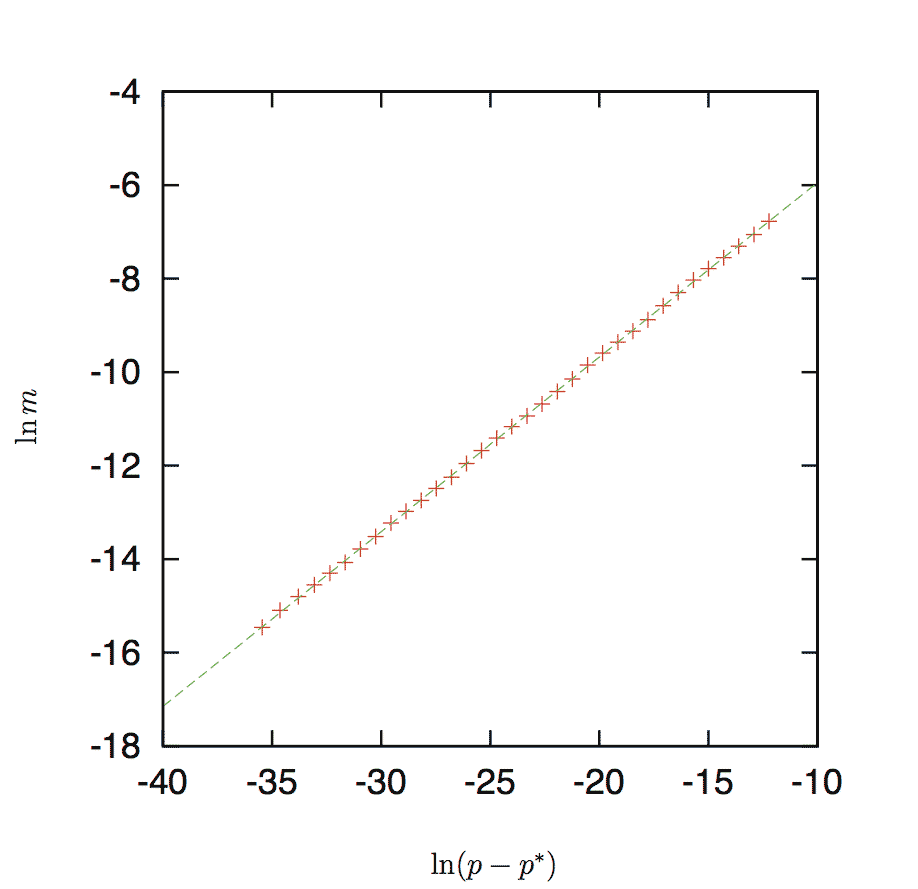}

  \caption{
    This figure shows mass scaling for a family of evolutions using 8000 gridpoints.
    The critical exponent is given by the slope of the fitted line, which yields 
    $\gamma \approx 0.373$.
    Note that this scaling is perfectly consistent with the analytical scaling law 
    and does not suffer from the leveling-off artifact present in mass scaling laws 
    generated from the compactified DICE code.
    }\label{fig:dn-mass-scaling}
\end{figure}

\begin{figure}
  \centering
  \includegraphics[width=0.6\textwidth]{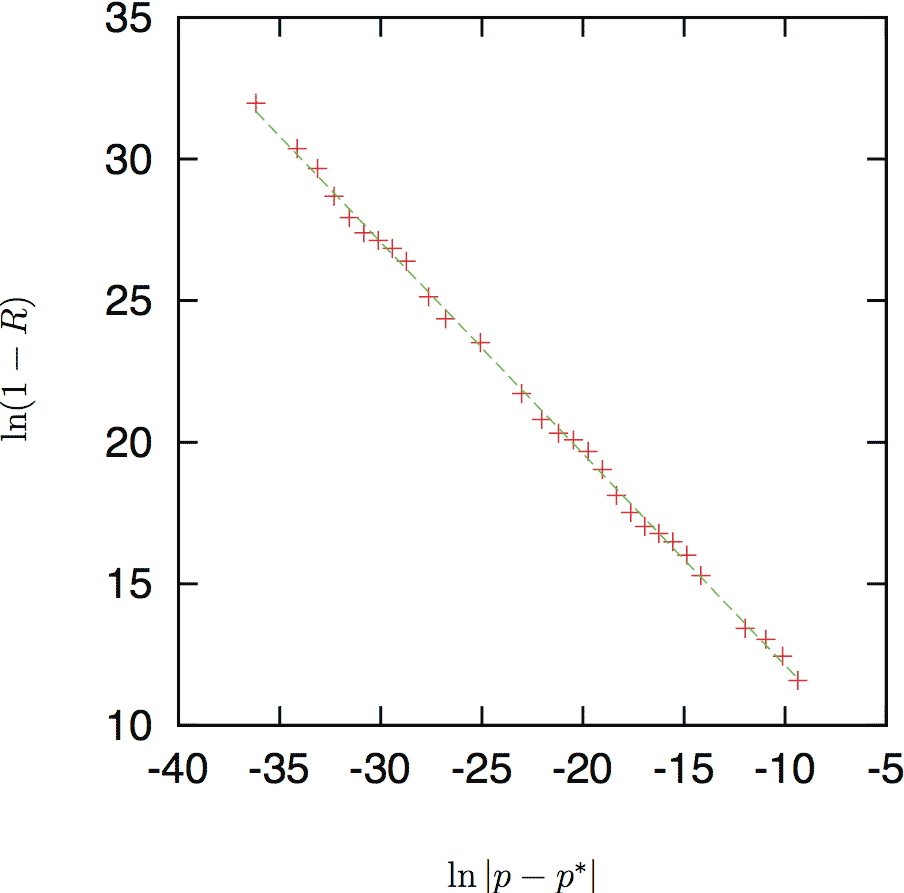}

  \caption{
    Here we show curvature scaling for a series of subcritical evolutions
    using 8000 gridpoints. The maximum of the Ricci scalar at the axis, plotted
    as $\ln (1-R)$, behaves as in equation \eqref{eq:curvature-scaling}. 
    The scaling exponent extracted from the fit is $\gamma \approx 0.374$.
  }\label{fig:dn-curvature-scaling}
\end{figure}

To complement the results from the DICE code, figures \ref{fig:dn-mass-scaling}
and \ref{fig:dn-curvature-scaling} show the expected scaling behavior for the
black hole mass and the Ricci scalar curvature from a one-parameter family of 
evolutions generated by the double null code, respectively.
According to its dimension being $1/L^2$, the scalar curvature scales
as $(p^*-p)^{-2\gamma}$, as discussed in section \ref{sec:scaling-law-derivation}.

\subsection{Horizons and $\mEXT$}

In section \ref{sec:mass_news} we mentioned that exterior matter, $\mEXT$, 
accumulates outside the past self-similarity horizon and we conjectured that, for late times,
the mass originates from backscattering of outgoing radiation during the self-similar
collapse. In a supercritical evolution, two scenarios are possible: Either $\mEXT$ 
eventually falls through the horizon, or it is radiated to infinity.
In the former case, the resulting black hole will have a tiny but finite 
Bondi mass, no matter how fine tuned the data are, whereas, in the latter case,
the Bondi mass can be made arbitrarily small.

We would like to explicitly verify whether this matter, $\mEXT$, falls 
through the event horizon. (A compactified code that could penetrate apparent
horizons would presumably still have to stop at the event horizon,
due to reaching future timelike infinity, $i^+$. Using an uncompactified 
code, we will have to settle for a reasonable local approximation of the event 
horizon.)

At first, it would seem that one could investigate the final fate of $\mEXT$  
with any evolution code that can penetrate apparent horizons, such as the
double null code of section \ref{sec:dn-code}. As it turns out, however, 
codes based on outgoing null slices are ill-suited to tackle this issue.

Only the immediate vicinity of the event horizon is relevant for the solution 
of this problem. Null slices too far to the future of the event horizon no 
longer contain $\mEXT$, since the slices will have ``bent'' towards the curvature 
singularity at $r=0$, and trapped gridpoints which come too close to the singularity
will have been excised, as has been discussed in section \ref{sec:horizon-detection-excision}. 
This leaves us with a slice where the very features we would like to study, 
namely the exterior matter, have probably already fallen into the singularity.

\begin{figure}
  \centering
  \includegraphics[width=0.9\textwidth]{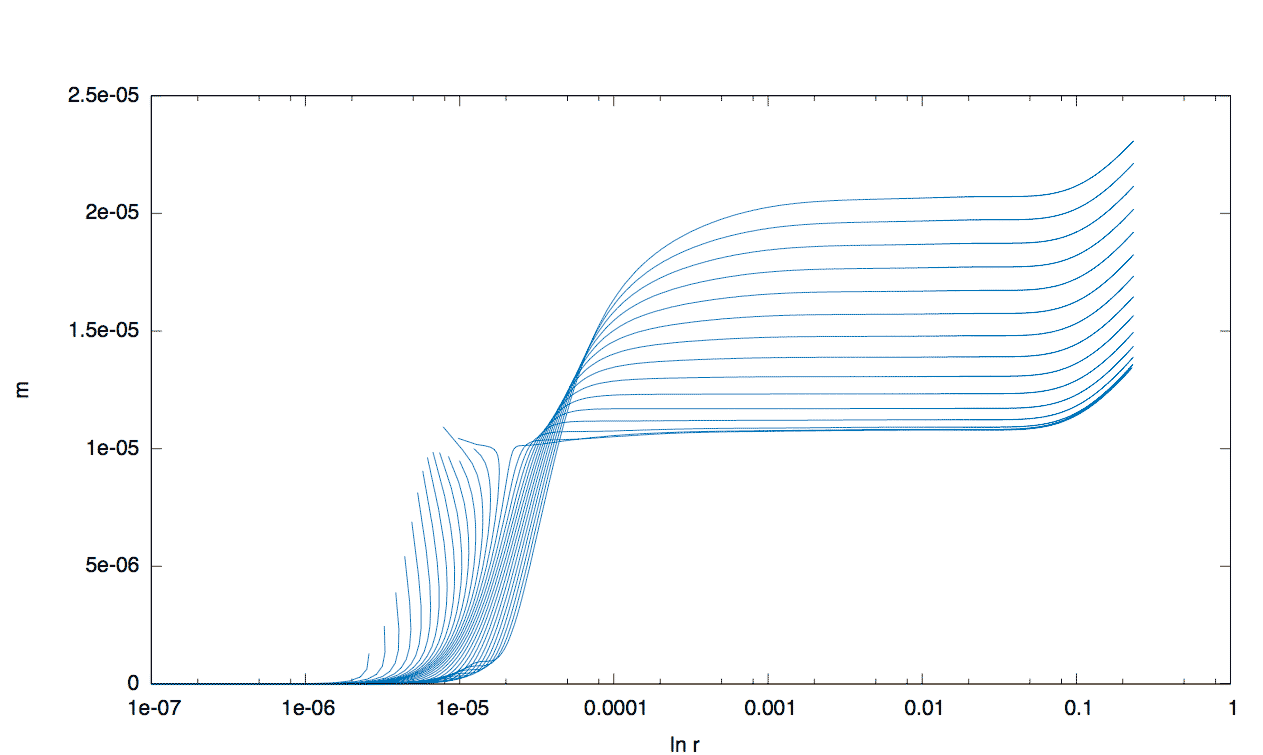}

  \caption{
    Shown is the mass-function of a series of null slices of a supercritical evolution close to
    criticality. The formation of an apparent horizon manifests itself in the bending over of the
    curves in the $r$-coordinate. After having reached its maximum $r$-value the slice bends back
    towards $r=0$ and reaches the curvature singularity. This necessitates the excision of
    gridpoints coming too close to $r=0$.}
  \label{fig:m-logr-series-supercritical}
\end{figure}

\begin{figure}
  \centering
  \includegraphics[width=0.9\textwidth]{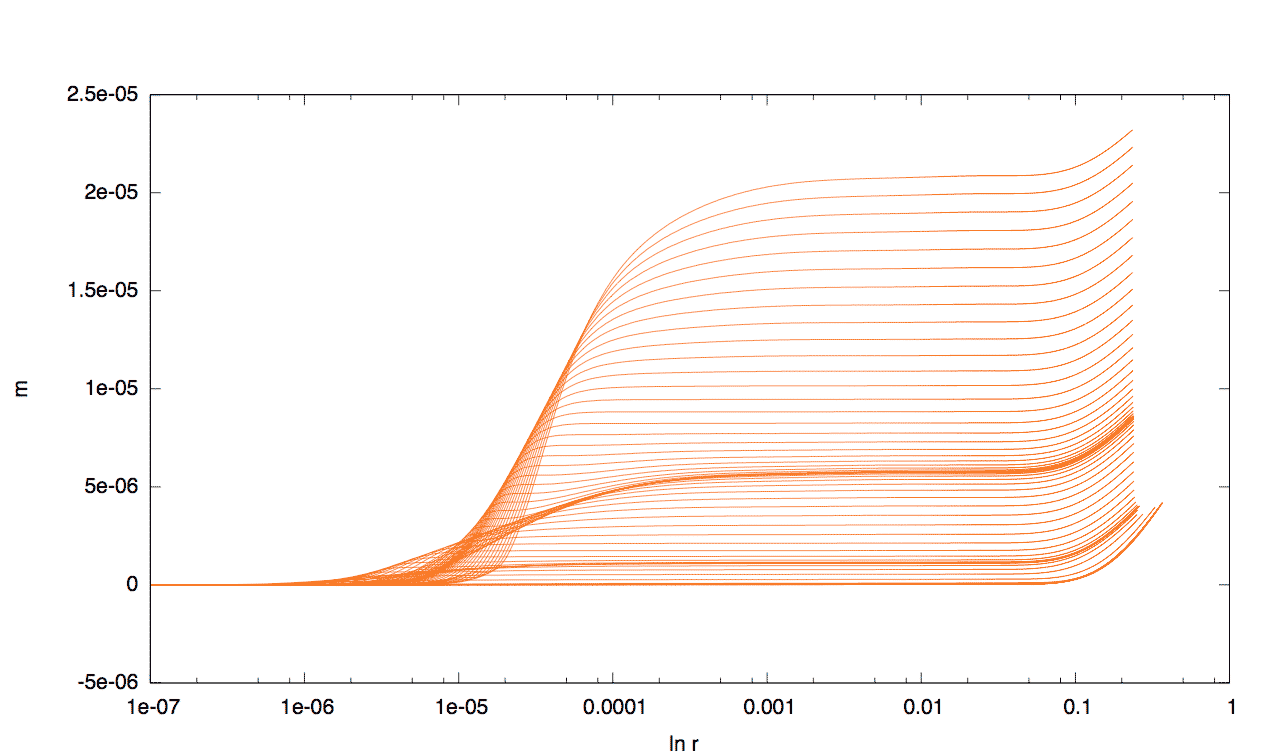}

  \caption{
    Shown is the mass-function of a series of slices of a subcritical evolution close to
    criticality. The initial data are very close to those in figure
    \ref{fig:m-logr-series-supercritical}, which ultimately forms an apparent horizon. In this
    evolution, the scalar field disperses, ultimately, reaching Minkowski space}
  \label{fig:m-logr-series-subcritical}
\end{figure}

\begin{figure}
  \centering
  \includegraphics[width=0.9\textwidth]{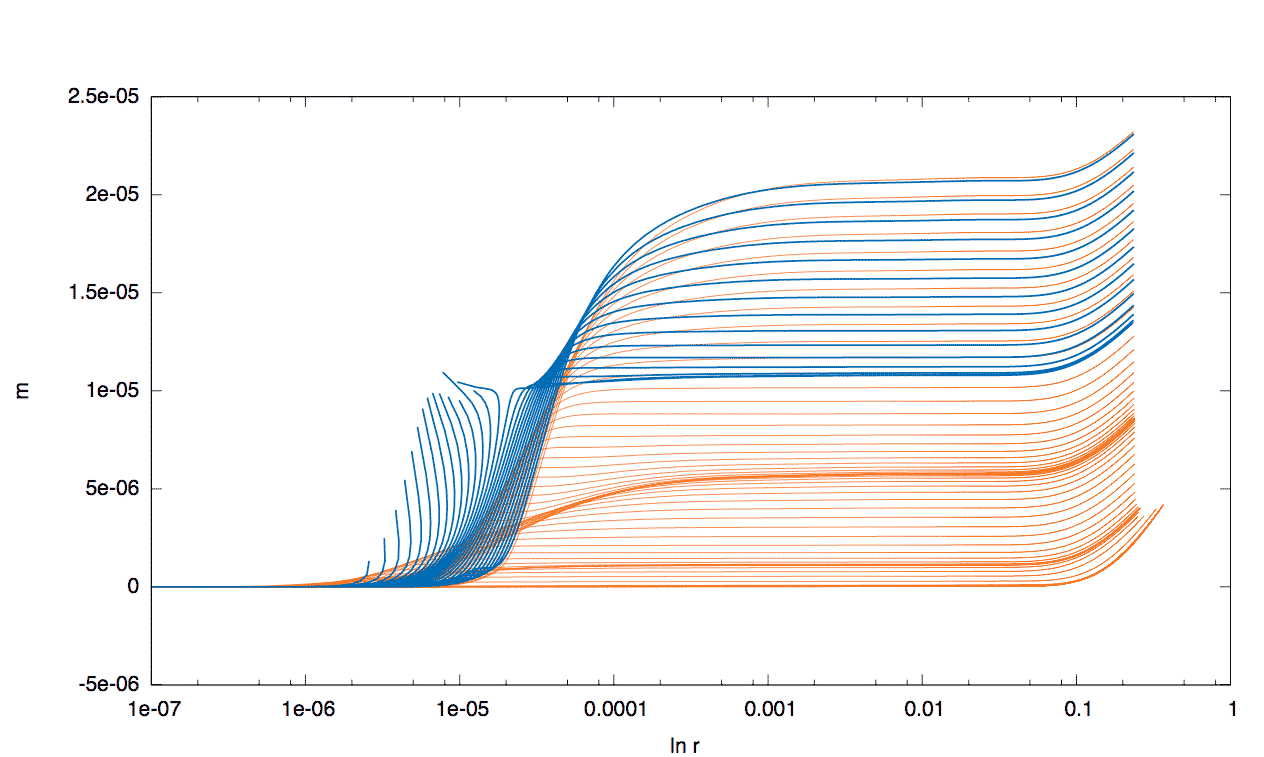}

  \caption{
    This figure shows figures \ref{fig:m-logr-series-supercritical} and
    \ref{fig:m-logr-series-subcritical} superimposed. The two evolutions almost coincide in the
    first few slices shown, when the apparent horizon has not yet formed. Gradually, the
    agreement gets worse for $r < 10^{-4}$, while it is still close for larger $r$. Then, an
    apparent horizon forms in the supercritical evolution, while the subcritical evolution
    continues to radiate scalar field that has reached the origin and ultimately disperses.}
  \label{fig:m-logr-series-superimposed}
\end{figure}

To illustrate the behavior of the mass function on outgoing null slices in the
vicinity of apparent horizon formation we compare a series of slices for a
near-critical evolution using 5000 gridpoints in figure 
\ref{fig:m-logr-series-supercritical}.
If a slice contains an apparent horizon, the function $r(v,u=const)$ 
increases monotonically from the origin outwards until the locus of the AH is 
reached; there, it reverses direction and decreases monotonically until it
reaches the singularity at $r=0$. Gridpoints which come too close to the 
singularity must be excised.
In contrast, figure \ref{fig:m-logr-series-subcritical} depicts a series of slices
for a near-critical evolution which ultimately is subcritical. The two plots
are superimposed in figure \ref{fig:m-logr-series-superimposed}.

\begin{figure}
  \centering
  \includegraphics[width=\textwidth]{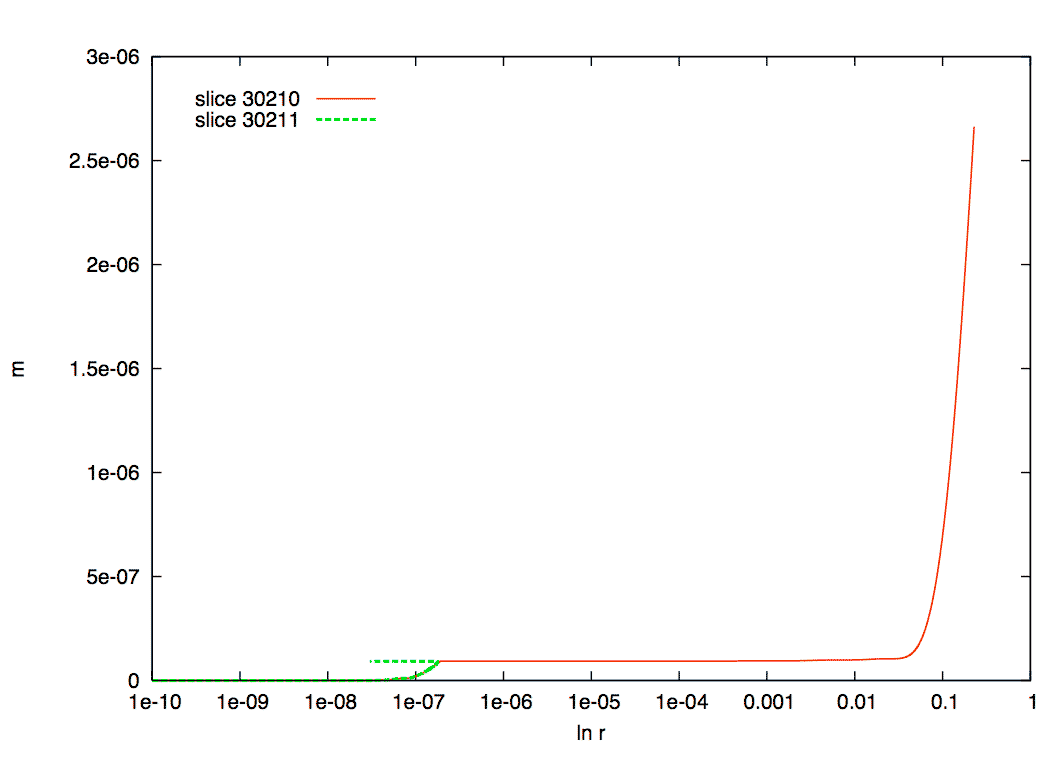}

  \caption{
    This figure shows the mass-function before and after the formation of an apparent horizon for
    a brute force evolution at 100000 gridpoints. The reduction of $\Delta u$ was still not able
    to resolve the bending over of the slices fast enough to have the trapped slice extend up to
    the exterior matter.
    }\label{fig:m-ext-brute-force}
\end{figure}

To improve the chance to observe the exterior mass falling into the horizon,
it is clearly beneficial to decrease the timestep $\Delta u$ as far as possible.
Unfortunately, the double null code relies on the condition $\Delta u = \Delta v$
for the approximation of derivatives close to the origin. We have tried to use
interpolation to get rid of this condition, but our efforts introduced 
instabilities that made evolutions unusable.
There is, in principle, another way. Namely, keep the condition $\Delta u = \Delta v$
and increase the overall spacetime resolution, i.e. both gridspacings. 
The result of such a brute force evolution using $100000$ gridpoints is shown
in figure \ref{fig:m-ext-brute-force}. The horizon mass is almost two orders
of magnitude lower, but the bending-over of the slices is still much too abrupt
and all the exterior mass has been excised on the first slice which contains
an apparent horizon in this evolution.

The behaviour discussed above is a deficiency intrinsic to our choice of slicing: 
Null-slices are inherently unstable near a horizon. Outgoing null-slices either 
extend till $\mathscr{I}^+$, if they are untrapped, or end at the singularity $r=0$
(with the exception of the event horizon which extends till $i^+$).
Thus, it would be seem to be appropriate to use a Cauchy code to investigate this issue further. This would allow us to scan the horizon for exterior infalling matter.


\subsection{Dependency of $m_{ext}$ on Initial Data}

In the following, we would like to discuss the dependance of the exterior matter,
$\mEXT$, on the initial data. 
First, choose a family of initial data and fix all free parameters except for the
parameter $p$ that will be used for fine-tuning. Within such a family, as we tune 
the parameter $p$ to criticality, we can reliably measure $\mEXT$ only if at some 
null slice in a given evolution, the interior mass is of the same order or smaller 
than $\mEXT$ or the interior and exterior mass are spatially well separated.
In practise, this condition is only fulfilled at late times for very near-critical 
evolutions which severely restricts the interval in the initial data parameter and
evolution time over which we can accurately measure the exterior mass $\mEXT$.

From the analysis of a typical single near-critical (but supercritical) evolution, 
$\mEXT$ oscillates from the time we can first detect it until the last null slice 
before horizon formation, while the interior mass
(contained in the past SSH) undergoes critical collapse and periodically sheds
mass. If we instead consider a family of supercritical evolutions which
become more and more near-critical, and measure $\mEXT$ at the last slice before
horizon formation, $\mEXT$ roughly stays constant.

It is also important to note that, using generic (e.g. Gaussian) initial data,
which have been fine-tuned, the solution evolves through a non-universal and family 
dependent stage before it approaches the self-similar regime. Since backscattering of
radiation is already present in this initial stage, we expect $\mEXT$ to also depend 
on the shape or family of initial data used.

\subsection{Double Gaussian Initial Data Results}\label{sec:dn-double-Gaussian}

In an effort to shed some light on the issue concerning the possible infall
of the external mass $\mEXT$, we have devised initial data to model 
this physical situation. Essentially, we are dealing with a near-critical
evolution where the interior part is in the DSS regime (for a certain time)
while further outside, there is another concentration of energy.
Therefore, it is sensible to chose initial data which consists of two Gaussian 
pulses of scalar field, which we call double Gaussian initial data for short.

As it turns out, $\mEXT$ is not dense enough to form an apparent horizon 
by itself. 
This can be guessed from the fact that there is no apparent peak in $2m/r$
at the locus of $\mEXT$. It is also possible to evolve data that are based
on the numerical scalar field solution on a null slice of a near-critical 
evolution where the presence of $\mEXT$ is the dominating contribution of mass.
Using such a solution, $s(\bar u,r) = s(\bar u,v(\bar u,r))$, 
(preferably the last available slice before the formation of an AH) as new 
initial data $s_0(u=0,v) = s(\bar u, 2r)$ is fairly straightforward.
In is necessary to employ interpolation to obtain equally spaced initial data
$s_0(0,v)$ for the scalar field . 
Here, we have also had to fix the null gauge, which determines how
the gridpoints are labeled by $u$ and $v$.
If necessary, taper the gridfunction such that the gradients in the 
interior part are smooth enough that they do not form trapped surfaces;
in practise, this is taken care of automatically by the interpolation.
We have carried out this procedure and found that the new spacetime
is subcritical.

The accumulation of exterior mass that occurs naturally via backscattering in 
critical collapse could be modeled by data where the outer pulse is chosen 
weak enough so that it does not collapse by itself.
Unfortunately, simulations of that type did not yield any new insights, since
we still run into the problem of the bending-over of slices and forced excision 
of the part of the slice we would like to follow further.
 
\begin{figure}
  \includegraphics[width=\textwidth]{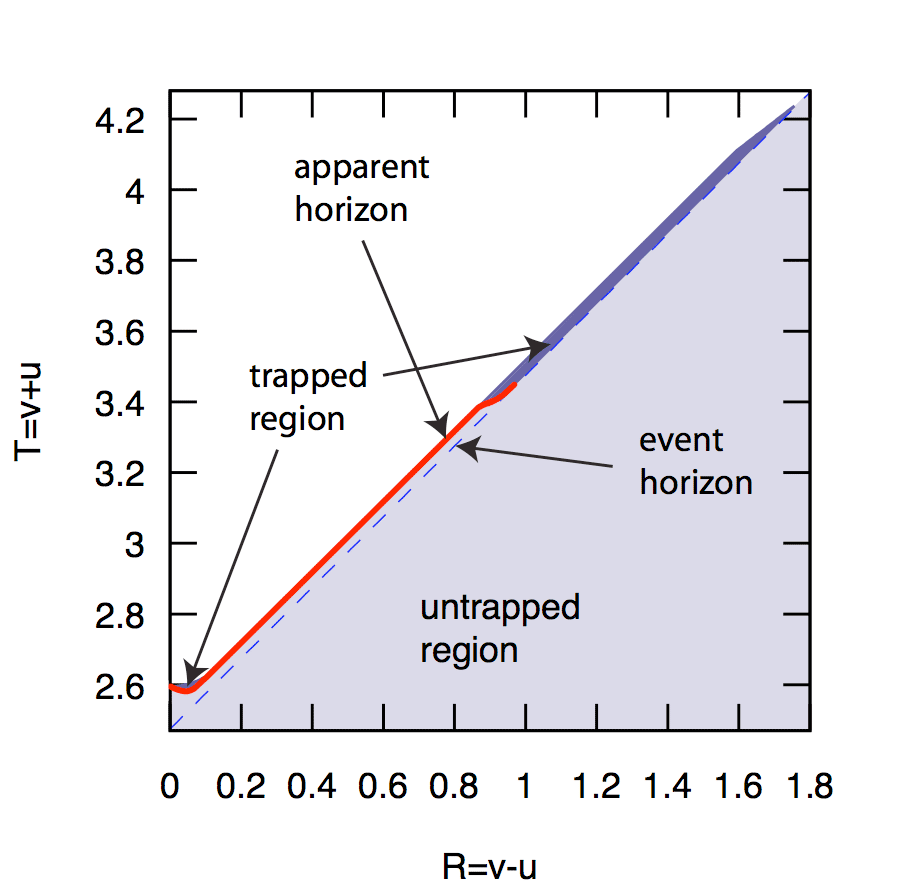}

  \caption{  
    Similar to figure \ref{fig:spacetime-diag-AH-T-R}, this spacetime diagram shows the formation
    of an apparent horizon and the region of spacetime that has been evolved. This time, we are
    dealing with an evolution using double Gaussian initial data, chosen such that the outer
    pulse forms the apparent horizon (first in $u$) while the inner one reaches $2m/r =1$ later
    in retarded time $u$. Then, excision kicks in and we are left with a very small grid. 
    }\label{fig:spacetime-diag-AH-dg-T-R}
\end{figure}

\begin{figure}
  \centering
  \includegraphics[width=0.78\textwidth]{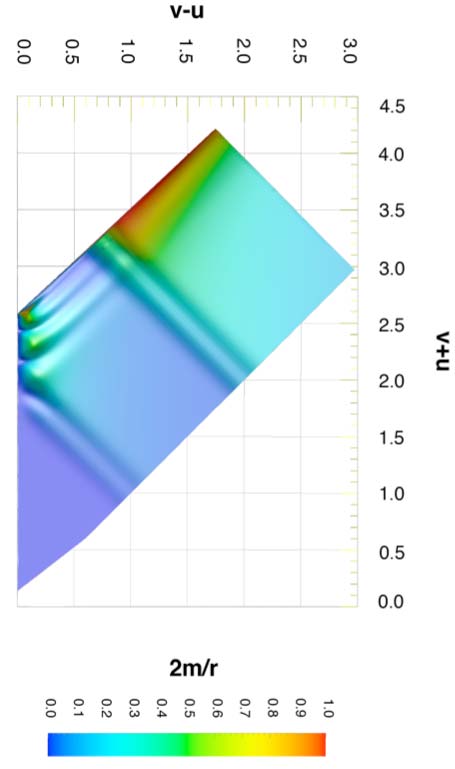}

  \caption{
    Collapse of double Gaussian initial data. Density plot of $2m/r$ shows the ingoing pulses, 
    contraction of matter and ultimate apparent horizon formation. 
    (exterior pulse first in $u$)
  \label{fig:double-gaussian-density-map} 
  }
\end{figure}

\begin{figure}
  \centering
  \includegraphics[scale=0.3]{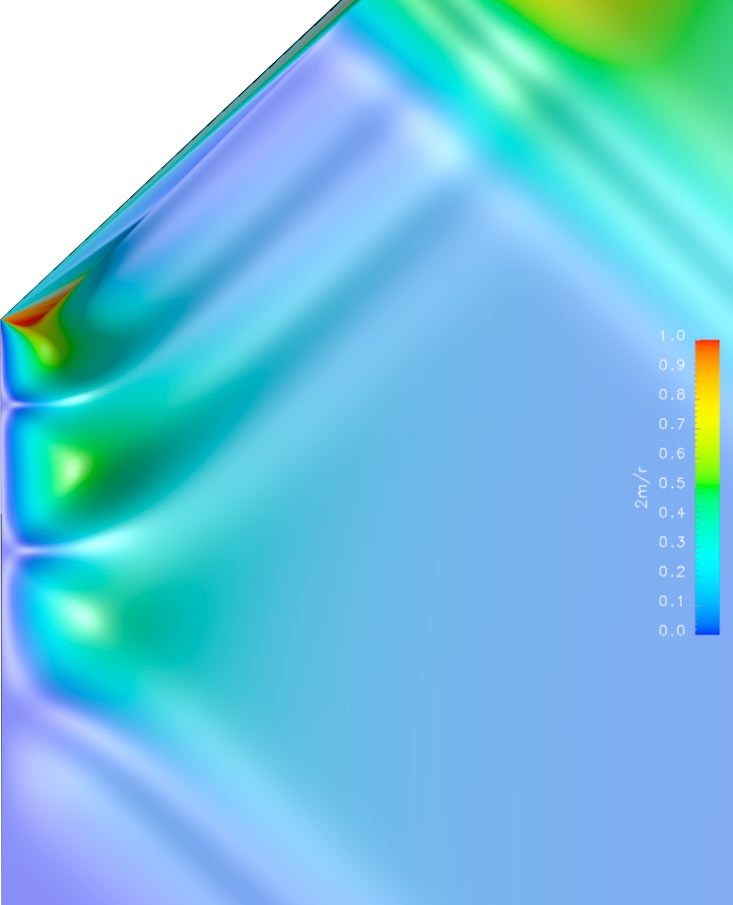}

  \caption{
    Shown is a detail of figure \ref{fig:double-gaussian-density-map}
  }\label{fig:double-gaussian-density-map-detail}
\end{figure}

\begin{figure}
  \includegraphics[width=\textwidth]{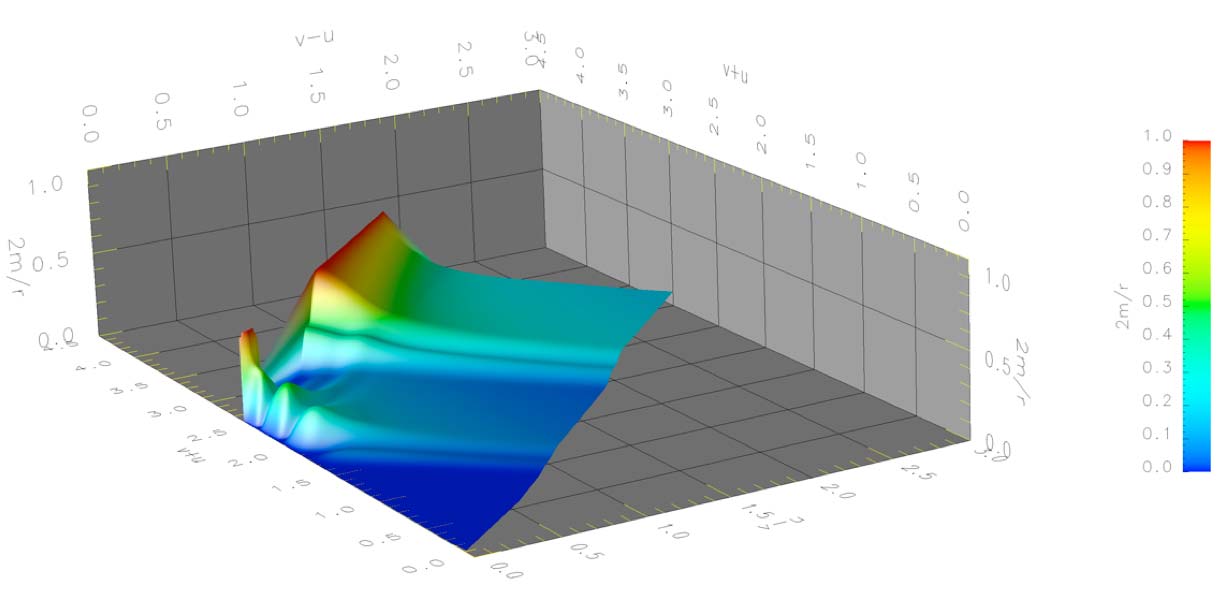}

  \caption{
    This figure shows a perspective view of the spacetime depicted in figure
    \ref{fig:double-gaussian-density-map}.
    }\label{fig:double-gaussian-3d}
\end{figure}

Although not directly related to the issue we would like to solve, 
we have investigated spacetimes where the outer pulse can form a horizon, 
while the interior pulse evolves, more or less unaffected from the outer pulse, 
as they are interesting in their own right.
In particular, we consider a configuration where the outer pulse froms an 
AH first (in retarded time $u$), while we can still follow the interior 
pulse in its critical collapse. Later (in retarded time), the interior pulse
may itself form trapped surfaces, which creates a jump in the trapping horizon.
A spacetime diagram of such an evolution is shown in figure 
\ref{fig:spacetime-diag-AH-dg-T-R}. Figure \ref{fig:double-gaussian-density-map}
presents a density plot of $2m/r$ for this evolution, (with a detail
shown in figure \ref{fig:double-gaussian-density-map-detail}), while figure
\ref{fig:double-gaussian-3d} shows a perspective view.


\subsection{Timelike Observers and $t$ versus $r$ Diagrams}

For timelike observers at $r=const$ we can compute the proper time $t(u,r=const)$
as discussed in sections \ref{sec:dn-diagnostics} and \ref{sec:numalg-timelike-observers}.
In doing so, we obtain a $(t(u,r),r)$ grid on which we can display quantities of
interest, such as the density function $2m/r$. Since this grid is neither
orthogonal, nor equispaced, we need to resort to interpolation for some plots.

\begin{figure}
  \centering
  \includegraphics[width=\textwidth]{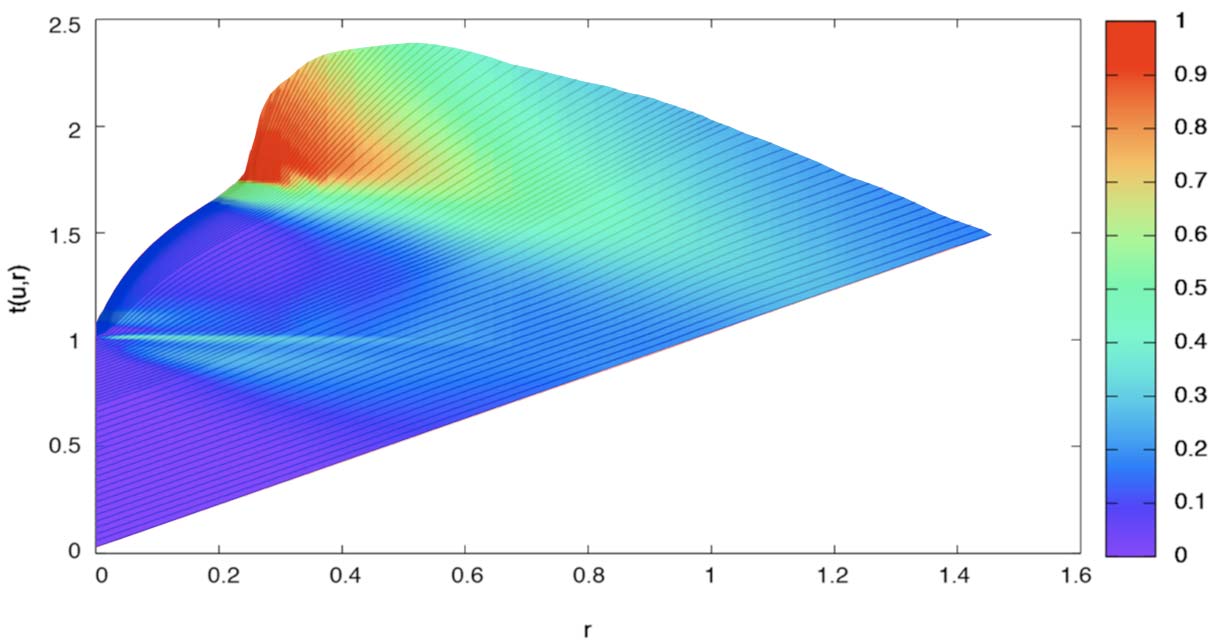}

  \caption{
    This figure shows a density plot of the diagnostic $2m/r$ on a $(t,r)$ grid
    for a supercritical evolution using double Gaussian initial data.
  }\label{fig:t_r_2m_r}
\end{figure}

In order to have more resolution in the vicinity of the origin, one can use a static
mapping in the distribution of the $r=const$ observers. A useful convex mapping is
to equally space the observers in a computational coordinate $\xi$ and 
define $r(\xi) = L(e^{a\xi} - 1)$ with $r(\xi = r_\text{max}) \overset{!}{=} r_\text{max}$,
so that $L = r_\text{max} / (e^{a r_\text{max}} - 1)$.
Figure \ref{fig:t_r_2m_r} shows a color-coded density plot of $2m/r$ on such a grid, where
the null slices $u=const$ have been overlaid. Grid refinements are clearly visible 
when the density of the slices suddenly doubles. The geometry is initially flat.
The evolution uses double Gaussian initial data, where the exterior pulse forms the
first (in retarded time $u$) apparent horizon and the interior collapses at a later
null slice after moving away from the DSS regime.

\begin{figure}
  \centering
  \includegraphics[width=0.8\textwidth]{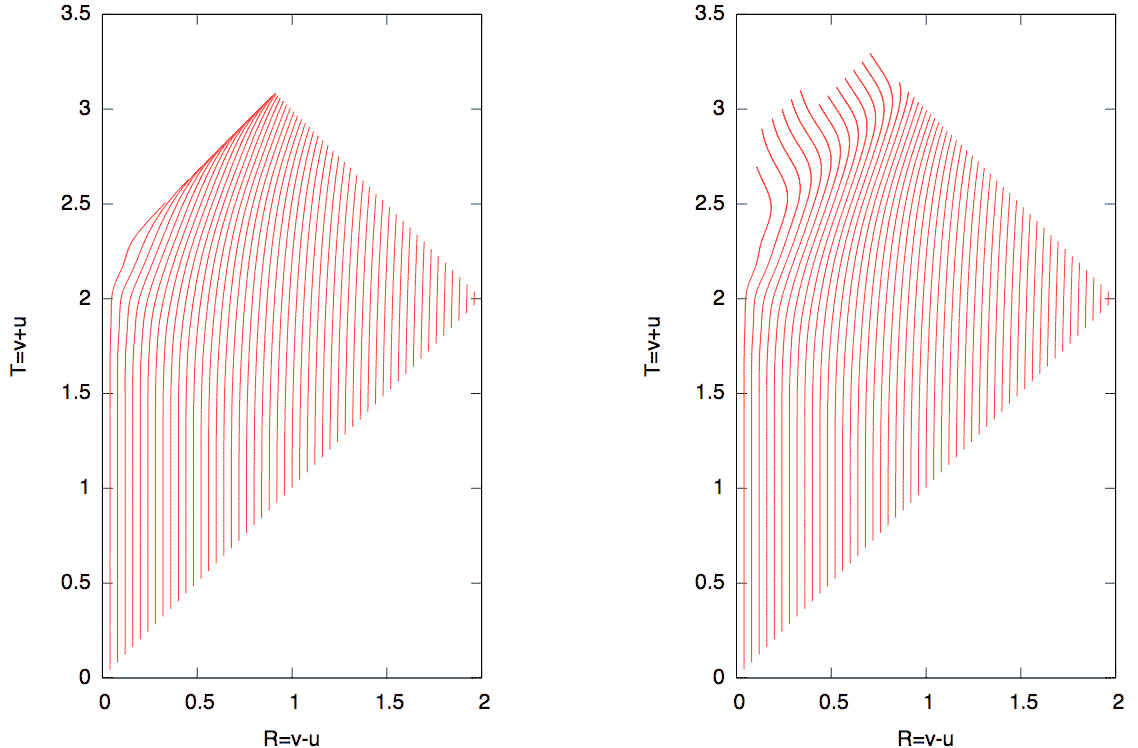}

  \caption{
    This figure shows spacetime diagrams for $r=const$ observers.
    In the left evolution, the scalar field configuration formed an apparent horizon as
    is evident from the $r=const$ curves approaching the null surface $r=2M$ of the 
    event horizon. In contrast, the evolution shown in the right frame turned out to be 
    subcritical.
    Compare this figure with the Penrose diagram shown in \ref{spacetime}
    }\label{fig:spacetime-diagrams-t_r-multi}
\end{figure}

In contrast, figure \ref{fig:spacetime-diagrams-t_r-multi} compares the fate of 
$r=const$ observers in a supercritical and a subcritical evolution.

\chapter{Discussion}
\label{chp:discussion}

\markright{\textbf{Discussion}}

In this thesis we have presented numerical constructions of portions of
near-critical spherically symmetric spacetimes that extend up to future null
infinity.
The simulations are based on a compactified approach, where the equations
have been regularized in a neighborhood of null infinity by introducing the
mass-function as an additional independent variable.
Further simulations have been carried out using a double-null code that can
penetrate apparent horizons. In this discussion, however, we focus on key results 
obtained from the compactified code as they represent the new developments in 
this thesis. 
The resolution necessary to resolve critical phenomena is gained through
letting our gridpoints fall along ingoing null geodesics, following Garfinkle
\cite{Garfinkle95}. The grid is repopulated with grid points, which are filled
with values through interpolations when half of the gridpoints have reached
the origin. This is not optimal, but ensures sufficient resolution for our 
current purposes.

We reproduce standard features of near-critical solutions, such as ``echoing'' 
and mass scaling with fine structure. The mass scaling law for compactified 
evolutions shows a ``leveling-off'' effect for very near-critical solutions
that is not present for uncompactified simulations which we attribute to 
numerical errors.
In addition, we extract radiation at null infinity by computing the news function 
and find a signal with rapidly increasing frequency as measured in Bondi time, 
which is the natural time coordinate associated with far away observers. 
In order to simplify the analysis of near-critical spacetimes, we
have defined an approximate adapted time in terms of Bondi time, in analogy
to the standard time coordinate which is adapted to the DSS discrete 
diffeomorphism \eqref{eq:DSS-def}.
This coordinate can be used by asymptotic observers to render the signal
from a near-critical collapse (almost) periodic.
The fact that such a definition actually works out, and makes DSS periodicity
manifest in quantities defined at null infinity is a non-trivial result, which
we explain in Sec. \ref{sec:crit_behavior}.

Note also that the amplitude of the news function, shown in Fig. 
\ref{fig:news} stays fairly constant after the initial transient.
This feature is clearly universal, as long as the radiation signal is dominated
by the DSS collapse, since the system can be approximated
in the DSS region by a perturbation of the critical solution.
Thus, in section \ref{sec:mass_news},
we neglect contributions from scalar field which is far outside the DSS region 
and does not contribute to the critical collapse dynamics.
Consequently, in this scenario the essential free parameters 
determining the radiation signal from an actual near critical solution
are the number of cycles the solution spends in the neighborhood of the critical 
solution and the length scale (e.g. the ``size'' of the past SSH) 
at which the solution comes close to the critical
solution. The robustness of this scenario, i.e. what happens if
the initial data are such that there is significant mass outside of
the SSH, is beyond the scope of this thesis and an issue for future research.

Perhaps the most surprising feature of the radiation signal has emerged
from our investigation of QNMs, which has been motivated by
\cite{Gundlach94b}, where  the first quasinormal mode is found in collapse
evolutions and the question is posed as to how QN ringing would change close
to criticality.
We find that even in very close-to-critical evolutions there is
a correlation of the radiation signal with the period of the first quasinormal 
mode, determined from the time-dependent value of the Bondi mass, as discussed 
in Sec. \ref{sec:QNMs}, Figs. \ref{c_QNM} and \ref{T0_QNM}. 
This correlation between the radiation of the highly dynamical near-critical 
solution and the quasinormal mode, that is defined in terms of perturbations of
a static spacetime, certainly deserves further investigation. 
This surprising feature might even turn out to be a key toward understanding 
the phenomenon of DSS behaviour in near-critical spacetimes.
Our results seem to suggest that the effective curvature potential for a DSS
self-gravitating field acts as a quasi-stationary background for
scattering processes which can be approximately described by quasinormal modes
of a one-parameter family of Schwarzschild black holes with exponentially
decreasing mass. 

This Schwarzschild background with decreasing mass can also be modeled by 
the Vaidya solution as discussed in section \ref{Vaidya-QNM}.
Abdalla, Chirenti, and Saa \cite{AbdallaChirentiSaa06} have computed
QNM for the Vaidya metric and have put forth a criterion to analyze whether
a spacetime is in the ``stationary adiabatic regime'' where the real part
of the QNM varies inversely with the mass function. We have been able to
verify that our critical collapse evolutions are indeed in this stationary 
regime.

The question of the applicability of QNM-motivated estimates
is quite relevant for numerical relativity, e.g. when extracting wave
forms from binary mergers.

For subcritical initial data we can evolve for very long times, and thus are
able to observe power-law tail behavior as shown in Figs. \ref{fig:power-law}
and \ref{fig:m_tail}.
Analytical calculations predict different falloff rates for radiation along
null infinity and along timelike lines, and the natural question arises, which
falloff rate would be seen by a hypothetical observer (in a realistic case,
observation of power-law tails would require an extremely large signal-to-noise 
ratio).
Accordingly, our results depicted in Fig. \ref{fig:power-law} show how
the rates at finite but large radius correspond to the value for null infinity
for a while before they approach the expected late-time value for finite
radius.
The interpretation of this phenomenon is suggested by
perturbative work of \cite{Barack99a,Barack99b}, 
where different tail falloff rates are computed.
There, different regions of spacetime are identified, where certain 
approximations hold. Within the perturbative regime, results obtained 
for null infinity are valid for what has been termed the
``astrophysical zone'' by Leaver \cite{LeaverJMP,LeaverPRD,Barack99a}, 
and which is defined as the region where $\Delta \uB \ll r$.
(Note, however, that $\uB \gg M$ must also be satisfied).
The physical idea is that the distance from observers of astrophysical
phenomena, e.g. gravitational wave detectors, to the radiation sources is 
very large compared to the time during which the source radiates at an observable rate.

We argue that our results illustrate that the relevant falloff from the point
of view of an astrophysical observer is the falloff rate at null infinity,
in accordance with the prediction from perturbation theory. 
We believe that this is a nice model calculation that exemplifies that
null infinity is indeed a useful approximation for observers at large distance
from the source, in the sense that such observers are located in the 
``astrophysical zone''.

Along the same lines, we would like to point out that by
taking appropriate limits in a conformally compactified manifold,
worldlines of increasingly distant geodesic observers converge to null
geodesic generators of future null infinity
and proper time converges to Bondi time \cite{Frauendiener98c}.
Note also that a naive correspondence between observers at large distance 
and spatial infinity would be problematic, e.g. compactification at spatial
infinity leads to ``piling up'' of waves, whereas at null infinity this effect
does not appear -- waves leave the physical spacetime through the boundary at 
null infinity.

Under practical circumstances, null infinity more realistically
corresponds to an observer that is sufficiently far away from the source
to treat the radiation linearly, but not so far away that cosmological
effects have to be taken into account. We are however not aware of
a discussion where this sloppy picture has been made more precise.

When looking at critical collapse from a global spacetime perspective,
as we have done here, one is confronted with some issues concerning
the mass-scaling, that we would like to comment on briefly:
When asking the question of whether infinitesimally small black holes can be formed 
-- the question which triggered the original work on critical collapse --
it could be phrased in two slightly different ways: (i) can we
form black-hole solutions where the final-state black hole has arbitrarily
small mass, or (ii) can we form arbitrarily small apparent horizons.
The question which has been answered in the affirmative by critical collapse
research is the second one. The first one still seems open. 
Our results for near-supercritical evolutions indicate that the mass outside 
the past self-similarity horizon, is significantly larger than the mass leading to scaling.
We conjecture that this external mass originates from backscattering of outgoing
radiation. If this mass falls into the black hole, then, clearly, one
cannot form arbitrarily small black holes, no matter how fine-tuned the data are.

The application of a double-null code that can penetrate apparent horizons has not been
able to decide whether the exterior matter really falls through the horizon; 
instead of a null slicing, which has proven inherently unstable close to the 
event horizon, we propose to use a Cauchy evolution code for future work.

Considering the observability of critical collapse from future null 
infinity in an astrophysical context, one might first object that the matter model
discussed here is unphysical. On the one hand, it was chosen for mathematical simplicity. 
On the other hand, the curved space wave equation of the simple scalar field model still shares 
some essential nonlinear features inherent in the full Einstein equations without symmetry.
Moreover, it would be possible to remedy this deficiency by using e.g. a perfect fluid 
or gravitational waves in axisymmetry.
A much more serious problem is the high degree of finetunedness required in the 
initial data to develop sufficient self-similar features in spacetime such that they could
be detected by far away observer. 


\appendix
\renewcommand{\chaptermark}[1]{\markboth{\textbf{Appendix \thechapter}:\ \textbf{#1}}{}}
\chapter{Tensor Components in Bondi Coordinates}\label{app:tensors-Bondi}

\begin{align}
\alignflush{R_{uu}} &= \frac{1}{2 r^3} \left[ 2V^2\left(\beta' + r\beta''\right)
  + 2r\dot V + r V\left(2\beta' V' V'' - 4\dot\beta - 4 r \dot\beta'\right)
  \right]\\
\alignflush{R_{ur}} &= \frac{1}{2r^2} \left[ 2V\left(\beta' + r\beta''\right)
  +r\left(2\beta' V' + V'' - 4r\dot\beta'\right) \right]\\
\alignflush{R_{rr}} &= \frac{4\beta'}{r}\\
\alignflush{R_{\theta\theta}} &= 1 - e^{-2\beta} V'\\
\alignflush{R_{\phi\phi}} &= \sin^2\theta \left( 1 - e^{-2\beta} V' \right)\\
\end{align}
\begin{equation}
R = \frac{e^{-2\beta}}{r^2} \left( 2 e^{2\beta} + 2V\beta' - 2 V'
  - 2r\beta' V' - 2r V\beta'' - r V'' + 4r^2\dot\beta' \right)
\end{equation}
\begin{align}
\alignflush{G_{uu}} &= \frac{1}{r^3} \left[ 2V^2\beta' + V\left( e^{2\beta} - V'
    - 2 r\dot\beta \right) + r\dot V \right]\\
\alignflush{G_{ur}} &= \frac{1}{r^2} \left( e^{2\beta} + 2 V\beta' - V' \right)\\
\alignflush{G_{rr}} &= \frac{4\beta'}{r}\\
\alignflush{G_{\theta\theta}} &= -\half e^{-2\beta} \left[ 2V\left(\beta' - r\beta''\right)
  + r \left( -2\beta' V' - V'' + 4r\dot\beta' \right) \right]\\
\alignflush{G_{\phi\phi}} &= -\half e^{-2\beta}\sin^2\theta
  \left[ 2V\left(\beta' - r\beta''\right)
  + r \left( -2\beta' V' - V'' + 4r\dot\beta' \right) \right]\\
\end{align}
Energy momentum tensor of a massless scalar field
\begin{align}
\alignflush{T_{ab}} &= \nabla_a \phi \nabla_b \phi - \half g_{ab} \nabla_c \phi
\nabla^c \phi\\
\alignflush{T_{uu}} &= \dot\phi^2 + \half\Vr \left( \Vr (\phi')^2 - 2 \dot\phi \phi' \right)\\
\alignflush{T_{ur}} &= \half \Vr (\phi')^2\\
\alignflush{T_{rr}} &= (\phi')^2\\
\alignflush{T_{\theta\theta}} &= -\half r^2 e^{-2\beta} \left( \Vr (\phi')^2 - 2\dot\phi
  \phi' \right)\\
\alignflush{T_{\phi\phi}} &= -\half r^2 \sin^2\theta e^{-2\beta}
\left( \Vr (\phi')^2 - 2\dot\phi \phi' \right)\\
\end{align}
\begin{equation}\label{eqn-trace-T}
T = g^{ab}T_{ab} = -e^{-2\beta} \left( \Vr (\phi')^2 - 2\dot\phi \phi' \right)
  = g^{AB}T_{AB}
\end{equation}
Curved space wave equation (matter field equation)
\begin{equation}\label{wave-eq}
0 = \square_g \phi = e^{-2\beta} \left[ \left(\frac{2V}{r^2} + (\Vr)'\right)
  \phi' - \frac{2}{r}\dot\phi - 2\dot\phi' + \Vr \phi'' \right]
\end{equation}
Christoffel symbols
\begin{align}
\alignflush{\Gamma^u_{uu}} &= \frac{1}{2r^2}\left(V - 2 r V \beta' - r V' + 4r^2 \dot\beta
\right)\\
\alignflush{\Gamma^u_{\theta\theta}} &= e^{-2\beta} r\\
\alignflush{\Gamma^u_{\phi\phi}} &= e^{-2\beta} r \sin^2\theta\\
\alignflush{\Gamma^r_{uu}} &= - \frac{1}{2r^3} \left(V^2 - 2 r V^2 \beta' - r V V' +
                   2 r^2 V\dot\beta - r^2 \dot V \right)\\
\alignflush{\Gamma^r_{ur}} &= \frac{1}{2r^2} \left( -V + 2 r V \beta' + r V' \right)\\
\alignflush{\Gamma^r_{rr}} &= 2\beta'\\
\alignflush{\Gamma^r_{\theta\theta}} &= - e ^{-2\beta}\\
\alignflush{\Gamma^r_{\phi\phi}} &= - e ^{-2\beta} \sin^2\theta\\
\alignflush{\Gamma^\theta_{r\theta}} &= \frac{1}{r}\\
\alignflush{\Gamma^\theta_{\phi\phi}} &= -\cos\theta \sin\theta\\
\alignflush{\Gamma^\phi_{r\phi}} &= \frac{1}{r}\\
\alignflush{\Gamma^\phi_{\theta\phi}} &= \cot\theta\\
\end{align}

\chapter{Tensor Components in Double Null Coordinates}\label{app:tensors-double-null}

Einstein Tensor
\begin{align}
G_{uu} &= \frac{4 \dot a \dot r - 2 a \ddot r}{a r}\\
G_{vv} &= \frac{4 a' r' - 2 a r''}{a r}\\
G_{uv} &= \frac{a^2/2 + 2r' \dot r + 2r {\dot r}'}{r^2}\\
G_{\theta\theta} &= \frac{4r^2\sin^2\theta \left( a' \dot a - a {\dot a}' - a^2 {\dot r}'/r \right)}{a^4}\\
G_{\phi\phi} &= \frac{4r^2 \left( a' \dot a - a {\dot a}' - a^2 {\dot r}'/r\right)}{a^4}
\end{align}
Christoffel symbols
\begin{align}
\Gamma^u_{uu} &= \frac{2 \dot a}{a}\\
\Gamma^u_{\theta\theta} &= \frac{2 r r'}{a^2}\\
\Gamma^u_{\phi\phi} &= \frac{2r r' \sin^2\theta }{a^2}\\
\Gamma^v_{vv} &= \frac{2 a'}{a}\\
\Gamma^v_{\theta\theta} &= \frac{2 r \dot r}{a^2} \\
\Gamma^v_{\phi\phi} &= \frac{2 r \dot r \sin^2\theta }{a^2}\\
\Gamma^\theta_{\theta u} &= \frac{\dot r}{r}\\
\Gamma^\theta_{\theta v} &= \frac{r'}{r}\\
\Gamma^\theta_{\phi\phi} &= -\cos\theta \sin\theta\\
\Gamma^\phi_{\phi u} &= \frac{\dot r}{r}\\
\Gamma^\phi_{\phi v} &= \frac{r'}{r}\\
\Gamma^\phi_{\phi\theta} &= \cot\theta
\end{align}

\chapter{Numerical Methods}\label{app:numerics}

\section{Finite Difference Methods for ODEs}\label{app:FD-ODEs}

Following \cite{Ascher98}, we consider numerical solutions to the initial value problem (IVP)
\begin{equation}\label{ODE}
\frac{dy}{dt} = f(t,y(t)), \quad y(t_0) = y_0, \quad 0 \le t \le T
\end{equation}
on a mesh
\begin{equation}
  0 = t_0 < t_1 < \dots < t_{N-1} < t_N = T
\end{equation}
and let $h_n = t_n - t_{n-1}$ the nth \emph{step size}.
We assume sufficient smoothness and boundedness on $f(t,y)$ so that
a unique solution $y(t)$ exists and has as many bounded derivatives as required.
Let $y_h$ be the mesh function which takes on the value $y_n$ at each $t_n$, where
the $y_n$ denote approximations to the exact solution $y(t_n)$.

For the sake of simplicity, consider the forward Euler method,
\begin{equation}
  y^{n+1} = y^n + h_{n+1} f(t^n, y^n).
\end{equation}
We define the difference operator $\mathcal N_h$ of the forward Euler method as
\begin{equation}
\mathcal N_h u(t_n) :=  \frac{u(t_n) - u(t_{n-1})}{h_n} - f(t_{n-1}, u(t_{n-1})),
\end{equation}
for some function $u(t)$ defined at the mesh points.
Then, $\mathcal N_h y_h = 0$ represents the discretized form of the ODE \eqref{ODE}.
The \emph{local truncation error} (LTE) (or \emph{local discretization error})
\footnote{
  This definition of the local truncation error can also be interpreted as
  the error generated by a single numerical step starting from exact data
  and dividing the result by $h_n$:
  i.e. for the forward Euler method denote the result of this step by
  $\tilde y_n := y(t_{n-1}) + h_n f(t_{n-1},y(t_{n-1}))$.
  Then $\tau_n = \left( y(t_n) - \tilde y_n \right) /h_n$.
  In contrast to this definition, the LTE is often defined by $y(t_n) - \tilde y_n$.
}
is the residual of the difference operator when it is applied to the exact solution:
\begin{equation}
  \tau_{n} := \mathcal N_h y(t_n).
\end{equation}
The difference method is said to be \emph{consistent} (or \emph{accurate}) of
\emph{order} $p$ if
\begin{equation}\label{eq:ODE-consistency-order-p}
  \tau_n = \bigO(h_n^p)
\end{equation}
for a positive integer $p$. 
For the forward Euler method we have
\begin{equation}
  \tau_n = \frac{y(t_n) - y(t_{n-1})}{h_n} - f(t_{n-1}, y(t_{n-1}))
  = \frac{h_n}{2} \ddot y(t_n) + \bigO(h_n^2),
\end{equation}
thus, it is consistent of order 1.

We define the \emph{global error}
\begin{equation}
  e_n := y_n - y(t_n),
\end{equation}
and assume that $e_0 = 0$.
Let $H = \max_{1\le n \le N} h_n$ and assume that $N H$ is bounded independent of $N$.
The difference method is said to be \emph{convergent} of \emph{order} $p$ if the
\emph{global error}, $e_n$, satisfies 
\begin{equation}
  e_n = \bigO(H^p),
\end{equation}
for $n=1,2, \dots, N$.

Theorem 3.1 of \cite{Ascher98} states that:
If a numerical method is consistent of order $p$ and fulfills a certain concept of 
stability (``0-stability''
\footnote{
  A difference method is called \emph{0-stable} if there exist positive constants
  $h_0$ and $K$ such that for any mesh functions $x_h$ and $z_h$ with $h \le h_0$,
  \begin{equation*}
    \lvert x_n - z_n \rvert \le K 
    \left\{ \lvert x_0 - z_0 \rvert   
          + \max_{1\le j\le N}  \lvert \mathcal N_h x_h(t_j) - \mathcal N_h z_h(t_j) \rvert 
    \right\},
    \quad 1 \le n \le N.
  \end{equation*}
}
), then it is convergent of order $p$.

In contrast to the local truncation error, the \emph{local error} is defined as the 
amount by which the numerical solution $y_n$ at each timestep differs from the solution 
$\bar y(t_n)$ to the initial value problem
\begin{equation}
  \dot {\bar y}(t) = f(t,\bar y(t)) \quad \bar y(t_{n-1}) = y_{n-1}.
\end{equation}
The local error is defined by
\begin{equation}
  l_n = \bar y(t_n) - y_n.
\end{equation}
For the numerical methods we consider here, the local error indicators $h_n \tau_n$ and 
$l_n$ are closely related:
\begin{equation}
  h_n \left( \lvert \tau_n \rvert + \bigO(H^{p+1})  \right) 
  = \lvert l_n \rvert \left( 1 + \bigO(h_n) \right).
\end{equation}

\section{Runge Kutta Methods}\label{app:RK-methods}

We discuss the \emph{Runge-Kutta methods} \cite{Ascher98} used to 
solve ODEs in our numerical codes. Runge-Kutta methods belong to the class of 
\emph{one-step methods}, i.e  methods that do not use any information 
from previous steps.
In a typical step of size $h=t_{n+1} - t_n$, we seek an approximation
$y_{n+1}$ to $y(t_{n+1})$, given the result of the previous step, $y_n$.

Given the scalar initial value problem
\begin{equation}\label{eq:scalarODE}
\frac{dy}{dt} = f(t,y(t)), \quad y(t_0) = y_0, \quad 0 \le t \le T
\end{equation}
we can approximate the area under the curve $\dot y(t)$ 
by applying one of the quadrature rules given in section \ref{app:quadrature-rules}
to the integral in
\begin{equation}\label{eq:ODE-integral}
  y(t_{n+1}) - y(t_n) = \int^{t_{n+1}}_{t_n} \dot y(t) dt
\end{equation}
on a mesh
\begin{equation}
  0 = t_0 < t_1 < \dots < t_{N-1} < t_N = T
\end{equation}
and let $h_n = t_n - t_{n-1}$ the nth \emph{step size}.
For instance, we can use the height at the midpoint of the interval,
i.e. $\dot y(t_{n+1/2})$, where $t_{n+1/2} = t_n + h/2$.
This choice leads to the \emph{implicit midpoint method}
\begin{equation}\label{eq:implicit-midpoint}
  y_{n+1} = y_n + h f\left( t_{n+1/2}, \frac{y_n + y_{n+1}}{2} \right).
\end{equation}
Approximating $y(t_{n+1/2})$ by the forward Euler method we obtain
the \emph{explicit midpoint method}:
\begin{align}\label{eq:explicit-midpoint}
  \hat y_{n+1/2} &= y_n + \frac{h}{2} f(t_n, y_n)\\
  y_{n+1} &= y_n + h f\left( t_{n+1/2}, \hat y_{n+1/2} \right).
\end{align}
According to equation \eqref{eq:ODE-consistency-order-p},
both midpoint methods have a local truncation error of $\bigO(h^2)$
and are consistent of order 2.

Applying the trapezoidal rule to \ref{eq:ODE-integral} yields
the \emph{implicit trapezoidal method}
\begin{equation}\label{eq:implicit-trapezoidal}
  y_{n+1} = y_n + \frac{h}{2} \left[  f(t_n,y_n) + f(t_{n+1},y_{n+1}) \right].
\end{equation}
To obtain an explicit method, we approximate $y_{n+1}$ using the forward 
Euler method, which yields
\begin{align}\label{eq:explicit-trapezoidal}
  \hat y_{n+1} &= y_n + h f(t_n, y_n)\\
  y_{n+1} &= y_n + \frac{h}{2} \left[  f(t_n,y_n) + f(t_{n+1},\hat y_{n+1}) \right].
\end{align}
This is an explicit 2-stage method of order 2, known as the 
\emph{explicit trapezoidal method}.

\subsection{Quadrature Rules}\label{app:quadrature-rules}
For the sake of reference we list some very well known approximations
to the integral $I = \int_a^b f(x) dx$ and their global errors:

\begin{itemize}
  \item \emph{Midpoint-Rectangle rule}:\\
  \begin{equation}\label{eq:midpoint-rule}
    I = h \sum_{i=1}^n f_{i-1/2} + \bigO(h^2)    
  \end{equation}
  \item \emph{Trapezoidal rule}:\\
  \begin{equation}\label{eq:trapezoidal-rule}
    I = h \left[ \frac{1}{2} f_0 + f_1 + \cdots + f_{n-1} + \frac{1}{2}f_n \right] + \bigO(h^2)    
  \end{equation}
  \item \emph{1/3 Simpson's rule}:\\\
  For the subinterval $[i-1,i+1]$:
  \begin{equation}\label{eq:Simpsons-rule-local}
    \int_{i-1}^{i+1}f(x)dx = \frac{h}{3}\left[ f_{i-1} + 4f_i + f_{i+1} \right] + \bigO(h^5),    
  \end{equation}
  while for the entire interval 
  \begin{equation}\label{eq:Simpsons-rule-global}
    I =  \frac{h}{3}\left[ f_0 + 4f_1 + 2 f_2 + \cdots + 4f_{m-1} + f_{m} \right] + \bigO(h^4),    
  \end{equation}
  where $m$ is even.
\end{itemize}

\section{Error Analysis of the NSWE-algorithm}\label{app:errors-NSWE}

\subsection{Local Errors}
\begin{figure}
  \centering
  \includegraphics[angle=37,totalheight=6cm]{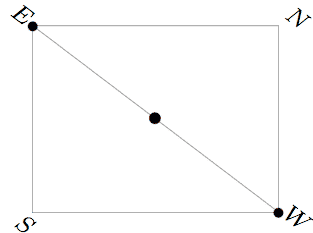}
  \caption{The diamond-shaped region in the NSWE-scheme.}\label{fig-NSWE-diamond}
\end{figure}

The local errors in the NSWE-scheme \eqref{eq:diamond-scheme} of section
\ref{sec:diamond-algorithm} stem from the approximation of the
integral by evaluating the integrand at the center of the small
null parallelogram (see figure \ref{fig-NSWE-diamond}) and from
the multiplication with the area $(\Delta u)(\Delta r)$ of this
diamond-shaped region. Evaluating the integrand at the center
amounts to the same as using the average between the points $E$
and $W$.

\begin{figure}
  \centering
  \includegraphics[angle=0,totalheight=2cm]{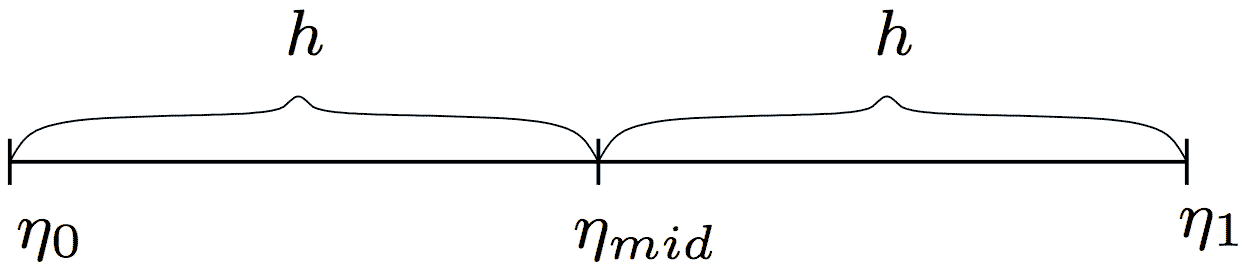}
  \caption{Illustration for the midpoint rule.}\label{fig-midpoint}
\end{figure}

Assume that we know a function $f$ at the endpoints $\eta_0$ and
$\eta_1$ of a given interval (see figure \ref{fig-midpoint}) and
want to evaluate it at the midpoint of the interval $\eta_\text{mid}$.
Using linear interpolation at the center of an interval $[\eta_0,\eta_1]$
we find
\begin{equation}\label{eqn-midpoint-rule}
  f_\text{mid} \approx \frac{1}{2} \left[ f(\eta_0) + f(\eta_1) \right].
\end{equation}
The approximation error for the linear interpolation
polynomial with pairwise distinct nodes is (see \cite{Deuflhard-Hohmann03})
\begin{equation}
  f(\eta_\text{mid}) - P(f| \eta_0, \eta_1)(\eta_\text{mid}) = 
  \frac{f^{(2)}(\xi)}{2!} \prod_{j=0}^1 \left( \eta - \eta_j \right)
  = \frac{h^2}{2} f''(\xi),
\end{equation}
for some $\xi \in (\eta_0,\eta_1)$.
Therefore, this polynomial approximation is second order accurate.

The approximation of the integral in \eqref{eq:diamond-scheme} applies the
midpoint-rule to the integrand which is multiplied by the area
$(\Delta u) (\Delta r)$ which is of the order $O(h^2)$.
(If $\Delta u \approx \Delta r$, then $(\Delta u)(\Delta r) \approx 2 h^2$.)
Since both factors are second order accurate, the local error of one
NSWE-integration step is $O(h^4)$, assuming that $\psi_W$, $\psi_E$ and $\psi_S$ 
are known at least to this order.

\subsection{Global Errors}

In the following, we give a heuristic argument for the global error of the NSWE-scheme.
First, we restrict analysis to the fixed background case.
This entails that the NSWE-scheme is no longer coupled to the geometry and
null-geodesic ODEs, which would in general have to be solved concurrently.
To simplify matters even more, we discard the Taylor expansion-region and
use just $\psi(r=0) = 0$ as a startup condition.
Note that for flat space the algorithm is exact, since $\left(\Vr\right)' = 0$.

\begin{figure}
  \centering
  \includegraphics[width=6cm]{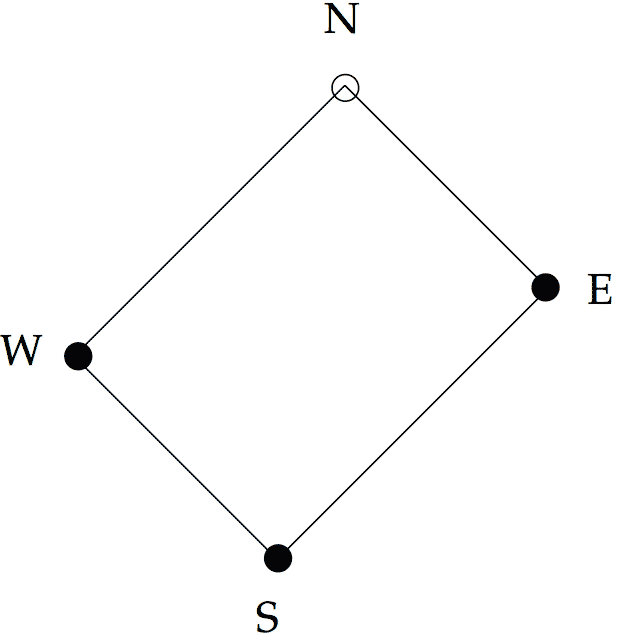}
  \caption{The computational molecule for the NSWE-scheme.}\label{fig-NSWE-cmol}
\end{figure}

\begin{figure}
  \includegraphics[width=14cm]{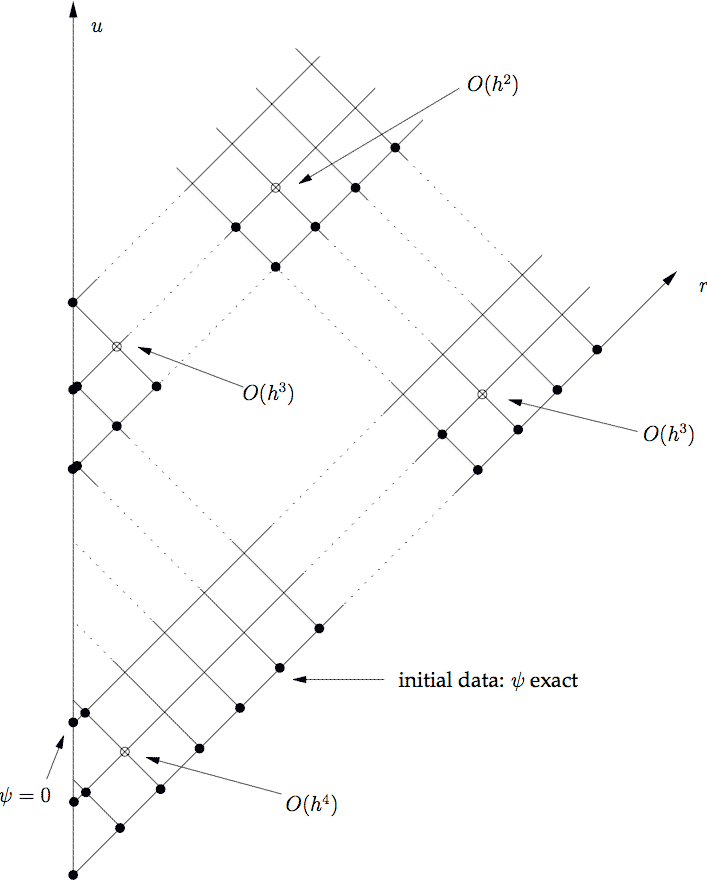}
  \caption{Error accumulation in the NSWE (diamond) - finite
    differencing scheme.}\label{fig-NSWE-global}
\end{figure}

The general idea is the following: We start out from the initial slice
where $\psi$ is known exactly. On the next slice, we march outwards from the 
origin using the NSWE-algorithm whose computational molecule is shown in 
figure \ref{fig-NSWE-cmol} and accumulate local error terms of order $O(h^4)$ 
as illustrated in figure \ref{fig-NSWE-global}. 
If we make $n$ steps and $h \approx 1/n$ we will loose one order of accuracy.
On the next slice, however, we are still using $\psi$-values of order $O(h^4)$ 
near the origin.
In a Cauchy code we would accumulate errors only along the time direction.
Here, we can distinguish two directions of error accumulation: one radially outward 
and another along the ingoing null-geodesics. Thus, we loose two orders in total 
and the diamond algorithm turns out to be globally second order accurate:
\begin{equation}
  \psi = \psi_{true} + \bigO(h^2).
\end{equation}

For the coupled NSWE-algorithm one has to choose the orders of
accuracy for the ODE solvers such that the diamond scheme stays globally 
second order accurate. In practise, second order accurate schemes
(see \ref{app:RK-methods}) have proven sufficient to ensure this.

\begin{figure}
  \centering
  \includegraphics[width=12cm]{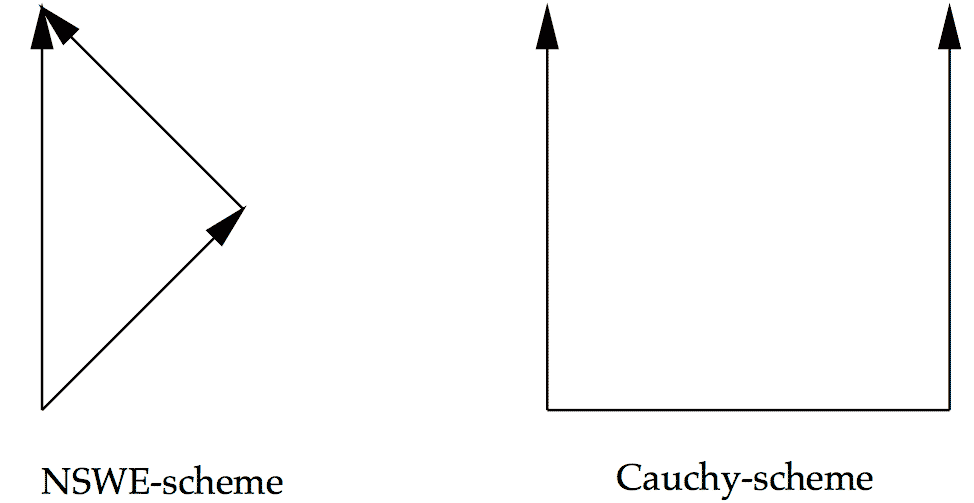}

  \caption{This figure shows the directions of error accumulation
  present in the NSWE-scheme and a prototypical Cauchy finite differencing scheme.
  In the Cauchy scheme, we loose only one order of accuracy from the accumulation 
  in the time direction.
  In the NSWE scheme, however, there is coupling in the radial direction,
  as we see from its computational molecule (figure \ref{fig-NSWE-cmol}).
  Combined with the ingoing null direction we therefore loose two orders of 
  accuracy, in total.
  }\label{fig-NSWE-leapfrog-comp}
\end{figure}

\chapter{Convergence Test Methodology}\label{app:convergence-methodology}
We briefly describe how we can check the accuracy of a numerical evolution 
code using the concept of convergence tests. These tests obtain error estimates by
subtracting numerical solutions on a grid which is shared by a set of  
evolutions using different (usually by powers of two) gridspacings.
Depending on whether the quantities to be tested are zero on the continuum
level or not, we take one of the following approaches:

In a so-called \emph{3-level convergence test}, 
  we use three different grid resolutions, with n being the number of
  gridpoints in the lowest resolution grid, and the higher ones
  having $2n$ and $4n$ gridpoints, respectively. 
  Without restriction of generality, we can assume that the gridspacing on a 
  grid of $n$ gridpoints is given by $h = 1/n$.
  Assume furthermore that the discretized field $\Psi_n$ is second order accurate,
  \begin{equation}
    \Psi_n = \Psi + h^2 e_2 + h^3 e_3 + \bigO(h^4),
  \end{equation}
  where the continuum error functions $e_2$ and $e_3$ do not depend on the gridspacing $h$.
  Then,
  \begin{align}
    \Psi_{2n} &= \Psi + (h/2)^2 e_2 + (h/2)^3 e_3 + \bigO(h^4)\\
    \Psi_{4n} &= \Psi + (h/4)^2 e_2 + (h/4)^3 e_3 + \bigO(h^4).
  \end{align}
  We may subtract these gridfunctions on the common gridpoints,
  \begin{equation}
    \Psi_{2n} - \Psi_n = -3/4 h^2 e_2 - 7/8 h^3 e_4 + \bigO(h^4),
  \end{equation}
  and
  \begin{equation}
    \Psi_{4n} - \Psi_{2n} = -3/4 (h/2)^2 e_2 - 7/8 (h/2)^3 e_4 + \bigO(h^4).
  \end{equation}
  Comparing the last two expressions we find that on the coarsest grid 
  \begin{equation}
    \Psi_{4n} - \Psi_{2n} = 1/4 \left( 1 + \bigO(h) \right) \left[\Psi_{2n} - \Psi_n\right].
  \end{equation}
  By taking a discrete spatial norm over the grid, we can compute a time-dependent convergence factor
  \begin{equation}
    Q(t) := \frac{\lVert \Psi_{4n} - \Psi_{2n} \rVert}{\lVert \Psi_{2n} - \Psi_n \rVert}.
  \end{equation}
  If the numerical scheme converges, then in the limit we should find
  \begin{equation}
    \lim_{h \to 0} Q(t) = 1/4.
  \end{equation}
  In general we have, for $h \to 0$,
  \begin{equation}
    \Psi_{\beta^2 n} - \Psi_{\beta n} = 1/\beta^\alpha (\Psi_{\beta n} - \Psi_n),
  \end{equation}
  where $\alpha$ denotes the exponent of the leading error term in the
  discretized field $\Psi_n = \bigO(h^\alpha)$ and $\beta$ the factor between
  grid resolutions.

A \emph{2-level convergence test}
is applicable to discretized quantities that are supposed to be zero
on the continuum level, i.e.
  \begin{equation}
    \Psi_n = 0 + h^2 e_2 + h^3 e_3 + \bigO(h^4).
  \end{equation}
It then follows that
  \begin{equation}
    \Psi_{2n} = 1/4 \left( 1 + \bigO(h) \right) \Psi_n.
  \end{equation}
More generally we have, for $h \to 0$,
  \begin{equation}
    \Psi_{\beta n} = 1/\beta^\alpha \Psi_n.
  \end{equation}


\bibliographystyle{alpha}
\bibliography{diss}

\end{document}